\documentclass[a4paper,12pt,twoside,openright]{book}
\usepackage[utf8]{inputenc}
\usepackage[english]{babel}

\usepackage[pass]{geometry}

\usepackage{appendix}
\usepackage[nottoc,notlot,notlof]{tocbibind}

\usepackage{fancyhdr}
\usepackage{emptypage}

\fancypagestyle{mypagestyle}{%
\fancyhf{}%
\fancyhead[LE,RO]{\thepage}
\fancyhead[LO]{\itshape\nouppercase{\rightmark}}
\fancyhead[RE]{\normalfont\nouppercase{\leftmark}}
}

\fancypagestyle{toc}{%
\fancyhf{}%
\fancyhead[LE,RO]{\thepage}
\fancyhead[LO]{}
\fancyhead[RE]{}
}

\makeatletter
  
\def\@makechapterhead#1{%
  \vspace*{20\p@}%
  {\parindent \z@
  \raggedright
  \normalfont
    \ifnum \c@secnumdepth >\m@ne
      \if@mainmatter
        \centering \reset@font \scshape \@chapapp\space \thechapter\par\nobreak
        \vspace*{20\p@}
      \fi
    \fi
    \interlinepenalty\@M
    \centering \reset@font \Large \bfseries #1\par\nobreak
    \vskip 90\p@
  }}
  
\def\@makeschapterhead#1{%
  \vspace*{20\p@}%
  {\parindent \z@ \raggedright
    \normalfont
    \interlinepenalty\@M
    \centering
    \Large \bfseries  #1\par\nobreak
    \vskip 80\p@
  }}
  
\renewcommand\section{\@startsection {section}{1}{\z@}%
                                   {-3.5ex \@plus -1ex \@minus -.2ex}%
                                   {2.3ex \@plus.2ex}%
                                   {\normalfont\large\bfseries}}
\renewcommand\subsection{\@startsection{subsection}{2}{\z@}%
                                     {-3.25ex\@plus -1ex \@minus -.2ex}%
                                     {1.5ex \@plus .2ex}%
                                     {\normalfont\normalsize   \bfseries}}
\renewcommand\subsubsection{\@startsection{subsubsection}{3}{\z@}%
                                     {-3.25ex\@plus -1ex \@minus -.2ex}%
                                     {1.5ex \@plus .2ex}%
                                     {\normalfont\normalsize\bfseries\centering}}
                                     
\makeatother

\usepackage{caption}
\usepackage{amssymb}
\usepackage{amsmath}
\usepackage{graphicx}
\usepackage{hyperref}
 \usepackage{dsfont}
\usepackage{mathrsfs}
\usepackage{url}
\usepackage[shortlabels]{enumitem}

\usepackage{amsthm}

\newcommand{\integers}{\ensuremath{\mathbb{Z}}}

\newcommand{\real}{\ensuremath{\mathbb{R}}}

\newcommand{\tr}{\ensuremath{\text{tr}}}
\newcommand{\id}{\ensuremath{\mathds{1}}}

\newcommand{\bigo}{\ensuremath{\mathcal{O}}}
\newcommand{\smallo}{\ensuremath{o}}
\newcommand{\vect}[1]{\ensuremath{\boldsymbol{#1}}}
\newcommand{\barr}[1]{\ensuremath{\overline{#1}}}

\newcommand{\ad}{\ensuremath{\text{ad}\,}}

\newcommand{\eqdef}{\ensuremath{:=}}
\newcommand{\defeq}{\ensuremath{=:}}
\newcommand{\su}{\ensuremath{\mathfrak{su}}}
\newcommand{\sprod}[2]{\ensuremath{{#1} \cdot {#2}}}
\newcommand{\extprod}[2]{\ensuremath{{#1} \times {#2}}}
\newcommand{\extderphase}{\ensuremath{\mathbf{d}}}

\newcommand{\insertion}{\ensuremath{\mathbf{i}}}
\newcommand{\weq}{\ensuremath{\approx}}
\newcommand{\lie}{\ensuremath{\mathcal{L}}}
\newcommand{\liephase}{\ensuremath{\boldsymbol{\mathcal{L}}}}
\newcommand{\Sym}{\mathsf{Sym}}
\newcommand{\Gau}{\mathsf{Gau}}
\newcommand{\Asym}{\mathsf{Asym}}

\newcommand{\Poi}{\mathsf{Poi}}
\newcommand{\SU}{\mathrm{SU}}
\newcommand{\U}{\mathrm{U}}
\newcommand{\eff}{\mathcal{F}}
\newcommand{\scri}{\mathscr{I}}
\newcommand{\realpart}{\ensuremath{\textrm{Re}}}
\newcommand{\imaginarypart}{\ensuremath{\textrm{Im}}}

\usepackage{tkz-graph}
 \usetikzlibrary{positioning,calc,arrows}
 \usetikzlibrary{decorations.pathmorphing}
 \usetikzlibrary{decorations.markings}
 \usetikzlibrary{decorations.text}

\newcommand{\arcThroughThreePoints}[4][]{
\coordinate (middle1) at ($(#2)!.5!(#3)$);
\coordinate (middle2) at ($(#3)!.5!(#4)$);
\coordinate (aux1) at ($(middle1)!1!90:(#3)$);
\coordinate (aux2) at ($(middle2)!1!90:(#4)$);
\coordinate (center) at ($(intersection of middle1--aux1 and middle2--aux2)$);
\draw[#1] 
 let \p1=($(#2)-(center)$),
      \p2=($(#4)-(center)$),
      \n0={veclen(\p1)},       
      \n1={atan2(\y1,\x1)}, 
      \n2={atan2(\y2,\x2)},
      \n3={\n2>\n1?\n2:\n2+360}
    in (#2) arc(\n1:\n3:\n0);
}

\begin{document}
\pagestyle{mypagestyle}
\title{Hamiltonian study of the asymptotic symmetries of gauge theories}
\author{Roberto Tanzi}


\pagenumbering{Alph}

\begin{titlepage}
\newgeometry{hmarginratio=1:1}

\center 

{\LARGE \bf DISSERTATION}\\[2.5cm]
submitted to the\\[8mm]
{\large Combined Faculties of Physics and Electrical Engineering}\\{\large of the University of Bremen, Germany}\\[8mm]
for the degree of\\[8mm]
{\large Doctor of Natural Sciences}

\vspace*{\fill}

Put forward by\\[8mm]
{\large \bf Roberto Tanzi}\\[8mm]
{\large born in Perugia, Italy}

\end{titlepage}
\restoregeometry
\cleardoublepage

\begin{titlepage}
\setcounter{page}{3}
\newgeometry{hmarginratio=1:1,bottom=2cm,top=2cm}

\center 
 

\includegraphics[width=8cm,keepaspectratio]{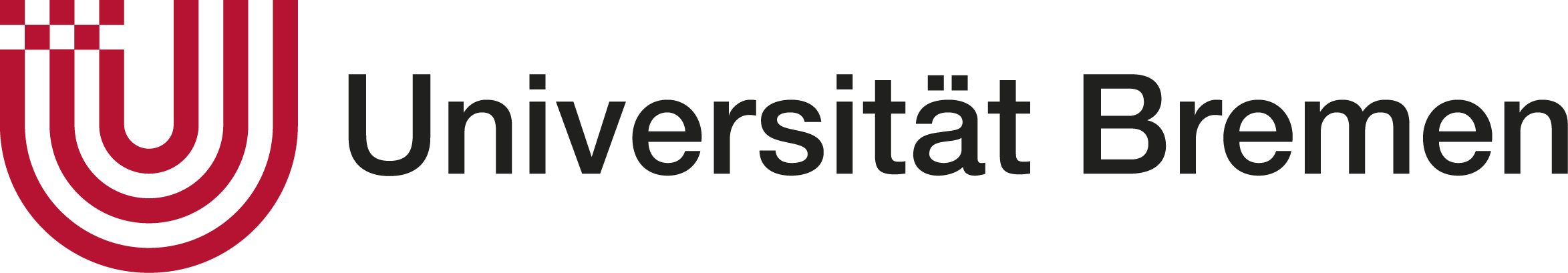}\\[3mm ]
\textsc{Zentrum f\"ur angewandte Raumfahrttechnik\\und Mikrogravitation (ZARM)}\\[4mm] 
\textsc{Research Training Group ``Models of Gravity''}\\ 
\rule{5cm}{0.3mm}\\[3mm]
{\small PhD in \textsc{Physics}}\\[3cm]

\textsc{ \Huge Hamiltonian study of}\\[7mm]
\textsc{ \Huge the asymptotic symmetries of}\\[7mm]
\textsc{ \Huge gauge theories}\\[3cm] 


\begin{minipage}[t]{0.4\textwidth}
\begin{center} 
\textbf{\large Candidate}\\[2mm]
Roberto \textsc{Tanzi} 
\end{center}
\end{minipage}
~
\begin{minipage}[t]{0.51\textwidth}
\begin{center} 
\textbf{\large Referees} \\[2mm]
\underline{Prof. Dr. Domenico \textsc{Giulini}}\\[3mm]
{\small  \it Leibniz University of Hannover\\ Institute for Theoretical Physics}\\[1mm]
{\small \it and}\\[1mm]
{\small  \it University of Bremen, Center of Applied Space Technology and Microgravity (ZARM)}
\\[8mm]
\underline{Prof. Dr. Jutta \textsc{Kunz}}\\[3mm]
{\small \it Carl von Ossietzky University of Oldenburg\\
Institute of Physics}
\end{center}
\end{minipage}

\vspace*{\fill}
\includegraphics[height=1.7cm,keepaspectratio]{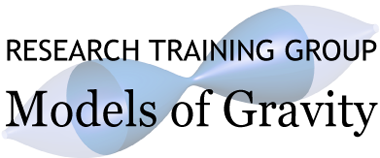}
\hspace*{\fill}
\includegraphics[height=1.7cm,keepaspectratio]{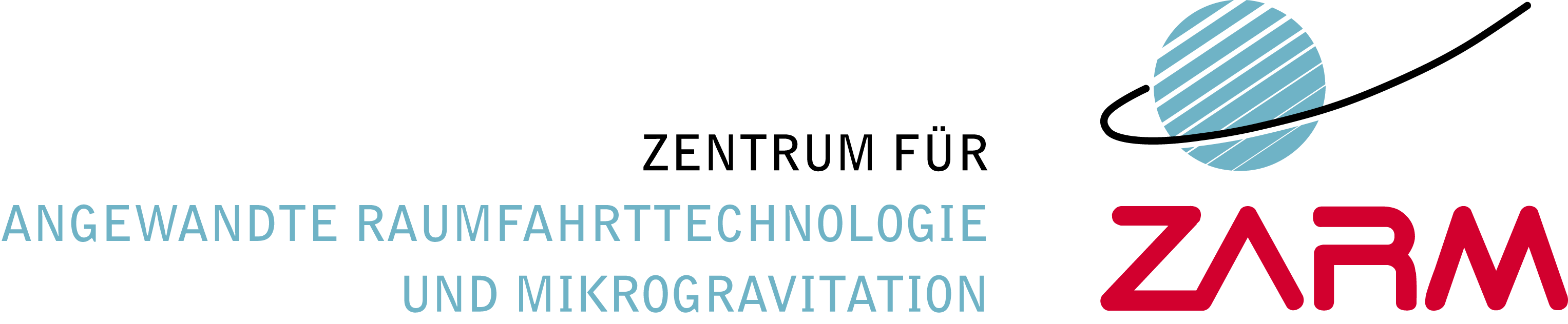}

\rule{8cm}{0.3mm}\\[1mm]
Date of oral examination: 26\textsuperscript{th} August 2021

\end{titlepage}
\restoregeometry


\cleardoublepage
\begin{flushright}
\thispagestyle{empty}
 \vspace*{\fill}
 \textit{Alla memoria di mio zio Dongio}
 \vspace*{\fill}
\end{flushright}
\cleardoublepage

\pagenumbering{roman}

\chapter*{Abstract}

Asymptotic symmetries are a general and important feature of theories with long-ranging fields, such as gravity, electromagnetism, and Yang-Mills.
They appear in the formalism once the analytic behaviour of fields near infinity is specified and have received a renewed interest in the last years
after a possible connection with the information-loss paradox has been conjectured.

One of the various methods used to study the asymptotic symmetries of field theories relies on the Hamiltonian formalism and was introduced in the seminal work of Henneaux and Troessaert, who successfully applied it to the case of gravity and electrodynamics, thereby deriving the respective asymptotic symmetry groups of these theories.
The main advantage of this approach is that the study of the asymptotic symmetries ensues from clear-cut first principles.
These include the minimal assumptions that are necessary to ensure the existence of Hamiltonian structures (phase space, symplectic form, differentiable Hamiltonian) and, in case of Poincar\'e invariant theories, a canonical action of the  Poincar\'e group.

In this thesis, after an extensive review of how the Hamiltonian approach to study asymptotic symmetries of gauge theories works, we apply these methods to two specific situations of physical interest.
First, we deal with the non-abelian Yang-Mills case and we show that the above principles lead to trivial asymptotic symmetries (nothing else than the Poincar\'e group) and, as a consequence, to a vanishing total colour charge.
This is a new and somewhat unexpected result. It implies that no globally colour-charged states exist in classical non-abelian Yang-Mills theory.

The second situation considered in this thesis is a scalar field minimally-coupled to an abelian gauge field, which can be used to study, at the same time, two specific cases: scalar electrodynamics and the abelian Higgs model.
We show that the situation in scalar electrodynamics amply depends on whether the scalar field is massive or massless, insofar as, in the latter case, one cannot canonically implement asymptotic symmetries.  
Furthermore, we illustrate that, in the abelian Higgs model, the asymptotic canonical symmetries reduce to the Poincar\'e group in an unproblematic fashion.

\chapter*{Acknowledgements}

My deepest gratitude goes to my supervisor Nico Giulini.
You have been a fantastic mentor during these years and I have learnt a lot from you.
It has always been a pleasure to discuss with you about my project, about physics, and about any other topic, especially during lunchtime.
I thank you for your advices, for your comments, and for your guidance.
Thank you also for the freedom you left me in developing my project:
Although it made me feel lost at the beginning, I can surely say now, when I am (almost) at the end of my journey as a PhD student, that this freedom greatly helped me in growing both as a scientist and as a person. 

I would like to thank the DFG Research Training Group 1620 ``Models of Gravity'' for having accepted me as one of their members and for having provided financial support.
Equally, I would like to thank the University of Bremen and, in particular, the Gravigruppe at ZARM for having hosted me and provided infrastructures.
It has been a pleasure to meet all the fantastic people of these two groups and I am thankful for all the conversations we have had in these years.

Special thanks go to Marc Henneaux and C\'edric Troessaert for insightful comments and useful suggestions that greatly helped me improve my work.
In addition, I am grateful to all the people who proposed improvements for this thesis and pointed out typos: in particular, Nico Giulini, Dennis Stock, Dennis Philipp, and Christian Pfeifer.
I am also grateful to Jutta Kunz for offering her time to be a referee.

Finally, I would like to thank my family and my girlfriend, Margarita, for their love and support during my studies and my life.
I would also like to thank my dear friends Peve and Alejandro, on whom I can always rely.
Even during this pandemic, I have never felt alone nor abandoned thanks to all of them.

\pagestyle{toc}
\begin{sloppypar}
 \tableofcontents 
\end{sloppypar}

\chapter{Introduction} \label{cha:introduction}
\pagenumbering{arabic} 
\setcounter{page}{1}
\pagestyle{mypagestyle} 

Symmetries have always been a central theme of investigation with an almost-unrivalled importance in physics.
A prominent example of their relevance can be seen in the role they play in the Standard Model of particle physics, whose architecture strongly relies on gauge symmetries and on the Poincar\'e group.
In particular, the Poincar\'e group of transformations --- consisting of spacetime translations, of rotations, and of Lorentz boosts combined together --- emerges as the symmetry group of a flat, empty spacetime, which provides the ideal background to study the dynamics and kinematics of matter in all those situations in which the gravitational interaction can be neglected.
The presence of the Poincar\'e group as a symmetry group has two important repercussions.
First, elementary particles are described by irreducible representations of the Poincar\'e group~\cite{Wigner-representations}.
Second, some physical quantities, such as energy and  angular momentum, have to be conserved as a consequence of Noether's theorem.
For instance, the conservation of energy follows from the symmetry under time translations, while the conservation of angular momentum is a consequence of the symmetry under rotations.

It is clear that the simple case employing the flat Minkowski spacetime as a background, despite finding its application in a numerous range of experimentally-relevant situations, is not well-suited to describe problems in which the gravitational interaction cannot be neglected.
Thus, it is often necessary to consider different and more general backgrounds, possibly featuring other symmetry groups.
For instance, in cosmology, one usually assumes that, on large scales, the Universe is described by the Friedmann-Lema\^itre-Robertson-Walker (FLRW) metric, whose symmetries amount to spatial translations and rotations.
Another example, which is of great relevance for this thesis, is provided by asymptotically-flat spacetimes.

Omitting formal definitions at this stage, asymptotically-flat spacetimes are those spacetimes that, at infinity, look like the flat Minkowski one and are very well-suited to describe isolated systems.
For this reason, they made their first appearance in the sixties in the study of gravitational radiation produced by a localised source and observed
at large ``infinite'' distance from the gravitating system~\cite{Bondi:waves-axisymmetric,Sachs:waves-asymptotically-flat}.
While the gravitational interaction can be extremely strong in the proximity of the source, one can assume the spacetime to be flat to good approximation at the position of the observer.
It is important to note that, since the gravitational radiation travels at the speed of light (neglecting non-linear self-interactions), the position of the observer has to be understood as being
at an ``infinite'' distance
along null geodesics, which can be described in more technical terms by saying that the observer is in a neighbourhood of (future) null infinity.

Inasmuch as the metric of an asymptotically-flat spacetime approaches the flat Minkowski one only near infinity but is otherwise generic, one does not expect to find any transformation that preserves the metric globally.
Thus, it is better to focus, rather than on global symmetries, on those transformations that preserve only the asymptotic form of the metric, commonly referred to as asymptotic symmetries.
Naively, one might expect the group of asymptotic symmetries of an asymptotically-flat spacetime to be the Poincar\'e group.
However, it turns out to be a much larger group named, after the people who discovered it, the ``Bondi-Metzner-Sachs (BMS) group''~\cite{Bondi:waves-axisymmetric,Sachs:waves-asymptotically-flat,Sachs:asymptotic-symmetries}.
As the Poincar\'e group, the BMS group contains rotations and boosts.
However, the four spacetime translations are replaced by an infinite number of transformations known as supertranslations, which include, but are not limited to, the usual translations.
Thus, the BMS group is an extension of the Poincar\'e group and  the former contains infinitely-many distinguished copies of the latter.

After more than half a century from these first studies, asymptotic symmetries have become a very active area of research. 
In particular, a huge interest in this subject has been generated after it has been conjectured by Hawking, Perry and Strominger~\cite{HPS} that asymptotic symmetries may be related to the solution of the long-standing black-hole information-loss paradox by encoding the ``supposedly-lost information'' into the asymptotic charges associated to asymptotic symmetries.
In connection to this possibility, it is important to mention that the recent efforts 
have unveiled that asymptotic symmetries and their charges are not an exclusive feature of gravity but, more generally, can appear in other theories with long-ranging fields, such as  electrodynamics.
Notably, over the last few years, several studies have analysed the asymptotic symmetries of electrodynamics~\cite{Strominger:QED1,Strominger:QED2,Strominger:QED3,Strominger:notes,Strominger:Soft-theorem,Campiglia-subleading-soft-photons,Conde:subleading-soft-photon,Strominger:Soft-theorem-magnetic} and of non-abelian gauge theories~\cite{Strominger-YM,Strominger-YM2,Barnich-YM}, mostly  focusing on the situation at null infinity and relying on the Lagrangian formulation.

Not long after, the method to perform analogous study at spatial infinity using the machinery of the Hamiltonian formulation of classical field theories was uncovered by Henneaux and Troessaert, whose analyses included a plethora of aspects.
They investigated, among others, the case of General Relativity~\cite{Henneaux-GR}, electrodynamics~\cite{Henneaux-ED}  and the coupled Maxwell-Einstein theory~\cite{Henneaux-ED-GR}.
Their seminal work showed that the Hamiltonian studies of asymptotic symmetries at spatial infinity can be complementary to the analogous studies at null infinity, which were performed earlier and are, perhaps, less demanding on the computational side.
The reason why one wishes, nevertheless, to pursue also the Hamiltonian treatment of the problem is not only that one expects to find an equivalence to the Lagrangian approach, but, more importantly, that the Hamiltonian tools are very well suited for a systematic characterisation of state spaces and the symmetries they support.
Needless to emphasise, it also  provides the basis for the canonical quantisation of the theory.

The purpose of the present thesis is to present the principles and the applications of the Hamiltonian approach to the study of asymptotic symmetries of gauge theories.
One of the main advantages of this approach is that it provides a systematic treatment of the study of asymptotic symmetries from clear-cut first principles.
We will present these principles in detail in section~\ref{sec:principles} after having reviewed the necessary mathematical tools.
For now, let us state broadly that these constitute the minimal requirements to make the Hamiltonian formulation well-defined and ensure the existence of the following four structures:
\begin{enumerate}[(i)]
\itemsep=-4pt
\item \label{item:phase-space} a phase space;
\item \label{item:symplectic-form} a well-defined symplectic form on phase space; 
\item a well-defined Hamiltonian;
\item \label{item:Poincare} a Hamiltonian action of the Poincar\'e group on phase space.
\end{enumerate}

In the case of classical field theories, which will be the subject of investigation in this thesis, the determination of the phase space consists of two steps.
First, one needs to find which fields are needed in a Hamiltonian description of the theory.
In the case of electrodynamics, for instance, the canonical fields amount to a spatial one-form $A_a$ (connected to the magnetic field) and to a vector density of weight one $\pi^a$ (connected to the electric field).
Second, one needs to impose conditions on the regularity and on the asymptotic behaviour of the canonical fields in order to ensure that the other conditions listed above are met.
More precisely, many physically-relevant quantities, such as the Hamiltonian and the symplectic form, will be found as formal expressions involving integrals over a three-dimensional manifold (the space).
Therefore, one needs to impose conditions on the fall-off behaviour of the fields at large distances to ensure that these integrals converge.
In addition, concerning the Hamiltonian, one also needs to make sure that it is differentiable with respect to the canonical fields, whose precise definition will be made clear in this thesis while reviewing the Hamiltonian methods.

Concerning point~\ref{item:Poincare} of the list above, let us mention that it amounts to the existence of a canonical generator, whose Poisson-bracket with the canonical fields returns the Poincar\'e transformations of the fields themselves.
This condition is necessary in order to recover the Poincar\'e transformations as a canonical symmetry of the theory, which is imperative for relativistic field theories on a flat background, such as the ones studied in this thesis.
An analogous but only asymptotic condition would have to be imposed on an asymptotically-flat background.
As for the Hamiltonian, also in this case, the canonical generator will be found as a formal expression involving integrals over space and we will need to make sure that this formal expression is finite and differentiable.
The greatest obstructions of this process, as we shall see, will come from the presence of the Lorentz boost and its behaviour at large distances.

Finally, let us mention that the correct characterisation of the phase space and, in particular, the second step described above is usually the most demanding task, which gets prolonged along the other points in the list.
Only once this step is completed and all the other requirements are met, one is allowed to study the (asymptotic) symmetries of the theory.
Due to point~\ref{item:Poincare}, the asymptotic-symmetry group will include the Poincar\'e group, but may be, in general, a non-trivial extension of it.
For this reason, we will say that the asymptotic symmetries of a theory are trivial if this extension is trivial, i.e., if the group of asymptotic symmetries is merely the Poincar\'e group.
One possibility that has to be taken into consideration is that, sometimes, one can find two different characterisations of the phase space meeting all the conditions~\ref{item:phase-space}--\ref{item:Poincare}, but having two different symmetry groups, e.g. the Poincar\'e group in the one case and an extension of it in the other.
Thus, it is important to consider all the possibilities meeting the conditions~\ref{item:phase-space}--\ref{item:Poincare} and select the one featuring the
biggest symmetry group.

\section{Results}

After providing an overview about the Hamiltonian study of the asymptotic symmetries of gauge theories, we will use it in order to study two situations of interests, which constitute the original results contained in this thesis.
Our overall plan is to apply the Hamiltonian strategy pioneered by Henneaux and Troessaert to other physically-relevant theories, starting with simple models and gradually including more fields of physical significance.

The first situation which we will consider is that of non-abelian gauge theories and, more precisely, of $\SU(N)$-Yang-Mills on a flat Minkowski background.
Previous studies of Yang-Mills theory in Hamiltonian formulation include~\cite{Teitelboim-YM1,Teitelboim-YM2} among others.
Although their focus is on the spherically-symmetric case, they nevertheless highlight some general and important features. 
We also mention the detailed discussion of boundary conditions allowing for globally-charged states in \cite{Chrusciel.Kondracki:1987}.

Based on the results obtained in the study of the asymptotic symmetries of Yang-Mills fields at null infinity~\cite{Strominger-YM,Strominger-YM2,Barnich-YM} and on the results obtained via the Hamiltonian approach in other gauge theories, such as electrodynamics~\cite{Henneaux-ED} and general
relativity~\cite{Henneaux-GR}, one would expect to find a well-defined Hamiltonian formulation of the non-abelian Yang-Mills theory, which features a canonical action of non-trivial asymptotic symmetries.
Quite surprisingly, we are not able to obtain this result.
Rather, we find a well-defined Hamiltonian formulation 
of the theory, but the group of asymptotic symmetries 
turns out to be trivially the Poincar\'e group and, accordingly, the total colour charge has do vanish.
Moreover, we find that if one tries to enlarge the phase space
in order to accommodate for a non-trivial asymptotic-symmetry group and for a non-vanishing value of the total  colour charge one either has to give up the existence of a symplectic form or looses the Hamiltonian action of the Poincar\'e transformations, i.e., one misses at least one of the conditions~\ref{item:symplectic-form} and~\ref{item:Poincare} in the aforementioned list.

The second situation, which is analysed in this thesis, is that of electromagnetism coupled to a scalar field.
More precisely, we deal with two main cases, of which the first contains two subcases.
In the first main case we consider what is commonly referred to as \emph{scalar electrodynamics}.
That is, a scalar field endowed with a potential, which, depending on its precise form, represents either a massless (first subcase) or a massive (second subcase) scalar field, minimally-coupled to electromagnetic fields.
Interestingly, the outcome of our analysis crucially depends 
on whether or not the scalar field  has a mass.
We show that a massive field has to decay at infinity faster than any power-like function in the affine coordinates, so that the behaviour of the electromagnetic fields, as well as the 
symmetry group, is the same as the one found by Henneaux and Troessaert in the case of free electrodynamics \cite{Henneaux-ED}.
On the other hand, a massless scalar field renders the boosts of the Poincar\'e transformations non-canonical in a way which is difficult to circumvent, leading either to a trivial asymptotic symmetry group or to a non-canonical action of the  Poincar\'e group.
We highlight a connection of this problem with the impossibility of a Lorenz gauge-fixing if the flux of  charge-current at null infinity is present, as  pointed out by Satishchandran and Wald \cite{Wald-Satishchandran}.

As our second main case we consider the \emph{abelian Higgs model}, i.e., a potential of the scalar field which leads to spontaneous symmetry breaking, thereby reducing the $\U(1)$ gauge-symmetry group to the trivial group.
We show that the asymptotic symmetry group  reduces in a straightforward way to the Poincar\'e  transformations without any complications.

These results have been already published in the following two papers:
\begin{center}
 \rule[6pt]{0.6\textwidth}{0.7pt} \\
 Roberto Tanzi and Domenico Giulini,\\
 \textit{Asymptotic symmetries of Yang-Mills fields in Hamiltonian formulation},\\
 \href{https://doi.org/10.1007/JHEP10(2020)094}{\textit{Journal of High Energy Physics} \textbf{10} (2020) 94},
 \href{https://arxiv.org/abs/2006.07268}{\textit{arXiv:} 2006.07268\textbf{[hep-th]}},\\
 October 2020.\\[20pt]
 Roberto Tanzi and Domenico Giulini,\\
 \textit{Asymptotic symmetries of scalar electrodynamics and of the abelian Higgs model in Hamiltonian formulation},\\
 \href{https://arxiv.org/abs/2101.07234}{\textit{arXiv:} 2101.07234\textbf{[hep-th]}},
 submitted to JHEP and under review,\\
 January 2021.\\
 \rule[0pt]{0.6\textwidth}{0.7pt}
\end{center}
In particular, in the former paper, the situations of $\SU(N)$-Yang-Mills is analysed, while, in the latter,  scalar electrodynamics and the abelian Higgs model are studied.
Parts of this thesis are taken and adapted from these two papers.

\section{Outline}

This thesis is structured as follows.
In order to provide some context to the specific investigations pursued here, we will begin with an overview of various aspects concerning the study of asymptotic symmetries.
After a historical survey, which includes the relevant definitions, we will point out the recent developments in the field.
In this way, the results of the thesis can be better interpreted as being part of a collective effort in the understanding of a wide and rich area of research.
All of this will be done in chapter~\ref{cha:asymptotic-symmetries}.

The following two chapters are designed to provide all the tools necessary to derive the findings of this thesis in a way, which is as much as possible self-contained.
To this end, we will need to illustrate the main features and methods concerning the Hamiltonian formulation of field theories.
In order to introduce these methods in an uncomplicated situation, we will show how the Hamiltonian formulation works in the simple case of classical mechanics, treated in chapter~\ref{cha:Hamiltonian-CM}.
Then, we will quickly generalise these methods to the relevant case of field theories in chapter~\ref{cha:Hamiltonian-ft}.
In addition, in this chapter, we will also show how to deal with gauge theories and with the Poincar\'e transformation.
All this preliminary work will allow us to state in greater detail the guiding principles in the Hamiltonian study of the asymptotic symmetries of gauge theories, done in section~\ref{sec:principles}.
These will be put immediately into action in section~\ref{sec:GR} to discuss briefly the situation in General Relativity, which concludes these first chapters.

The consecutive chapters are dedicated to the detailed analysis of those two situations which constitute the original contribution of this thesis.
Specifically, we will begin with the study of the asymptotic symmetries of non-abelian gauge theories and, more precisely, the $\SU(N)$-Yang-Mills case in chapter~\ref{cha:Yang-Mills}.
In this chapter, we will present the results already published in~\cite{Tanzi-Giulini:YM} and, in addition, we will show how these results change in higher dimensions, in order to highlight how special the physically-relevant four-dimensional case is.
The second situation --- which includes the study of scalar electrodynamics and of the abelian Higgs model --- will be discussed in chapter~\ref{cha:scalar-electrodynamics}, in which we present the results already published in~\cite{Tanzi-Giulini:abelian-Higgs}.

Finally, chapter~\ref{cha:conclusions} contains ample discussions and concluding remarks about the topics treated in this thesis.
Moreover, in this chapter, we will also point out some possibilities for the future short-term and medium-term developments of the findings of this thesis.

\newpage
\section{Conventions and notation}

Throughout this thesis, the speed of light is set to the value of $1$.
The spacetime manifold will be denoted by $M$.
This is a four-dimensional Lorentzian manifold, except in the few sections dealing with the situation in higher dimensions, in which case the dimension of $M$ will be $n+1$.
The spacetime Lorentzian metric, denoted by ${}^4 g$, is chosen accordingly to the mostly-plus convention with signature $(-+++)$.
Only in chapter~\ref{cha:asymptotic-symmetries}, we will denote the spacetime metric with $g$, since there is no risk of confusion with the spatial metric appearing in the $3+1$ decomposition.

The Hamiltonian formulation requires the spacetime to be ``split into space and time'' according to the $3+1$ decomposition.
In this case, $\Sigma$ will be a three-dimensional manifold --- representing abstractly the space --- and will be provided with a three-dimensional positive-definite metric $g$.
In the few sections dealing with the situation in higher dimensions, the dimension of $\Sigma$ will be $n$.

Points in $M$ will be denoted by lower-case letters, such as $x \in M$, whereas points in $\Sigma$ will be denoted by lower-case bold letters, such as $\vect{x} \in \Sigma$.
When there is no risk of confusion, we may eventually relax this rule and denote points in $\Sigma$ with non-bold letters, such as $x \in \Sigma$.

We will often use radial-angular coordinates on $M$ and on $\Sigma$, i.e. $(t,r,\barr{x})$ and $(r,\barr{x})$, respectively.
In this case $\barr{x}$ will denote suitable coordinates on the unit two-sphere, which we will often choose to be the usual $(\theta,\varphi)$.

When coordinates are employed, lower-case Greek indices will run over spacetime components, e.g. $\alpha = 0,1,2,3$, lower-case Latin indices over spatial components, e.g. $a = 1,2,3$, and lower case barred Latin indices will run over angular components, e.g. $\barr{a} = \theta,\varphi$.
In addition, capital Latin letters will be used as indices in other situations, e.g. $A = 1, \dots, N^2-1$ for the components of the $\su(N)$ Lie algebra.
As usual, the sum of repeated indices (one upstairs and one downstairs) has to be understood, unless differently specified.

The symbols $\lie$ and $d$ will be respectively the Lie derivative and the exterior derivative, either on $M$ or on $\Sigma$ depending on the situation.
In addition, $\liephase$ and $\extderphase$ will be respectively the Lie derivative and the exterior derivative on phase space.


Finally, $A \setminus B$ denotes the set difference, i.e., the set consisting of those elements of $A$ that do not belong to $B$.
In addition, we will write $A \subset B$ when $A$ is strictly contained in $B$ and $A \subseteq B$ when the inclusion is not in the strict sense.

\chapter{An introduction to asymptotic symmetries} \label{cha:asymptotic-symmetries}

We wish to start this thesis with an overview of asymptotic symmetries, a very rich field of research, which has been studied to a great extend for several decades.
The first studies of the asymptotic symmetries in General Relativity date back to 1962, when three papers about the topic were published in short sequence.
In the first one~\cite{Bondi:waves-axisymmetric}, Bondi, van der Burg, and Metzner studied the radiation of gravitational waves from an axisymmetric isolated system.
To this end, suitable coordinates were introduced and the spacetime metric was required to have some precise asymptotic behaviour while reaching (null) infinity or, in other words, to satisfy some fall-off conditions at (null) infinity.
The transformations preserving these fall-off conditions were identified and are part of what we now refer to as asymptotic symmetries.
In addition, the authors provided a useful definition of mass for the isolated system measured at null infinity, commonly referred to as ``Bondi mass''.
The main feature of the Bondi mass is that its value is constant so long as no gravitational radiation is present, while it monotonically decreases if there is gravitational radiation emitted from the system, thus providing a good measure of how the mass of an isolated system changes due to gravitational radiation.

The other two papers appeared in 1962 are both by Sachs.
The first one of the two~\cite{Sachs:waves-asymptotically-flat} generalises the findings of~\cite{Bondi:waves-axisymmetric} by studying the radiation of gravitational waves in asymptotically-flat spacetimes, which look like flat Minkowski spacetime at infinity and include, but are not limited to, the axisymmetric isolated systems considered in~\cite{Bondi:waves-axisymmetric}.
The second paper by Sachs~\cite{Sachs:asymptotic-symmetries} focused on the asymptotic transformations preserving the asymptotic form of the metric in a similar fashion to the  discussion by Bondi and Metzner, thus finding the asymptotic symmetries of asymptotically-flat spacetimes.

The intuitive expectation at the time was that, since asymptotically-flat spacetimes look at infinity like the Minkowski spacetime, the asymptotic-symmetry group should have been the same as the symmetry group of Minkowski, i.e., the Poincar\'e group.
We remind that this is constituted by the combination of ten transformations: one time translation, three spatial translations, three rotations, and three Lorentz boosts.
More precisely, in mathematical language, we say that the Poincar\'e group is the semidirect product of the Lorentz group (rotations and boosts) with the four spacetime translations.
To great surprise, the group of asymptotic symmetries that was identified ended up to be a much larger, infinite-dimensional group, which is the semidirect product of the Lorentz group with the so-called super-translations, infinitely-many transformations that generalise the four spacetime translation.
Sachs proposed the name ``Generalised Bondi-Metzner'' (GBM) group.
The ensuing literature gave it the name ``Bondi-Metzner-Sachs'' (BMS) group, which remained up to today.
We will provide a brief overview about the features of asymptotically-flat spacetimes and the BMS group in section~\ref{sec:AFST}.

After the initial enthusiasm, the BMS group lived in a niche area of research for quite some time.
This was until Hawking, Perry and Strominger published their renowned paper~\cite{HPS}, in which they conjecture that the BMS group could provide a solutions to the long-standing black-hole information-loss paradox.
The intuitive idea of how this solution should work is that the information of an object which collapses into a black hole is not lost, but gets somehow transferred to the asymptotic charges of the BMS group.
Connected to this possibility, it was also realised that asymptotic symmetries and their charges are not an exclusive property of the gravitational field in asymptotically-flat spacetimes, but they actually seem to be a common feature of those field theories with long-ranging interactions.
Independently on whether or not the proposal of~\cite{HPS} can actually solve the information-loss paradox, it has certainly the merit of having brought a great momentum in the study of asymptotic symmetries.
Over the last years, many new studies appeared and analysed, for instance, the situation of electrodynamics~\cite{Strominger:QED1,Strominger:QED2,Strominger:QED3} and that of Yang-Mills~\cite{Strominger-YM,Strominger-YM2,Barnich-YM} at null infinity finding a large group of asymptotic symmetries.
We will briefly discuss this topic in section~\ref{sec:Strominger}.
A more detailed review about the topics of this section can be found in~\cite{Alessio-Esposito:review}.

\section{Asymptotically-flat spacetimes} \label{sec:AFST}

Intuitively, an asymptotically-flat spacetime is a spacetime which looks like Minkowski at infinity.
Of course, this is not a definition at all, since we have to specify what ``looks like'' and ``infinity'' mean in mathematical terms.
We will provide a mathematical definition and references with detailed discussions later on in this section, but let us for now proceed in a more down-to-earth way.

To this end, let us begin by considering the line element of the flat Minkowski spacetime in radial coordinates $(t,r,\theta,\varphi)$, i.e.,
\begin{equation} \label{Minkowski-metric-tr}
 ds^2 = -dt^2 +dr^2 + r^2 \bigl( d\theta^2 + \sin^2 \theta \, d \varphi^2 \bigr) \,.
\end{equation}
Broadly speaking, there are five ``infinities'' that could be introduced by considering some limit.
Future and past \emph{timelike infinity} are reached by taking the limit $t \rightarrow + \infty$ or $t \rightarrow - \infty$, respectively, while $r$ stays finite.
The former is denoted by $i^+$ and represents the distant future of observers, while the latter is denoted by $i^-$ and represents their distant past.
Future and past \emph{null infinity}, denoted by $\scri^+$ and $\scri^-$, are reached by following, respectively, a future-directed or a past-directed null geodesics up to an infinite value of the affine parameter.
This corresponds to take the limit $r \rightarrow +\infty$ while the \emph{retarded time} $u \eqdef t -r$ converges to a finite value (future null infinity) or while the \emph{advanced time} $v \eqdef t + r$ does (past null infinity).
Finally, \emph{spacelike infinity} or \emph{spatial infinity}, denoted by $i^0$, is reached by taking the limit $r \rightarrow + \infty$ while $t$ stays finite.

\begin{figure}[t]
\centering
\begin{tikzpicture}[scale=1.5]
   \coordinate (A) at (0.5,0);
   \coordinate[label=right:{$i^0$}] (i0) at (2,0);
   \coordinate[label=above:{$i^+$}] (i+) at (0,2);
   \coordinate[label=below:{$i^-$}] (i-) at (0,-2);
   \draw (i-) -- node[below right] {$\scri^-$} (i0) -- node[above right] {$\scri^+$} (i+);
   \draw[dashed] (i-) -- node[above,sloped] {\footnotesize $r=R=0$} (i+);
   \draw[color=gray,
   very thick,
    decoration={
    markings,
    mark=at position 0.7 with {\arrow{latex}},
    mark=at position 0.35 with {\arrow{latex}}
    },
   postaction={decorate}] (i-) to[bend right=25] node[pos=0.7,above,sloped] {\scriptsize massive} (i+);
   \begin{scope}[very thick,decoration={
    markings,
    mark=at position 0.25 with {\arrow{latex}},
    mark=at position 0.8 with {\arrow{latex}}
    }
    ]
   \draw[densely dotted,color=gray,postaction={decorate}] (0.6,-1.4) -- (0,-0.8) -- node[pos=0.7,below,sloped] {\scriptsize light ray} (1.4,0.6);
   \end{scope}
   \draw[thick,->,>=latex] (-1,-1) -- node[fill=white] {\it \footnotesize time} (-1,1);
   \fill[black] (i0) circle(1.5pt);
   \fill[black] (i+) circle(1.5pt);
   \fill[black] (i-) circle(1.5pt);
\end{tikzpicture}
  \caption{The Penrose diagram of Minkowski spacetime, suppressing the angular coordinates.
  Neglecting the known coordinate singularity at $r=0$, the physical spacetime is contained in the interior of a triangle, whose boundary consists of timelike, spacelike, and null infinities.
  The trajectories of massive particles originate at $i^-$ and terminate at $i^+$, whereas the trajectories of light rays originates at $\scri^-$ and terminates at $\scri^+$.
  }
  \label{fig:Penrose-diagram}
\end{figure}
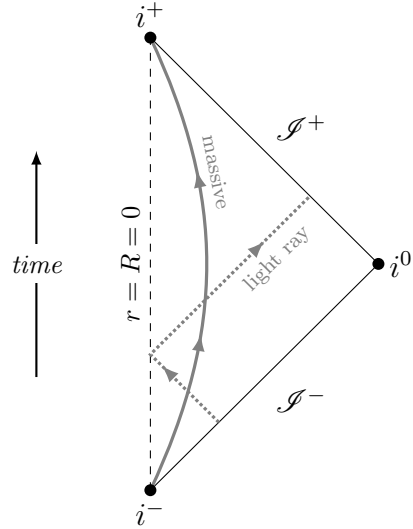

As noted by Penrose, it is actually useful to make use of a conformal transformation and define these infinities not as limits, but rather as actual points of an (unphysical) manifold, where they are at a finite distance with respect to an (unphysical) metric.
The idea works concretely as follows. 
First, let us replace $t$ and $r$ with the new coordinates
\begin{subequations}
 \begin{align}
  T &\eqdef \frac{\arctan (t + r) + \arctan (t - r)}{\pi} 
  \quad \text{and} \\
  R &\eqdef \frac{\arctan (t + r) - \arctan (t - r)}{\pi} \,,
 \end{align}
\end{subequations}
which sweep the entire Minkowski spacetime while their values range on the ``triangle'' obtained by the three conditions
\begin{equation}
 R \ge 0 \,, \qquad
 | T + R | < 1 
 \qquad \text{and} \qquad
 | T - R | > 1 \,.
\end{equation}
Then, the limits described above corresponds to the borders of said triangle.
In particular, in terms of the pair $(T,R)$ and neglecting for a moment the angular coordinates, $i^\pm$ correspond to the two points $(\pm 1,0)$, $i^0$ to the point $(0,1)$, and $\scri^\pm$ to the two open segments connecting $i^\pm$ with $i^0$.
Second, we extend the Minkowski spacetime $M$ to an unphysical manifold $\widetilde{M}$, which include the boundary of this triangle.
Third, we introduce an unphysical metric $\tilde{g}$ by means of the conformal factor
\begin{equation} \label{AFST:conformal-factor-Minkowski}
 \Omega^2 =
 \frac{4}{\pi^2(1+u^2)(1+v^2)}
 = \frac{4}{\pi^2}
 \cos^2 \left[ \frac{\pi}{2} (T - R) \right]
 \cos^2 \left[ \frac{\pi}{2} (T + R) \right] \,,
\end{equation}
so that $\tilde{g}$ has the line element
\begin{equation}
 d\tilde{s}^2 \eqdef \Omega^2 ds^2 =
 -dT^2 + dR^2 + C (T,R) \bigl( d \theta^2 + \sin^2 \theta \, d \varphi^2 \bigr) \,,
\end{equation}
where $C$ is equal to $r^2 \Omega^2$ expressed in terms of $T$ and $R$.
The situation can be actually visualised in the so-called Penrose diagram depicted in figure~\ref{fig:Penrose-diagram}.

A few things can be noted.
First, with this construction, the infinities are the actual boundary of an unphysical manifold and they are at a finite distance with respect to the unphysical metric.
Second, the trajectory of every massive particle begins at $i^-$ and ends at $i^+$.
Moreover, the trajectory of every light ray begins at $\scri^-$ and ends at $\scri^+$.
Third, taking into account also the angular coordinates, future null infinity is topologically a cylinder $\real \times S^2$ described by the retarded time $u \in \real$ and the angles.
The same holds true for past null infinity replacing $u$ with $v$.
On the contrary, timelike and spacelike infinities are all single points (see e.g. the discussion in~\cite[Chap.~11]{Wald:GR}).

The Penrose's procedure described above allows the conformal treatment of infinity, in which timelike, spacelike and null infinities are introduced as a conformal boundary of the spacetime.
As we shall see in the next subsection, asymptotically-flat spacetimes will allow a similar treatment of conformal infinity.
Before we do that, let us remark two things.
First, since we are following a more-historical line of reasoning in this chapter, we will focus on the situation at null infinity.
On the contrary, the following chapters will be more focused on the situation at spacelike infinity, due to the way the Hamiltonian formulation is built.
Secondly, in order to derive the results of this thesis, it would suffice to consider the flat Minkowski spacetime, on which the field theories that we analyse are defined.
Nevertheless, we will include a discussion about the more general asymptotically-flat spacetimes to make a connection between the situation in these field theories and in gravity.
Moreover, one possible generalisation of the findings of this thesis consists in considering these field theories on an asymptotically-flat spacetime rather than on flat Minkowski.

\subsection{Definition}

We now provide a definition of asymptotically-flat spacetimes.
The reader who is not interested in the mathematical definition can skip to the next subsections in which we will work using suitable coordinates.
Since we focus on the situation at null infinity in this section, we provide a definition for the so-called \emph{asymptotically-flat spacetimes at null infinity}, following~\cite{Geroch:structure-spacetime}.
For a treatment that includes spacelike infinity, see e.g.~\cite[Chap.~11]{Wald:GR}

Proceeding in a way similar to the case of the flat Minkowski spacetime, let $M$ be a smooth four-dimensional manifold with a smooth Lorentzian metric $g$, which we will refer to as the \emph{physical spacetime} and the \emph{physical metric}, respectively.
In order to introduce null infinity as a conformal boundary of the physical spacetime, let us consider a second manifold $\widetilde{M}$ with a smooth Lorentzian metric $\tilde{g}$,  which we will refer to as the \emph{unphysical spacetime} and the \emph{unphysical metric}, respectively.

The unphysical spacetime is called an \emph{asymptote} of the physical spacetime if the following conditions are satisfied.
\begin{enumerate}[(i)]
\itemsep=-4pt
 \item The unphysical spacetime has a non-empty boundary $\scri \eqdef \partial \widetilde{M}$ and there is a diffeomorphism between $M$ and $\widetilde{M} \setminus \scri$.
 By means of this diffeomorphism, we will identify the physical spacetime $M$ with the interior of the unphysical spacetime, $\widetilde{M} \setminus \scri$, and we will regard all the tensor fields of the former as tensor fields on the latter.
 \item There is a smooth function $\Omega \colon \widetilde{M} \rightarrow \real$, such that $\tilde{g} = \Omega^2 g$ on the interior of the unphysical spacetime.
 Thus, we see that the two metrics are conformally related on $\widetilde{M} \setminus \scri$ and, since they are both smooth Lorentzian metrics, it must be that $\Omega \ne 0$ on $\widetilde{M} \setminus \scri$.
 \item At the boundary $\scri$, the conformal factor satisfies the three conditions
 \begin{equation} \label{AFST:conformal-factor-conditions}
  \Omega = 0 \,, \qquad
  d\, \Omega \ne 0 \,,
  \qquad \text{and} \qquad
  \tilde{g}^{\alpha \beta} \, \partial_\alpha \Omega \, \partial_\beta \Omega = 0 \,.
 \end{equation}
\newcounter{saveenumi}
\setcounter{saveenumi}{\value{enumi}}
\end{enumerate}
It is straightforward to check that the conformal factor~(\ref{AFST:conformal-factor-Minkowski}) satisfies all the three conditions stated in~(\ref{AFST:conformal-factor-conditions}).
In general, the first of the conditions in~(\ref{AFST:conformal-factor-conditions}), together with the fact that $\tilde{g}$ is smooth on $\widetilde{M}$ and conformally related to $g$ on the interior, tells us that $\scri$ is very far away with respect to the physical metric $g$.
The second condition tells us, broadly speaking, that the fall-off behaviour of $\Omega$ is that of $1/r$.
Finally, the third condition tells us that $\scri$ is a null hypersurface or, in other words, that we are dealing with null infinity.

The definition above does not lead to a unique asymptote for each physical spacetime.
To this end, one needs to note that, up to diffeomorphisms, there is a maximal asymptote, which will be the only one considered from now on.
Several general statements about the asymptotic structure of spacetimes can be proven already with this weak definition, as it is done, e.g., in~\cite{Geroch:structure-spacetime}.
However, we wish to focus our attention to asymptotes that look similar to the situation described for the Minkowski spacetime.
To this end, we will add the further condition:
\begin{enumerate}[(i)]
\setcounter{enumi}{\value{saveenumi}}
 \item Each maximally-extended null geodesic intersect $\scri$ exactly twice.
\end{enumerate}
With this further requirement, one can show that $\scri$ consists of two parts: a future part ($\scri^+$) and a past part ($\scri^-$)~\cite{Penrose:future-past-null-infinity}.
In addition, each of these two parts has the topology of $\real \times S^2$.
Finally, we will say that a spacetime is \emph{asymptotically-flat at null infinity} if it admits a maximally extended asymptote satisfying all the conditions (i)--(iv).

More detailed discussions can be found in the contribution by Geroch~\cite{Geroch:structure-spacetime}, in the book by Wald~\cite[Chap.~11]{Wald:GR}, and in the one by Hawking and Ellis~\cite[Chap.~6.8]{hawking-ellis}.
We now turn to the more-explicit analysis of asymptotically-flat spacetimes at null infinity in suitable coordinates.

\subsection{Bondi-Sachs coordinates}
The two original papers by Bondi, van der Burg and Metzner~\cite{Bondi:waves-axisymmetric} and by Sachs~\cite{Sachs:waves-asymptotically-flat} used suitable coordinates in order to study the gravitational radiation emitted from axisymmetric isolated systems and in asymptotically-flat spacetimes, respectively.
Let us see how this works starting from the familiar line-element of Minkowski~\ref{Minkowski-metric-tr}.

To begin with, let us point out that, since the focus was on the situation very far away from the source or, more precisely, in a neighbourhood of future null infinity, it is more convenient to replace $t$ with the retarded time $u$.
In this case, the line element~(\ref{Minkowski-metric-tr}) becomes
\begin{equation} \label{Minkowski-metric-tu}
 ds^2 = -du^2 -2 du dr +r^2 \bigl( d\theta^2 + \sin^2 \theta d \varphi^2 \bigr) \,.
\end{equation}

In the general situation, as it is shown in~\cite{Sachs:waves-asymptotically-flat}, it is possible to choose a set of coordinates $(u,r,\theta,\varphi)$ covering at least a region of the spacetime, so that the line element takes the form 
\begin{subequations}\label{Sachs-metric}
\begin{equation}
 ds^2 = \frac{e^{2\beta} V}{r} du^2
 - 2e^{2\beta} du dr 
 + r^2 h_{\bar m \bar n} \bigl(dx^{\bar m} -U^{\bar m} du \bigr) \bigl( dx^{\bar n} -U^{\bar n} du \bigr) \,,
\end{equation}
where indices with a bar above run over the angular components and the two-metric $h$ is such that
\begin{equation}
 2 h_{\bar m \bar n} dx^{\bar m} dx^{\bar n} =
 \bigl( e^{2\gamma} + e^{2\delta} \bigr) d\theta^2 + 4 \sin\theta \sinh (\gamma - \delta) d\theta d\varphi
 +\sin^2 \theta \bigl( e^{-2\gamma} + e^{-2\delta} \bigr) d\varphi^2 \,,
\end{equation}
\end{subequations}
so that $\det h = \sin^2 \theta$.
These coordinates, introduced by Sachs in~\cite{Sachs:waves-asymptotically-flat} --- and very closely related to those introduced by Bondi in~\cite{Bondi:waves-axisymmetric} --- will be referred to as \emph{Bondi-Sachs coordinates}.
Note that the metric~(\ref{Sachs-metric}) is written in terms of six arbitrary functions of the coordinates: $V$, $\beta$, $U^{\bar m}$, $\gamma$, and $\delta$.
For the physical meaning of these coordinates see Fig.~1 of~\cite{Sachs:waves-asymptotically-flat}, Fig.~1 of~\cite{Sachs:asymptotic-symmetries}, and the related discussions.

Now, if the metric~(\ref{Sachs-metric}) is to be that of an asymptotically-flat spacetime, some conditions on the range in which these coordinates can change and on the behaviour of the six functions must be imposed.
In particular, as it is done in~\cite{Bondi:waves-axisymmetric,Sachs:waves-asymptotically-flat,Sachs:asymptotic-symmetries}, one can assume that the coordinates cover at least the range
\begin{equation} \label{Sachs-metric-range}
 u_0 < u < u_1 \,, \quad
 r_0 < r <  \infty \,, \quad
 0 \le \theta \le \pi \,, 
 \quad \text{and} \quad
 0 \le \varphi \le 2 \pi \,,
\end{equation}
where the points $\varphi = 0$ are identical to those $\varphi = 2\pi$.
In addition, the six functions are such that the metric~(\ref{Sachs-metric}) approaches asymptotically the flat Minkowski metric~(\ref{Minkowski-metric-tu}).
In other words, in the limit $r \rightarrow +\infty$,
one has
\begin{equation} \label{Sachs-metric-limit}
 \lim \bigl( V/r \bigr) = -1
 \qquad \text{and} \qquad
 \lim \bigl( rU^{\bar m} \bigr) =
 \lim \beta =
 \lim \gamma =
 \lim \delta = 0 \,.
\end{equation}
To be more precise, using the field equations, it was shown in~\cite{Bondi:waves-axisymmetric,Sachs:waves-asymptotically-flat} that the six function have the fall-off behaviour
\begin{subequations} \label{Sachs-metric-expansion}
 \begin{align}
 \label{Sachs-metric-expansion-V}
  V &= -r +2M(u,\theta,\varphi) + \bigo ( 1/r ) \,, \\
  \label{Sachs-metric-expansion-c}
  \beta &= - \frac{ c(u,\theta,\varphi)  c^*(u,\theta,\varphi) }{4r^2}
  + \bigo \bigl( 1/r^4 \bigr) \,, \\
  h_{\bar m \bar n} dx^{\bar m} dx^{\bar n} &=
  \bigl( d \theta^2 + \sin^2 \theta \, d \varphi^2 \bigr)
  + \bigo ( 1/r ) \,, \\
  U^{\bar m} &= \bigo \bigl( 1/r^2 \bigr) \,,
 \end{align}
\end{subequations}
where the function $M(u,\theta,\varphi)$ is called the \emph{Bondi mass aspect}, $c(u,\theta,\varphi)$ is a complex function and $c^*(u,\theta,\varphi)$ is its complex conjugate.

This result led Sachs to define in~\cite{Sachs:asymptotic-symmetries} an asymptotically-flat spacetime as one in which there are coordinates $(u,r,\theta,\varphi)$ ranging as in~(\ref{Sachs-metric-range}), such that the metric was of the form~(\ref{Sachs-metric}) in terms of six functions with asymptotic behaviour~(\ref{Sachs-metric-expansion}).
Before we introduce the symmetry group of asymptotically-flat spacetimes, let us briefly discuss the physical importance of the two functions $M(u,\theta,\varphi)$ and $c(u,\theta,\varphi)$ appearing in the expansion~(\ref{Sachs-metric-expansion}).
In particular, one can define the \emph{Bondi mass} as the average over the two sphere of the Bondi mass aspect, i.e.,
\begin{equation} 
 M_B (u) \eqdef \langle M (u,\theta,\varphi) \rangle_{S^2} =
 \frac{1}{4 \pi} \int_{S^2} d \cos \theta \, d \varphi \, M(u,\theta,\varphi) \,.
\end{equation}
Then, one can show~\cite{Bondi:waves-axisymmetric,Sachs:waves-asymptotically-flat} that
\begin{equation} \label{Bondi-mass-loss}
 \frac{d M_B}{du}  =
 - \Bigl\langle
 \left| \frac{\partial c}{\partial u} \right|^2
 \Bigr\rangle_{S^2}
\end{equation}
As a consequence, the Bondi mass decreases so long as the \emph{Bondi news function} $N \eqdef \partial c / \partial u$ is non-zero and stays the same if $N=0$.
The Bondi mass can be used to define the mass of an isolated system, while a non-vanishing news represent the presence of emitted gravitational radiation.
In this way, we see that, due to equation~(\ref{Bondi-mass-loss}), the mass of the isolated system decreases when it emits gravitational radiation and stays the same otherwise.
See~\cite{Bondi:waves-axisymmetric,Sachs:waves-asymptotically-flat} for a detailed discussion about this topic.\
\enlargethispage{-\baselineskip}

\subsection{The BMS group}

As we have seen, an asymptotically-flat spacetime allows coordinates, which satisfies the conditions~(\ref{Sachs-metric-range}) and in which the metric takes the form~(\ref{Sachs-metric}) with asymptotic behaviour~(\ref{Sachs-metric-expansion}).
There is a group transformations that preserves the three conditions~(\ref{Sachs-metric}), (\ref{Sachs-metric-range}), and~(\ref{Sachs-metric-expansion}), as it was first noted by Bondi and Metzner for the axisymmetric case~\cite{Bondi:waves-axisymmetric} and by Sachs for the more-general case~\cite{Sachs:waves-asymptotically-flat,Sachs:asymptotic-symmetries}.
This group of transformations is precisely the \emph{Bondi-Metzner-Sachs group} or, more simply, the \emph{BMS group}.

Let us assume that $u$ is not restricted to an interval, but can range in the whole real line.
The transformation $(u , \theta,\varphi) \mapsto (\bar u , \bar \theta, \bar \varphi)$ is a BMS transformation if it satisfies the following two conditions.
\begin{enumerate}[(i)]
 \item The transformation of the angular coordinates is a conformal transformation of the sphere.
 In other words,  writing
 \begin{subequations} \label{BMS:transformations}
 \begin{equation} \label{BMS:transformations-angles}
  \bar \theta = H (\theta,\varphi) 
  \qquad \text{and} \qquad
  \bar \varphi = I (\theta,\varphi) \,,
 \end{equation} 
 the metric of the two-sphere transforms as
 \begin{equation} \label{BMS:transformations-conformal-sphere}
  d \theta ^2 + \sin^2 \theta \, d \varphi^2 = 
  K^2 (\bar \theta, \bar \varphi)
  \bigl(
  d \bar \theta ^2 + \sin^2 \bar \theta \, d \bar \varphi^2
  \bigr) \,,
 \end{equation}
 where the conformal factor $K (\bar \theta, \bar \varphi)$ is a positive function.
 \item The retarded time transforms as
 \begin{equation} \label{BMS:transformations-u}
  \bar u (u,\theta,\varphi) = 
  \frac{u + \alpha (\theta , \varphi)}{K (\theta , \varphi)} \,,
 \end{equation}
 \end{subequations}
 where $\alpha (\theta, \varphi)$ is an arbitrary function and $K (\theta, \varphi)$ is the conformal factor appearing in~(\ref{BMS:transformations-conformal-sphere}) expressed in terms of the old coordinates.
\end{enumerate}
In the definition above, we have not specified how regular the various functions appearing in the BMS transformation~(\ref{BMS:transformations}) should be, since numerous options have been considered in the literature.
In the original treatment by Sachs~\cite{Sachs:asymptotic-symmetries}, the conformal transformation~(\ref{BMS:transformations-angles}) had to be regular everywhere --- which make this subgroup of transformations isomorphic to the Lorentz group --- and the function $\alpha$ had to be, at least, twice differentiable.
For example, one generalisations considered in the literature is that of McCarthy~\cite{McCarthy:1}, who assumed the function $\alpha$ to be merely a square-integrable function on the two-sphere, while studying the representation theory of the BMS group, which we will briefly mention at the end of this subsection.
Another, more recent example is that of Barnich and Troessaert~\cite{Barnich-Troessaert}, who have proposed to weaken the condition that~(\ref{BMS:transformations-angles}) is regular everywhere, obtaining a much larger group than that originally discussed by Sachs.
\enlargethispage{-\baselineskip}

Let us now discuss briefly the structure of the group of BMS transformations~(\ref{BMS:transformations}) and see how it generalises the Poincar\'e group.
If we consider the case if which the angles are not transformed, i.e. $\bar \theta = \theta$ and $\bar \varphi = \varphi$, we are left with a non-trivial transformation of the retarded time~(\ref{BMS:transformations-u}) parametrised by the arbitrary function $\alpha (\theta,\varphi)$.
This transformation is called a \emph{supertranslation} and reads
\begin{equation} \label{BMS:supertranslation}
 \bar u = u + \alpha (\theta,\varphi) \,,
\end{equation}
which consists simply in shifting $u$ by an angle-dependent function $\alpha$.
A few things can be said about the supertranslations.
First, from~(\ref{BMS:supertranslation}), one immediately see that the combination of any two supertranslations $\alpha_1$ and $\alpha_2$ is the supertranslation $\alpha_1 + \alpha_2$.
Thus, the supertranslations form an abelian subgroup $\mathcal{A}$ of the BMS group.
Second, decomposing $\alpha$ in terms of spherical harmonics, one sees that this subgroup is actually infinite-dimensional.
Third, one can show that $\mathcal{A}$ is a normal subgroup of the BMS group.
Moreover, there is a four-dimensional normal subgroup $\mathcal{T}$ of the supertranslations parametrised by those $\alpha$ which can be written as linear combinations of the $\ell=0$ and $\ell=1$ spherical harmonics.
The four-dimensional subgroup $\mathcal{T}$ constitutes the four spacetime translations.

Thus, we have seen that it is possible to identify the spacetime translations inside the BMS group and that these are a special case of the more-general supertranslations.
Now, one can ask whether it is possible to identify not just the translations, but the entire Poincar\'e group inside the BMS group.
The answer is yes, but not unequivocally.
To be more precise, one can show that the BMS group is the semi-direct product of $\mathcal{A}$ with the Lorentz group.
Therefore, we have all the elements to find a copy of the Poincar\'e group as a subgroup BMS group.
However, since this subgroup is not normal, it is possible to find infinitely-many other distinct copies of the Poincar\'e group.

Before we conclude this section, let us briefly mention that, among the early contributions to the study of the BMS group, there was a certain effort in the study of its (projective irreducible) representations.
The motivation was that, since particles on a flat Minkowski spacetime are classified according to the irreducible projective representations of the Poincar\'e group as noted by Wigner~\cite{Wigner-representations}, particles on an asymptotically-flat spacetime should have been classified according to the representations of the BMS group.
The first analyses were pursued by McCarthy~\cite{McCarthy:1,McCarthy:2} and gave quite promising results.
Indeed, McCarthy showed that, contrary to the case of the Poincar\'e group, all the representations of the BMS group had discrete spin.
To be completely fair, McCarthy showed that all the representations, found applying Mackey's theory of induced representations~\cite{Mackey}, had discrete spin.
Despite Mackey's theory can be successfully applied to obtain all the irreducible representation of the Poincar\'e group, there was no guarantee that it could provide all, and not just a part, of the irreducible representations of the BMS group, too.
Luckily, a few years later, it was shown by Piard to be true that Mackey's theory provides all the representations of the BMS group~\cite{Piard:1,Piard:2}.

The original enthusiasm of the first papers and of the subsequent studies~\cite{McCarthy:3,McCarthy:4,McCarthy:5} was partially lost when it was noted by Girardello and Parravicini~\cite{Girardello-Parravicini} and by McCarthy himself~\cite{McCarthy:nuclear} that, by choosing a different topology for the BMS group, continuous-spin representations can appear.
Thus, the use of the BMS group instead of the Poincar\'e group  cannot get rid unequivocally of the continuous-spin representation, in contrast to the original hope.
Nevertheless, the representation theory of the BMS group can find different applications, e.g. in the context of holography~\cite{Dappiaggi}.
\enlargethispage{-\baselineskip}

\section{Recent developments} \label{sec:Strominger}

After the momentum brought in the subject by the first studies, which were mentioned in the previous section, the BMS group and asymptotic symmetries remained a secondary area of research for some decades.
However, this situation has changed radically in recent years, the turning point being the well-known paper ``Soft hair on black holes'' by Hawking, Perry and Strominger~\cite{HPS}.
In this paper, the three authors conjectured a connection between asymptotic symmetries and a possible solution to the long-standing black-hole information-loss paradox.
From that moment onward, asymptotic symmetries have been a very rich and active field of investigation.
\enlargethispage{-\baselineskip}

To provide some context, the information-loss paradox dates back to 1976~\cite{Hawking:info-loss}, when it was theorised by Stephen Hawking as a natural consequence of Hawking radiation~\cite{Hawking:radiation1,Hawking:radiation2}.
In short, let us assume that we have an isolated system consisting in an astrophysical object, such as a star, which collapses and forms a black hole.
Before the collapse, the state of the system is described by very many parameters and contains a lot of information.
However, after the collapse, the black hole is expected to be  described only by a handful of parameters (the mass, the angular momentum, and the charge), a result which is often referred to by saying that a black hole has no hair.\footnote{
For details and references, see Chap.~33 of~\cite{Gravitation}, Chap.~9 of~\cite{hawking-ellis}, and Chap.~12 of~\cite{Wald:GR}.
}
Classically, the situation described does not present any issue.
We are merely noting that almost all the information about the original object falls into the singularity, which acts as a sink for the universe, but this happens in a region hidden behind the horizon and forever inaccessible to us.

However, this situation changes completely due to Hawking's finding that a black hole emits thermal radiation when semi-classical effects are taken into considerations~\cite{Hawking:radiation1,Hawking:radiation2}.
Indeed, emitting thermal radiation and, thus, loosing energy, the black hole can completely evaporate after some (long) time.
After the evaporation process is complete, we would end up with a universe that does not contain a horizon and a singularity any-more, but only thermal radiation.
Thus, overall, the initial state consisting of a fairly-complex astrophysical object will eventually evolve into a final state consisting of mere thermal radiation, the net effect being a loss of information from the initial to the final state, despite neither singularities nor horizons are present in these states.
Several possible solutions have been proposed in the literature, although none has been completely accepted as a solution to the paradox.
For a list of some of the most prominent proposal, see~\cite{Wald:info-loss}, although the authors are very critical with each one of them.

Coming back to our starting point, the proposal by Hawking, Perry, and Strominger~\cite{HPS} aimed at solving the paradox by showing that a black hole can have hair and can be described by more-than-just-three parameters, namely, the charges associated to asymptotic symmetries, which could be used to store the information without a loss.
In this context, it was also noted that asymptotic symmetries and their associated charges are not an exclusive aspect of gravity, but are actually a feature of many long-ranging field theories, such as electrodynamics.
Independently on whether or not these methods can be actually used to solve the long-standing paradox, the work of~\cite{HPS} produced undoubtedly a huge interest in the study of asymptotic symmetries, so that many new discoveries have been made in this field and a greater understanding of the subject has been reached.

Over the last few years it has been shown that electrodynamics possesses a large symmetry group at null infinity, consisting of angle-dependent gauge transformations that do not vanish at $\scri$, and that the Ward identities associated to this symmetry coincide with Weinberg's soft photon theorems~\cite{Strominger:QED1,Strominger:QED2,Strominger:QED3}.\footnote{
For the results by Weinberg, see~\cite{Weinberg:infrared} and Chap.~13 of~\cite{Weinberg:QFT1}.
}
These results concerning electrodynamics have been extended at the subleading order~\cite{Strominger:Soft-theorem,Campiglia-subleading-soft-photons,Conde:subleading-soft-photon} and in the presence of magnetic charges~\cite{Strominger:Soft-theorem-magnetic}.\footnote{
See the lecture notes~\cite{Strominger:notes} for a gentle introduction about these recent developments with details and references.
}
Moreover, similar results have been derived in the case of non-abelian gauge theories~\cite{Strominger-YM,Strominger-YM2,Barnich-YM} and in the case of higher-spin fields~\cite{Francia:spin}.
Other recently-pursued investigations try to find the asymptotic symmetries of physically-relevant spacetimes that are not asymptotically flat.
These studies include, for instance, the asymptotically-(A)dS case~\cite{BMS-ADS1,BMS-ADS2} and the asymptotically-FLRW (Friedmann-Lema\^itre-Robertson-Walker) case~\cite{BMS-FRLW1,BMS-FRLW2}.

It is important to note that the recent developments have not been limited to purely-theoretical aspects, but also potentially-detectable effects related to asymptotic symmetries have been considered.
This is the case of the memory effect~\cite{Wald:memory,Wald:memory-FRLW,Compere1,Compere2,Francia:color-memory}, which, in the simplest case, consists in a permanent displacement of the relative distance of test particles due to the passage of a burst of gravitational waves and could be detected in gravitational-wave experiments.

Finally, let us conclude this survey about the recent developments in the study of asymptotic symmetries by mentioning that, for a long time, it was falsely believed that the BMS group was an exclusive feature of null infinity, which was not present at spatial infinity.
Ideally, the study of asymptotic symmetries at spatial infinity are best performed using the Hamiltonian formulation, which is going to be thoroughly reviewed in the next two chapters.
In the first paper dealing with this subject~\cite{RT}, Regge and Teitelboim were able to recover the Poincar\'e group as the asymptotic-symmetry group of asymptotically-flat spacetimes using Hamiltonian methods.
However, the BMS group was not recovered and this fact lead to the aforementioned wrong conclusion about the absence of the BMS group at spatial infinity.
This apparent conflict between analyses at null and spatial infinity was only recently resolved by Henneaux and Troessaert in~\cite{Henneaux-GR} and we will discuss briefly their solution in section~\ref{sec:GR}.
Let us mentioned that, a few years after the paper by Regge and Teitelboim~\cite{RT} and many years before the one by Henneaux and Troessaert~\cite{Henneaux-GR}, Ashtekar and Hansen performed a purely-kinematical study of the structure of spacelike infinity, finding a (rather big) symmetry group known as SPI~\cite{SPI}.
In this study, however, the dynamics of the fields was not taken into consideration, as it is done in~\cite{Henneaux-GR}.

The success of the analysis by Henneaux and Troessaert~\cite{Henneaux-GR} in recovering the BMS group using the Hamiltonian formulation of General Relativity has led to similar analyses of the asymptotic structure of many field theories of interest.
In particular, it has been already studied the situation of electrodynamics in four dimensions~\cite{Henneaux-ED} (which has been related to the mentioned studied at null infinity by means of~\cite{Campiglia-U1}), in higher dimensions~\cite{Henneaux-ED-higher}, and coupled to gravity~\cite{Henneaux-ED-GR}.\footnote{
See also the review~\cite{Henneaux-review}.
}
Further studies include the case of the electromagnetic duality~\cite{Henneaux:duality}, of a massless scalar field (and its dual two-form field)~\cite{Henneaux-scalar}, of the Pauli-Fierz theory~\cite{Henneaux:Pauli-Fierz}, and of the Rarita-Schwinger theory~\cite{Henneaux:Rarita-Schwinger}.
This thesis adds to the list two cases.
The first one consists in $\SU(N)$-Yang-Mills, discussed in chapter~\ref{cha:Yang-Mills}, where differences with the corresponding studies at null infinity~\cite{Strominger-YM,Strominger-YM2,Barnich-YM} are found.
The second one is the case of scalar electrodynamics and of the abelian Higgs model, discussed in chapter~\ref{cha:scalar-electrodynamics}.

\chapter{Hamiltonian methods in classical mechanics} \label{cha:Hamiltonian-CM}

In the previous chapter, we have provided the basic insight into various aspects of asymptotic symmetries.
In order to discuss in an exhaustive way the finding of this thesis, we also need to introduce the methods of the Hamiltonian formulation of classical field theories.
To this end, we will provide a brief review of the Hamiltonian methods and techniques in this and the next chapter.
Specifically, we will begin with the simpler case of classical mechanics in this chapter and, then, generalise the obtained results to the case of classical field theories in the next chapter.

This review is meant to introduce all the tools that are needed in order to derive the main results of this thesis.
Therefore, we will not need to deal with every single aspect and mathematical subtlety of the Hamiltonian formulation of classical field theories.
For a detailed analysis of the subject and for the technicalities, we redirect the reader to the book by Marsden and Ratiu~\cite{Marsden-Ratiu} and references therein.
Some basic knowledge of differential geometry is assumed and we refer again to~\cite{Marsden-Ratiu} for an introduction on the subject, although we are going to try to keep the discussion as self contained as possible.

Most of the mathematical difficulties arise due to the fact that, in field theories, there is an infinite number of degrees of freedom.
Therefore, we will begin by reviewing the classical-mechanical case, in which the number of degrees of freedom is finite.
Although this would not have a direct application to the  theories we wish to study, it will provide a simple setup in which all of the tools and equations of the Hamiltonian formulation can be introduced and understood without complications.
Thus, we will devote this chapter exclusively to the case of classical mechanics.
Starting from the principle of least action and the Lagrangian picture, we will derive the relevant quantities of the Hamiltonian picture, from Hamilton equations to the analysis of canonical symmetries.

\section{General considerations}

Let us begin with a very simple situation, that is a physical system consisting of one point particle in three dimensions.
In Cartesian coordinates with respect to some origin, we can describe the position of the particle (at a given time) using a triple of real numbers $(x^1,x^2,x^3) \in \real^3$ and, as a consequence, $Q = \real^3$  is the set of all possible positions.
In general, one needs not rely on Cartesian coordinates to describe the system, but can use any useful triple of real numbers $(q^1,q^2,q^3)$ to describe the points in $Q$, such as radial coordinates, cylindrical coordinates and much more.
The fact that the concept of position should exist independently of coordinates leads naturally to the idea that the set of all possible positions $Q$ should actually be treated as a (smooth) manifold.
Then, Cartesian coordinates are merely a choice of chart on $Q$.
Although this description might seem somehow not needed to study the case of one particle in three dimensions, it will prove to be of great use in the discussion of more-complicated systems and, in particular, in the case of classical field theories, which we analyse in the next chapter.

In the simple case of one point particle, the problem one would like to solve is to determine the position of the particle at any time.
Actually, it would be enough for any practical application to restrict the time interval to any $t$ between an initial time $a$ and a final one $b$.
In other words, we wish to determine the trajectory of the particle, which is the sufficiently-regular curve $\gamma : [a,b] \rightarrow Q$.
By ``sufficiently-regular curve'', we mean anything that makes the following mathematical manipulations possible.
A $C^2$ curve is enough and we will not discuss in great detail how regular a curve must and can be.
Note that, in principle, we could have denoted the trajectory of the particle as $q(t)$, but we prefer to keep separated, at the moment, the notations concerning a single point $q \in Q$ and the curve $\gamma:  [a,b] \rightarrow Q$, in order to avoid possible misunderstandings and inaccuracies.

In classical mechanics, the trajectory $\gamma$ is usually found as one of the solutions of well-posed second-order differential equations, e.g. Newton's second law.\footnote{
We do not discuss the case of higher-order equations of motion, as this is irrelevant to the derivation of the results of this thesis.
}
Thus, in order to find the unique $\gamma$ describing the position of the particle as a function of time, one needs to know the initial position and velocity of said particle, i.e., $\gamma (a)$ and $\dot \gamma (a)$, respectively.
These are known as the initial conditions and are actually six conditions, since $\gamma (a)$ belongs to the three-dimensional manifold $Q$ and $\dot \gamma (a)$ to the three-dimensional tangent space $T_{\gamma (a)} Q$.

Similar considerations can be made for any system in classical mechanics.
In general, the configuration of a system can be described by a point in a finite-dimensional smooth manifold $Q$, which we call the \emph{configuration space}.
We will denote by $N$ the dimension of $Q$.
Points in $Q$ are a generalisation of the concept of position and can be referred to as generalised positions.
The dynamics of the system is then described by the sufficiently-regular curve $\gamma \colon [a,b] \rightarrow Q$, which can be found as the solution of well-posed second-order differential equations, known as the equations of motion.
Also in this case, one needs to specify the $2N$ initial conditions $\gamma(a)$ and $\dot\gamma (a)$, in order to have a unique solution.

So far, we have not said how to find the equations of motion of a given system.
In principle, one could simply postulate their form for the system under consideration and then verify the validity of this postulation experimentally.
In practice, it is very often more convenient to postulate the action functional associated to a system and to derive the equations of motion following the principle of least action.
We will review briefly this concepts and the derivation of the equations of motion in the next sections.
Specifically, we will begin with the Lagrangian mechanics in which one assumes the knowledge of a function, the Lagrangian, with the use of which the action functional is built.
The equations of motion ensuing from the application of the principle of least action are written in terms of the Lagrangian and take the name of Euler-Lagrange equations, which are obtained from the action functional by calculus of variation and are, in general, second-order differential equations.

From there, we will see how to convert these second-order equations into equivalent first-order equations, which will lead us to the basics of Hamiltonian mechanics.
Finally, we will analyse the structure and the tools that the Hamiltonian mechanics introduces, in order to study, for instance, the symmetries of a system.

\section{The principle of least action and the Lagrangian formulation} \label{sec:least-action-Lagrangian-CM}

The principle of least action (or, more precisely, of stationary action) is a method to derive the equations of motion starting from the action functional $S[\gamma;a,b]$ which maps every sufficiently-regular curve $\gamma : [a,b] \rightarrow Q$ to a real number.
In the Lagrangian formulation of classical mechanics, the action is written as
\begin{equation} \label{action-Lagrangian-classical-mechanics}
 S[\gamma ;a,b] = \int_a^b dt \, L[\gamma(t),\dot \gamma(t)] \,,
\end{equation}
in term of the Lagrangian $L$, which is a sufficiently smooth function from the tangent bundle of $Q$ to the real numbers, i.e. $L \colon TQ \rightarrow \real$.

Let us put this in simpler words by considering again the case of one particle in three dimensions, whose trajectory is described by the curve $\gamma$ in $Q = \real^3$.
At a given time $t$, the position of the particle is the point $q \eqdef \gamma (t) \in \real^3$, while its velocity is the tangent vector to the curve at that point, i.e. $\dot q \eqdef \dot \gamma (t)$.
Then, the Lagrangian assigns to the position $q$ and to the velocity $\dot q$ the real number $L[q,\dot q]$.
In the general situation, the evolution of the system is described by a curve $\gamma$ in the configuration space $Q$.
At a given time $t$, the system is in the generalised position $q = \gamma (t) \in Q$ and has the generalised velocity $\dot q = \dot \gamma (t) \in T_q Q$.
Then, the Lagrangian assign to every position $q \in Q$ and to every tangent vector $\dot q \in T_q Q$ a real number, which is exactly the meaning of $L : TQ \rightarrow \real$.\footnote{
In general, the Lagrangian $L$ must not be defined on all of $TQ$, but possibly only on a subbundle.
Anyway, we will not pay much attention to this subtlety in this thesis.
}
Note that, in principle, we could allow the Lagrangian to depend explicitly on time and write $L[\gamma(t),\dot \gamma(t);t]$ in the above integral, but this would be a complication without any benefit in the ensuing discussion, so that we will neglect this possibility.

In many physically-relevant situations, the Lagrangian of one particle takes the form
\begin{equation} \label{Lagrangian-CM-point-particle}
 L[q,\dot q] = \frac{1}{2} m \| \dot q \|^2 - V(q) \,, 
\end{equation}
where $m$ is the mass of the particle,
$\| \dot q \|^2 = (\dot q^1)^2 + (\dot q^2)^2 + (\dot q^3)^2$ in Cartesian coordinates and $\| \dot q \|^2 = g_{ij} \dot q^i \dot q^j$ in general coordinates, while $V(q)$ is the potential.
We are going to use this Lagrangian to provide explicit examples in the remainder of this chapter.

The principle of stationary action states the following.
Given the time interval $[a,b] \subset \real$ and given $q_1,q_2 \in Q$, let us consider all the sufficiently-regular curves $\gamma : [a,b] \rightarrow Q$, such that $\gamma (a) = q_1$ and $\gamma (b) = q_2$.
Then, $\bar \gamma$ is a solution to the equations of motion if, and only if, it is a critical point of $S[\gamma;a,b]$.
Intuitively, this means that, if $\gamma$ differs only by ``a little'' from $\bar \gamma$, then the value of the action at $\gamma$ does not vary from the value at $\bar \gamma$ at first order.
A bit more rigorously, let us consider some coordinates on $Q$ and let $c$ be any sufficiently-regular curve such that $c(a)=0=c(b)$ in these coordinates.
Let us consider $\gamma = \bar \gamma + \lambda c$ in these coordinates, being $\lambda \in \real$.
It is clear that $\gamma(a)=\bar \gamma(a) = q_1$ and $\gamma(b)=\bar \gamma(b) = q_2$.
Thus, $\bar \gamma$ is a solution to the equations of motion if, and only if, $S[\gamma;a,b] - S[\bar \gamma;a,b] = o(\lambda)$ for any $c$, which fact is usually referred to by saying that the variation of the action is zero.\footnote{
We write $f(\lambda) = o \bigl( g(\lambda) \bigr)$ in the limit $\lambda \rightarrow 0$ if $\lim_{\lambda \rightarrow 0} |f(\lambda)/g(\lambda)| = 0$.
Generalisations to other values for the limit are obvious.
}

Let us see what this means in terms of the action~(\ref{action-Lagrangian-classical-mechanics}).
To this end, let us insert $\gamma = \bar \gamma + \lambda c$ in~(\ref{action-Lagrangian-classical-mechanics}) and expand in $\lambda$, finding
\begin{equation}
\begin{aligned}
 S[\gamma;a,b] ={}& \int_a^b dt \, \left\{
 L[\bar \gamma(t) ,\dot{\bar \gamma}(t) ] +
 \left( \frac{\partial L}{\partial q^I} [\bar \gamma(t) ,\dot{\bar \gamma}(t)] \right) \lambda c^I (t) + \right. \\
 & \left.  +\left( \frac{\partial L}{\partial {\dot q}^I} [\bar \gamma(t) ,\dot{\bar \gamma}(t)] \right) \lambda {\dot c}^I (t)
 + o(\lambda)
 \right\} \,,
\end{aligned}
\end{equation}
where the sum over $I=1,\dots,N$ has to be understood.
The condition $S[\gamma;a,b] = S[\bar \gamma;a,b] + o(\lambda)$ is then satisfied if
\begin{equation} \label{variation-S-0-CM}
\begin{aligned}
 0 ={}& \int_a^b dt \, \left\{
 \left( \frac{\partial L}{\partial q^I} [\bar \gamma(t) ,\dot{\bar \gamma}(t)] \right) c^I (t)
 + \left( \frac{\partial L}{\partial {\dot q}^I} [\bar \gamma(t) ,\dot{\bar \gamma}(t)] \right) {\dot c}^I (t)
 \right\}= \\
 ={}& \int_a^b dt \, c^I (t) \left\{
 \left( \frac{\partial L}{\partial q^I} [\bar \gamma(t) ,\dot{\bar \gamma}(t)] \right) 
 - \frac{d}{dt} \left( \frac{\partial L}{\partial {\dot q}^I} [\bar \gamma(t) ,\dot{\bar \gamma}(t)] \right)
 \right\} \,,
\end{aligned}
\end{equation}
where we have integrated by parts on the last step.
Note that, since $c(a)=c(b)=0$, the boundary term coming from the integration by parts is actually zero.
Since the above integral needs to vanish for any $c$, we must conclude that the expression in graph brackets needs to be actually zero, i.e.,
\begin{equation} \label{Euler-Lagrange-equations-CM}
 \frac{d}{dt} \left( \frac{\partial L}{\partial {\dot q}^I} [\bar \gamma(t) ,\dot{\bar \gamma}(t)] \right) =
 \frac{\partial L}{\partial q^I} [\bar \gamma(t) ,\dot{\bar \gamma}(t)] \,.
\end{equation}
The above equations are called the Euler-Lagrange equations and provide the equations of motion of a classical mechanical system once the Lagrangian is known.
Note that the left-hand side of equations above depends in general on $\bar \gamma$, $\dot{\bar \gamma}$, and $\ddot{\bar \gamma}$, whereas the right-hand side on $\bar \gamma$ and $\dot{\bar \gamma}$.
Therefore, due to the left-hand side the Euler-Lagrange equations are in general a system of second-order ordinary differential equations and, as a consequence, they need to be complemented with initial conditions providing the values of $\bar \gamma (a)$ and $\dot{\bar \gamma} (a)$ at a given initial time $a$, usually taken to be zero.
Thus, one needs to specify $2N$ conditions --- $N$ for $\bar \gamma (a) \in Q$ and other $N$ for $\dot{\bar \gamma} (a) \in T_{\bar \gamma (a)} Q$ --- in order to find a unique solution.

These considerations hold true if the equations are in fact second order for every $\bar \gamma^I$.
Specifically, let us expand the left-hand side of~(\ref{Euler-Lagrange-equations-CM}), obtaining
\begin{equation} 
 \frac{d}{dt} \left( \frac{\partial L}{\partial {\dot q}^J} [\bar \gamma(t) ,\dot{\bar \gamma}(t)] \right) =
 \ddot{\bar \gamma}^I (t) \frac{\partial^2 L}{\partial {\dot q}^I \partial {\dot q}^J} [\bar \gamma(t) ,\dot{\bar \gamma}(t)]
 + \dot{\bar \gamma}^I (t) \frac{\partial^2 L}{\partial {q}^I \partial {\dot q}^J} [\bar \gamma(t) ,\dot{\bar \gamma}(t)] \,,
\end{equation}
which shows that the Euler-Lagrange equations are linear in the second-order terms $\ddot{\bar \gamma}^I (t)$.
Therefore, the rank of the Hessian $\partial^2 L/ \partial \dot q^I \partial \dot q^J$ is going to determine on how many of the second-order terms the equations actually depend.
We say that the Lagrangian $L(q,\dot q)$ is \emph{regular} if the Hessian $\partial^2 L/ \partial \dot q^I \partial \dot q^J$ is not singular, which ensure that the Euler-Lagrange equations are actually $N$ second-order ordinary differential equations.
In this thesis, when dealing with field theories, we will have to analyse some situations in which this condition is not satisfied: in particular, in the case of gauge theories.
We are going to neglect this possibility for now and come back to this topic when discussing gauge theories in the next chapter.

Let us conclude this section by applying the derived results to one specific example.
To this end, let us consider the Lagrangian~(\ref{Lagrangian-CM-point-particle}), describing the dynamics of a particle in three dimensions in the presence of a potential.
It is easy to verify that this Lagrangian is regular since $\partial^2 L/ \partial \dot q^i \partial \dot q^j = g_{ij}$ which is not singular.
In addition, in this simple example, the Euler-Lagrange equations in Cartesian coordinates are
\begin{equation} \label{Newton-law}
 m  {\ddot {\bar \gamma} }^i (t) =
 - \left( \frac{\partial V}{\partial q^i} \right) [\bar \gamma (t)]
 \qquad (i=1,2,3) \,,
\end{equation}
which are easily recognised as Newton's Second Law.

\section{The principle of least action more rigorously} \label{subsec:action-Lagrangian-formal}
In the previous section, we have stated the principle of stationary action and derived from that the Euler-Lagrange equations.
To do so, we first fixed coordinates on $Q$ in order to be able to write $\gamma = \bar \gamma + \lambda c$ for any $c$ satisfying $c(a)=c(b)=0$.
The principle was then stated as $\bar \gamma$ is a solution to the equations of motion if, and only if, $S[\gamma;a,b] - S[\bar \gamma;a,b] = o(\lambda)$ for any $c$.
Actually, this statement can be made more rigorous.
Specifically, let us consider the space $\mathcal{Q}(q_1,q_2,[a,b])$, consisting of all the sufficiently-regular curves $\gamma \colon [a,b] \rightarrow Q$ satisfying $\gamma(a) = q_1$ and $\gamma(b) = q_2$.
Thus, $\mathcal{Q}$ consists of all the paths in $Q$ with fixed endpoints and the action $S$ is then a map $S \colon \mathcal{Q} \rightarrow \real$.

It is possible to show that $\mathcal{Q}$ is a smooth infinite-dimensional manifold.
Hence, given $\gamma \in \mathcal{Q}$, one can consider the tangent space $T_\gamma \mathcal{Q}$.
In details, according to Proposition 8.1.2 of~\cite{Marsden-Ratiu}, this consists of all the $C^2$ maps $V \colon [a,b] \rightarrow TQ$ satisfying the following two properties.
First, they make the following diagram commutative
\begin{center}
 \begin{tikzpicture}[scale = 1.9]
 \node (A) at (0,0) {$[a,b]$};
 \node (B) at (2,0) {$TQ$};
 \node (C) at (2,-1) {$Q$};
 \draw[->] (A) -- (B) node [midway, above] {$V$};
 \draw[->] (A) -- (C) node [midway, below] {$\gamma$};
 \draw[->] (B) -- (C) node [midway, right] {$\pi_Q$};
\end{tikzpicture}
\end{center}
Here, $\pi_Q \colon TQ \rightarrow Q$ is the canonical projection, which takes the element $(q,\dot q) \in TQ$, being $\dot q \in T_q Q$, and maps it to $q$.
This first property implies that $V(t)=\big( \gamma(t), v(t) \big)$, where $v(t)$ belongs to the tangent space $T_{\gamma (t)} Q$.
Second, $v(a) = 0$ and $v(b) = 0$.\footnote{
Note that, despite both the vectors $v(a)$ and $v(b)$ are zero, they belong to different tangent spaces.
In particular, the former belongs to $T_{q_1} Q$, while the latter to $T_{q_2} Q$.
}
This second property follows from the fact that all the curves in $\mathcal{Q}$ have the same endpoints, $q_1$ and $q_2$.
One usually says that a vector $V \in T_{\gamma} \mathcal{Q}$ is the (infinitesimal) variation of $\gamma \in \mathcal{Q}$ and often writes $V = \delta \gamma$, which we are also going to do.

At this point, we are ready to reformulate the principle of stationary action formally.
To this end, let us denote with $\extderphase_\mathcal{Q}$ the exterior derivative on the manifold $\mathcal{Q}$ and let us consider the one-form $\extderphase_\mathcal{Q} S$, which we refer to as the variation of the action.
Note that the one-form $\extderphase_\mathcal{Q} S$ assigns at each $\gamma \in \mathcal{Q}$ and at each $\delta \gamma \in T_\gamma \mathcal{Q}$ a real number. 
Then, $\bar \gamma \in \mathcal{Q}$ is a solution to the equations of motion if, and only if, the one-form $\extderphase_\mathcal{Q} S = 0$ at $\bar \gamma$.
This means that, for every $\delta \bar \gamma \in T_{\bar \gamma} \mathcal{Q}$, we have $\extderphase_\mathcal{Q} S (\delta \bar \gamma) = 0$, which coincides with equation~(\ref{variation-S-0-CM}) if one uses coordinates and identifies $\delta \bar \gamma$ with $c$.

Before we conclude this subsection and move to the Hamiltonian formulation, let us stress one point in the notation introduced in this subsection.
In particular, we have used different symbols to denote the variation of the action $\extderphase_\mathcal{Q} S$ and the infinitesimal variation $\delta \gamma$ of $\gamma$.
The reason for that is that these are very different objects.
Specifically, on the one hand, $\extderphase_\mathcal{Q} S$ is the one-form obtained by applying the exterior derivative $\extderphase_\mathcal{Q}$ to $S$ and is a well-defined object by itself.
On the other hand, $\delta \gamma$ denotes simply a generic vector in $T_{\gamma} \mathcal{Q}$ and does not refer to a specific vector, nor it needs to, since $\extderphase_\mathcal{Q} S (\delta \bar \gamma)$ has to vanish for every $\delta \bar \gamma$.
In the literature, the variation of the action $\extderphase_\mathcal{Q} S$ is often denoted also as $\delta S$.

So far, we have been rather pedantic in keeping a different notation for points, e.g. $q \in Q$, and curves, e.g. $\gamma \colon [a,b] \rightarrow Q$.
In the following, where there is no risk of confusion, we are going to denote the curves simply by $q(t)$, often omitting the dependence on time.
Thus, for instance, we would write the equations~(\ref{Euler-Lagrange-equations-CM}) and~(\ref{Newton-law}) simply as 
\begin{equation}
 \frac{d}{dt} \left( \frac{\partial L}{\partial {\dot q}^I} \right) =
 \frac{\partial L}{\partial q^I}
\end{equation}
and
\begin{equation} \label{EL-equation-point-particle}
 m  {\ddot  q }^i =
 -  \frac{\partial V}{\partial q^i}   \qquad (i=1,2,3) \,,
\end{equation}
respectively.
Note that we are also omitting the point at which the partial derivatives of $L$ and $V$ are evaluated.

Before we introduce the Hamiltonian formulation, let us note that, so far, we have always limited the integration in the action principle to a finite interval $[a,b]$, but the derived equations of motion were not depending on $a$ nor on $b$ (the initial conditions, however, were specified at $t=a$).
Assuming that the variation of the action is well-defined for any finite time interval $[a,b] \subset \real$ and always leads to the same equations of motion, it is customary to write the action as an integral over the entire real line.
The principle of least action, in this case, works by first selecting any finite time interval, which is big enough for one's purposes, and then proceeding as in the previous sections.

\section{From the Lagrangian to the Hamiltonian}
In the Lagrangian picture, one needs to have a function $L \colon TQ \rightarrow \real$, from which one derives the Euler-Lagrange equations~(\ref{Euler-Lagrange-equations-CM}).
If one assumes that the Lagrangian is regular as we shall do in this subsection, these are a system of $N$ second-order differential equations.
The basic idea of the Hamiltonian picture is to convert this system of $N$ second-order  equations into a system of $2N$ first-order equations.

Let us first see how this work in the simple example provided by the Lagrangian~(\ref{Lagrangian-CM-point-particle}), whose equations of motion are given, in Cartesian coordinates, by~(\ref{EL-equation-point-particle}).
Let us define the quantities
\begin{equation} \label{linear-momentum-PP-CM}
 p_i \eqdef m \, \delta_{ij} \, \dot q^j \qquad (i=1,2,3) \,,
\end{equation}
which are the components of the well-known linear momentum.
Note that we are writing the index of the momentum $p$ downstairs, contrary to that of the velocity $\dot q$, since the momentum is a covector rather than a vector, technically speaking.
This will be manifest when discussing the general case.
In addition, in the expression above, we have used the Euclidean metric in Cartesian coordinates $\delta = \text{diag} (1,1,1)$ to lower indices and we will use its inverse to raise them.
The three second-order equations~(\ref{EL-equation-point-particle}) are equivalent to
\begin{equation} \label{Hamilton-equation-CM-PP-Cartesian}
 \left\{ 
 \begin{aligned}
  \dot q^i &= \frac{\delta^{ij} \, p_j}{m} \\
  \dot p_i &= - \frac{\partial V}{\partial q^i}
 \end{aligned}
 \right.
 \qquad (i = 1,2,3) \,,
\end{equation}
which is a system of six first-order equations in the variables $q$ and $p$.
These equations need to be complemented by the initial conditions specifying the values of $q^i(0)$ and $p^i(0)$.
The former were already needed for the Euler-Lagrange equations, while the latter can be obtained as $p_i(0) = m \, \delta_{ij} \, \dot q^j(0)$ easily.
The above equations are a specific example of the Hamilton equations.
Note that they are written in terms of the position $q$ and the momentum $p$, but not of the velocity $\dot q$.
In principle, we would like these equation to ensue from the principle of least action by rewriting the action $S$ in a suitable manner.
But we postpone this discussion to the next sections, where we will do it directly in the general case.

Let us consider the more general case of a system described by the configuration space $Q$ and a regular Lagrangian $L \colon TQ \rightarrow \real$.
The first step in the example above, was to define the linear momentum $p$ from the velocity $v = \dot q$.
In the general case, this is done by the \emph{fibre derivative} (or \emph{Lagrange transform}) $\mathbb{F} L \colon TQ \rightarrow T^*Q$.
Fixing a point $q \in Q$ the fibre derivative assigns to every generalised velocity $v \in T_q Q$ a generalised momentum $p \in T_q^*Q$, called the \emph{canonical momentum}.
Schematically, we can write 
\begin{equation*}
 \begin{array}{rccl}
  \mathbb{F} L \colon & TQ    & \longrightarrow & T^*Q \\
                      & (q,v) & \longmapsto     & \mathbb{F}L (q,v)=\big(q,\mathbb{F}_q L (v) \big)
 \end{array} \,,
\end{equation*}
where $v \in T_q Q$ and $\mathbb{F}_q L (v) \in T_q^* Q$.
In practice, if we fix a point $q \in Q$, we can interpret $\mathbb{F}_q L$ as a map that assigns to every vector $v \in T_q Q$ a covector $\mathbb{F}_q L (v)$. 
This latter, in turn, can be unequivocally defined by the way it maps any  vector $w \in T_q Q$ to a real number.
Specifically, this map is defined as
\begin{equation}
 \big[ \mathbb{F}_q L (v) \big] (w) \eqdef
 \frac{d}{d \lambda} L(q,v + \lambda w) \,,
\end{equation}
where $\lambda \in \real$.
Thus, the momentum $p$ associated to a velocity $v \in T_q Q$ is nothing else than $p = \mathbb{F}_q L (v) \in T_q^* Q$.

It is useful to work out the above definition when a chart $(q^I)$ is chosen, that is, when points of a subset $U \subseteq Q$ are described by the $N$ real numbers $(q^1, \dots q^N)$.
The chart $(q^I)$ induces a basis on the tangent spaces naturally, so that a vector $v = \dot q$ can be written as $\dot q = \dot q^I \partial/\partial q^I$ in terms of its components $(\dot q^I)$.
Analogously, a covector $p$ can be written as $p = p_I dq^I$ in terms of its components $(p^I)$.
Then, the definition is equivalent to
\begin{equation} \label{canonical-momentum-CM}
 p_I \eqdef \frac{\partial L}{\partial \dot{q}^I} (q, \dot q) \,.
\end{equation}
Note that the expression above becomes exactly~(\ref{linear-momentum-PP-CM}) when the Lagrangian is~(\ref{Lagrangian-CM-point-particle}) in Cartesian coordinates.

At this point, we would like to describe the theory by using the position $q$ and the canonical momentum $p$ instead of the velocity $\dot q$.
Now, it is clear that, if we wish to get rid completely of the velocity $\dot q$ replacing it with the canonical momentum $p$, the relation between $p$ and $\dot q$ needs to be invertible.
Let us work, for simplicity, using the chart $(q^I)$.
At each point $q \in Q$, equation~(\ref{canonical-momentum-CM}) allows us to express the components of the momentum $(p_I) \in \real^N$ as a function of the components of the velocity $(\dot q^I) \in \real^N$.
The inverse-function theorem ensures that we can invert the relation locally and express $(\dot q^I)$ as a function of $(p_I)$, so long as the Jacobian $\partial p_I / \partial \dot q^J$ is invertible for every $\dot q$ (and at each $q \in Q$), whose condition is easily seen to be equivalent to the requirement of $\partial^2 L / \partial \dot q^I \partial \dot q^J$ being invertible.
Thus, we see that we can use the momenta instead of the velocities, so long as the Lagrangian is regular.

Actually, in order to avoid mathematical issues, we need to make a slightly stronger requirement.
Specifically, we need to ask that the Legendre transform $\mathbb{F} L \colon TQ \rightarrow T^* Q$ is not only invertible, but also a diffeomorphism.
When this happens, we say that the Lagrangian is \emph{hyper-regular}, to which case we shall limit our analysis from now on, except in the case of gauge theories discussed in the next chapter.

Finally, we would like to replace the Euler-Lagrange equations~(\ref{Euler-Lagrange-equations-CM}) with equivalent equations written in terms of $q$ and $p$.
To this end, let us first introduce the \emph{energy function} $E \colon T Q \rightarrow \real$ as
\begin{equation} \label{energy-function-CM}
 E (q,\dot q) = p (q,\dot q) \cdot \dot q - L(q,\dot q) \,,
\end{equation}
where the velocity $\dot q \in T_q Q$, the momentum $p (q, \dot q) \eqdef \mathbb{F}_q L (\dot q) \in T_q^* Q$, and $p \cdot q$ denotes the contraction of a covector and a vector, i.e., $p_I q^I$ in coordinates.
Second, let us define the \emph{Hamiltonian} $H \colon T^*Q \rightarrow \real$, so that $H \eqdef E \circ (\mathbb{F} L)^{-1}$ as in the following diagram.
\begin{center}
 \begin{tikzpicture}[scale = 3]
 \node (A) at (0,0) {$T^* Q$};
 \node (B) at (1,0) {$TQ$};
 \node (C) at (2,0) {$\real$};
 \draw[->] (A) -- (B) node [midway, above] {$(\mathbb{F} L)^{-1}$};
 \draw[->] (A) to[bend right] node [midway, below] {$H$} (C);
 \draw[->] (B) -- (C) node [midway, above] {$E$};
\end{tikzpicture}
\end{center}
In other words, the Hamiltonian is obtained by taking the energy function~(\ref{energy-function-CM}) and replacing every appearance of the velocity $\dot q$ with the corresponding momentum $p$.
Thus, in the Hamiltonian formulation, the theory is described by using the generalised position $q$ and the canonical momenta $p$. 
As one can check by direct computation, the Euler-Lagrangian equation are equivalent to the \emph{Hamilton equations}
\begin{equation} \label{Hamilton-equation}
 \left\{ 
 \begin{aligned}
  \dot q^I ={}& \frac{\partial H}{\partial p_I} \\
  \dot p_I ={}& - \frac{\partial H}{\partial q^I}
 \end{aligned}
 \right. \,,
\end{equation}
where the functions $\big(p_I(t)\big)$ and $\big(q^I(t)\big)$ need to be considered as independent. 
The equations above are indeed $2N$ first-order differential equations, which need to be complemented with the $2N$ initial conditions specifying the values $q^I$ and $p_I$ at a given initial time.
To be more precise, the first of the equations above is equivalent to the requirement that the canonical momentum is related to the Lagrangian by~(\ref{canonical-momentum-CM}) and, after this fact has been taken into consideration, the second equation reduces to the Euler-Lagrange equations.

We are going to show in the next sections that these equations  can be rewritten in a more compact and efficient way taking advantage of the geometrical structure of the cotangent bundle $T^* Q$.
This formulation of the Hamilton equations is going to prove especially convenient when dealing with classical field theories and with symmetries.
But, before we do that, let us conclude this section with the example of one point particle described by the Lagrangian~(\ref{Lagrangian-CM-point-particle}), which we have already partially described at the beginning of this subsection.
Using the definition~(\ref{canonical-momentum-CM}), the components of the momentum are
\begin{equation}
 p_i = m \, g_{i j} \, \dot q^j \,.
\end{equation}
The above expression, clearly reduces to the already discussed~(\ref{linear-momentum-PP-CM}), when Cartesian coordinates are employed.
Then, the energy function~(\ref{energy-function-CM}) becomes
\begin{equation}
 E(q,\dot q) = \frac{1}{2} m \| \dot q \|^2 + V(q) \,,
\end{equation}
which is easily recognised as the mechanical energy of the system, consisting of the sum of the kinetic and potential energies.
From this expression, one easily finds the Hamiltonian
\begin{equation} \label{Hamiltonian-PP}
 H(q,p) = \frac{\| p \|^2}{2m} + V(q) \,,
\end{equation}
where $\| p \|^2 \eqdef g^{i j} \, p_i \, p_j$, being $g^{ij}$ the inverse of $g_{ij}$, i.e., $g^{ik} g_{kj} = \delta^i_j$.
Finally, one can easily compute the Hamilton equations
\begin{equation}
 \left\{
 \begin{aligned}
  \dot q^i ={}& \frac{g^{i j} \, p_j}{m} \\
  \dot p_i ={}& - \frac{\partial V}{\partial q^i}
 \end{aligned}
 \right. \,,
\end{equation}
which reduce to~(\ref{Hamilton-equation-CM-PP-Cartesian}) when Cartesian coordinates are employed.

\section{Phase space and symplectic form}

Let us now see how to cast the Hamilton equations in a geometric way.
This will lead us to the introduction of the phase space and of the symplectic form, both of which will play a fundamental role in the derivation of the results contained in this thesis.

To begin with, let us stress one crucial difference between the Lagrangian and Hamiltonian formulations.
Specifically, in the Euler-Lagrange equations~(\ref{Euler-Lagrange-equations-CM}), the position and the velocity are not independent from each other, since the former is the described by the curve $q (t)$ and the latter is the tangent vector $\dot q (t)$ to the same curve at each given time $t$.
As a consequence, the Euler-Lagrange equations are differential equation for the function $q(t)$.
The reason for this can be tracked down to the fact that, in the action~(\ref{action-Lagrangian-classical-mechanics}), the Lagrangian is evaluated at the specific velocity $\dot q (t)$.

On the contrary, as we have already mentioned, in the Hamilton equations~(\ref{Hamilton-equation}), the position $q(t)$ and $p(t)$ are to be considered as independent.
They become bound  to one another only by virtue of the Hamilton equations themselves and not before.
For this reason, rather than considering a curve $q \colon [a,b] \rightarrow Q$, we need to consider a curve $ z \colon [a,b] \rightarrow T^*Q$, i.e., $z(t) = \big(q(t),p(t) \big)$, so that $q (t)$ and $p (t)$ can be treated as independent.
In addition, it is useful to note that the curve $z$ takes values in the cotangent bundle $T^*Q$ and that this latter is by itself a smooth manifold of dimension $2N$, i.e., twice the dimension of the configuration space $Q$.
As a consequence, we can take advantage of the rich differential structure of the  manifold $Z \eqdef T^*Q$, when describing the Hamiltonian formulation.

The manifold $Z$, which we have introduced as the cotangent bundle of the configuration space, is called the \emph{phase space} and plays an important role in the characterisation of a physical system, as we shall see later.
For now, let us note that the choice of a chart on the configuration space $Q$ naturally induces coordinates on $Z = T^* Q$.
Specifically, if $(q^I)$ are coordinates on $Q$, then $(q^I,dq^I)$ are coordinates on $T^*Q$, which we call \emph{canonical coordinates}.
We will see that these are very useful when writing explicit equations.
However, for reasons which will be made clear in the ensuing discussion, it is better to formulate the theory so that, at the end of the day, it does not rely too much on canonical coordinates (or on any other coordinates).

At this point, we are ready to write the Hamilton equations~(\ref{Hamilton-equation}) in terms of geometric objects on the manifold $Z$.
To this end, let us note that, if we use canonical coordinates, the curve $z(t)$ takes the form $\big( q^I (t),p_I (t)\big)$ and, thus, we recognise the left-hand sides of the equations~(\ref{Hamilton-equation}) as the components $\big( \dot q^I (t), \dot p_I (t)\big)$ of the tangent vector to the curve $z(t)$.
This tangent vector is an element of $T_{z(t)} Z$.
Thus, let us write
\begin{equation}
 \dot z (t) = X_H \big(z (t) \big)\,,
\end{equation}
where $X_H$ is a vector field on $Z$, so that $X_H \big(z (t) \big)$ belongs to $T_{z(t)} Z$.
In order for the above equation to be equivalent to~(\ref{Hamilton-equation}), the components of the vector field needs to be
\begin{equation} \label{Hamiltonian-vf-canonical}
 X_H = \left( \frac{\partial H}{\partial p_I} ,
            - \frac{\partial H}{\partial q^I}\right) \,,
\end{equation}
which is called \emph{Hamiltonian vector field} and looks closely related to the components of the one form
\begin{equation} \label{Hamiltonian-one-form-canonical}
 \extderphase H = \frac{\partial H}{\partial q^I} \extderphase q^I +
             \frac{\partial H}{\partial p_I} \extderphase p^I \,.
\end{equation}
In the above expression, we have introduce the symbol $\extderphase$ to denote the exterior derivative on the phase space $Z$.

The link between the vector field $X_H$ and the one-form $\extderphase H$ can be achieved by the introduction of a two-form $\Omega$ on $Z$, which is called the \emph{symplectic form}.
By being a two-form, $\Omega$ defines, at each point $z \in Z$, a skew-symmetric linear map from $T_z Z \times T_z Z $ to the real numbers, i.e., $\Omega_z \colon T_z Z \times T_z Z \rightarrow \real$.
In canonical coordinates, we will define $\Omega = \extderphase p_I \wedge \extderphase q^I$, that is an exact two-form since $\Omega = \extderphase \Theta$, where $\Theta \eqdef  p_I \extderphase q^I$ is called the \emph{symplectic potential}.
Then, one can check directly that, with this definition, the symplectic form is \emph{weakly non-degenerate}, i.e., at each $z \in Z$, if $\Omega_z(X,Y) = 0$ for every $Y \in T_z Z$, then it must be that $X = 0$.

At this point, we can build a map $\Omega_z^\flat$ from the tangent space $T_z Z$ to the cotangent space $T_z^* Z$.
Specifically, at each $z \in Z$, we define
\begin{equation}
 \begin{array}{rccc}
  \Omega_z^\flat \colon & T_z Z & \longrightarrow &  T_z^* Z\\
                        & X     & \longmapsto & \Omega_z^\flat (X)
 \end{array} \,,
\end{equation}
where $\big[ \Omega_z^\flat (X) \big] (Y) \eqdef \Omega_z (X,Y) $ for any $Y \in T_z Z$.
We will write $\Omega^\flat$ to refer to the map from the tangent bundle to the cotangent bundle, which reduces to $\Omega_z$ at each $z \in Z$.
In a straightforward way, $\Omega^\flat$ can be used to map vector fields to one-forms.
Furthermore, the weak non-degeneracy of $\Omega$ is equivalent to the fact that $\Omega^\flat$ is injective.
But, at each $z \in Z$, the linear map $\Omega_z^\flat$ is injective if, and only if, it is surjective, since $T_z Z$ and $T^*_z Z$ are linear spaces of the same finite dimension.
Thus, we conclude that $\Omega_z^\flat$ is an isomorphism, whose inverse is $\Omega^\sharp_z \colon T_z^* Z \rightarrow T_z Z$.
Equivalently, we denote with $\Omega^\sharp$ the inverse of $\Omega^\flat$.

The fact that $\Omega^\flat$ is invertible and, thus, that $\Omega^\sharp$ exists is commonly referred to as the \emph{strong non-degeneracy} of the symplectic form $\Omega$.
As we have just seen, if the dimension of $Z$ is finite, the weak non-degeneracy is sufficient to ensure the strong one, so that there is no need to specify whether we are talking about a strong or a weak symplectic manifold.
However, this will no longer be the case if the dimension of $Z$ is infinite.
While a strongly-non-degenerate symplectic form is also weakly non-degenerate, the converse is not always true, so that it will be important to distinguish between weak and strong symplectic manifolds when discussing field theories in the next chapter.

Finally, we can make the connection between the Hamiltonian vector field $X_H$ given in~(\ref{Hamiltonian-vf-canonical}) and the one form $\extderphase H$ given in~(\ref{Hamiltonian-one-form-canonical}).
Specifically, one can check that $X_H = - \Omega^\sharp (\extderphase H)$.
Note that, by means of the map $\Omega^\flat$ the relation $X_H = - \Omega^\sharp (\extderphase H)$ can be rewritten as $\extderphase H = -\Omega^\flat (X_H) = - \insertion_{X_H} \Omega$, where $\insertion$ denotes the contraction (or insertion) operator of a vector field with a differential form on the phase space.
The Hamilton equations~(\ref{Hamilton-equation}) can be written as
\begin{equation} \label{Hamilton-equation-vf}
 \dot z (t) = X_H \big( z(t) \big)
  = -\Omega^\sharp_{z(t)} \Big( \extderphase H \big( z(t) \big) \Big) \,.
\end{equation}
These are, as expected, first-order equations in the curve $z(t)$ and need to be complemented with the initial condition
$z(a) = z_1 \in Z$.
Thus, we see that points in $Z$ are enough to specify the physical state of a system at a given time completely, in the sense that, if we know that at a given time $t_0$ the system is at the point $z_0 \in Z$, then we can find the entire path $z(t)$ at any time as the unique solution to the above equation satisfying the initial condition $z(t_0) = z_0$.
The points in the configuration space $Q$ do not share the same property, since one needs to specify, other than the position $q \in Q$, the momentum $p \in T^*_q Q$  in order to have a unique solution to the Hamilton equations (or the velocity $\dot q \in T_q Q$ for the Euler-Lagrange equations).\footnote{
Since $Z$ has twice the dimension of $Q$, but the same as $TQ$, it is no wonder that specifying only elements of $Q$ is not sufficient to describe the state of the system.
}
For this reason, the phase space $Z$ plays a central role in the characterisation of the possible states of a physical system.

To conclude, let us rewrite the principle of least action using the elements introduced in this subsection.
Proceeding as in subsection~\ref{subsec:action-Lagrangian-formal}, let us consider the space $\mathcal{Z} (z_1,z_2,[a,b])$ of all the curves $z \colon [a,b] \rightarrow Z$ with fixed endpoints, i.e., satisfying $z(a)=z_1 \in Z$ and $z(b)=z_2 \in Z$.
Also in this case, one can show that $\mathcal{Z}$ is an infinite-dimensional manifold.
The action needs to be a function $S \colon \mathcal{Z} \rightarrow \real$, whose stationary points coincide with the solutions to the Hamilton equations.
So, let us define it as
\begin{equation}
 S \big[ z(t) \big] \eqdef
 \int_a^b dt \Big[ \dot z(t) \cdot \Theta \big( z(t) \big) - H \big( z(t) \big) \Big] \,,
\end{equation}
where the one-form $\Theta$ is the symplectic potential and $\dot z(t) \cdot \Theta \big( z(t) \big)$ is the contraction of the vector $\dot z (t)$ and the covector $\Theta\big( z(t) \big)$.
In canonical coordinates, the above expression reads
\begin{equation} \label{action-Hamiltonian-canonical-CM}
 S \big[ z(t) \big] =
 \int_a^b dt \Big[ p_I(t) \dot q^I(t) - H \big( q(t), p(t) \big) \Big] \,,
\end{equation}
where $z(t) = \big( q^I(t) , p_I(t) \big)$.
With the same methods described in subsection~\ref{subsec:action-Lagrangian-formal}, one can show that $\bar z (t)$ solves the Hamilton equations if, and only if, the variation of the action $\extderphase_\mathcal{Z} S (\delta \bar z) = 0$ for every $\delta \bar z \in T_z \mathcal{Z}$, where the symbol $\extderphase_\mathcal{Z}$ is the exterior derivative on $\mathcal{Z}$.
From now on, we will denote the variation of the action simply as $\extderphase S$ and drop the subscripts $\mathcal{Z}$ (used in this subsection) or $\mathcal{Q}$ (used in subsection~\ref{subsec:action-Lagrangian-formal}). 

\section{Symplectomorphisms} \label{sec:symplectomorphisms}
In the previous section, we have introduced the phase space $Z$ as the cotangent bundle $T^*Q$ of some configuration space $Q$.
Although this description of the phase space is valid in many situations of interest, it does fail in other equally-important cases.
For instance, this is the case for the reduced phase space of gauge theories, discussed in section~\ref{subsec:reduced-ps}, and it also happens theories whose equations of motions contain time derivatives higher than second order.
Neglecting for now the details, it important to know that in both cases the phase space $Z$ is a manifold which \emph{cannot} be obtained as the cotangent bundle of some configuration space $Q$.
Luckily, the formalism can be adapted and applied to these situations as well.

Specifically, let us consider the phase space $Z$ as a finite-dimensional smooth manifold, whose points identify completely the state of the physical system.
The phase space is equipped with a closed, weakly non-degenerate two-form $\Omega$, which plays the role of the symplectic form and is needed to write the Hamilton equations.
Two things should be pointed out concerning the symplectic form.
First, in general, it is not required to be an exact form as it was in the last subsection, but only a closed one.
Second, due to a simple theorem of linear algebra, the non-degeneracy of $\Omega$ implies that the dimension of $Z$ is even.
In addition, one can show that at any point $z \in Z$ there are local coordinates $(q^I,p_I)$ where the symplectic form takes the canonical form $\Omega = \extderphase p_I \wedge \extderphase q^I$.
These coordinates are called canonical coordinates.
If $Z$ is infinite dimensional, as it will be in the case of field theories, this fact is no longer true, but we leave the details about this discussion to the next chapter.

The symplectic form is not enough to determine the time evolution of the system.
For this, indeed, we need also the Hamiltonian $H$ which is a function from $Z$ to the real numbers.
For physical reasons, this function needs to be bounded from below (positivity of energy).
To the Hamiltonian $H$, one can associate a Hamiltonian vector field $X_H$, which satisfies the equation
\begin{equation} \label{X_H-Hamiltonian}
 \extderphase H = - \insertion_{X_H} \Omega \,.
\end{equation}
The fact that $X_H$ exists is guaranteed by the fact that, since $Z$ is finite dimensional, the map $\Omega^\sharp$ exists so long as $\Omega$ is non-degenerate, so that one can find $X_H = - \Omega^\sharp (\extderphase H)$.
Then, the equations of motion are simply
\begin{equation}
 \dot z = X_H (z) \,, 
\end{equation}
from which we see that vector field $X_H$ tells us how the points of the phase space are displaced at first order under the action of the time evolution.
In other words, the vector field $X_H$ tells us the infinitesimal change of a point $z \in Z$ under the time evolution.

In general, one may wish to study how the system changes under the action of other (continuous) transformations and not only of the time evolution.
For instance, one may wish to study the action of translations, rotations or, more generally, of a Lie group.
To this end, we can follow a similar strategy to that employed for the study of time evolution and relate continuous transformations to vector fields in the phase space.
Thus, let us consider a vector field $X$ on the phase space and the differential equation
\begin{equation} \label{eom-symplectic-transf}
 z'(\lambda) = X \big( z(\lambda) )
\end{equation}
with initial condition $z(0) = z_0 \in Z$.
The solution to this equation is a path $z(\lambda)$ in phase space passing through $z_0$ at $\lambda=0$.
We can interpret this path as the transformation of the point $z_0$ under the action of a one-parameter family of continuous transformations, being $\lambda \in \real$ the parameter.
By varying the point $z_0 \in Z$ in the specification of the initial condition, we get a one-parameter family of transformations
\begin{equation*}
 \begin{array}{rccc}
  \varphi \colon & \real \times Z  & \longrightarrow &  Z\\
                 & (\lambda,z_0) & \longmapsto & \varphi_\lambda (z_0)
 \end{array} \,,
\end{equation*}
where $\varphi_\lambda (z_0)$ is obtained by taking the solution to $z' = X(z)$ with initial condition $z(0) = z_0$ and evaluating it at $\lambda$.\footnote{
Let us neglect problems arising if the vector field $X$ is not globally integrable.
In general, the following observations are valid for all the value of $\lambda$ for which $\varphi_\lambda$ is well-defined.
}
The function $\varphi$ is called the flow of the vector field $X$ and, in the case $X = X_H$, Hamiltonian flow.
Fixing the value of $\lambda$, the map $\varphi_\lambda \colon Z \rightarrow Z$ is in general bijection\footnote{
The injectivity follows from the fact that integrable curves do not intersect, while the surjectivity from the fact that, for every point in the target space, we can solve the differential equation for negative values of $\lambda$ and find the corresponding point in the domain.
} and a diffeomorphism if $X$ is smooth, to which case we limit our analysis.

After applying the transformation $\varphi_\lambda$, in general, the symplectic form does not stay invariant and is mapped to $(\varphi_{\lambda})_* \Omega$, where $(\varphi_{\lambda})_*$ is the push forward.
We say that $\varphi_{\lambda}$ is a \emph{symplectomorphism} if $(\varphi_{\lambda})_* \Omega = \Omega$ or, equivalently, if $\Omega = (\varphi_{\lambda})^* \Omega$ which is written in term of the pull back $(\varphi_{\lambda})^*$.
In other words, a symplectomorphism does not change the symplectic form.
Due to the fact that $\varphi$ is the flow of the smooth vector field $X$, we can find out whether or not $(\varphi_\lambda)_{\lambda \in \real}$ is a family of symplectomorphisms by checking whether or not its Lie derivative along $X$ is zero.
Therefore, the vector field $X$ induces a family of symplectomorphisms if, and only if, $\liephase_X \Omega = 0$, where $\liephase$ is the Lie derivative on $Z$.
In this case, we say that $X$ is a \emph{symplectic vector field}.
Using Cartan's magic formula, we find
\begin{equation}
\begin{aligned}
 0 &= \liephase_X \Omega = \extderphase (\insertion_X \Omega) + \insertion_X (\extderphase \Omega) = \\
 &= \extderphase (\insertion_X \Omega) \,,
\end{aligned}
\end{equation}
where the last term of the first line vanishes since $\Omega$ is closed.
Thus, we see that the vector field $X$ is symplectic if, and only if, the one-form $\insertion_X \Omega$ is closed.

If the one-form $\insertion_X \Omega$ is not only closed but also exact, we say that $X$ is a \emph{Hamiltonian vector field}.
Thus, for a Hamiltonian vector field $X$, one can find a function $F \colon Z \rightarrow \real$, such that $\extderphase F = -\insertion_X \Omega$, where the minus sign is just a convention.
This equation is exactly of the same form of~(\ref{X_H-Hamiltonian}), so that we see that the vector field $X_H$ associated to the Hamiltonian $H$ is indeed Hamiltonian in the sense now defined, as one could have guessed from the terminology.
So, we see that $X_H$ is only one special case of Hamiltonian vector fields, namely the one providing the (infinitesimal) time evolution.
From now on, we will write $X_F$ to denote the Hamiltonian vector field associated to $F$, i.e., the one satisfying
\begin{equation} \label{Hamiltonian-vf-condition}
 \extderphase F = -\insertion_{X_F} \Omega \,,
\end{equation}
and we will refer to the function $F$ as the \emph{canonical generator} of the transformation.
Note that, if the first de Rham cohomology group of $Z$ is trivial, then every symplectic vector field is Hamiltonian.
This is for instance the case if $Z$ is an open simply-connected subset of $\real^{2N}$, due to Poincar\'e's lemma.
But, in general, this will not be the case, especially in the case in which $Z$ is not finite dimensional, such as in classical field theories.

So far, we have started from a vector field $X_F$ and inferred its canonical generator $F$, at the condition that $X_F$ is Hamiltonian.
In principle, one could also attempt to take the inverse path.
Thus, let us consider a scalar function $F$ on $Z$.
In order to find a Hamiltonian vector field $X_F$ associated to $F$, we need to find a solution to equation~(\ref{Hamiltonian-vf-condition}), which can also be written as $\extderphase F = - \Omega^\flat (X_F)$.
Therefore, we see that, if $Z$ is finite dimensional, we can always solve the equation and find $X_F = - \Omega^\sharp (\extderphase F)$.
However, this situation will change in classical field theories since $Z$ will no longer be finite dimensional and many problems discussed in this thesis are linked to this possible obstruction.

\section{Poisson brackets}
We are now ready to introduce the Poisson brackets, which play an important role in Hamiltonian mechanics.
Let us consider two canonical generators $F$ and $G$ with their respective vector fields $X_F$ and $X_G$.
The Poisson bracket of $F$ and $G$ is defined as
\begin{equation} \label{Poisson-bracket-def}
 \{ F , G \} \eqdef - \insertion_{X_G} (\insertion_{X_F} \Omega) \,,
\end{equation}
which is a function from $Z$ to $\real$.

The Poisson brackets satisfy several interesting properties and we are going to mention some of them without providing proofs, which can be found e.g. in~\cite{Marsden-Ratiu}.
First, the Poisson brackets are skew-symmetric, i.e.,
\begin{equation} \label{Poisson-brackets-skew-symmetric}
 \{ F , G \} = - \{ G , F \} \,,
\end{equation}
which follows from the above definition and the fact that $\Omega$ is a two-form, i.e. a skew-symmetric covariant tensor of rank two. 
Second, they are bilinear, i.e.,
\begin{equation} \label{Poisson-brackets-bilinear}
 \{ c \, F , G \} = c \, \{  F , G \}
 \qquad \text{and} \qquad
 \{ F + G , K \} = \{ F , K \} + \{ G , K \} \,,
\end{equation}
where $c \in \real$ and the analogous statements for the second argument of the brackets follows from the statements above combined with the skew-symmetry. 
Third, they satisfy the Jacoby identity
\begin{equation} \label{Poisson-brackets-Jacobi}
 \big\{ F , \{G , K\} \big\} +
 \big\{ G , \{K , F\} \big\} +
 \big\{ K , \{F , G\} \big\} = 0 \,.
\end{equation}
Fourth, as we have said, $\{ F, G \}$ is a function from $Z$ to $\real$, so that it is certainly the canonical generator of some vector field $X_{\{ F, G \}}$ if $Z$ is finite dimensional.
This statement holds true even if $Z$ is not finite dimensional and one can show that
\begin{equation} \label{Poisson-brackets-closure}
 X_{\{ F, G \}} = -[X_F,X_G] \,.
\end{equation}
where $[X_F,X_G]$ is the commutator (or Lie-Jacobi bracket) of the two vector fields $X_F$ and $X_G$.\footnote{\label{footnote:commutator-vf}
The formal definition of the \emph{commutator} (or \emph{Lie-Jacobi bracket}) of two vector fields can be found in the already-provided literature.
Let us simply remind that, in coordinates, its components are simply given by $[X,Y]^i = X^m \partial_m Y^i - Y^m \partial_m X^i$ and that it satisfies three important properties.
First, it is skew-symmetric, i.e. $[X,Y] = - [Y,X]$ for any two vector fields $X$ and $Y$.
Second, it is bilinear, i.e. $[\lambda X, Y] = \lambda [X,Y]$ and $[X+Y,Z] = [X,Z] + [Y,Z]$ for all scalars $\lambda$ and vector fields $X$, $Y$ and $Z$ (the linearity in the second argument follows from the combination of these first two properties).
Third, it satisfies the Jacobi identity
$\big[X,[Y,Z]\big] + \big[Y,[Z,X]\big]+ \big[Z,[X,Y]\big] =0$ for all vector fields $X$, $Y$,and $Z$.
Thus, the vector fields on a manifold with the Lie-Jacobi bracket form a Lie algebra (see footnote
~\ref{footnote:Lie-algebra} on page~\pageref{footnote:Lie-algebra} for the definition). 
}
The properties so-far mentioned imply that the canonical generators --- which are all functions in $C^\infty (Z)$ if $Z$ is finite dimensional or, more generally, strongly symplectic --- together with the Poisson brackets form a Lie algebra and that Hamiltonian vector fields form a Lie subalgebra of the vectors fields on $Z$.\footnote{\label{footnote:Lie-algebra}
A \emph{Lie algebra} can be abstractly defined as a linear space $V$ with a bilinear map $[\; , \;] \colon V \times V \rightarrow V$, called the Lie bracket, which is skew-symmetric, bilinear, and satisfies the Jacobi identity.
These are precisely the properties discussed in footnote~\ref{footnote:commutator-vf} on page~\pageref{footnote:commutator-vf} for the commutator of vector fields and in~(\ref{Poisson-brackets-skew-symmetric}), (\ref{Poisson-brackets-bilinear}), and~(\ref{Poisson-brackets-Jacobi}) for the Poisson brackets of canonical generators.
Moreover, we say that a Lie algebra is \emph{abelian} if $[X,Y] = 0$ for all $X,Y \in V$.
Furthermore, let us assume that there is a subspace $W \subset V$ which is closed under Lie brackets, i.e., such that $[X,Y] \in W$ for all $X,Y \in W$.
Then, $W$ equipped with the restriction of the brackets on itself $[\;,\;]_{|W}$ is a Lie algebra and we call it a \emph{Lie subalgebra} of $(V,[\;,\;])$.
Note that, in order to see if a subspace $W$ of $V$ is a subalgebra, we merely need to verify that it is closed under the Lie brackets since all the other properties follow immediately from the properties of $[\;,\;]$.
Finally, a linear-space homomorphism (isomorphism) $V_1 \rightarrow V_2$ is said to be a Lie-algebra homomorphism (isomorphism) if it preserves the Lie brackets.
}
In addition to these properties, the Poisson brackets satisfy the Leibniz rule, that is
\begin{equation} \label{Poisson-brackets-Leibniz}
 \{ F G , K \} = F \, \{ G , K \} + \{ F , K \} \, G \,.
\end{equation}

The actual computation of Poisson brackets is often carried more efficiently if canonical coordinates $(q^I,p_I)$ are employed, in which the Poisson brackets reduce to the well-known expression 
\begin{equation} \label{Poisson-brackets-canonical}
 \{ F , G \} = \frac{\partial F}{\partial q^I} \frac{\partial G}{\partial p_I} - \frac{\partial F}{\partial p_I} \frac{\partial G}{\partial q^I} \,,
\end{equation}
from which one infers the canonical brackets
\begin{equation}
 \{ q^I , q^J \} = 0 \,, \qquad
 \{ p_I , p_J \} = 0 \,, \qquad
 \{ q^I , p_J \} = \delta^I_J \,.
\end{equation}
These works very well in combination with the bilinearity and the Leibniz rule when computing the Poisson brackets of analytic functions of $(q^I,p_I)$.
Note that, if we set $G = H$ and we take either $F=q^I$ or $F = p_I$, the expression~(\ref{Poisson-brackets-canonical}), once evaluated at the point $z(t)$, reduces to the Hamilton equations, so that we can write the equations of motion using the Poisson brackets.

Indeed, the strength of the Poisson brackets is that they can be used to compute how a canonical generator $F$ changes under the time evolution.
To see this, let us first note that, using the definition of Hamiltonian vector field and its relation with its canonical generator, we can write the Poisson brackets~(\ref{Poisson-bracket-def}) also as
\begin{equation} \label{Poisson-brackets-alternative}
 \{ F , G \} = \extderphase F \cdot X_G = - \extderphase G \cdot X_F \,,
\end{equation}
where the dot represents the contraction of a covector and a vector.
Let us now consider a canonical generator $F \colon Z \rightarrow \real$ and a solution to the equations of motion $z \colon [a,b] \rightarrow Z$.
We wish to compute the change of the function $F$ along the path $z(t)$, that is the quantity $\dot F \big( z(t) \big)$.
Fixing coordinates $(z^I)_{I=1,\dots,2N}$ on $Z$ and using the chain rule, we find
\begin{equation}
 \dot F \big( z(t) \big) =
 \frac{\partial F}{\partial z^I} \big( z(t) \big) \, \dot z^I (t) = \extderphase F \big( z(t) \big) \cdot X_H \big( z(t) \big) \,,
\end{equation}
where, in the last step, we have used the Hamilton equations~(\ref{Hamilton-equation-vf}) and obtained an expression which is coordinate-independent.
Using~(\ref{Poisson-brackets-alternative}), we finally arrive at the desired expression
\begin{equation} \label{time-evo-canonical-generator}
 \dot F \big( z(t) \big) =
 \{ F , H \} \big( z(t) \big) \,,
\end{equation}
which we will often write as $\dot F = \{ F , H \} $ omitting the $\big( z(t) \big)$.

One simple application of the expression above is the conservation of energy.
Indeed, given a solution to the equations of motion $z(t)$, its energy is defined as the value taken by the Hamiltonian on that solution, i.e. $E_z(t) \eqdef H \big( z(t) \big)$.
Therefore,
\begin{equation}
 \dot E_z (t) =
 \{ H , H \} \big( z(t) \big) = 0 \,,
\end{equation}
where the last equality trivially follows from the skew-symmetry of the Poisson brackets.
Since the energy is constant on each solution of the equations of motion, we will simply denote it by $E_z$.

The above considerations about the time evolution of $F$ can be applied also to the case in which we want to study how $F$ changes under transformations other than time evolution.
Thus, if $\varphi^G_\lambda$ is a one-parameter family of symplectomorphisms induced by the vector field $X_G$ associated to the canonical generator $G$, for every $z \in Z$, we can define the path $z(\lambda) = \varphi^G_\lambda (z)$.
The change of $F$ along this path, i.e. under the transformation induced by $G$, is given by
\begin{equation} \label{transf-canonical-generator}
  \delta_{G} F \big( z(\lambda) \big) \eqdef
  F' \big( z(\lambda) \big) =
 \{ F , G \} \big(  z( \lambda ) \big) \,,
\end{equation}
which we will often write simply as $\delta_G F = \{ F , G \}$.
The above expression can be easily derived in the same way as~(\ref{time-evo-canonical-generator}) using the results of section~\ref{sec:symplectomorphisms}.

Before we conclude this section and turn to the study of symmetries, let us make two final observations.
First, Poisson brackets are preserved by the flow of symplectic vector fields.
This means that if $X$ is a symplectic vector field and $(\varphi_\lambda)$ the family of induced symplectomorphisms, then, for all canonical generators $F$ and $G$, we have
\begin{equation}
 (\varphi_\lambda)^* \{ F , G \} =
 \{ (\varphi_\lambda)^* F , (\varphi_\lambda)^* G \} \,.
\end{equation}
In terms of canonical coordinates $(q^I,p_I)$, this statement means that the canonical brackets~(\ref{Poisson-brackets-canonical}) have the same form before and after the symplectomorphism is applied.

Second, although we have defined Poisson brackets $\{\;,\;\}$ starting from a symplectic form $\Omega$, Poisson brackets can be defined and used also in situations in which a symplectic form does not exist.
Specifically, a manifold $Z$ is called a \emph{Poisson manifold} if, on $C^{\infty} (Z)$, there is the operation
\begin{equation*}
 \{ \; , \; \} \colon C^{\infty} (Z) \times C^{\infty} (Z) \longrightarrow C^{\infty} (Z) \,,
\end{equation*}
called Poisson brackets, which makes $\big( C^{\infty}(Z), \{ \; , \;\} \big)$ a Lie algebra and satisfies the Leibniz rule.
In other works, the Poisson brackets need to satisfy the properties~(\ref{Poisson-brackets-skew-symmetric}), (\ref{Poisson-brackets-bilinear}), (\ref{Poisson-brackets-Jacobi}), and~(\ref{Poisson-brackets-Leibniz}).
As we have seen, every strong symplectic manifold is a Poisson manifold, but the converse is not true.
We will not provide more information about this topic, which can be found e.g. in~\cite{Marsden-Ratiu}, since, in this thesis, we will study situations in which we have a (weak) symplectic manifold.
In this case, Poisson brackets are defined only for those functions, which we have called canonical generators, for which a Hamiltonian vector field exists.

\section{Symmetries}

Let us conclude this section about Hamiltonian methods in classical mechanics with one of the central aspects of this thesis, i.e., the study of symmetries.
To make things simpler, we are going to present an example in the first subsection and leave the general discussion to the next one.
Specifically, we are going to discuss the very simple case of translations of a point particle in three dimensions, introducing step by step the relevant quantities and definitions.

\subsection{A simple example}

Let us study the case of translations of a point particle in three dimensions and let us begin by reminding that the Hamiltonian is given by~(\ref{Hamiltonian-PP}), from which the action easily ensues using~(\ref{action-Hamiltonian-canonical-CM}), and let us work in Cartesian coordinates, where the equations of motion are~(\ref{Hamilton-equation-CM-PP-Cartesian}).
The phase space $Z$ is the cotangent bundle of the configuration space $Q = \real^3$, so that we can use the canonical coordinates $(q^i,p_i)$ for $i=1,2,3$.

The first thing one needs in order to discuss a possible symmetry is that of a transformation of phase space, which derives from a group action.
Leaving the precise definition to the next subsection, let us directly consider the case of translations.
Each translation can be parametrised using three real parameters $(a^1,a^2,a^3) \eqdef a \in \real^3$ representing the displacement along the three Cartesian axes, so that its action $T_a$ on the phase space, if Cartesian coordinates on $Q$ are employed, is simply the  map
\begin{equation*}
 \begin{array}{rccc}
  T_a \colon & Z         & \longrightarrow &  Z\\
             & (q^i,p_i) & \longmapsto & (q^i + a^i,p_i)
 \end{array} \,,
\end{equation*}
which is a diffeomorphism.
Three important things should be noted.
First, the trivial translation $T_0$ is the identity map on $Z$.
Second, the combination of two translations is again a translation and in particular $T_a \circ T_b = T_{a + b}$.
From this, it follows that any translation is, as already pointed out, bijective since $T_a \circ T_{-a} = T_0$ and that two translations commutes, i.e., $T_a \circ T_b = T_b \circ T_a$, although the latter property will not be shared by the majority of the transformations that we will analyse.
Third, for each value of the parameter $a\in \real$, we have a different transformation of the phase space, i.e., $T_a \ne T_b$ if $a \ne b$.
These three facts imply that the map $T \colon \real^3 \times Z \rightarrow Z$ is a faithful (left) action of the abelian group $(\real^3,+)$ on the phase space, as we shall see in the next subsection.

Now, the weakest possible definition of a symmetry is that of a transformation that maps solutions of the equations of motion to solutions of the equations of motion.
In this case, we will say that the transformation is a \emph{symmetry of the equations of motion}.
So, let us take a curve $z(t) = \big( q(t), p(t) \big)$ which solves the equations of motion~(\ref{Hamilton-equation-CM-PP-Cartesian}).
Under the action of a generic translation $T_a$, this solution is mapped to the curve
\begin{equation}
 \tilde z(t) \eqdef T_a \big(z (t) \big) = \big( \tilde q(t) , \tilde p(t) \big) \,,
\end{equation}
where $\tilde q(t) = q(t) +a$ and $\tilde p(t) = p(t)$.
To see whether or not the curve $\tilde z(t)$ is still a solution of~(\ref{Hamilton-equation-CM-PP-Cartesian}), let us compute
\[
 \dot{\tilde q} (t) = \dot q (t) = \frac{p (t)}{m} = \frac{\tilde p (t)}{m} \,,
\]
where, other than omitting the index, we have used the first one of the equations of motion~(\ref{Hamilton-equation-CM-PP-Cartesian}) and the fact that $\dot a = 0$.
This let us conclude that $z'(t)$ satisfies the first one of the Hamilton equations.
In addition,
\[
 \dot{\tilde p}(t) = \dot p(t) = -\frac{\partial V}{\partial q} \big( q(t) \big) \,.
\]
Thus, in order for $\tilde z(t)$ to solve the second one of the equations of motion~(\ref{Hamilton-equation-CM-PP-Cartesian}), it must be that the right hand-side of the above equation is such that
\[
 \frac{\partial V}{\partial q} \big( q(t) \big) = \frac{\partial V}{\partial q} \big( \tilde q(t) \big) =
 \frac{\partial V}{\partial q} \big( q(t) +a \big)
\]
In order for this to hold for every $a\in \real^3$, it must be that $\partial V / \partial q$ does not depend on the position $q \in Q$, which implies that the potential $V$ is an affine function in Cartesian coordinates, i.e., $V(q) = V_0 - F_i q^i$.
The system which we have just described is that of a point particle in a uniform force.
As we have seen it possesses translations as symmetries of the equations of motion.
Thus, if we know one solution, we can find a continuous of distinguished solutions simply by translating the original one.

Let us now use this very simple example and move one step forward.
As we have seen throughout this section, solutions to the equations of motion are those paths in phase space that are stationary points of the action.
Thus, let $z(t)$ be a path (not necessarily a solution) and let us see how the value of the action changes under the translation $T_a$.
It is easy to verify that, in the case of a uniform force, we have
\begin{equation} \label{action-translation-CM}
 S\big[ T_a\big( z (t) \big) \big] = S[z (t)] + a^i F_i \, \Delta t \,, 
\end{equation}
where $\Delta t$ is the size of the time interval under consideration.
The second summand of the right-hand side depends on the time interval $\Delta t$ (which is fixed before varying the action), on the force $F$ (which does not depend on the position), and on the parameter $a$ of the translation.
In particular, this second summand does \emph{not} depend on the path $z(t)$.
As a consequence, if $z_0 (t)$ is a stationary point of the action, so is ${\tilde z}_0 (t) \eqdef T_a \big( z_0 (t) \big)$.
Indeed, since $T_a$ is a diffeomorphism, we know that every path close to ${\tilde z}_0 (t)$ can be obtained from a path close to $z_0 (t)$ by applying $T_a$.
But, since the second summand on the right-hand side is the same for all these paths, we conclude that the variation of the left-hand side is zero if, and only if, the variation of the first summand of the right-hand side is zero, which proves our statement.

The above considerations apply --- and are actually easier to show --- in the special case in which the action functional $S$ is invariant under the action of the translations, that is if the second summand of the right-hand side of~(\ref{action-translation-CM}) is actually zero.
In order for this to happen for all $a \in \real^3$, we need to require $F = 0$ or, equivalently, that the potential $V (q) = V_0$ is uniform.
As a consequence, we see that the second of the equations of motion~(\ref{Hamilton-equation-CM-PP-Cartesian}) reduces to $\dot p (t) = 0$, i.e., that the solution of the equations of motion have constant momentum.
This is the well-known fact that the linear momentum is conserved if the action is invariant under spatial translations and is a special example of the Noether's theorem.
In this case, we will say that the transformation is a \emph{symmetry of the action}.
As we have seen, this definition is stronger than the previous one and has stronger consequences, for it leads in general to conserved quantities by means of Noether's theorem.

At this point, let us see how to deal with this simple example using the Hamiltonian tools that we have developed so far.
Thus, let us introduce a real parameter $\lambda \in \real$ and let us consider the one-parameter family of transformations $T_{\lambda a}$, which we can interpret as the flow of a vector field $X_a$, as discussed in section~\ref{sec:symplectomorphisms}.
To find the desired vector field $X_a = (\delta_a z)$, let us write in canonical coordinates
\begin{equation}
 \begin{aligned}
  T_{\lambda a} (z) &= z + \lambda \delta_a z + o (\lambda) \\
  &= z + \lambda (a,0) \,,
 \end{aligned}
\end{equation}
so that we easily read that the desired vector field is $X_a = (a,0)$.
From now on, we will denote the components of a vector field using the notation $X = (\delta_X z)$ or $X= (\delta_X q, \delta_X p)$ in canonical coordinates.
If the vector field $X$ is dependent on some parameter, such as $X_a$ for the translations, we will often write $\delta_a$ instead of $\delta_{X_a}$.
Thus, instead of $X_a = (a,0)$, we will often write $\delta_a q = a$ and $\delta_a p = 0$.

As one can easily verify, $X_a$ is a symplectic vector field and, since the phase space is finite dimensional, is also Hamiltonian.
In particular, it is generated by the function
\begin{equation}
 P_a = a^i p_i \,,
\end{equation}
which satisfies $\extderphase P_a = - \insertion_{X_a} \Omega$.
It is actually useful to write a canonical generator for the three translations along each Cartesian axis.
To this end, let us define $(\theta_i)_{i=1,2,3}$ as the standard basis of $\real^3$, i.e.
\[
 \theta_1 = (1,0,0) \,, \qquad
 \theta_2 = (0,1,0) \,, \qquad \text{and} \qquad
 \theta_3 = (0,0,1) \,, 
\]
so that $a = a^i \theta_i$, and let us define the three vector fields $X_i \eqdef X_{\theta_i}$, with $i =1,2,3$.
It is easy to see that the vector field $X_i$ induces translations along the $i$-th axis in Cartesian coordinate and that the respective canonical generator is $P_i \eqdef P_{\theta_i} = p_i$.
Thus, we see that $i$-th component of the linear momentum $p_i$ is the generator of the translation along the $i$-th axis.
We will say that the transformation is a \emph{canonical symmetry} if it ensues from a canonical generator which Poisson-commutes with the Hamiltonian.\footnote{
Two canonical generators $F$ and $G$ are said to  \emph{Poisson-commute} if $\{ F,G \} = 0$.
}
Since the analyses of this thesis rely on the Hamiltonian formulation of classical field theories, this definition will be the most relevant one to us.

In Hamiltonian formulation, the conservation of the linear momentum can be shown quite easily by computing the Poisson brackets
\begin{equation}
 \{ P_i , H \} = \left\{ p_i , \frac{\| p \|^2}{2m} + V_0 \right\} = 0 \,,
\end{equation}
from which follows immediately $\dot P_i = 0$.
Note that the equation $\{ P_i , H \} = 0$ can be interpreted in two ways.
The first is that the $P_i$ is invariant under time evolution (i.e. the action of H) and the second, by inverting the order in the Poisson brackets using the skew-symmetry, is that $H$ is invariant under translations (i.e. the action of $P_i$).
It is worth noting that a canonical symmetry is a symmetry of the equations of motion and leads to conserved quantities.

Finally, let us conclude this subsection by computing the Poisson brackets of the three canonical generators $(P_i)_{i=1,2,3}$, obtaining
\[
 \{ P_i , P_j \} = 0 \qquad (i,j = 1,2,3) \,.
\]
As we shall see in the next subsection, this is a direct consequence of the fact that translations commutes among each other.

\subsection{General discussion} \label{subsec:symmetries-general}

Let us now move from the special case of translations of a point particle to the general case.
To begin with, we need a transformation of the phase space, which ensues from a faithful (left) action of a group.
So, let $Z$ be the phase space and $G$ a group.
A left action of the group $G$ on the phase space $Z$ is a map
\[
 \begin{array}{rccc}
  \Phi \colon & G \times Z & \longrightarrow &Z \\
                 & (g,z)      & \longmapsto     &\Phi_g(z)
 \end{array}
 \,,
\]
which is compatible with the group structure.
With ``compatible with the group structure'', we mean that the action $\Phi$ satisfies two properties.
First, if $e\in G$ is the group identity, then $\Phi_e = \text{id}_Z$, i.e., $e \cdot z = z$ for every $z \in Z$, where we have used the standard notation for a left action $g \cdot z \eqdef \Phi_g (z)$.
Second, for every $g,h \in G$, we have $\Phi_g \circ \Phi_h = \Phi_{gh}$ or, equivalently, $g \cdot ( h \cdot z ) = (gh) \cdot z$ for all $g,h \in G$ and $z\in Z$.
Note that these two properties imply that the map $\Phi_g \colon Z \rightarrow Z$ is a bijection for every $g\in G$, the inverse map being given by $(\Phi_g)^{-1} = \Phi_{g^{-1}}$.
These two properties constitute in general the definition of a left action of a group on a set.
In addition to these, we wish the action to be \emph{faithful} (or \emph{effective}), which means that, if $g$ and $h$ are two distinct elements of the group $G$, then $\Phi_g \ne \Phi_h$.\footnote{
Analogously, one can define a right action $\Phi^R$  by imposing the same properties of a left action except the behaviour under composition that becomes $\Phi^R_g \circ \Phi^R_h = \Phi^R_{hg}$.
Using the standard notation for a right action $z \cdot g \eqdef \Phi^R_g (z)$, this becomes simply $(z \cdot h) \cdot g = z \cdot (hg)$.
Note that one can convert a right action into a left action and vice versa by means of the group inversion.
Specifically, if $\Phi^R$ is a right action, then $\Phi^L$ defined by $\Phi^L_g \eqdef \Phi^R_{g^{-1}}$ is a left action.
Analogously, if $\Phi^L$ is a left action, $\Phi^R_g \eqdef \Phi^L_{g^{-1}}$ defines a right action.
}

This definition clearly applies to the case of the previous subsection, where $G = (\real^3, +)$ and $\Phi_a = T_a$.
In that case, in addition, the transformations $T_a$ were depending continuously on the parameter $a \in \real^3$, so that we could speak of continuous transformation and continuous symmetry.
In general, we say that $G$ is a \emph{Lie (or continuous) group} if, other than a group, $G$ is also a smooth manifold, whose topology is compatible with the group structure.
This means that both the group operation $(g,h) \mapsto g \cdot h$ and the inversion $g \mapsto g^{-1}$ are smooth maps.
Of course, it is possible use the same definition as above for the action $\Phi$ of the group $G$ on the set $Z$, i.e., $\Phi_e = \text{id}_Z$ and $\Phi_g \circ \Phi_h = \Phi_{gh}$.
But, since both $G$ and $Z$ are now manifolds, it is possible to provide a stronger definition.
Specifically, we say that $\Phi$ is a \emph{Lie-group action} on $Z$ if it is an action and satisfy two further properties.
First, the map $\Phi_g \colon Z \rightarrow Z$ is a diffeomorphism for all $g \in G$.
Second, the map $G \rightarrow \text{Diff} (Z)$ defined by $g \mapsto \Phi_g$ is smooth.\footnote{
Note that, from the definition of action, it follows that $g \mapsto \Phi_g$ is a group homomorphism.
Since it is also smooth, it is a Lie-group homomorphism.
}
The definition of faithful action is unchanged.
From now on, unless stated otherwise, when we say that there is an action of a Lie group on the phase space, we will assume that it is actually a Lie-group action.

We say that $\Phi$ is a \emph{symmetry of the equations of motion} if it maps solutions of the equations of motion to solutions of the equations of motion or, in other words, for every solution $z(t)$ of the equations of motion and for every $g \in G$, the curve $\Phi_g\big(z(t)\big)$ is again a solution to the equations of motion.
In addition, we say that $\Phi$ is a \emph{symmetry of the action} if it leaves the action functional $S$ invariant, i.e., for all $g \in G$, $S \big[ \Phi_g \big( z(t) \big) \big] = S [z(t)]$.
In this case, due to Noether's theorem, there are some conserved quantities.
We will not be more specific about this topic now, since we will be interested in studying continuous symmetries using the Hamiltonian methods.
Thus, we postpone the discussion of conserved quantities directly to the case of canonical symmetries.

In order to deal with continuous symmetries with the Hamiltonian methods, let us focus on the case in which $G$ is a Lie group.
As we have done in the previous subsection, we would like relate the action of $G$ on phase space with (Hamiltonian) vector fields on phase space.
In order to do that, rather than the group $G$, we would need to consider its Lie algebra $\mathfrak{g}$, which we define formally below.
The reason for this is that, as we have already mentioned, a vector field in phase space is related to the idea of an infinitesimal transformation, whereas its flow is related to the full transformation.
Now, since the action of the group is the full transformation, we will need something like an ``infinitesimal group element'' to relate to the infinitesimal transformation.
And this will be exactly the role played by the Lie algebra $\mathfrak{g}$ associated to $G$.

Precisely, the Lie algebra $\mathfrak{g}$ of the Lie group $G$ can be identified, as a set, with the tangent space to the group identity $T_e G$, which intuitively explains why its elements should be related to infinitesimal transformations.
In order to define the Lie brackets on $\mathfrak{g}$, let us first define, for every $g \in G$, the left translation map
\begin{equation}
 \begin{array}{rcccl}
  L_g \colon &G & \longrightarrow &G      & \\
             &h & \longmapsto     &L_g(h) &\eqdef gh
 \end{array} \,.
\end{equation}
One can easily check that $L$ is a left action of the Lie group $G$ on itself.
The right translations maps $R_g$ could be defined in an analogous way, but we will not make use of it.
A vector field $X$ on $G$ is said to be \emph{left-invariant} if $(L_g)_* X = X$ for all $g \in G$, where $(L_g)_*$ is the push forward of $L_g$.
Let us denote by $\mathfrak{X}_L$ the set of all the left-invariant vector fields on $G$.
This set, equipped with the Lie-Jacobi bracket, is a Lie algebra.
Indeed, if $X, Y \in \mathfrak{X}_L$,
\[
 (L_g)_* [X,Y] = \big[ (L_g)_* X , (L_g)_* Y \big] = [X,Y] \,,
\]
which shows that also $[X,Y] \in \mathfrak{X}_L$.
Thus, $\mathfrak{X}_L$ is a subalgebra of the vector fields on $G$ (see footnote~\ref{footnote:commutator-vf} on page~\pageref{footnote:commutator-vf} and footnote~\ref{footnote:Lie-algebra} on page~\pageref{footnote:Lie-algebra}).
At this point, we merely need to note that $T_e G$ and $\mathfrak{X}_L$ are isomorphic and use this isomorphism to define Lie brackets on $T_e G$ starting from the Lie-Jacobi brackets on $\mathfrak{X}_L$.

The isomorphism works as follows.
Consider a vector $\xi \in T_e G$.
For each $g$, the left translation $L_g$ maps trivially the group identity $e$ to $g$.
Thus, its push forward $(L_g)_{*}$ maps $\xi \in T_e G$ to a vector $X_\xi (g) \in T_g G$.
Varying $g \in G$, $X_\xi (g)$ defines a vector field, which can be easily verified to be left-invariant.
The map $\xi \mapsto X_\xi$ is linear (since the push forward is linear) and invertible (since its inverse is easily recognised as $X \in \mathfrak{X}_L \mapsto X(e) \in T_e G$), which shows that $T_e G $ is isomorphic to $\mathfrak{X}_L$ as a linear space.
Thus, we can define the Lie algebra $\mathfrak{g}$ associated to the Lie group $G$ as the tangent space to the group identity $T_e G$ by imposing that the Lie brackets are
\[
 [\xi,\eta] \eqdef [X_\xi, X_\eta] \,,
\]
which are defined in terms of the Lie-Jacobi brackets on $\mathfrak{X}_L$.
Note that, if $G$ is a linear space $V$ with the addition as group operation, as in the case discussed in the previous subsection where $G = (\real^3,+)$, then the above definition implies that the associated Lie algebra $\mathfrak{g}$ coincides with $V$ itself and the Lie brackets are trivially equal to zero, i.e., $\mathfrak{g}$ is an abelian Lie algebra.
This last statement is true, more generally, if $G$ is an abelian Lie group.\footnote{
Given a Lie group $G$, we have seen that one can build a Lie algebra $\mathfrak{g}$ associated to it.
Thus, one could ask the question on whether the converse is true, i.e., given a Lie algebra $\mathfrak{g}$ defined abstractly as in footnote~\ref{footnote:Lie-algebra} on page~\pageref{footnote:Lie-algebra}, one can find a Lie group $G$, whose associated Lie algebra is $\mathfrak{g}$.
If $\mathfrak{g}$ is finite-dimensional, then the answer is always positive due to Lie's third theorem, but the theorem does not generalise to the infinite-dimensional case.
Note that, when the group exists, it is not unique in general.
Indeed, for instance, a group and its universal cover share the same Lie algebra and so do a non-connected Lie group and its connected subgroup.
}
\enlargethispage{\baselineskip}

Before we make the connection with symmetries, let us discuss one last topic about Lie groups and Lie algebras, namely, the exponential map.
This is a map from the Lie algebra to the Lie group, i.e., $\exp \colon \mathfrak{g} \rightarrow G$.
In order to define it, let us consider an element $\xi \in \mathfrak{g} = T_e G$ and the left-invariant vector field $X_\xi$ associated to it.
One can show that there is a unique curve $\gamma_\xi \colon \real \rightarrow G$ which solves $\gamma'_\xi (\lambda) = X_\xi \big(\gamma_\xi (\lambda) \big)$ with the initial condition $\gamma_\xi (0) = e$.
Furthermore, one can also show that
\[
 \gamma_\xi (\lambda_1 + \lambda_2) = \gamma_\xi (\lambda_1) \gamma_\xi (\lambda_2)
\]
for all $\lambda_1,\lambda_2 \in \real$.
Thus, $\langle \xi \rangle \eqdef \{ \gamma_\xi (\lambda) | \lambda \in \real \}$ is a smooth one-parameter subgroup of $G$.
At this point, we can simply define the exponential map as
\[
 \exp(\xi) \eqdef \gamma_\xi (1) \,,
\]
which is smooth and satisfies $\exp(\lambda\xi) \eqdef \gamma_\xi (\lambda)$
As a consequence the one-parameter subgroup $\langle \xi \rangle = \{ \exp(\lambda \xi) |\lambda \in \real \}$, i.e., it is generated by the exponential map.\footnote{
The exponential map is \emph{injective} and, as a consequence, every element of the group in $U \eqdef \exp(\mathfrak{g}) \subseteq G$ can be uniquely written as the exponential of an element in the Lie algebra $\mathfrak{g}$.
Since $e \in U$, we see that there is at least a  neighbourhood of the group identity where every element can be written as the exponential of an element in $\mathfrak{g}$.
This certainly suffices to describe those ``infinitesimal group elements'' which we mentioned in our non-rigorous discussion.
Note that the exponential map is \emph{not surjective in general}, since e.g. $\exp(\mathfrak{g})$ is always connected, whereas $G$ is not in general.
Nevertheless, there are some notable cases in which the exponential map is known to be surjective, e.g., if the Lie group is abelian and connected or if it is compact and connected.
In any case, even if the exponential map is not surjective, one can still write every element of a connected Lie group $G$ as the finite product of elements in $\exp(\mathfrak{g})$, also in the case in which $G$ is infinite dimensional.
}
If $G$ is a linear space, as in the previous subsection where $G = (\real^3,+)$, then the exponential map is trivially the identity map.
The one-parameter family of transformations $T_{\lambda a}$ which we considered in the previous subsection were nothing else than the transformations generated by the action of the one-parameter subgroup $\langle a \rangle$ on the phase space.
\enlargethispage{\baselineskip}

Finally, let us consider the action $\Phi$ of the Lie group $G$ on the phase space $Z$.
We say that the action $\Phi$ is \emph{symplectic} if $\Phi_g^* \Omega = \Omega$ for all $g \in G$.
Given an element $\xi \in \mathfrak{g}$, let us consider the one-parameter family of transformations $\varphi_\lambda^{(\xi)} \colon z \mapsto \exp(\lambda \xi) \cdot z$, which is easily recognised as the flow of the vector field $X_\xi (z) \eqdef  \big[ \exp(\lambda \xi) \cdot z \big]'_{\lambda = 0}$ (not to be confused with the left-invariant vector fields discussed before, although we are using the same symbol).
One can show that any two of such vector fields satisfy the identity
\begin{equation} \label{lie-algebra-to-vf}
 [X_\xi, X_\eta] = - [\xi,\eta] \,,
\end{equation}
which means that $\xi \mapsto X_\xi$ is an anti-homomorphism (due to the minus sign) from $\mathfrak{g}$ to the vector fields on $Z$. 
If $\Phi$ is symplectic, then the vector fields $X_\xi$ are symplectic for all $\xi \in \mathfrak{g}$, i.e., $\liephase_{X_\xi} \Omega = 0$.
We say that the action is \emph{canonical} (or \emph{Hamiltonian}) if, in addition, there is a family of canonical generators $P_\xi$, such that $\extderphase P_\xi = - \insertion_{X_\xi} \Omega$.
Note that the map $P\colon \xi \mapsto P_\xi$ can always be redefined to be linear and we will assume that such a redefinition has been made.
The combination of~(\ref{Poisson-brackets-closure}) and~(\ref{lie-algebra-to-vf}) shows that the Poisson brackets of these canonical generators satisfy the identity
\begin{equation} \label{Poisson-representation}
 \{ P_\xi , P_\eta \} = P_{[\xi,\eta]} \,,
\end{equation}
which implies that the map $P \colon \xi \mapsto P_\xi$ is a Lie-algebra isomorphism or, as we shall say, that the canonical generators $P_\xi$ form a Poisson-representation of the Lie algebra $\mathfrak{g}$.\footnote{
The identity~(\ref{Poisson-representation}) shows that the map $P \colon \xi \mapsto P_\xi$ is a homomorphism from the Lie algebra $\mathfrak{g}$ to the canonical generators.
Since we have imposed the group action $\Phi$ to be faithful, it follows that $P$ is injective.
As a consequence, $P$ is a isomorphism between $\mathfrak{g}$ and $P(\mathfrak{g})$.
}

It is often useful in practical situations to fix a basis $(T_A)_{A=1,\dots,\dim \mathfrak{g}}$ in $\mathfrak{g}$, so that a generic element $\xi$ can be written as $\xi = \xi^A T_A$.
The Lie bracket of two elements of the basis is an element of $\mathfrak{g}$ and, therefore, it can be written as a linear combination of the $(T_A)$.
Thus, we can write
\begin{equation}
 [T_A,T_B] = F^{M}{}_{AB} T_M \,,
\end{equation}
where the coefficients $F^{M}{}_{AB}$ are called the \emph{structure constants} of the Lie algebra and are skew-symmetric in the lower indices, i.e., $F^{M}{}_{AB} = -F^{M}{}_{BA}$.
At this point, let us define $P_A \eqdef P_{T_A}$, so that
\[
 P_{\xi} = \xi^A P_A 
\]
due to the linearity of $P_\xi$.
Then, the identity~(\ref{Poisson-representation}) can be rewritten equivalently as
\begin{equation}
 \{ P_A , P_B \} = F^M{}_{AB} P_M \,,
\end{equation}
which displays again that the canonical generators $(P_A)$ form a Poisson-representation of the Lie algebra $\mathfrak{g}$.

The \emph{momentum map} of a canonical action is the map $\boldsymbol{P} \colon Z \rightarrow \mathfrak{g}^*$ defined by $\big[ \boldsymbol{P} (z) \big] (\xi) \eqdef P_\xi (z)$.
We say that a canonical action $\Phi$ is a \emph{canonical symmetry} if it leaves the Hamiltonian $H$ invariant.
If $P_\xi$ are the canonical generators of the action, then this statement translates into
\begin{equation}
 \{ H , P_\xi \} = 0 \qquad \forall \xi \in \mathfrak{g} \,,
\end{equation}
from which it immediately follows that $\dot P_\xi = 0$, i.e., the value of the canonical generators of the symmetries are conserved quantities along the solutions of the equations of motion.
Analogously, if $\varphi_t$ is the flow associated to the Hamiltonian $H$, we could have said that $\boldsymbol{P} \circ \varphi_t = \boldsymbol{P}$, i.e., that the momentum map $\boldsymbol{P}$ is conserved under time evolution.

This concludes our review of the Hamiltonian methods in classical mechanics, where we have introduced the basic concepts from the equations of motion to the study of symmetries.
In order to be able to present the results of this thesis about the asymptotic symmetries of classical field theories, we will have to generalise the results of this section to this case, which is going to be the topic of the next chapter.
The main difference will be that the phase space will be an infinite-dimensional manifold.

\chapter{Hamiltonian methods in field theories} \label{cha:Hamiltonian-ft}

In this chapter, we will discuss the Hamiltonian formulation of classical field theories, thus generalising the results of the previous chapter.
The findings of this chapter are going to constitute the basis of the investigations contained in this thesis.
As in the case of classical mechanics, the starting point is going to be the action functional in Lagrangian formulation.
From this, we will show how to derive the Hamiltonian and discuss the asymptotic symmetries of the theory under consideration.
Also in this case, we are not going to discuss every formal aspect, nor provide the mathematical proof of every statement.
For this, we redirect the reader to the detailed discussion contained in~\cite{Marsden-Ratiu} and to the other references provided along this chapter.

The main difference with respect to the previous section is that, in the case of field theories, the degrees of freedom are the fields, that is, quantities whose value can change from one spacetime point to another.
In the cases which we will analyse, the fields will be tensor fields and tensor densities on a flat Minkowski spacetime, but different types of objects, such as spinor fields, and more general spacetimes, such as  asymptotically-flat spacetimes, can be taken into considerations.
The action and other important physical quantities will be defined as integrals of some combination of the fields.
Therefore, for these quantities to be well-defined, we will need to restrict the allowed fields to those belonging to some suitable function space.
Specifically, an important role in making the integrals finite is played by restricting the possible behaviour of fields at (spatial) infinity.
In particular, these restrictions amount to imposing fall-off conditions on the fields, i.e., specifying how quickly the fields have to vanish at (spatial) infinity, often complemented with parity conditions, as we shall discuss in details in specific examples.
\enlargethispage{-\baselineskip}

One thing that should be already mentioned about fall-off and parity conditions is that there are two competing aspects, which should be taken into consideration when imposing restriction on the asymptotic behaviour of the fields.
On the one hand, as we have just mentioned, these conditions must be strong enough to ensure that physically-relevant quantities are well-defined.
On the other hand, they should be weak enough, so that solutions of physical interest are not excluded and that the symmetry group is as large as possible.

Another aspect to consider is that, independently on the specific fall-off and parity conditions, the space of allowed field configurations --- i.e. the phase space --- will be built starting from some function spaces, to which the fields have to belong.
As a consequence, in general, the phase space will be an infinite-dimensional manifold, so that we will need to adapt the results of the previous chapter to this situation.
We will do this in the first section of this chapter, where the discussion will be kept as general as possible.

Immediately after, in the second section, we will begin the specific discussion about relativistic field theories.
In this case, the fundamental objects are fields defined on the four-dimensional spacetime $M$.
However, in order to set up the Hamiltonian formalism, we will need to ``split'' the spacetime into space and time, by means of the so-called $3+1$ decomposition.
Indeed, if we set aside for a moment the possible mathematical issues, we could infer intuitively the equations for the field-theoretical case from the equations of the previous chapter by replacing the finitely-many $q^I$ and $p_I$ with the infinitely-many position-dependent $q(\vect{x})$ and $p(\vect{x})$, where $\vect{x}$ represents the position in space, while time enters in the equations as an evolution parameter in both cases.
Thus, we see that the position in space $\vect{x}$ behaves similarly to the label $I$, whereas time has the profoundly different role of parametrising the evolution of the system.
In addition, sums over the index $I$ will be replaced by integrals over $\vect{x}$, so that issues concerning the convergence of these integrals will arise, as we have already mentioned.

In order to derive the results of this thesis, we will need to deal with two further aspects of classical field theories using the Hamiltonian formalism.
First, we will need to derive the action of the Poincar\'e transformations on the phase space, which will play an important role in the discussion about the asymptotic symmetries of the theories under analysis.
Second, we will need to introduce the concept of constraints and gauge transformations, which we chose not to discuss in the classical-mechanical case.
In order not to leave the discussion about gauge transformations too abstract, we are going to work with the specific example of Yang-Mills, which is neither a too-trivial nor a too-complicated example of a gauge theory.
Furthermore, it will also find a direct application in the discussion of chapter~\ref{cha:Yang-Mills}.
\enlargethispage{-2\baselineskip}

\section{Hamiltonian methods on an infinite-dimensional phase space} \label{sec:Hamiltonian-inf-dim}

As we have said, the analysis of field theories using the Hamiltonian formalism will require to introduce in general  a phase space which is an infinite-dimensional manifold.
Therefore, in this section, we are going to generalise the results of the previous chapter, which concerned classical-mechanical systems described by a finite-dimensional phase space $Z$, to the case in which $Z$ is indeed an infinite-dimensional manifold.
We will try to leave the discussion as general as possible and we will not provide the proofs of the statements, which can be found again in~\cite{Marsden-Ratiu} and in the literature therein.
\enlargethispage{-2\baselineskip}

As in this case, it is often useful to start from an action 
\begin{equation}
 S [q(t)] = \int dt L[q(t), \dot q (t)]
\end{equation}
written in terms of a Lagrangian $L \colon TQ \rightarrow \real$.
Note that we are not writing explicitly the boundary of the integration, but it has to be understood that $t$ belongs to a finite interval $I \subset \real$.
The difference with respect to the previous chapter is that now $Q$ is an infinite-dimensional manifold.

As in the previous case, we can introduce the canonical momenta by means of the Legendre transform $\mathbb{F} L \colon TQ \rightarrow T^* Q$.
If the Lagrangian $L$ is hyper-regular, i.e. if $\mathbb{F} L$ is a diffeomorphism, we can replace the velocities with the momenta and define a Hamiltonian $H$, in the same way used in the previous chapter.
Thus, we obtain an infinite-dimensional symplectic manifold $Z = T^* Q$ equipped with the canonical symplectic form $\Omega$, which is exact and (weakly) non-degenerate, and a Hamiltonian $H$.

More generally, although the phase is introduced in many cases the cotangent bundle $T^* Q$ of a the configuration space $Q$, we do not have to rely on this fact.
Thus, we will assume that we are given a phase space $Z$, which is simply a smooth manifold.
In addition, we are given the symplectic form $\Omega$, which is a close weakly-non-degenerate two-form on the phase space.
We remind that $\Omega$ is \emph{weakly non-degenerate} if it is such that, at each $z \in Z$, if $\Omega_z(X,Y) = 0$ for every $Y \in T_z Z$, then it must be that $X = 0$.
The weak non-degeneracy of $\Omega$ is equivalent to the fact that $\Omega^\flat \colon TZ \rightarrow T^*Z$ is injective.
However, since $TZ$ and $T^*Z$ are now infinite-dimensional linear space, we \emph{cannot} conclude any more that the map is also surjective.
Thus, we say that $\Omega$ is \emph{strongly non-degenerate} if $\Omega^\flat \colon TZ \rightarrow T^*Z$ is invertible and we denote its inverse with $\Omega^\sharp \colon T^*Z \rightarrow TZ$.
In this case, Darboux's theorem ensures that local canonical coordinates can always be found, but this is not always the case if $\Omega$ is only weakly non-degenerate.
Depending on whether $\Omega$ is weakly or strongly non-degenerate, we refer to $Z$ as a weak or strong symplectic manifold.

Finally, in order to complete the minimal structure needed to set up the Hamiltonian formulation, we need the Hamiltonian, i.e., a canonical generator of the time evolution $H \colon Z \rightarrow \real$.
Concerning the Hamiltonian, two things should be said at this point.
First, in the cases of our interest, $H$ will be given as the integral over a spatial slice of a Hamiltonian density $\mathscr{H}$, i.e.
\begin{equation} \label{Hamiltonian-density}
 H = \int_\Sigma d^3 \vect{x} \, \mathscr{H} (\vect{x}) \,, 
\end{equation}
where $\mathscr{H}(x)$ is a local function of the canonical fields and their (spatial) derivatives.
Second, in order for $H$ to be a canonical generator, there must be a vector field $X_H$ such that $\extderphase H = - \insertion_{X_H} \Omega$.
Then, the equations of motion are simply $\dot z = X_H (z)$, as in the case of classical mechanics.
Note that if $\Omega$ is strongly non-degenerate, we are sure of the existence of such a vector field, which is simply found as $X_H = - \Omega^\sharp(\extderphase H)$.
However, when $\Omega$ is weakly non-degenerate, the existence of $X_H$ is not guaranteed since $\extderphase H$ might lie outside the image of $\Omega^\flat$.
As a consequence, some extra care in specifying the Hamiltonian of a theory is needed in order to ensure the existence of a Hamiltonian vector field.
In our analysis, after specifying the correct fall-off and parity conditions of the fields (i.e. in choosing the right phase space), most of this extra care will consist in complementing the expression~(\ref{Hamiltonian-density}) with the correct boundary terms, finding
\begin{equation} \label{Hamiltonian-density-boundary}
 H = \int_\Sigma d^3 x \, \mathscr{H} (x) +
 \oint_{\partial \Sigma} d^2 \barr{x} \, \mathscr{B} (\barr{x}) \,, 
\end{equation}
where $\barr{x}$ are coordinates on the boundary and $\mathscr{B} (\barr{x})$ is a local function of the fields and their derivatives.
In the situations which we will analyse in this thesis, the boundary $\partial \Sigma$ will be actually a boundary at infinity, but we will come back to this point in these specific situations.

As in the classical-mechanical case, we will be interested in considering vector fields other than $X_H$.
Also in this case, we say that $X$ is \emph{symplectic} if $\liephase_X \Omega = 0$ or, equivalently, if $\extderphase (\insertion_X \Omega) = 0$.
In addition, $X$ is Hamiltonian if there is a smooth function $F \in \mathcal{C}^\infty (Z)$ such that $\extderphase F = -\insertion_X \Omega$.
In this case, we write $X_F$ instead of $X$ and we say that $F$ is the canonical generator of $X_F$.
Note that, if $\Omega$ is \emph{not} strongly symplectic, not every smooth function $F \in \mathcal{C}^\infty (Z)$ is the canonical generator of some vector field, but only those for which $\extderphase F$ belongs to $\Omega^\flat (TZ)$.
The rest of the discussion the previous chapter concerning Poisson brackets and symmetries goes unchanged for the canonical generators. 

With this, we conclude this general section and we move to the discussion of relativistic field theories.
We will first discuss the $3+1$ decomposition and then move to the Poincar\'e transformations and gauge theories, introducing all the tools needed to derive the results of this thesis.

\section{3+1 decomposition} \label{sec:3+1}

From this section, we turn our attention to relativistic field theories.
In this case, the theory is written in terms of fields living on the spacetime manifold $M$, which is four dimensional and possesses a Lorentzian metric ${}^4 g$, where the superscript $4$ empathises that it is the metric of the four-dimensional spacetime $M$.
The equations of motion ensue from the variational principle of an action
\begin{equation} \label{spacetime-action-general}
 S[\phi;{}^4 g] = \int_M d^4 x \, \mathscr{L} \big( \phi(x) , \partial \phi (x) ,\dots, \partial^k \phi(x) ; {}^4 g (x) \big) \,,
\end{equation}
where the Lagrangian density $\mathscr{L}$ is a local function of the fields --- which we have collectively denoted with $\phi$ --- and of a finite number of their derivatives.
In addition, $\mathscr{L}$ depends parametrically on ${}^4 g$, with the exception of General Relativity where ${}^4 g$ is part of the collection $\phi$, either alone (empty space) or with other fields as well (gravity coupled to matter). 

Some facts should be noted about this action.
First, in order for the integral to make sense, $\mathscr{L}$ must be a scalar density of weight one, for otherwise the expression would be dependent on the choice of coordinates on $M$.
Secondly, as we have seen in the previous chapter, the principle of least action requires the variation to be considered among all the curves with fixed endpoints and, for this reason, the integration was limited to a time interval $t \in [a,b]$.
However, not only does the expression above not have a similar limitation being an integral over the entire spacetime $M$, but there is not even a clear choice of time.
The action principle in this case has to be understood as taking place on a sandwich between two hypersurfaces (see e.g. Chap. 21 of \cite{Gravitation} for more details).
Thirdly, during the variation, it is usually necessary to integrate some expression by parts, as we shall see.
Therefore, some boundary terms might be present at the end of the variation and it might be necessary to include some boundary terms in the definition~(\ref{spacetime-action-general}) of the action as well, in order to compensate for them.
But, we will come back to this point with all the details when dealing with specific theories.
Lastly, in the cases which we will analyse, $\mathscr{L}$ will depend only on the $\phi$ and on its first derivatives $\partial \phi$, with the only exception of General Relativity where second-order derivatives will be included.
In any case, the equations of motion will always turn out to be, at most, second-order partial-differential equations.

The main idea of the Hamiltonian formulation is to convert the above set-up into an equivalent theory formulated in terms of fields living on a three-dimensional manifold $\Sigma$ --- the ``space'' --- where time appears as a parameter, similarly to the situation of the previous chapter.
This is the goal of the $3+1$ decomposition, which allows us to split the spacetime into space and time, as well as to decompose tensor fields on $M$ into tensor fields on $\Sigma$ carrying the same amount of information.
From now on, we are going to denote points of $\Sigma$ with bold letters, such as $\vect{x} \in \Sigma$, and points in $M$ with non-bold letters, e.g. $x \in M$.

Before we begin with the technical part, let us mention that the information contained in this section can be found, among others, in the already-mentioned Chap.~21 of~\cite{Gravitation}, in the seminal work by Kucha\v{r}~\cite{Kuchar:hyperspace1,Kuchar:hyperspace2,Kuchar:hyperspace3}, in the book by Henneaux and Teitelboim~\cite{Henneaux-Teitelboim}, and in the review by Giulini~\cite{Giulini:Hamiltonian-GR}.
Additional information can be found also in the papers by Isham and Kucha\v{r}~\cite{Kuchar-Isham1,Kuchar-Isham2} and in the one by Hojman, Kucha\v{r} and Teitelboim~\cite{Geometrodynamics-regained}.

Let us begin by introducing the concepts of embedding and of foliation, which will play a fundamental role in the discussion of this and of the next sections.
An \emph{embedding} is a smooth injection $e \colon \Sigma \hookrightarrow M$, where $\Sigma$ is a three-dimensional smooth manifold and $e(\Sigma) \subset M$ is a \emph{space-like hypersurface}, i.e., the pull-back of the four-dimensional metric ${}^4 g$ on $e(\Sigma)$ is a Riemannian metric.
The three-dimensional manifold $\Sigma$ will act as our ``space'' and will be here that the Hamiltonian dynamics will take place.
A smooth one-parameter family of embeddings $(e_t)_{t\in I}$, where $I \subseteq \real$ is an interval, is said to be a \emph{foliation} of the spacetime $M$ if the hypersurfaces $\Sigma_t \eqdef e_t (\Sigma)$ form a partition of $M$.\footnote{
With ``smooth one-parameter family'' of embeddings, we mean that the map $\real \times \Sigma \rightarrow M$, defined by $(t,\vect{x}) \mapsto e_t(\vect{x})$, is smooth.
In addition, we remind that $\{\Sigma_t : t\in I\}$ is a partition of $M$ if it satisfies the following three properties.
First, $\Sigma_t$ is not empty for every $t \in I$.
Second, $\Sigma_t \cap \Sigma_{t'} = \emptyset$ if $t \ne t'$.
Third, $\cup_{t \in I} \Sigma_t = M$.
}
In this case, the hypersurfaces $\Sigma_t$ are called spatial slices or leaves of $M$.

Thus, if we have a foliation of $M$, the parameter $t \in I$ would play the role of ``time'', whereas the three dimensional manifold $\Sigma$ would play the role of ``space''.
However, two things should be said to avoid possible misunderstandings.
First, not every spacetime $M$ allows the existence of a foliation.
Indeed, this is only possible if the spacetime is globally hyperbolic, to which case we will restrict our attention.
Second, even when a foliation exist, it is not unique.
Actually, there is a plethora of distinguished foliations for each globally-hyperbolic spacetime, as it should be on physical grounds since there is no absolute space and no absolute time.

In the next subsections, we will first show how to convert the tensors fields on the spacetime $M$ into tensors fields on the spacetime $\Sigma$ without loosing any information and, second, we will re-express the dynamics of the former ones in terms of the latter ones.
The so-found theory on $\Sigma$ will be precisely the Hamiltonian formulation of the original field theory and will be the starting point to the analysis of the specific theories treated in this thesis.

\subsection{Decomposition of tensor fields}
Let us begin with the first step, that is the decomposition of tensor fields on $M$ in terms of tensor fields on $\Sigma$.
Specifically, we wish to convert a tensor field on $M$ into tensor fields on $\Sigma$ carrying the same amount of information.
This is possible provided that there is a foliation $e_t \colon \Sigma \hookrightarrow M$.
In order not to make any confusion we will write momentarily a superscript $4$ on the right of the tensor fields on $M$, which we wish to decompose.
This notation, which was already used for the metric ${}^4 g$ is extended to the other relevant tensor fields in the next few subsections.
On the contrary, tensor fields on $\Sigma$ will not have any superscript.

Let us first consider the simplest case, i.e., a scalar field ${}^4 \phi(x)$.
For each $t \in I$, the pull-back of the map $e_t \colon \Sigma \hookrightarrow M$ can be used to define a scalar field $\phi_t$ on $\Sigma$ by the expression
$\phi_t \eqdef  e^*_t ({}^4 \phi)$.
In other words, since $(e_t)_{t \in I}$ is a foliation, for each point $x \in M$ there are unique $t \in I$ and $\vect{x} \in \Sigma$ such that $x = e_t (\vect{x})$; vice-versa, for each $t \in I$ and for each $\vect{x} \in \Sigma$ there is a unique $x \in M$ such that $x = e_t (\vect{x})$.
Thus, we simply define $\phi_t (\vect{x}) \eqdef {}^4 \phi (x)$ and obtain a one-parameter family $(\phi_t)_{t \in I}$ of scalar fields on $\Sigma$ by varying the parameter $t \in I$.
It is easy to see that the one-parameter family $(\phi_t)_{t \in I}$ contains the same amount of information as the scalar field ${}^4 \phi$.
We will often write (one-parameter families of) tensor fields on $\Sigma$ omitting the $t$, where there is no risk of misunderstanding.

The decomposition of other tensor fields is a bit more complicated.
Indeed, let us consider the case of a Lorentzian metric ${}^{4}g $.
Simply using the pull-back of the one-parameter family of embeddings $(e_t)_{t\in I}$, we can define a one-parameter family of three-dimensional metrics $g_t \eqdef e^*_t ({}^4 g)$, which are Riemannian due to the fact that every $\Sigma_t = e_t(\Sigma)$ is space-like.
But, the one-parameter family of three-dimensional metrics does \emph{not} carry the same amount of information of the four-dimensional metric ${}^4 g$.
Indeed, at each spacetime point $x \in M$, the four-dimensional metric ${}^4 g$ has ten independent components, whereas, at the corresponding pair $(t, \vect{x})$, the three metric $g_t$ has only six.
The reason is that the embedding $e_t$ is merely injective and induces a bijection only between $\Sigma$ and $\Sigma_t \subset M$.
Thus, when we use it to pull-back tensor fields from $M$ to $\Sigma$, it only takes care of the components tangent to $\Sigma_t$ neglecting all the others.

To solve this issue, we need to decompose tensor fields into vertical components (which are normal to the hypersurfaces) and horizontal components (which are tangent to it) as follows.
Let us introduce the one-form $\underline{n}(x)$ defined by the following three properties.
First, it annihilates all the vectors tangent to any $\Sigma_t$, i.e., if $x \in \Sigma_t$ and $v \in T_x \Sigma_t$, we have $\underline{n} (x) \cdot v = 0$.
Second, it is future-directed.
Third, it is normalised such that ${}^4 g^{-1} (\underline{n},\underline{n}) = -1$.
These three properties unequivocally identify one, and only one, $\underline{n}(x)$ since $T_x \Sigma_t$ is a three-dimensional linear subspace of the four-dimensional $T_x M$ at each $x \in M$.
Note that we write $\underline{n}$ with a bar below in order to remember that it is a one-form and, if coordinates $(x^\alpha)$ are employed, it has an index below.
In particular in this case, we would write the components as $\underline{n} = n_{\alpha} dx^{\alpha}$, without the bar since it would be superfluous.
In addition, the normalisation condition in terms of the components would simply be ${}^4 g^{\alpha \beta} n_\alpha n_\beta = -1$.

At the same time, we can introduce the vector field $\overline{n} \eqdef {}^4 g^{\sharp} (\underline{n})$, which is written with a bar above since it is a vector field and is clearly normalised as ${}^4 g(\overline{n},\overline{n}) = -1$.
In coordinates, we write $\overline{n} = n^\alpha \partial/ \partial x^\alpha $ and we have the relations $n^\alpha = {}^4 g^{\alpha \beta} n_\beta$ and ${}^4 g_{\alpha \beta} n^\alpha n^\beta = -1$.
It is clear from the definition, that $\overline{n}(x)$ spans the one-dimensional linear subspace of $T_x M$ normal to $T_x \Sigma_t$.
Note that $\underline{n}$ and $\overline{n}$, despite being tensor field on $M$, are written without the superscript $4$ since we do not wish to decompose them.
Rather, we wish to decompose other tensor fields by means of them.

\begin{figure}[b]
\centering
\begin{tikzpicture}[scale=.50]
   \coordinate[label=below:{\small $x$}] (x) at (0,0);
   \coordinate[label=left:$\Sigma_t$] (L) at (-6, -1);
   \coordinate (R) at (6, -1);
   \coordinate[label=below left:{\footnotesize ${}^4 v^\perp \overline{n}$}] (A) at (0,2.5);
   \coordinate[label=above left:{\footnotesize ${}^4 v^\parallel$}] (B) at (3.5,0);
   \arcThroughThreePoints{R}{x}{L};
   \fill[black] (x) circle(3pt);
   \draw[->,thick=2pt] (x) -- (A);
   \draw[->,thick=2pt] (x) -- (B);
   \draw [dashed] (B) -- ($(A)+(B)$) -- (A);
   \draw[->,thick=2pt] (x) -- node[above,sloped] {\footnotesize ${}^4 v$} ++ ($(A)+(B)$);
   \draw (0.4,0) -- (0.4,0.4) -- (0,0.4);
\end{tikzpicture}
  \caption{The decomposition of a vector field ${}^4 v$ into horizontal (${}^4 v^\parallel$) and vertical (${}^4 v^\perp$) components at a point $x \in \Sigma_t \subset M$.}
  \label{fig:horizontal-vertical-vf}
\end{figure}
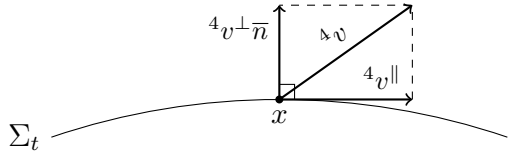

We will say that a vector field ${}^4 v (x)$ is \emph{horizontal} if $\underline{n} (x) \cdot v(x) = 0$ everywhere and that it is \emph{vertical} if it is proportional to $\overline{n}$.
The decomposition of a generic vector field ${}^4 v (x)$ into horizontal and tangent components can then be achieved quite easily.
Indeed, let us write
\begin{equation}
 {}^4 v(x) = {}^4 v^\perp (x) \, \overline{n} (x) + {}^4 v^\parallel (x) \,, 
\end{equation}
where $\underline{n} \cdot {}^4 v^\parallel = 0$, whereas ${}^4 v^\perp (x)$ is a scalar field.
Thus, ${}^4 v^\parallel$ and ${}^4 v^\perp \overline{n}$ are respectively a horizontal and a vertical vector field, called the horizontal and vertical components of ${}^4 v$ (see Fig.~\ref{fig:horizontal-vertical-vf}).

Specifically, they can be found unequivocally in terms of the vector field ${}^4 v$ simply as
\begin{equation} \label{vector-field-decomp}
 {}^4 v^\perp  = - \underline{n} \cdot {}^4 v
 \qquad \text{and} \qquad
 {}^4 v^\parallel  = {}^4 v + (\underline{n} \cdot {}^4 v) \, \overline{n} 
\end{equation}
using the relation $\underline{n} \cdot \overline{n} = -1$.
As we have already seen, the scalar field ${}^4 v^\perp$ can be pulled back on $\Sigma$ obtaining a one parameter family  $v^\perp_t (\vect{x})$ of scalar fields on $\Sigma$.
In addition, also the vector field ${}^4 v^\parallel$, due to the fact that ${}^4 v^\parallel (x)$ belongs to $T_x \Sigma_t$ and not merely to $T_x M$, can be pulled back to the one parameter family of vector fields $v^\parallel_t (\vect{x})$ on $\Sigma$.
Thus, we have decomposed one vector field ${}^4 v$ on $M$ into a one-parameter family $v^\perp_t$ of scalar fields on $\Sigma$ and a one-parameter family $v^\parallel_t$ of vectors field on $\Sigma$.
Only the two-parameter families $v^\perp_t$ and $v^\parallel_t$ considered together contain the same information of ${}^4 v$.

Before discussing the decomposition of different types of tensor fields, let us make a quick remark.
Since we have a Lorentzian metric ${}^4 g$ on $M$, we could have use it to convert the vector field ${}^4 v$ into a one-form ${}^4 \underline{v} \eqdef g^\flat (v)$.
This one-form could have been then pulled back directly to the one-parameter family $\underline{v}_t \eqdef e^*_t ({}^4 \underline{v})$ of one-forms on $\Sigma$, since the pull-back of one-forms is always defined as long as we have an injection.
Now, using the one-parameter family of Riemannian metrics $g_t$ defined before, we would have obtained a one-parameter family $v_t \eqdef g_t^\sharp (\underline{v}_t)$ of vector fields on $\Sigma$.
One can show that this one-parameter family of vector fields is the same as the $v^\parallel_t$ discussed above, so that we will often write $v_t$ instead of $v^\parallel_t$, sometime even omitting the $t$ if there is no risk of confusion.\footnote{
One way to show this fact is by doing the explicit computation in the foliation-induced coordinates introduced in the next subsection.
}
Thus, we see that the pull-back, which can be defined for a tensor field of any rank eventually using the Lorentzian metric ${}^4 g$ and the one-parameter family of Riemannian metrics $g_t$, only takes care of the parallel components and neglects the normal ones.

The decomposition of one-forms behaves very similarly to that of vector fields.
Specifically, if ${}^4 A$ is a one-form, such as the four-potential of electrodynamics, we can write
\begin{equation}
 {}^4 A (x) = {}^4 A_\perp (x) \, \underline{n} (x) + {}^4 A_{\parallel} (x) \,,
\end{equation}
where ${}^4 A_{\parallel} \cdot \overline{n} = 0$, so that the normal and parallel components are easily found to be
\begin{equation} \label{one-form-decomp}
 {}^4 A_\perp = - {}^4 A \cdot \overline{n}
 \qquad \text{and} \qquad
 {}^4 A_\parallel = {}^4 A + ({}^4 A \cdot \overline{n}) \, \underline{n} \,.
\end{equation}
The pull-back of the scalar field ${}^4 A_\perp$ and of the one-form ${}^4 A_\parallel$ defines the one-parameter family $A_\perp^t$ of scalar fields on $\Sigma$ and the one-parameter family of one-forms $A_\parallel^t$ on $\Sigma$, respectively.
Note that $A_\parallel^t = e^*_t ({}^4 A)$ and, in the following, we will often write $A^t$ instead of $A_\parallel^t$, sometime even omitting the $t$ if there is no risk of confusion.

The decomposition of tensor fields of higher rank follows the same scheme, but is rather more cumbersome.
Therefore, we will limit our analysis to the decomposition of the four-dimensional metric ${}^4 g$, since we will not need explicitly any of the other higher-rank tensors.
It is possible to proceed in the same way as we did for a one-form, but introducing also some mixed components.
In other words, let us write
\begin{equation} \label{decomp-metric-n}
 {}^4 g = {}^4 g_{\perp \perp} \, \underline{n} \otimes \underline{n} +
 ({}^4 g_{\parallel \perp} \otimes \underline{n} +
 \underline{n} \otimes {}^4 g_{\perp \parallel} ) +
 {}^4 g_{\parallel \parallel} \,,
\end{equation}
where ${}^4 g_{\perp \perp}$ is a scalar field, ${}^4 g_{\perp \parallel}$ and ${}^4 g_{\parallel \perp}$ are two one-forms, and ${}^4 g_{\parallel \parallel}$ is a second-rank covariant tensor.
The terms in brackets are precisely the mixed components mentioned above.
In order for the decomposition to be well-defined, we need to impose the conditions
\begin{equation} \label{decomp-metric-conditions}
 {}^4 g_{\parallel \perp} \cdot \overline{n} =
 {}^4 g_{\perp \parallel} \cdot \overline{n} = 0
 \qquad \text{and} \qquad
 {}^4 g_{\parallel \parallel} (\overline{n},\cdot) =
 {}^4 g_{\parallel \parallel} (\cdot,\overline{n}) = 0 \,, 
\end{equation}
where ${}^4 g_{\parallel \parallel} (\overline{n},\cdot)$ is the contraction of the first index of ${}^4 g_{\parallel \parallel}$ with $\overline{n}$ and ${}^4 g_{\parallel \parallel} (\cdot,\overline{n})$ the contraction of the second index.
The decomposition~(\ref{decomp-metric-n}) with the conditions~(\ref{decomp-metric-conditions}) constitutes the starting point for the decomposition of any second-rank covariant tensor field.
In addition, for the specific case of the spacetime metric, we also have three further pieces of information.
First, since the metric is symmetric, we also know that ${}^4 g_{\parallel \perp} = {}^4 g_{\perp \parallel}$ and that ${}^4 g_{\parallel \parallel}$ is symmetric.
Second, due to the chosen normalisation of $\overline{n}$, we infer the further condition 
${}^4 g (\overline{n}, \overline{n}) = -1$.
This latter condition, combined with~(\ref{decomp-metric-conditions}), let us conclude that
\[
 {}^4 g_{\perp \perp} = -1 \,.
\]
Third, since by definition $\underline{n} = {}^4 g^\flat (\overline{n}) = {}^4 g (\overline{n}, \cdot)$, we also infer that
\[
 {}^4 g_{\parallel \perp} = {}^4 g_{\perp \parallel} = 0 \,.
\]
Thus, the original decomposition~\ref{decomp-metric-n}) reduces simply to
\begin{equation} \label{decomp-metric-simplified}
 {}^4 g = - \, \underline{n} \otimes \underline{n} +
 {}^4 g_{\parallel \parallel} \,.
\end{equation}
The parallel components ${}^4 g_{\parallel \parallel}$ can be found in terms of the spacetime metric as ${}^4 g_{\parallel \parallel} = {}^4 g + \, \underline{n} \otimes \underline{n} $ and, if pulled back on $\Sigma$, give rise to the one-parameter family of Riemannian metrics $g_t$.

Let us know see how these results can be expressed using a particular choice of coordinates, which will turn out to be extremely convenient in explicit computations.

\subsection{Foliation-induced coordinates}

Although we have worked until now without using coordinates, it is of great use to rewrite the various expressions in some special coordinates.
In particular, if $(\vect{x}^a)$ are coordinates on $\Sigma$, we can define the foliation-induced coordinates $(x^\alpha)$ on $M$ simply by demanding that the point $x = e_t (\vect{x})$ has coordinates $x^0 = t$ and $x^a = \vect{x}^a $.\footnote{
If the coordinates $(\vect{x}^a)$ cover only a subset $\mathcal{U} \subset \Sigma$, then the foliation-induced coordinates would cover only the subset $\cup_{t \in I} \, e_t(\mathcal{U}) \subset M$.
}
In principle, it would be possible to work with unrelated coordinates on $\Sigma$ and $M$, see e.g.~\cite{Kuchar:hyperspace1}, but we will stick to the simple case of foliation-induced coordinates.

Note that, keeping $t$ constant, the point $(t, \vect{x}^a)$ varies on the spatial slice $\Sigma_t$.
Thus, the tangent subspace $T_x \Sigma_t$ is spanned by the vectors \mbox{$\partial/\partial x^a \defeq \partial_a$} and it is annihilated by the one-form $dt$.
The pull-back of the relevant tensor fields on $\Sigma$ is trivial in these coordinates.
Indeed, the pull-back of ${}^4 A (t,\vect{x}) = {}^4 A_\alpha (t,\vect{x}) dx^\alpha$ is $A_t (\vect{x}) =  A^t_a (\vect{x}) d\vect{x}^a$, whose components satisfy $A^t_a (\vect{x}) = {}^4 A_a (t,\vect{x})$ or, in lighter notation, $A_a = {}^4 A_a$.
Thus, we see that only the spatial components labelled by $a = 1,2,3$ are carried by the pull-back.
Analogous results hold for \emph{parallel} vector fields and for the metric.
In particular for the latter, we have ${}^4 g_{ab} = g_{ab}$.

Before we can decompose tensor fields in coordinates, we need to introduce the lapse function and shift vector.
To this end, let us first define the four-dimensional vector field $\overline{\mathrm{N}}$, defined at the point $x = e_{t_0} (\vect{x})$ as
\begin{equation} \label{definition-N-vf}
 \overline{\mathrm{N}} (x) \eqdef
 \left[ \frac{d e_t (\vect{x})}{dt} \right]_{t = t_0} \,.
\end{equation}
In other words, if we fix $\vect{x} \in \Sigma$, the foliation $(e_t)_{t \in I}$ defines a curve in $M$ by $\gamma_{\vect{x}} (t) \eqdef e_t (\vect{x})$.
Since $(e_t)_{t \in I}$ is a foliation, each point $x \in M$ has a unique curve of the collection $\{\gamma_{\vect{x}} \}_{\vect{x} \in \Sigma}$ passing through it. 
Thus, the tangent vector to that curve at that point defines unequivocally the value of the vector field $\overline{\mathrm{N}}$ at that point, which precisely the meaning of~(\ref{definition-N-vf}).
Decomposing $\overline{\mathrm{N}}$ in normal and parallel components, we get
\begin{equation} \label{N-vf-decomp}
 \overline{\mathrm{N}} (x) = {}^4 N(x) \overline{n} (x) + {}^4 \vect{N} (x) \,, 
\end{equation}
where ${}^4 N = -\underline{n} \cdot \overline{\mathrm{N}}$ is a scalar field and ${}^4 \vect{N} = \overline{\mathrm{N}} - {}^4 N \overline{n} $ is a parallel vector field.
Thus, the latter, can be written in components as ${}^4 \vect{N} = {}^4 N^m \partial_m$.
The pull-back of the scalar field ${}^4 N$ and of the parallel vector field $ {}^4 \vect{N}$ defines, on $\Sigma$, a one-parameter family $N_t$ of scalar fields and a one-parameter family of vector fields $\vect{N}_t$, respectively.
In components, we have $N_t (\vect{x}) = {}^4 N(t,\vect{x})$ and $N^m_t (\vect{x}) = {}^4 N^m (t,\vect{x})$.
We will call $N$ the \emph{lapse function} and $\vect{N}$  the \emph{shift vector}.
Note that, by its definition, the vector field $\overline{\mathrm{N}}$ depends on the chosen foliation and so do, as a consequence, both the lapse and the shift.

At this point, we can finally rewrite the $3+1$ decomposition of the various tensor fields in foliation-induced coordinates.
The results will be expressed using the components of the corresponding tensor fields on $\Sigma$, the lapse, and the shift.
To begin with, from the definition~(\ref{definition-N-vf}), it follows that $\overline{\mathrm{N}} (x) = \partial/\partial t \defeq \partial_t$ in foliation-induced coordinates.
Thus, using~(\ref{N-vf-decomp}), we find
\begin{equation}
 \overline{n} = \frac{1}{N} \left(
 \partial_t - N^m \partial_m \right)
\end{equation}
In addition, since $\underline{n}$ must annihilates every parallel vector, it must be proportional to $dt$.
Thus, from the normalisation $\underline{n} \cdot \overline{n} = -1$, we infer
\begin{equation}
 \underline{n} = -N dt \,.
\end{equation}

The decomposition of the one-form ${}^4 A = {}^4 A_\alpha dx^\alpha$ is then straightforward.
Reminding that the one-parameter family $(A_t)_{t \in I}$ of one-forms on $\Sigma$ has components $A_a = {}^4 A_a$ and using~(\ref{one-form-decomp}), we find
\begin{equation}
 {}^4 A_\perp = - \frac{1}{N} ({}^4 A_0 - N^m {}^4 A_m) \,,
\end{equation}
which can be used to express ${}^4 A_0$ in term of $A_m$ and ${}^4 A_\perp = A_\perp$.
Thus, we find the decomposition
\begin{equation} \label{3+1-decomp-A}
 {}^4 A = (N^m A_m - N A_\perp) dt + A_m dx^m \,.
\end{equation}

The situation for a vector field does not differ much.
So, let us consider ${}^4 v = {}^4 v^\alpha \partial_\alpha$.
From~(\ref{vector-field-decomp}), we find
\begin{equation}
 {}^4 v^\perp = {}^4 v^0 N
 \qquad \text{and} \qquad
 {}^4 v^\parallel = ({}^4 v^a - {}^4 N^a) \partial_a \,.
\end{equation}
Thus, the components of the pulled-back parallel vector field $v^\parallel_t = v^m \partial/ \partial \vect{x}^m$ need to satisfy the equation $v^m = {}^4 v^m + {}^4 N^m$.
Using the relations ${}^4 N^m = N^m$ and ${}^4 v^\perp = v^\perp$, we arrive at the wished expression
\begin{equation} \label{3+1-decomp-v}
 {}^4 v = \frac{v^\perp}{N} \partial_t + (v^m - N^m) \partial_m \,.
\end{equation}

Finally, the Lorentzian metric can be expressed in coordinates as follows.
First, from ${}^4 g_{\parallel \parallel} (\overline{n},\cdot) = 0$ and $({}^4 g_{\parallel \parallel})_{ab} = g_{ab}$, it follows that
\begin{equation}
 ({}^4 g_{\parallel \parallel})_{00} = N^m N_m
 \qquad \text{and} \qquad
 ({}^4 g_{\parallel \parallel})_{0a} =
 ({}^4 g_{\parallel \parallel})_{a0} = N_a \,,
\end{equation}
where we have defined $N_a \eqdef g_{am} N^m$.
Second, using~(\ref{decomp-metric-simplified}), we reach the wanted expression
\begin{equation} \label{4-metric-decomposition}
 {}^4g_{\alpha \beta}=
 \left(
 \begin{array}{c|c}
  -N^2+g^{ij} N_i N_j	& N_b	\\ \hline
  N_a	& g_{ab}
 \end{array}
 \right) \,,
\end{equation}
which is the well-known decomposition of the four-dimensional Lorentzian metric ${}^4 g$ in terms of the three-dimensional metric $g$, of the lapse $N$, and of the shift $\vect{N}$.
From the above expression, we can compute the inverse metric
\begin{equation} \label{4-inverse-metric-decomposition}
 {}^4g^{\alpha \beta}=
 \left(
 \begin{array}{c|c}
  -1/N^2	& N^b/N^2	\\ \hline\\[-11pt]
  N^a/N^2	& g^{ab} - N^a N^b / N^2
 \end{array}
 \right)
\end{equation}
and infer the relation $\det {}^4 g = - N^2 \det g$ for the determinant of the metric.
From both these expression, we see that the lapse $N$ must be everywhere non-zero.

Actually, the fact that $N \ne 0$ is a consequence of the fact that $(e_t)_{t \in I}$ is a foliation.
In order to see this, let us first note that the lapse and the shift have a nice geometrical interpretation.
Working in foliation-induced coordinates,let us consider a freely-falling observer initially located at the point $(t,\vect{x}) \in M$ with initial four-velocity perpendicular to the hypersurface $\Sigma_t$ and let us say that it is described by a curve $\gamma (\tau)$ parametrised using the proper time $\tau$.
The four-velocity of an observer must be a time-like future-directed vector and, in the proper-time parametrisation, it must also be unit in module.
Therefore, the initial four-velocity must coincide with the vector $\overline{n} (t, \vect{x})$.
Thus, the initial position and four-velocity of the observer are
\[
 \gamma(0) = (t,\vect{x})
 \qquad \text{and} \qquad
 \dot \gamma (0) = \overline{n} (t, \vect{x}) =
 \left( \frac{1}{N_t (\vect{x})} \; , \;
 -\frac{\vect{N}_t (\vect{x})}{N_t (\vect{x})} \right) \,.
\]
After an infinitesimal proper time $\Delta \tau$, the new position of the observer is
\[
 \gamma(\Delta \tau) =
 \left(
 t +\frac{\Delta \tau}{N_t (\vect{x})} \; , \;
 \vect{x} -\frac{\Delta \tau \vect{N}_t (\vect{x})}{N_t (\vect{x})}
 \right) \,,
\]
which belongs to the hypersurface $\Sigma_{t + \Delta t}$, being $\Delta t = \Delta \tau / N_t (\vect{x})$.
Hence, we see that $N_t (\vect{x}) \Delta t$ is the proper time needed by an observer whose initial velocity is perpendicular to the hypersurface $\Sigma_t$ to reach the hypersurface $\Sigma_{t + \Delta t}$.
In addition, we also see that, if our observer wanted to land at the point of $\Sigma_{t + \Delta t}$ labelled by the same spatial coordinate $\vect{x}$ of its initial position, he would need to travel beforehand on $\Sigma_t$ with a displacement $\vect{N}_t (\vect{x}) \Delta t$.\footnote{
This displacement has to be understood in a mathematical way, since it is impossible for a physical observer to move on a  space-like hypersurface.
}

The relation $\Delta \tau = N_t (\vect{x}) \Delta t$, let us infer two further pieces of information.
First, if $N > 0$, a positive increment of the parameter $t$ corresponds to a positive increment of the proper time $\tau$.
Now, the parameter $t$, despite having being called ``time'' is actually a mere label used in the definition of a foliation and it is \emph{not}, in general, the time measured by some physical clock, although there is a simple relation $\Delta \tau = N_t (\vect{x}) \Delta t$ between the parameter $t$ and the time measured by the specific observer described above.
When the lapse is positive, the parameter $t$ and the time measured by any physical clock are increasing simultaneously, although possibly by a different rate.
As a consequence, for instance, if $t_2 > t_1$, the hypersurface $\Sigma_{t_2}$ is in the causal future of the hypersurface $\Sigma_{t_1}$.
From now on, we will assume without loss of generality that every foliation is such that $N>0$.

\begin{figure}[t]
\centering
\begin{tikzpicture}[scale=.70]
   \coordinate[label=below:{\footnotesize $x$}] (x) at (0,0);
   \coordinate (L) at (-6, 0);
   \coordinate[label=right:$\Sigma_t$] (R) at (6, 0);
   \coordinate[label=above:{\footnotesize $x''$}] (x2) at (3.5,3);
   \coordinate (L2) at (-6,2.5);
   \coordinate[label=right:$\Sigma_{t+\Delta t}$] (R2) at (6,2.5);
   \coordinate[label=above left:{}] (A) at (x -| x2);
%
   \draw (L) -- (R);
   \arcThroughThreePoints{R2}{x2}{L2};
   \fill[black] (x) circle(3pt);
   \fill[black] (x2) circle(3pt);
   \draw[->,thick=2pt,>=latex] (x) -- node[below] {\footnotesize ${}^4 \vect{N} \Delta t$} (A);
   \draw[->,thick=2pt,>=latex] (A) -- node[right] {\footnotesize $\overline{n} N \Delta t$}  (x2);
   \draw[->,thick=2pt,>=latex] (x) -- node[above,sloped] {\footnotesize $\overline{\textrm{N}} \Delta t$} (x2);
   \draw[dotted] (x) -- node[above,pos=1] {\footnotesize $x'$} (0,3.25);
   \fill[black] (0,3.25) circle(3pt);
   \draw ($(A)+(-0.4,0)$) -- ($(A)+(-0.4,0.4)$) -- ($(A)+(0,0.4)$);
   \draw ($(x)+(0.4,0)$) -- ($(x)+(0.4,0.4)$) -- ($(x)+(0,0.4)$);
\end{tikzpicture}
  \caption{
  The geometrical interpretation of the lapse and the shift.
  An observer at the position $x = (t,\vect{x}) \in \Sigma_t$ with initial four-velocity perpendicular to $\Sigma_t$ would reach the hypersurface $\Sigma_{t + \Delta t}$ at the point $x' = (t + \Delta t, \vect{x} + \Delta \vect{x})$ in an infinitesimal proper-time interval ${}^4 N \Delta t = N \Delta t$.
  In order to reach the point $x'' = (t+\Delta t, \vect{x})$ on the second hypersurface $\Sigma_{t + \Delta t}$, he would first need to travel along the first hypersurface $\Sigma_{t}$ with a displacement ${}^4 \vect{N} \Delta t$ and then move perpendicularly to $\Sigma_t$ for a proper-time interval $N \Delta t$.
  Equivalently, $x''$ could have been reached from $x$ by moving along $\overline{\textrm{N}} \Delta t$, since $\overline{\textrm{N}}$ is the tangent vector to the curve $\gamma_{\vect{x}} (t) = (t,\vect{x})$ at fixed $\vect{x}$.
  We also recognise ${}^4 \vect{N}$ and $\overline{n} N$ as the horizontal and vertical components of $\overline{\textrm{N}}$, respectively.
  See also Figure~21.2 of~\cite{Gravitation} and the discussion there for further details.
  }
  \label{fig:lapse-shift}
\end{figure}
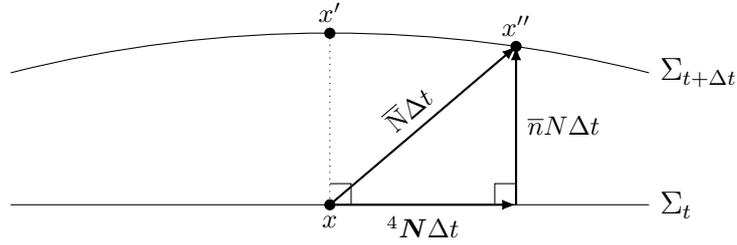

Second, let us take the limit $N_t (\vect{x}) \rightarrow 0$ for some $t$ and $\vect{x}$.
In this case, a finite increment $\Delta t > 0$ in the parameter $t$ corresponds to a proper-time increment $\Delta \tau = 0$.
But $\Delta t > 0$ means that we are moving from the hypersurface $\Sigma_t = e_t (\Sigma)$ to the hypersurface $\Sigma_{t + \Delta t} = e_{t + \Delta t} (\Sigma)$, while $\Delta \tau = 0$ means that our observer is not physically moving from its initial position.
Thus, in this case, we would have two hypersurfaces, $\Sigma_t $ and $\Sigma_{t + \Delta t}$ sharing at least a point or, in other words, $(e_t)_{t_I}$ would not be a foliation.
Thus, we see that a foliation needs to satisfy $N \ne 0$, while a generic one-parameter family of embeddings $(e_t)_{t \in I}$ does not.
It is good to remember this fact since, in the following discussion, we will want to consider also one-parameter families of embeddings that are not foliations.

Let us now turn our attention to the dynamics of the $(3+1)$-decomposed fields, which will lead us to the Hamiltonian formulation of the relativistic field theories.

\section{Dynamics of the fields}

After having performed the $3+1$ decomposition, the strategy to go from the Lagrangian to the Hamiltonian formulation of relativistic field theories is, at least conceptually, quite simple.
Let us begin from the action~(\ref{spacetime-action-general}), written as the integral over the spacetime $M$ of a Lagrangian density $\mathscr{L} [{}^4\phi; {}^4 g]$.\footnote{
The dependence of $\mathscr{L}$ on the derivatives of the fields $\phi$ is not written explicitly for simplicity.
}
Given a foliation $(e_t)_{t \in I}$, we can write the action in terms of the normal and parallel components of the fields, obtaining
\begin{equation} \label{action-3+1-begin}
\begin{aligned}
 S[\phi;{}^4 g] ={}& \int_M d^4 x \, \mathscr{L} \big( {}^4 \phi(x) ; {}^4 g (x) \big) = \\
 ={}& \int_{I} dt \int_{\Sigma_t} d^3 x \, \mathscr{L} \big( {}^4 \phi^{\perp \cdots} (x),{}^4 \phi^{\parallel \cdots} (x) ; {}^4 g_{\parallel \parallel} (x) , {}^4 N (x) , {}^4 \vect{N} (x) \big) \,,
\end{aligned}
\end{equation}
where we have split the integral over $M$ into integrals over the spatial slices $\{ \Sigma_t \}_{t \in I}$.
In the case in which the fields $\phi$ contains higher-rank tensors, all the mixed components, other than the purely-normal and purely-parallel ones, will in general appear in the decomposition and this fact is reminded by the ellipses on the superscripts of ${}^4 \phi$ in the second line of the expression above.

The expression above for the action can be equivalently written in terms of the (one-parameter families of) fields living on the space manifold $\Sigma$.
Indeed, due to the properties of the pull-back, it follows immediately that
\begin{equation} \label{action-3+1-end}
 S[\phi;{}^4 g] = \int_{I} dt \int_{\Sigma} d^3 \vect{x} \, \mathscr{L} \big( \phi^{\perp \cdots} (\vect{x}), \phi^{\parallel \cdots} (\vect{x}) ; g (\vect{x}) , N (\vect{x}) , \vect{N} (\vect{x}) \big) \,,
\end{equation}
where we have omitted the label $t$ on the right-hand side and, with an abuse of notation, we have denoted with the same symbol the (spacetime) Lagrangian density $\mathscr{L} (x)$ appearing in~(\ref{action-3+1-begin}) --- which is a local function of the four-dimensional fields and, thus, depends on the spacetime points $x \in M$ --- and the (spatial) Lagrangian density $\mathscr{L} (\vect{x})$ appearing in~(\ref{action-3+1-end}) --- which is a local function of the (one-parameter families of) fields on $\Sigma$ and, thus, depends on the spatial point $\vect{x} \in \Sigma$.
The relation between the two is obvious.
Note that the Lagrangian density $\mathscr{L} (\vect{x})$ depends parametrically on the three-dimensional metric $g$, on the lapse $N$, and on the shift $\vect{N}$.
This situation is slightly different in the case of General Relativity, since the geometry is not any more a parameter but a true degree of freedom.
We will briefly comment about this case in section~\ref{sec:GR}

The action written in the form~(\ref{action-3+1-end}) is precisely what we need to set up the machinery of section~\ref{sec:Hamiltonian-inf-dim} starting from the Lagrangian
\begin{equation} \label{Lagrangian3+1}
 L[\phi^{\perp \cdots}, \phi^{\parallel \cdots}; g,N,\vect{N}] = \int_{\Sigma} d^3 \vect{x} \, \mathscr{L} \big( \phi^{\perp \cdots} (\vect{x}), \phi^{\parallel \cdots} (\vect{x}) ; g (\vect{x}) , N (\vect{x}) , \vect{N} (\vect{x}) \big) \,,
\end{equation}
where now the expression above depend on fields on $\Sigma$ and \emph{not} on one-parameter families of fields.
Indeed, the one-parameter families of fields corresponds to curves in the configuration space and are needed when writing the action functional~(\ref{action-3+1-end}) from the Lagrangian~(\ref{Lagrangian3+1}).\footnote{
Compare with the situation in classical mechanics of section~\ref{sec:least-action-Lagrangian-CM}, where $L$ is a function of $q$ and $\dot q$, while $S$ is a function of the curve $q(t)$.
}
In general, when considering the variation, it is necessary to limit the first integral of~(\ref{action-3+1-end}) to a finite, close time interval $[a,b] \subset I$.
This is equivalent to restrict the original spacetime action~(\ref{action-3+1-begin}) to an integral on the spacetime region contained between the hypersurfaces $\Sigma_a$ and $\Sigma_b$, i.e., on a spacetime sandwich.

If the Lagrangian~(\ref{Lagrangian3+1}) is hyper-regular, then the methods of section~\ref{sec:Hamiltonian-inf-dim} returns us a phase space $Z = T^* Q$ equipped with the canonical symplectic form
\begin{equation} \label{symplectic-form-field-theories}
 \Omega = \int_\Sigma d^3 \vect{x} \, \extderphase \pi_{\cdots} (\vect{x}) \wedge \extderphase \phi^{\cdots} (\vect{x})
\end{equation}
written in terms of the fields simply denoted with $\phi^{\cdots}$ and on their conjugated momenta $\pi_{\cdots}$.\footnote{
Explicitly, the conjugated momenta can be found as $\pi_{\cdots} = \delta L / \delta \dot \phi^{\cdots}$.
Thus, the easiest way to compute the momenta is to consider the variation of the Lagrangian $\delta L$ and focus on the terms containing $\delta \dot \phi^{\cdots}$, which may appear also with a finite number of spatial derivatives.
After integrating by parts to move the spatial derivatives away from $\delta \dot \phi^{\cdots}$ and collecting together the similar terms, we can directly read the momenta from the coefficient in front of $\delta \dot \phi^{\cdots}$.
Note that, due to the integration by parts, there may be some boundary terms depending on the asymptotic part of $\delta \dot \phi^{\cdots}$.
If this happen, one has to include a boundary in the canonical symplectic form, so that the asymptotic part of $\delta \dot \phi^{\cdots}$ has the momentum given by the boundary term of $\delta L$.
In practice, we will neglect the boundary terms when deriving the Hamiltonian formulation and introduce them back later in the formalism, as we shall see in detail in the specific situations discussed in this thesis.
However, in section~\ref{subsec:Lorentz-Lorenz-free-ED}, we will see an explicit example of how boundary terms in $\delta L$ can lead to boundary terms in the symplectic form.
}
In addition, we obtain a Hamiltonian
$H [\phi^{\cdots},\pi_{\cdots} ; g, N , \vect{N}]$, which, other than on the fields and their conjugated momenta, also depends parametrically on the lapse, on the shift, and on the three-dimensional metric.
Specifically, the Hamiltonian takes the form\footnote{
See e.g.~\cite{Kuchar:hyperspace3} for more details.
}
\begin{equation} \label{Hamiltonian-lapse-shift}
 H [\phi^{\cdots},\pi^{\cdots} ; g, N , \vect{N}] = 
 \int_\Sigma d^3 \vect{x} \Big[
 N (\vect{x}) \mathscr{H} (\vect{x}) +
 N^m (\vect{x}) \mathscr{H}_m (\vect{x})
 \Big] \,,
\end{equation}
where the functions $\mathscr{H} (\vect{x})$ and $\mathscr{H}_m (\vect{x})$ are built from the fields $\phi^{\cdots}$, their conjugated momenta $\pi^{\cdots}$, and the three-dimensional metric $g$, \emph{but not} from the lapse and the shift.
This fact will have some non-trivial consequences, as we shall see.

Before we continue, it is necessary to make some remarks about boundary terms.
So far, the Lagrangian~(\ref{Lagrangian3+1}), the symplectic form~(\ref{symplectic-form-field-theories}), and the Hamiltonian~(\ref{Hamiltonian-lapse-shift}) were all written as integrals over the space manifold $\Sigma$.
This situation does not suffice to describe the theories which we wish to analyse and we will need to complement the given expression with some boundary terms, i.e., with some integrals over $\partial \Sigma$.
Note that $\partial \Sigma$ will actually be a boundary at infinity in the cases discussed in this thesis.
This has to be interpreted as follows.
The space manifold $\Sigma$ will be (at least asymptotically) equivalent to $\real^3$.
If we replace (at least asymptotically) $\Sigma \sim \real^3$ with \mbox{$\Sigma_R \sim B_R$} --- being $B_R$ a ball of sufficiently-large, finite radius $R$ --- we obtain a manifold with a true boundary $\partial \Sigma_R \sim S^2_R$, i.e., the surface of a two-sphere of radius $R$.
At this point, we can derive the wished expressions (such as the symplectic form, the Hamiltonian, or the generators of some symmetries) working on the manifold $\Sigma_R$ with boundary $\partial \Sigma_R$ and, only at the end, take the limit $R \rightarrow \infty$.
As we shall see in the explicit computations of the next sections and of the next chapters, some of the boundary terms will not vanish in this limit and, hence, need to be properly taken into account in the correct formulation of the theory.

Independently of whether $\partial \Sigma$ is an actual boundary or a boundary at infinity, the three expressions for the Lagrangian~(\ref{Lagrangian3+1}), for the symplectic form~(\ref{symplectic-form-field-theories}), and for  the Hamiltonian~(\ref{Hamiltonian-lapse-shift}) need to be complement, in general, with boundary terms.
The strategy which is usually followed at this point works as follows.
First, one neglects the possible presence of boundary terms in the Lagrangian, in the symplectic form, and in the Hamiltonian.
Then, one checks if the so-obtained Hamiltonian theory is well-defined and has the wished properties.
Specifically, in our case, we will check that the Hamiltonian admits the existence of a vector field $X_H$ and of a canonical generator of the Poincar\'e transformations~(see the next section).
If this is not the case, one usually tries to add boundary terms to the Hamiltonian~(\ref{Hamiltonian-lapse-shift}), but sometimes also to the symplectic form~(\ref{symplectic-form-field-theories}), in order to find a well-defined Hamiltonian theory.
We shall see explicit examples of this in the next sections and chapters.

Note, that the addition of boundary terms to the symplectic form and to the Hamiltonian corresponds to the addition of some boundary term to the Lagrangian~(\ref{Lagrangian3+1}) and, as a consequence, to the spacetime action~(\ref{spacetime-action-general}).
Since we are working on a $(3+1)$-decomposed globally-hyperbolic spacetime, the boundary term in the spacetime action would be eventually an integral over $\real \times \partial \Sigma$.
In this regards, already in the Lagrangian formulation of electrodynamics, the need for a boundary term was noted~(see e.g. \cite{Campiglia-U1}), although the boundary term was an integral over the hyperboloid at infinity in that case.
This shows that the appearance of boundary term is a general feature of (long-ranging) field theories and is not merely limited to the Hamiltonian formulation.

Finally, let us compute the equations of motion ensuing from the symplectic form~(\ref{symplectic-form-field-theories}) and the Hamiltonian~(\ref{Hamiltonian-lapse-shift}).
To this end, we need to find the vector field
$X_H = \big( \delta_H \phi^{\cdots} (\vect{x}),
\delta_H \pi_{\cdots} (\vect{x}) \big)$
satisfying $\extderphase H = - \insertion_{X_H} \Omega$.
Then, the equations of motion are simply
$\dot \phi^{\cdots} (\vect{x}) = \delta_H \phi^{\cdots} (\vect{x})$ and
$\dot \pi_{\cdots} (\vect{x}) = \delta_H \pi_{\cdots} (\vect{x})$.
So, let us first note that
\begin{equation} \label{i-Omega-general}
 - \insertion_{X_H} \Omega = \int_\Sigma d^3 \vect{x} \Big[
 \delta_H \phi^{\cdots} (\vect{x}) \, \extderphase \pi_{\cdots} (\vect{x})
 -\delta_H \pi_{\cdots} (\vect{x}) \, \extderphase \phi^{\cdots} (\vect{x})
 \Big]
\end{equation}
if the symplectic form~(\ref{symplectic-form-field-theories}) does not contain any boundary term, which we are going to assume for now, since it is a common situation.
The expression above must be equal to $\extderphase H$, which can be written in general as
\begin{equation} \label{dH-general}
 \extderphase H = \int_\Sigma d^3 \vect{x} \Big[
 A_{\cdots} (\vect{x}) \extderphase \phi^{\cdots} (\vect{x}) +
 B^{\cdots} (\vect{x}) \extderphase \pi_{\cdots} (\vect{x})
 \Big]
 + \oint_{\partial \Sigma} d^2 \barr{x} \, C (\barr{x}) \,,
\end{equation}
where $A_{\cdots}(\vect{x})$ and $B^{\cdots}(\vect{x})$ are written in terms of the canonical fields, the lapse and the shift.
As we shall see in the explicit examples provided in the next sections, the reason why we obtain the expression above is that, after that we have computed the $\extderphase$ of the integrand, we usually end up with terms containing partial derivatives, e.g. $A_1(\vect{x}) \partial^{k} \big(\extderphase \phi^{\cdots} (\vect{x})\big)$.
Integrating by parts this expression, we get a contribution $(-1)^k  \partial^{k} A_1(\vect{x}) \,\extderphase \phi^{\cdots} (\vect{x})$ to the integral on the bulk and, in general, a boundary term.
The former, together with the other terms of this kind, form the first summand in the square brackets of the expression above, while the latter contributes to the boundary term $C(\barr{x})$.
Thus, we see that, due to the need of integrating by parts, $\extderphase H$ contains in general non-vanishing boundary terms, even if $H$ does not.
However, comparing~(\ref{i-Omega-general}) and~(\ref{dH-general}), we see that the equation $\extderphase H = - \insertion_{X_H} \Omega$ can be satisfied if, and only if, $C(\barr{x})$ vanishes upon integration on $\partial \Sigma$.

We say that the Hamiltonian $H$ is \emph{differentiable \`a la Regge-Teitelboim} if~(\ref{dH-general}) does not contain any boundary term, i.e., if $C(\barr{x})$ vanishes upon the integration on $\partial \Sigma$.
This condition ensures the existence of a vector field $X_H$, whose components are explicitly found to be
\begin{equation}
\delta_H \phi^{\cdots} (\vect{x}) = B^{\cdots} (\vect{x})
\qquad \text{and} \qquad
\delta_H \pi_{\cdots} (\vect{x}) = - A_{\cdots} (\vect{x})
\end{equation}
from the direct comparison between~(\ref{i-Omega-general}) and~(\ref{dH-general}), if the symplectic form is exactly as in~(\ref{symplectic-form-field-theories}) \emph{without} boundary terms.
In this case, the equations of motion are simply
\begin{equation} \label{eoms-field-theory}
\dot \phi^{\cdots} (\vect{x}) = B^{\cdots} (\vect{x})
\qquad \text{and} \qquad
\dot \pi_{\cdots} (\vect{x}) = - A_{\cdots} (\vect{x})
\end{equation}
where $A_{\cdots}(\vect{x})$ and $B^{\cdots}(\vect{x})$ are written in terms of the canonical fields, the lapse and the shift.

Let us mention two common situations in which the boundary term in $\extderphase H$ vanishes.
The former is the case in which $\Sigma$ is a closed manifold, i.e., it is compact and without a boundary.
Therefore, $\Sigma$ neither has an actual boundary, nor an asymptotic one.
Despite this situation does not apply to the cases of our interest, it is still worth mentioning that it is a common trick used in the literature, including in many of the papers which we have cited, in order to avoid issues with the boundary terms.
The latter situation is the case discussed e.g. in the seminal paper by Regge and Teitelboim~\cite{RT}, from which the name ``differentiability \`a la Regge-Teitelboim'' follows.
In this case, the canonical fields,\footnote{
In~\cite{RT}, the authors analysed the case of asymptotically-flat spacetimes using the Hamiltonian formulation of General Relativity and we will briefly mention their findings in section~\ref{sec:GR}.
Nevertheless, the same ideas can be applied to other fields theories with long-ranging interactions.
} the lapse, and the shift are required to satisfy some fall-off and parity conditions in a neighbourhood of the asymptotic boundary $\partial \Sigma$.
Then, either the boundary term in~(\ref{dH-general}) vanishes directly due to this conditions or it can be written as a total derivative, i.e.,
\[
 \oint_{\partial \Sigma} d^2 \barr{x} \, C (\barr{x}) =
 - \extderphase \oint_{\partial \Sigma} d^2 \barr{x} \, \mathscr{C} (\barr{x}) \,.
\]
Thus, redefining the Hamiltonian as
\begin{equation}
 H' [\phi^{\cdots},\pi^{\cdots} ; g, N , \vect{N}] = 
 H [\phi^{\cdots},\pi^{\cdots} ; g, N , \vect{N}]
 + \oint_{\partial \Sigma} d^2 \barr{x} \, \mathscr{C} (\barr{x})\,,
\end{equation}
we see that $H'$ is differentiable \`a la Regge-Teitelboim.

Finally, if the symplectic form~(\ref{symplectic-form-field-theories}) needs to be complemented with some boundary terms, the differentiability \`a la Regge-Teitelboim does not guarantees the existence of the vector field $X_H$.
One common strategy in this case is to use the equation $\extderphase H = - \insertion_{X_H} \Omega$ first neglecting all the boundary terms.
In this way, we find a candidate for the vector field $X_H$, which satisfies the general expression up to boundary terms.
Then, inserting the candidate $X_H$ into the general expression $ - \insertion_{X_H} \Omega$ now including boundary terms, we can check whether or not it equates $\extderphase H$, eventually adding a boundary term to $H$.
We will discuss this situation in chapter~\ref{cha:scalar-electrodynamics} for instance.

In this subsection, we have discussed how to obtain the equations of motion of a $(3+1)$-decomposed field theory, thus finding the time evolution of the canonical fields.
This does not suffices for the purposes of this thesis.
Indeed, other than the behaviour of the fields under time evolution, we will need to know also their behaviour under the Poincar\'e transformations.
So, let us turn our attention to this topic.

\section{Poincar\'e transformations} \label{sec:Poincare-general}

The Poincar\'e transformations play an important role in relativistic field theories on a flat Minkowski background and on asymptotically-flat spacetimes, in the latter case appearing as asymptotic transformations.
Notably, one of the requirements which we impose when setting up the Hamiltonian formulation of these field theories is that the Poincar\'e transformations are present as canonical symmetries, as defined in section~\ref{subsec:symmetries-general}.



\subsection{One-parameter families of embeddings}
We restrict our analysis to the case of field theories on a flat Minkowski spacetime and only mention how to generalise the results to the asymptotically-flat case.
Thus, we can assume that the spacetime manifold $M = \real^4$, the space manifold $\Sigma = \real^3$, and that the spacetime metric ${}^4 g = \text{diag} (-1,1,1,1)$ in some global chart of normal coordinates.
In addition, we complement the space manifold with an asymptotic boundary $\partial \Sigma = S^2_{\infty}$.
For better clarity, let us fix some coordinates $(\vect{x}^a)$ on $\Sigma$ and $(x^\alpha)$ on $M$, which do not need to be in any particular relation, contrary to the foliation-induced coordinates used in section~\ref{sec:3+1}.
We could assume for simplicity that $(x^\alpha)$ are global normal coordinates on $M$ such that the four-metric takes the simple form ${}^4 g = \text{diag} (-1,1,1,1)$, although we would make use also of radial coordinates at some point in the following discussion.

In order treat the Poincar\'e transformations in the Hamiltonian framework, let us consider them as transformations acting on hypersurfaces by means of one-parameter families of embeddings.
Thus, let us consider the one-parameter family of embeddings $(e_\lambda)_{\lambda \in I}$, such that the open interval $I \supset [0,1]$.
For simplicity, let us further assume that $\Sigma_0 \eqdef e_0 (\Sigma)$ consists of the hyperplane in $M$ determined by the condition $x^0 = 0$.
A unit time translation of $\Sigma_0$, for instance, would be described by the one-parameter family $(e_\lambda)_{\lambda \in I}$, whose expression in coordinates $(x^\alpha)$ is
\begin{equation} \label{time-translation-unit}
 \big[ e_\lambda (\vect{x}) \big]^0 = \lambda
 \qquad \text{and} \qquad
 \big[ e_\lambda (\vect{x}) \big]^a = \vect{x}^a \,.
\end{equation}
As the embedding parameter $\lambda$ varies continuously from $0$ to $1$, these family of embeddings transform the hypersurface $\Sigma_0$ continuously to the hypersurface $\Sigma_1$, where the former is describe by the condition that the ``time'' $x^0 = 0$ and the latter by $x^0 = 1$, so that the meaning of unit time translation becomes clear. 
One can easily verify that, if we extend the interval $I$ to the entire real numbers, the family $(e_\lambda)_{\lambda \in \real}$ is actually a foliation and that the chosen coordinates are foliation induced.

However, this is not the case for the other Poincar\'e transformations: spatial translations, rotations, and Lorentz boosts.
For instance, spatial translations and spatial rotations map points of the hypersurface $\Sigma_0$ to (other) points of the very same hypersurface.
It is clear that these cannot be described by a foliation and, as a matter of facts, neither can the Lorentz boosts.
Therefore, we need to generalise some facts presented in section~\ref{sec:3+1} to the case in which $(e_\lambda)_{\lambda \in I}$ is \emph{not} a foliation.
To this end, let us follow the general strategy pursued by Kucha\v{r}~\cite{Kuchar:hyperspace1,Kuchar:hyperspace2,Kuchar:hyperspace3} and define $\mathscr{E}$ as the space of all embeddings $e \colon \Sigma \hookrightarrow M$.
In this language, a foliation is a special curve on $\mathscr{E}$, but more general curves can be considered as well.\footnote{
In~\cite{Kuchar:hyperspace1,Kuchar:hyperspace2,Kuchar:hyperspace3}, Kucha\v{r} introduces also the \emph{hyperspace} $\mathcal{H}$ as the quotient of $\mathscr{E}$ with respect to the equivalence relation $e_1 \equiv e_2 \iff e_1 (\Sigma)= e_2 (\Sigma)$, i.e., two embeddings are equivalent if they map $\Sigma$ to the same hypersurface in $M$.
For this reason, $\mathcal{H}$ is the space of hypersurfaces in $M$.
In this thesis, we will not discuss all the technicalities about Kucha\v{r}'s very-important analysis, since they go beyond the purpose of this thesis.
Nevertheless, all the details can be found in the already-mentioned  papers. 
}

Now, if a generic path $(e_\lambda)_{\lambda \in I}$, which is not a foliation, is considered in $\mathscr{E}$ many equations discussed in section~\ref{sec:3+1} become problematic.
Indeed, for instance, the expression for the inverse four-metric~(\ref{4-inverse-metric-decomposition}) would be divergent if the lapse $N = 0$ and so would be, in general, the expression for the Lagrangian~(\ref{Lagrangian3+1}).
These issues are mostly related to the fact that those expressions are derived in foliation-induced coordinates, which are not well-defined if $(e_\lambda)_{\lambda \in I}$ is not a foliation.
Nevertheless, the symplectic form~(\ref{symplectic-form-field-theories}) and the Hamiltonian~(\ref{Hamiltonian-lapse-shift}) would be well-defined, non-divergent quantities, written in terms of the canonical fields $(\phi^{\cdots},\pi_{\cdots})$ living on $\Sigma$, on the Riemannian metric $g$ (which is well-defined even for a single embedding), on the lapse and the shift (on which the Hamiltonian density depends linearly).
Furthermore, also the action --- when defined in the Hamiltonian formulation --- is well-defined, leading to the concept of \emph{hyperspace action} discussed in~\cite{Kuchar:hyperspace3}.
We will briefly mention this in subsection~\ref{subsec:Poincare-algebra}.

The equations of motion~(\ref{eoms-field-theory}), too, would be well-defined.
To see this, let us remind that~(\ref{eoms-field-theory}) depends 
on the lapse $N$ and the shift $\vect{N}$, since $A_{\cdots} (\vect{x})$ and $B^{\cdots} (\vect{x})$ depend on them.
Let us write $A_{\cdots} [N,\vect{N}]$ and $B^{\cdots} [N,\vect{N}]$ to stress this point.
Then, from the Hamiltonian~(\ref{Hamiltonian-lapse-shift}), it follows that $A_{\cdots} [N,\vect{N}]$ and $B^{\cdots} [N,\vect{N}]$ depends linearly on $N$, on $\vect{N}$, and on a finite number of their spatial derivatives.\footnote{
For the cases which we will consider, the spatial derivatives will be only of first order.
}
As a consequence, we do not have any divergence even if $N=0$ in some region or everywhere.

The transformation of the canonical fields under the one-parameter family of embeddings $(e_\lambda)_{\lambda \in I}$ is then given by the equations of motion~(\ref{eoms-field-theory}), where the lapse and the lapse and the shift are the one associated to a generic family of embeddings and not necessarily to a foliation.
Therefore, in order to find the Poincar\'e transformations of the canonical fields, we need to find the lapse and shift associated to them.

\subsection{Poincar\'e transformations of the fields}

The lapse and shift associated to a unit time translation can be found quite trivially to be $N = 1$ and $\vect{N} = 0$ by applying the results of section~\ref{sec:3+1} to the specific foliation~(\ref{time-translation-unit}).
Fort this reason, in the following chapters, we will restrict to this choice of lapse and shift when we will be interested merely in determining the behaviour of the fields under time evolution.
Let us derive, as an example, the lapse and shift associated to a Lorentz boost.
The derivation for the ones of a generic Poincar\'e transformations is along the same lines and, thus, we will present only the final result.

For simplicity, let us consider a boost along the $x^1$-axis starting from the hyperplane $\Sigma_0$ of points satisfying $x^0 = 0$ as in the previous subsection.
In other words, the transformation is give in terms of the one-parameter family of embeddings $(e_\lambda)_{\lambda \in I}$, whose expressions in coordinates $(x^\alpha)$ is
\begin{subequations} \label{boost-along-x1}
\begin{align}
 \big[ e_\lambda (\vect{x}) \big]^0 &= -\vect{x}^1 \sinh (\lambda b_1) \,,   &
 \big[ e_\lambda (\vect{x}) \big]^1 &= \vect{x}^1 \cosh (\lambda b_1) \,, \\
 \big[ e_\lambda (\vect{x}) \big]^2 &= \vect{x}^2 \,,
 \qquad \text{and}
 &
 \big[ e_\lambda (\vect{x}) \big]^3 &= \vect{x}^3 \,,
\end{align} 
\end{subequations}
where $b_1 \in \real$ is the boost parameter and the usual part of the boost containing time is absent since we are boosting the hyperplane at $x^0 = 0$.
The embedding parameter $\lambda$ is required to vary at least in an open interval $I \supset [0,1]$, but we see from the expression above that there is no harm to extend its range to all the real numbers.

To begin with, let us note that, fixing $\lambda$, the hyperplane $\Sigma_0$ is mapped to the hypersurface $\Sigma_\lambda$, which is again a hyperplane and, more precisely, the one described by the equation $x^0 = - x^1 \tanh (\lambda b_1)$.
Thus, we see that the hyperplane $\Sigma_0$ is tilted under the action of a boost.
Note that the tilting angle is, in absolute value, always lesser than $45^\circ$.

At this point, it is not difficult to find the covector $\underline{n}_\lambda (x)$ normal to the hypersurface $\Sigma_\lambda$ at the point $x = e_\lambda(\vect{x}) \in \Sigma_\lambda$.
Explicitly, one finds
\begin{equation} \label{boost-normal-covector}
 \underline{n}_\lambda \big( e_\lambda(\vect{x}) \big) =
 - \cosh (\lambda b_1) d x^0
 - \sinh (\lambda b_1) d x^1 \,.
\end{equation}
Note that we have included a subscript $\lambda$ since the covector normal to the hypersurface $\Sigma_\lambda$ cannot be extended to a one-form on $M$, due to the fact that $(e_\lambda)_{\lambda \in \real}$ is not a foliation.
As a consequence, the respective vector
\begin{equation} \label{boost-normal-vector}
 \overline{n}_\lambda \big( e_\lambda(\vect{x}) \big) =
 \cosh (\lambda b_1) \frac{\partial}{\partial x^0}
 - \sinh (\lambda b_1) \frac{\partial}{\partial x^1}
\end{equation}
does not extend to a vector field on $M$ as well.
Similarly, $\overline{\mathrm{N}}$ defined in~(\ref{definition-N-vf}) can be computed at each point in $\Sigma_\lambda$ but not globally, thus finding
\begin{equation} \label{boost-N-vector}
 \overline{\mathrm{N}}_\lambda \big( e_\lambda(\vect{x}) \big) =
 - \vect{x}^1 b_1 \cosh (\lambda b_1) \frac{\partial}{\partial x^0}
 + \vect{x}^1 b_1 \sinh (\lambda b_1) \frac{\partial}{\partial x^1}
\end{equation}

Finally, decomposing~(\ref{boost-N-vector}) according to~(\ref{N-vf-decomp}) by means of~(\ref{boost-normal-covector}) and taking the pull-back on $\Sigma$, we find the lapse
\begin{equation}
 N (\vect{x}) = - (e_\lambda)^* \big( \underline{n} \cdot \overline{\mathrm{N}} \big) 
 = - b_1 \vect{x}^1 \,. 
\end{equation}
Note that this lapse vanishes on the plane defined by the equation $\vect{x}^1 = 0$, which implies that $(e_\lambda)_{\lambda \in \real}$ is not a foliation, as expected.
In addition, using~(\ref{boost-normal-vector}), we also find the shift
\begin{equation}
 \vect{N} = (e_\lambda)^* \big(\overline{\mathrm{N}} - N \overline{n} \big)
 = 0 \,,
\end{equation}
where the pull-back is well-defined since the vector is horizontal.
This conclude the derivation of the lapse and shift associated to a (particular) Lorentz boost.

For a generic Poincar\'e transformation, the lapse $N = \xi^\perp$ and shift $\vect{N} = \xi$, whose values in Cartesian coordinates are respectively given by the expressions
\begin{align} \label{poincare-xi-cartesian}
 \xi^\perp =a^\perp + b_i \vect{x}^i
 \qquad \text{and} \qquad
 \xi^i= a^i +\omega^{i}{}_j \vect{x}^j \,,
\end{align}
where $a^\perp$ is responsible for the time translation, $a^i$ for the spatial translations, $b_i$ for the Lorentz boost (we changed the sign for convenience with respect to the expression derived above), and the antisymmetric $\omega_{ij} \eqdef g_{i\ell}\omega^{\ell}{}_j$ for the spatial rotations.
Note that, following~\cite{Henneaux-GR,Henneaux-ED}, we have absorbed the contribution  of the boost $x^0 \,b^i$, which would appear in $\xi^i$, into the parameters $a^i$.
The reason for doing so is that these two terms have the same dependence on the radial distance in the asymptotic expansion at spatial infinity 
and this will be the relevant fact in the following discussion.


For the following discussion, it is actually more convenient to move to spherical coordinates $(x^0,r,\barr{x})$, where $\barr{x}$ are coordinates on the unit two-sphere, such as the usual $\theta$ and $\varphi$.
The flat three-metric is
\begin{equation} \label{3-metric-radial-angular}
 g_{ab}=
 \left(
 \begin{array}{c|c}
  1	& 0	\\
  \hline
  0	& r^2 \, \barr{\gamma}_{\bar a \bar b}
 \end{array}
 \right) \,,
\end{equation}
where $\barr{\gamma}_{\bar a \bar b}$ is the metric of the unit round sphere and indices with bars above, such as $\bar a$, run over the angular components.
Using these coordinates, the components of the vector field~(\ref{poincare-xi-cartesian}) corresponding to Poincar\'{e} transformations are
\begin{equation} \label{poincare-xi}
 \xi^{\perp}=r b +T \,, \qquad
 \xi^r = W \,, \qquad
 \xi^{\bar a} = Y^{\bar a} + \frac{1}{r} \barr{\gamma}^{\bar a \bar m}\, \partial_{\bar m} W \,.
\end{equation}
In the above expression, $b$, $Y^{\bar a}$, $T$, and $W$ are functions on the sphere satisfying the equations 
\begin{equation}
 \barr{\nabla}_{\bar a} \barr{\nabla}_{\bar b} W +\barr{\gamma}_{\bar a \bar b} W=0 \,, \qquad
 \barr{\nabla}_{\bar a} \barr{\nabla}_{\bar b} b +\barr{\gamma}_{\bar a \bar b} b=0 \,, \qquad
 \mathcal{L}_Y \barr{\gamma}_{\bar a \bar b}=0 \,, \qquad
 \partial_{\bar a} T=0 \,,
\end{equation}
where $\barr{\nabla}$ is the covariant derivative on the unit round two-sphere.
Moreover, $b$, $Y^{\bar a}$, $T$, and $W$ are related to the parameters $a^\perp$, $a^i$, $m^i \eqdef - \epsilon^{i j k} \omega_{jk} /2$, and $b^i$ by the expressions
\begin{subequations}
\begin{align}
 b(\theta,\varphi) &=b_1 \sin \theta \cos \varphi +b_2 \sin \theta \sin \varphi+b_3 \cos\theta \,,\\
 Y(\theta,\varphi) &=
 m_1 \left( -\sin \varphi \frac{\partial}{\partial \theta} - \frac{\cos \theta}{\sin \theta} \cos \varphi \frac{\partial}{\partial \varphi} \right)\nonumber\\
 &+m_2 \left( \cos \varphi \frac{\partial}{\partial \theta} - \frac{\cos \theta}{\sin \theta} \sin \varphi \frac{\partial}{\partial \varphi} \right)\nonumber\\
 &+m_3 \frac{\partial}{\partial \varphi} \,,\\
 W(\theta,\varphi) &=a_1 \sin \theta \cos \varphi +a_2 \sin \theta \sin \varphi+a_3 \cos\theta \,, \\
 T(\theta,\varphi) &=a^\perp \,,
\end{align}
\end{subequations}
where we have used explicitly the usual $\theta$ and $\varphi$ as angular coordinates.

Finally, in order to find the Poincar\'e transformations of the canonical fields, we only need to use the vector field $X_P = (\delta_P \phi^{\cdots},\delta_P \pi_{\cdots})$, which provides the infinitesimal transformations under the action of the Poincar\'e group and is such that
\begin{equation} \label{Poincare-fields-general}
 \delta_P \phi^{\cdots} (\vect{x}) = B^{\cdots} \big[ \xi^\perp (\vect{x}) , \vect{\xi}(\vect{x}) \big]
 \qquad \text{and} \qquad
 \delta_P \pi_{\cdots} (\vect{x}) = - A_{\cdots} \big[ \xi^\perp (\vect{x}) , \vect{\xi} (\vect{x}) \big] \,,
\end{equation}
obtained from the equations of motion~(\ref{eoms-field-theory}) by formally replacing the generic lapse $N$ and shift $\vect{N}$ with $\xi^\perp$ and $\vect{\xi}$, respectively.

Note that, if we are not on a flat Minkowski background, the Poincar\'e transformations are not expected to be symmetries, nor to be defined by the above procedure at all.
Nevertheless, in the case of asymptotically-flat spacetimes, we expect the Poincar\'e transformations to be part of the asymptotic-symmetry group of the theory, so that the equations above are expected to hold in some sense asymptotically, i.e., when the lapse and shift reduces to the one of the Poincar\'e transformations only at infinity.
We will come back to this point with more details in section~\ref{sec:GR}, when discussing the case of General Relativity.

\subsection{Poisson-representation of the Poincar\'e algebra} \label{subsec:Poincare-algebra}

Let us conclude this section by showing that the described  procedure for finding the Poincar\'e transformations leads to a Poisson-representation of the Poincar\'e Lie algebra.
To do this, let us remind that the action in Hamiltonian formulation can be defined also in the case of generic one-parameter families of embeddings, as we mentioned at the beginning of this section. 
To be completely fair, if one takes the lapse and shift of the Poincar\'e transformations some problems may emerge in the definition of the Hamiltonian $H[N,\vect{N}]$ and, thus, of the action.
Specifically, these potential issues are due to the boundary terms arising when $N-1$ and $\vect{N}$ do not vanish quickly enough at infinity, which could make $H[N,\vect{N}]$ divergent or could prevent a Hamiltonian vector field $X_H$ from existing.\footnote{
Instead of $N -1$, we could have written $N$ minus a constant.
}
But, let us neglect this problem for now (and set to zero any boundary term), as this would be part of the thorough analyses of the next sections and chapters.

The next step is to turn the (one-parameter family of) embeddings $(e_t)_{t \in I}$ into canonical variables and assign canonical momenta $p$ to them~\cite{Kuchar:hyperspace3,Kuchar-Isham1}.
Note that, in terms of coordinates $(x^\alpha)$ on $M$ and omitting to write the label $t$, the embedding $e$ can be decomposed in four functions $e^\alpha$ and, thus, the canonical momenta associated to it can be written as $p_\alpha$.
The dynamics of the system is then described by the so-called \emph{parametrised action} of the field theory
\begin{equation}
 S [\phi^{\cdots},\pi_{\cdots},e,p;N,\vect{N}] =
 \int_I dt \int_\Sigma d^3 \vect{x} \Big[
 p_\alpha \dot e^\alpha
 + \pi_{\cdots} \dot \phi^{\cdots}
 - N  \tilde{\mathscr{H}} 
 - N^m  \tilde{\mathscr{H}}_m 
 \Big] \,,
\end{equation}
where
\begin{equation}
 \tilde{\mathscr{H}} \eqdef -p_\perp + \mathscr{H}
 \qquad \text{and} \qquad
 \tilde{\mathscr{H}}_m \eqdef p_m + \mathscr{H}_m
\end{equation}
are called the \emph{super-Hamiltonian} and the \emph{super-momentum} of the parametrised theory, respectively.
The action above needs to be varied independently by the canonical fields $\phi^{\cdots}$ and $\pi_{\cdots}$, by the embedding variables $e^\alpha$ and their momenta $p_\alpha$, and by $N$ and $\vect{N}$.
Note that $N$ and $\vect{N}$ are now \emph{independent} from the embedding variables $e$.
They become dependent on them only through the equations of motion ensuing from the variation of the parametrised action and, if $e$ is a foliation, they reduce to the lapse and shift discussed in the $3+1$ decomposition.
For this reason, they are denoted with the same symbols. 
In addition, the variation with respect to $N$ and $\vect{N}$ does not provide any dynamical equation, but simply imposes the two constraints\footnote{
See section~\ref{sec:gauge-theories} for the detailed discussion about constraints and the notation.
}
\begin{equation}
 \tilde{\mathscr{H}} (\vect{x}) \approx 0
 \qquad \text{and} \qquad
 \tilde{\mathscr{H}}_m (\vect{x}) \approx 0 \,,
\end{equation}
which the super-Hamiltonian and the super-momentum need to satisfy.

One can show, as it is done in~\cite{Kuchar:hyperspace3} and in~\cite{Kuchar-Isham1}, that the super-Hamiltonian and the super-momentum satisfy the algebra
\begin{subequations} \label{algebra:super-Hamiltonian-momentum}
\begin{align}
 \label{algebra:HH}
 \big\{ \tilde{\mathscr{H}} (\vect{x}), \tilde{\mathscr{H}} (\vect{x}') \big\} ={}& - g^{ab} (\vect{x}) \, \tilde{\mathscr{H}}_a (\vect{x}) \, \partial_b \delta_{\vect{x}} (\vect{x}') - (\vect{x}\leftrightarrow \vect{x}') \\
 \label{algebra:HM}
 \big\{ \tilde{\mathscr{H}}_a (\vect{x}), \tilde{\mathscr{H}} (\vect{x}') \big\} ={}&  -\tilde{\mathscr{H}} (\vect{x}) \, \partial_a \delta_{\vect{x}} (\vect{x}') \\
 \label{algebra:MM}
 \big\{ \tilde{\mathscr{H}}_a (\vect{x}), \tilde{\mathscr{H}}_b (\vect{x}') \big\} ={}&  - \tilde{\mathscr{H}}_b (\vect{x}) \, \partial_a \delta_{\vect{x}} (\vect{x}') - (\vect{x}\leftrightarrow \vect{x}',a \leftrightarrow b)
\end{align}
\end{subequations}
where $\delta_{\vect{x}} (\vect{x}')$ is the $\delta$-distribution.\footnote{
We define the $\delta$-distribution of a point $\vect{x} \in \Sigma$, written as $\delta_{\vect{x}}$, by its action on a smooth test function $f$ as
\[
 \langle \delta_{\vect{x}} , f \rangle = \int_\Sigma d^3 \vect{x}' \, \delta_{\vect{x}} (\vect{x}') f(\vect{x}') \eqdef f(\vect{x}) \,,  
\]
where the expression in the middle merely shows the usual notation used instead of $\langle \delta_{\vect{x}} , f \rangle$.
The derivatives of the distribution are then defined as $\langle \partial_a \delta_{\vect{x}} , f \rangle \eqdef - \langle \delta_{\vect{x}} , \partial_a f \rangle$.
}
Equivalently, the algebra above could have been written in a smeared version as
\begin{equation} \label{algebra:super-Hamiltonian-momentum-smeared}
 \big\{ \tilde{H} [N_1, \vect{N}_1], \tilde{H} [N_2, \vect{N}_2] \big\} = \tilde{H} [\hat{N}, \hat{\vect{N}}]
\end{equation}
where we have defined the generator
\begin{equation}
 \tilde{H} [N, \vect{N}] \eqdef \int_{\Sigma} (N \tilde{\mathscr{H}} + N^a \tilde{\mathscr{H}}_a)
\end{equation}
and the multipliers on the right-hand-side of equation~(\ref{algebra:super-Hamiltonian-momentum-smeared}) are given by
\begin{subequations}
\begin{align}
 \hat{N} ={}& \, N_1^a \partial_a N_2 - N_2^a \partial_a N_1 
 \\
 \hat{N}^a ={}& \, g^{ab} (N_1 \partial_b N_2 - N_1 \partial_b N_2 ) + [\vect{N}_1,\vect{N}_2]^a  \,,
\end{align}
\end{subequations}
being $[\vect{N}_1,\vect{N}_2]$ the Lie-Jacobi commutator of the two vector fields.
In the case of gauge theories, as we shall see better later on, the algebra~(\ref{algebra:super-Hamiltonian-momentum}) and its smeared version~(\ref{algebra:super-Hamiltonian-momentum-smeared}) may be modified by the presence of constraints on the right-hand side, so that they hold up to gauge transformations, but this does not constitute a problem in general.
It is now a matter of mere computation to verify that the algebra~(\ref{algebra:super-Hamiltonian-momentum}) or its smeared version~(\ref{algebra:super-Hamiltonian-momentum-smeared}) reduce to the Poincar\'e algebra if we replace $N$ and $\vect{N}$, respectively, with the $\xi^\perp$ and the $\vect{\xi}$, which are given in Cartesian coordinates by~(\ref{poincare-xi-cartesian}) and in radial coordinates by~(\ref{poincare-xi}).

Thus, we have seen how to obtain the Poincar\'e transformations of the fields.
Note that, in this section, we have neglected any possible issue coming from boundary terms and, actually, we have neglected boundary terms altogether.
In fact, these will turn out to be the greatest possible obstruction to a canonical realisation of the Poincar\'e transformations.
The strategy that we will follow in this thesis is to, first, proceed as in this section neglecting the boundary terms and obtaining a candidate for the Poincar\'e transformations of the fields by means of~(\ref{Poincare-fields-general}).
Secondly, we will check whether or not these transformations are symplectic, which, in the case of the Poincar\'e transformations, is sufficient to conclude that they are also canonical.\footnote{
We will discuss this in section~\ref{sec:principles}.
}
When this is the case, the canonical generator of the Poincar\'e transformations will be $\tilde{H}[\xi,\vect{\xi}]$, eventually complemented with a boundary term.

Actually, in the following sections and chapters, we will keep the discussion simpler by working with the generator $H[\xi,\vect{\xi}]$, i.e., we will not do explicitly the passage to the parametrised theory.
Nevertheless, the formal procedure highlighted in this section has to be implicitly understood.
This concludes the general discussion about the Hamiltonian formulation of relativistic field theories.
Let us now discuss the last general topic left open, that is, gauge theories.

\section{Gauge theories} \label{sec:gauge-theories}

So-far, we have always assumed the Legendre transformation to be invertible, always specifying, however, that this fact should have been changed in the case of some important field theories.
Indeed, this happens in gauge theories, in which category fall all the cases analysed in this thesis.\footnote{
The only exception is the complex scalar field, which, however, is mostly studies in order to analyse scalar electrodynamics and the abelian Higgs case, both of which are gauge theories.
}

In order not to make the discussion overcomplicated, we will analyse one, specific case and infer the general rules from this.
Namely, we will take into consideration the case of $\SU (N)$-Yang-Mills on a flat Minkowski background, which we prefer over the simpler and often chosen case of free electrodynamics for two reasons.
First, as we shall see, some results can be generalised more easily after analysing the Yang-Mills situation, since they are too trivial in electrodynamics to see their natural generalisation, as in the case of the constraints' algebra.
Second, the case of electrodynamics can be found in great detail in many textbooks, including~\cite[Chap.~2]{Dirac-book}.
In addition, the results of this section are propaedeutical to the discussion of chapter~\ref{cha:Yang-Mills}.
In any case, a detailed discussion of the general situation can be found in the book by Henneaux and Teitelboim~\cite{Henneaux-Teitelboim} and in the references therein.

Since, in this section, we wish to focus only on gauge symmetries, we will further simplifying the discussion by working in a foliation in which the lapse and the shift are trivial, i.e., $N = 1$ and $\vect{N} = 0$ and postpone the discussion about the Poincar\'e transformations to chapter~\ref{cha:Yang-Mills}.
After a small summary of the features of $\SU (N)$, we will discuss the Hamiltonian formulation of Yang-Mills and introduce the concept of gauge transformations.
This section is taken and adapted from~\cite{Tanzi-Giulini:YM}.

\subsection{SU(N): group, algebra and conventions}

The group $\SU (N)$ can be defined as the group of $N \times N$ complex matrices satisfying the two properties
\begin{equation} \label{SU(N)-properties}
 \mathcal{U}^{-1} = \mathcal{U}^\dagger
 \qquad \text{and} \qquad
 \det \mathcal{U} = 1 \,,
\end{equation}
where $\mathcal{U}^\dagger$ is the complex-conjugated and transposed matrix of $\mathcal{U}$.
In this case, the group operation on $\SU (N)$ is simply the matrix multiplication and the topology is inherited by $\real^{2 N^2}$.\footnote{
$\SU (N)$ is a subset of the $(2 N^2)$-dimensional real linear space of $N \times N$ complex matrices.
The topology of this linear space is the one induced by a norm and the topology of $\SU (N)$ is the induced topology on a subset of a topological space.
Which norm is chosen is not important, since they all lead to the same topology on a finite-dimensional linear space.
One common choice for the norm of a linear operator $\Lambda \colon V \rightarrow W$ between two normed linear spaces is
\[
 \| \Lambda \| \eqdef
 \sup_{v \in V} \frac{\|\Lambda v\|_{W}}{\|v\|_V}
\]
where $\| \cdot \|_V$ and $\| \cdot \|_W$ are the norms of $V$ and $W$, respectively, and the supremum is taken excluding $v=0$.
But any other norm would lead to the same topology.
}
In addition, one can show that $\SU(N)$ is compact.

The associated Lie-algebra is denoted by $\su (N)$ and can be obtained from the group $\SU(N)$ with the procedure described in section~\ref{subsec:symmetries-general}.
Doing so, we find out that $\su (N)$ is the linear space of trace-free anti-hermitian $N \times N$ matrices, i.e., a generic element $M \in \su (N)$ must satisfy the two properties
\begin{equation} \label{su(N)-properties}
 M^\dagger = - M
 \qquad \text{and} \qquad
 \tr M = 0 \,,
\end{equation}
which follows, respectively, from the first and the second properties of~(\ref{SU(N)-properties}) expanding $\mathcal{U} = \exp M \simeq \id + M$.
It is easy to check that $\su(N)$ is a $(N^2-1)$-dimensional real linear space.
Thus, we will denote with $\{ T_A\}_{A=1,\dots,N^2-1}$ a basis and use upper-case Latin indices to denote the components of an element $M \in \su(N)$ with respect to it, e.g. $M = M^A T_A$, where the sum over $A$ ranging from $1$ to $N^2-1$ is understood.

To be completely fair, the definition of $\su (N)$ could have been achieved abstractly.
In this regards, the definition above corresponds to identify $\su (N)$ with the image of its fundamental (also called ``defining'') representation.
In this fashion, we embed the abstract Lie algebra into the associative algebra of endomorphisms with (associative) product being matrix multiplication.
In this way, the Lie product becomes the associative product's commutator and, moreover, we may speak of (associative) products of elements of the Lie algebra, like, e.g., in formulae \eqref{eq:DefKillingFormFundRep} and \eqref{eq:CompModifiedKillingProduct} below, which is very useful --- though not necessary --- for many later calculations and which would not make sense on an abstract level of Lie algebras.
Note that the matrix product of elements in $\su (N)$ will generally yield matrices outside $\su (N)$.

The structure constants $f^A{}_{BC}$ are defined by the 
relation
\begin{equation}
 [T_B,T_C]=f^A{}_{BC} T_A \,.
\end{equation}
On $\su(N)$, we consider a positive-definite inner product, which we obtain from the Killing form, $\kappa$, through multiplication with $(-2N)^{-1}$. This will turn out to be a convenient normalisation in later calculations.
To explain this in slightly more detail, we recall that the Killing form itself is a symmetric bilinear form on the Lie algebra, defined by 
\begin{equation}
 \label{eq:DefKillingForm}
\kappa(T_A,T_B):=
\tr\bigl(\ad_{T_A}\circ \ad_{T_B}\bigr)
=f^N{}_{AM}f^M{}_{BN}\,,
\end{equation}
where $\circ$ denotes the operation of composition (of endomorphisms).
On $\su(N)$, the Killing form defines a negative-definite inner product (like for any compact Lie algebra).
Moreover, through our identification of $\su(N)$ with its image under the fundamental representation, we can eliminate the occurrence of the adjoint representation in the definition of the inner product and express it directly trough traces of products of Lie algebra elements in a form that is only valid for $\su(N)$: 
\begin{equation}
 \label{eq:DefKillingFormFundRep}
\kappa(T_A,T_B)=
2N\,\tr(T_AT_B)\,.
\end{equation}
Here, juxtaposition of matrices in $\su(N)$ refers to matrix multiplication.
Now, the inner product we shall be using is 
\begin{equation}
 \label{eq:DefModifiedKillingForm}
S:=-\frac{1}{2N}\, \kappa\,.
\end{equation}
Its components with respect to the basis $\{ T_A\}_{A=1,\dots,N^2-1}$ are therefore
\begin{equation}
 \label{eq:CompModifiedKillingProduct}
S_{AB}= - \tr (T_A T_B)\,.
\end{equation}
Its inverse has components $S^{AB}$ 
and satisfies 
\begin{equation}
 \label{eq:InvModifiedKillingProduct}
S^{AM}S_{BM}=\delta^A_B\,.
\end{equation}
In this section and in chapter~\ref{cha:Yang-Mills}, we shall exclusively use $S$ and hence continue, for simplicity, to refer to it as ``Killing inner product'', keeping in mind that it is actually a negative multiple of $\kappa$.
  
We use $S_{AB}$ and $S^{AB}$ to raise and lower indices in the standard fashion, e.g., in order to define the index-lowered structure constants
\begin{equation}
\label{eq:DefIndexLoweredSC}
 f_{ABC} \eqdef S_{AA'} f^{A'}{}_{BC}\,,
\end{equation}
which are easily seen to be completely antisymmetric, using the equation  $f_{ABC}=-\tr\bigl(T_A[T_B,T_C]\bigr)$ and the cyclicity of the trace.

Finally, given two Lie-algebra-valued functions 
$\phi (x) \eqdef \phi^A (x) T_A$ and $\psi (x) \eqdef \psi^A (x) T_A$, we denote their positive-definite inner product
by a dot, like
\begin{equation}
\label{eq:DefSprod}
 \sprod{\phi}{\psi} \eqdef \phi^A S_{AB} \psi^B\,,
\end{equation}
and the commutators by 
\begin{equation}
\label{eq:DefExtprod}	
 \extprod{\phi}{\psi} \eqdef [\phi,\psi] \,.
\end{equation}
With this notation, inner product and commutator then obey the 
familiar rule
\begin{equation}
\label{eq:SprodExtprod}	
 \sprod{\phi}{(\extprod{\psi}{\chi})}=
 \sprod{\psi}{(\extprod{\chi}{\phi})}=
 \sprod{\chi}{(\extprod{\phi}{\psi})} \,,
\end{equation}
with the same cyclic property of the triple product.
In this notation, the Jacobi identity reads
\begin{equation}
\label{eq:JacobiIdenity}	
 \extprod{\phi}{(\extprod{\psi}{\chi})}+
 \extprod{\psi}{(\extprod{\chi}{\phi})}+
 \extprod{\chi}{(\extprod{\phi}{\psi})}=0 \,.
\end{equation}
In addition, by means of the positive-definite 
inner product, we may and will identify (as vector 
spaces) the Lie-algebra and its dual and this we 
extend to functions. So, if $\hat\phi$ is 
dual-Lie-algebra-valued function, we assign it to the 
unique Lie-algebra-valued function $\phi$ 
satisfying $\hat\phi(\psi)=\sprod{\phi}{\psi}$ 
for all $\psi$. Examples of such dual-Lie-algebra-valued 
functions that we will encounter in the 
following sections and identify with their 
corresponding Lie-algebra-valued functions
are the conjugated momenta $\pi^\alpha$ and 
the Gauss constraint~$\mathscr{G}$.

\subsection{From the action to the Hamiltonian} 
\label{subsec:Lagrangian-Hamiltonian-YM}

Let us now begin the discussion about the $\SU(N)$-Yang-Mills theory.
The spacetime action in Lagrangian picture is
\begin{equation} \label{YM:action-Lagrangian}
 S[A_\alpha,\dot A_\alpha;g] = -\frac{1}{4} \int d^4 x \sqrt{-{}^4g} \, {}^4g^{\alpha \gamma} \, {}^4g^{\beta \delta} \,  \sprod{F_{\alpha \beta}}{F_{\gamma \delta}}
 +(\text{boundary})\,,
\end{equation}
where $A_\alpha$ is the $\su(N)$-valued one-form potential, i.e., $A_\alpha (x) = A_\alpha^I (x) T_I$ and the $\su(N)$-components $A_\alpha^I (x)$ are $(N^2 -1)$-many one-forms on the spacetime $M$.
The field $A_\alpha$ is the fundamental object in the $\SU(N)$-Yang-Mills theory and it enters in the action above though the \emph{curvature two-form}
\begin{equation}
 F_{\alpha \beta} \eqdef \partial_\alpha A_\beta -\partial_\beta A_\alpha + \extprod{A_\alpha}{A_\beta} \,.
\end{equation}
In addition, the action also contains the four-dimensional flat spacetime metric  ${}^4 g$ and, possibly, a boundary term necessary to make the Lagrangian functionally-differentiable and to make the following manipulations meaningful.
For now, we just assume its existence and postpone a thorough discussion about it to the chapter~\ref{cha:Yang-Mills} as we wish to focus on gauge transformations in this section.

To this end, let us also set $N = 1$ and $\vect{N} = 0$ so that the spacetime four-metric ${}^4g$ is $(3+1)$-decomposed into
\[
 {}^4g_{\alpha \beta}=
 \left(
 \begin{array}{c|c}
  -1	& 0	\\ \hline
  0	& g_{ab}
 \end{array}
 \right) \,.
\]
Although we are dealing with flat Minkowski spacetime, it is more convenient to leave the three-metric $g$ in general coordinates for now, so that the derived equations will be valid both in Cartesian and in other coordinates.
In particular, later on, we will be interested in expressing the results in radial-angular coordinates, but there is no advantage in doing it at this stage.
From now on, spatial indices are lowered and raised using the three-metric $g$ and its inverse.
In addition, contrary to what we did in section~\ref{sec:3+1}, we will not denote any more points of $\Sigma$ with bold letters and it will be clear from the context whether $x$ is a point of the spacetime manifold $M$ or of the space manifold $\Sigma$.
Basically, after the $3+1$ decomposition is completed, we will almost exclusively deal with points of $\Sigma$.
In addition, we will not denote the space on which integrations take place, unless there is some risk of confusion.

The $3+1$ decomposition of the fields in the case of $N = 1$ and $\vect{N} = 0$ is trivial.
Indeed, using foliation-induced coordinates and up to a sign, the vertical components are those having the index $0$ and the horizontal ones are those having the index $a = 1,2,3$.
Thus, we will use directly them instead of introducing the normal ($\perp$) and parallel ($\parallel$) notation.
The action becomes $S=\int dt L[A,\dot A;g]$, where the Lagrangian is
\begin{equation} \label{YM:Lagrangian}
  L[A_\alpha,\dot A_\alpha;g] =\int d^3 x \sqrt{g} \left[
  \frac{1}{2} g^{ab} \sprod{F_{0a}}{F_{0b}}
  -\frac{1}{4} \sprod{F_{ab} }{F^{ab}}
  \right] + (\text{boundary}) \,.
\end{equation}
The variation of the Lagrangian above with respect to the spatial components $\dot A_a$ yields the conjugated three-momenta
\begin{equation} \label{YM:momenta}
 \pi^a \eqdef \frac{\delta L}{\delta \dot A_a} =\sqrt{g} \, g^{ab} F_{0b} \,,
\end{equation}
which are vector densities of weight $+1$.
However, the variation with respect to $\dot A_0$ vanishes.
Following Dirac, let us write this fact as
\begin{equation} \label{YM:primary-constraints}
 \pi^0 \eqdef \frac{\delta L}{\delta \dot A_0} \weq 0 \,.
\end{equation}
Thus, we see that the Legendre transform is not invertible.
Indeed, while $\dot A_a$ can be expressed in terms of the momenta $\pi^a$ by means of~(\ref{YM:momenta}), $\dot A_0$ cannot.

The way to proceed in this situation is the one illustrated by Dirac in~\cite{Dirac-book}.
Specifically, one begins by taking the phase space as being $Z = T^* Q$ and considers, on the phase space, a hypersurface defined by the equation~(\ref{YM:primary-constraints}).
For this reason, we use the symbol $\weq$ instead of $=$ to remember that the equation is satisfied by a subset of points on $Z$.
We will refer to~(\ref{YM:primary-constraints}) and to similar equations as \emph{constraints}.
Note that~(\ref{YM:primary-constraints}) contains actually $N^2-1$ independent constraints since $\pi^0$ has $N^2-1$ independent components.
To be clear, the physical solutions to the equations of motion need to be found among the point on the hypersurface defined by the constraints~(\ref{YM:primary-constraints}), since they  follow directly from the variation of the action~(\ref{YM:action-Lagrangian}).
Thus, what we are doing here is to consider a phase space $Z$ that is much bigger than what strictly needed and we are doing so for mathematical convenience.

Since we are considering $Z = T^*Q$, we can take on it the canonical symplectic form
\begin{equation} \label{YM:symplectic-form_0}
 \Omega_0[A_\alpha,\pi^\alpha]=
 \int d^3 x \, \sprod{\extderphase \pi^\alpha \wedge}{\extderphase A_\alpha} \eqdef
 \int d^3 x \, \extderphase \pi^\alpha_A \wedge \extderphase A_\alpha^A \,,
\end{equation}
where the bold $\extderphase$ and $\wedge$ are, respectively, the exterior derivative and the wedge product in phase space.
Moreover, the symbol $\sprod{\wedge }{}$ means that, at the same time, we are doing the wedge product in phase space and (the negative of) the Killing inner product in the $\su(N)$ degrees of freedom.

Finally, the Hamiltonian is obtained in two steps.
First, one proceeds as usual using the formula $H\eqdef \int d^3 x \, \sprod{\pi^\alpha}{\dot A_\alpha}-L$ and replaces $\dot A_a$ with $\pi^a$ by means of~(\ref{YM:momenta}).
Second, in order for the constraint~(\ref{YM:primary-constraints}) to ensue from the action principle in Hamiltonian formulation we include it in the Hamiltonian --- and, thus, in the action --- multiplied by a Lagrange multiplier $\mu$.
In this way, we obtain the Hamiltonian
\begin{equation} 
\label{YM:Hamiltonian_0}
\begin{aligned}
 H_0[A,\pi;g;\mu]=
 \int d^3 x &\left[ 
 \frac{\sprod{\pi^a}{\pi_a}}{2\sqrt{g}} 
 +\frac{\sqrt{g}}{4} \sprod{F_{ab}}{F^{ab}}
 -\sprod{A_0}{(\partial_a \pi^a+\extprod{A_a}{\pi^a})} + \right. \\
 & + \sprod{\mu}{\pi^0}
 \Big] + (\text{boundary}) \,.
\end{aligned}
\end{equation}
The Lagrange multiplier $\mu$ is an arbitrary (Lie-algebra-valued) function which need to be varied along with canonical fields in the action principle, but it does not appear in the symplectic form~(\ref{YM:symplectic-form_0}).
Note that, in deriving the expression above, we have absorbed $\dot A_0$ in the Lagrange multiplier $\mu$.

\subsection{Secondary constraints and constraints' algebra} \label{subsec:YM:secondary-constraints}

The Hamiltonian~(\ref{YM:Hamiltonian_0}) is not yet the correct Hamiltonian of the $\SU(N)$-Yang-Mills theory.
The reason is that the solutions to the equations of motion need to satisfy the constraint~(\ref{YM:primary-constraints}).
However, the Hamilton equations ensuing from the symplectic form~(\ref{YM:symplectic-form_0}) and the Hamiltonian~(\ref{YM:Hamiltonian_0}) do not preserve the constraint $\pi^0 \weq 0$.
Thus, a field configuration, that is initially on the hypersurface defined by the constraint, will in general move away from it while evolving in time.
But the constraint $\pi^0 \weq 0$ needs to be satisfied by physical solutions at any time.

To solve this issue, we proceed as follows.
First, let us compute
\begin{equation}
 \dot \pi^0 = \{ \pi^0, H_0 \} = \partial_a \pi^a + \extprod{A_a}{\pi^a} \,,
\end{equation}
which is, in general, different from zero.\footnote{
Equations involving the Poisson brackets of position-dependent quantities, such as $\pi^0 (x)$, have to be understood as being valid when smearing the left- and right-hand sides of the equation by any test function.
Some carefulness about boundary terms of the smeared equations is in general needed, but we are neglecting boundary terms altogether in this section.
}
Second, let us enforce the further constraint
\begin{equation} \label{YM:secondary-constraints}
 \mathscr{G} \eqdef \partial_a \pi^a + \extprod{A_a}{\pi^a} \weq 0 \,,
\end{equation}
so that the original constraint~(\ref{YM:primary-constraints}) is now preserved under time evolution by all the field configurations which satisfy both~(\ref{YM:primary-constraints}) and~(\ref{YM:secondary-constraints}).
Note that the expression in~(\ref{YM:secondary-constraints}) is precisely the term multiplied by $A_0$ in the Hamiltonian~(\ref{YM:Hamiltonian_0}) and that it is build using the gauge-covariant derivative $D_b \pi^a \eqdef \partial_b \pi^a + \extprod{A_b}{\pi^a}$.

At this point, one needs to ensure that also the constraints~(\ref{YM:secondary-constraints}) are preserved by time evolution.
This is indeed the case since
\begin{equation}
 \dot{\mathscr{G}} = \{ \mathscr{G}, H_0 \} = -\extprod{A_0}{\mathscr{G}} \weq 0 \,.
\end{equation}
Thus, if a field configuration satisfies at the initial time the constraints~(\ref{YM:primary-constraints}) and~(\ref{YM:secondary-constraints}), it will satisfy them at any time.
The highlighted procedure takes the name of the \emph{Dirac-Bergmann algorithm} and is used to find all the constraints of a Hamiltonian system.

The constraints derived directly from the Lagrangian, such as~(\ref{YM:primary-constraints}) in this case, are called \emph{primary constraints}, while the constraints imposed to make the primary constraints preserved by the time evolution, such as~(\ref{YM:secondary-constraints}) in our case, are called \emph{secondary constraints}.
In general, it may happen that the secondary constraints are not preserved by time evolution.
In this case, one introduces further constraints (still referred to as secondary constraints) to impose this condition and continues so on until all the constraints are preserved by time evolution.

At the level of the Hamiltonian formulation, there is not a substantial difference between primary and secondary constraints.
Indeed, as we shall see in the next subsection, they will all enter the Hamiltonian in the same way and the only distinction will be that they were derived in a different order.
However, there is a substantial difference between first-class and second-class constraints.
In general, we say that a canonical generator is \emph{first-class} if its Poisson-bracket with the constraints vanishes on the constraint-hypersurface.
Otherwise it is \emph{second-class}.

In our case, the constraints are first class.
To see this fact, let us decompose them into components, $\pi^0_A \eqdef \sprod{\pi^0}{T_A}$ and
\mbox{$\mathscr{G}_A \eqdef \sprod{\mathscr{G}}{T_A}$}, and compute their Poisson brackets.
We get
\begin{equation}
\label{YM:constraint-algebra}
\begin{split}
 \{ \pi^0_A (x), \pi^0_B (x') \}&=0 \,, \\
 \{ \pi^0_A (x), \mathscr{G}_B (x') \}&=0 \,,\\ 
 \{ \mathscr{G}_A (x), \mathscr{G}_B (x') \}&=f^M{}_{AB} \, \mathscr{G}_M (x) \delta_x (x') \weq 0 \,.
\end{split}
\end{equation}
Notably, the last one of the expressions above shows that the constraints  $\{ \mathscr{G}_A\}_{A=1,\dots,N^2-1}$ form a Poisson-representation of the $\su (N)$ algebra.
For a full discussion about first- and second-class constraints, see~\cite{Henneaux-Teitelboim} and the references therein.
We will just mention that only first-class constraints are related to gauge symmetries.

\subsection{Hamiltonian of free Yang-Mills theory}
As well as the primary constraints~(\ref{YM:primary-constraints}), also the secondary ones~(\ref{YM:secondary-constraints}) need to be included in the  Hamiltonian~(\ref{YM:Hamiltonian_0}) multiplied by a Lagrange multiplier $\lambda$, for otherwise they will not ensue from the action principle in Hamiltonian formulation.
Doing so and reabsorbing $A_0$ in the definition of $\lambda$, one obtains the extended Hamiltonian of free Yang-Mills theory
\begin{equation} 
\label{YM:extended-Hamiltonian}
\begin{aligned} 
H_{\text{ext}}[A_\alpha,\pi^\alpha;g;\mu,\lambda]=
 \int  d^3 x & \left[ 
 \frac{\sprod{\pi^a}{\pi_a}}{2\sqrt{g}} 
 +\frac{\sqrt{g}}{4} \sprod{F_{ab}}{F^{ab}}
 +\sprod{\mu}{\pi^0}
 +\sprod{\lambda}{\mathscr{G}}
 \right] +\\
 +&(\text{boundary})\,.
\end{aligned}
\end{equation}
We see, indeed, that there is no distinction in the Hamiltonian between primary and secondary constraints.
They both appear in the Hamiltonian multiplied by an arbitrary function.

As discussed by Dirac in the case of electrodynamics~\cite{Dirac-book}, one can remove the degrees of freedom corresponding to $\pi^0$ and $A_0$, since they do not contain any physical information.
Indeed, their equations of motion are
\begin{equation} \label{YM:eom-zero-comp}
 \dot A_0=\mu \,, \qquad
 \dot \pi^0 \weq 0 \,,
 \qquad \text{and} \qquad
 \pi^0 \weq 0 \,,
\end{equation}
so that the time derivative of $A_0$ is the completely arbitrary Lagrange multiplier $\mu$ and $\pi^0$ is identically zero for physical solutions.
Therefore, we discard completely these degrees of freedom obtaining the symplectic form
\begin{equation} 
\label{YM:symplectic-form}
 \Omega[A,\pi]=\int d^3 x \, \sprod{\extderphase \pi^a \wedge}{\extderphase A_a}
\end{equation}
and the Hamiltonian of free Yang-Mills theory
\begin{equation} 
\label{YM:Hamiltonian}
 H[A,\pi;g;\lambda]=
 \int d^3 x \left[ 
 \frac{\sprod{\pi^a}{\pi_a}}{2\sqrt{g}} 
 +\frac{\sqrt{g}}{4} \sprod{F_{ab}}{F^{ab}}
 +\sprod{\lambda}{\mathscr{G}}
 \right] +(\text{boundary}) \,,
\end{equation}
where the only constraints left are the $(N^2-1)$ first-class Gauss-like constraints
\begin{equation} \label{YM:gauss-constraint}
 \mathscr{G} \eqdef 
 \partial_a \pi^a  + \extprod{A_a}{\pi^a}
 =D_a \pi^a
 \weq 0\,.
\end{equation}

Finally, the knowledge of the symplectic form~(\ref{YM:symplectic-form}) and of the Hamiltonian~(\ref{YM:Hamiltonian}) allows one to compute the equations of motion by finding the Hamiltonian vector field $X_H$ satisfying $\extderphase H = - \insertion_{X_H} \Omega$ and equating its components to the time derivative of the canonical fields.
In this way, we find
\begin{subequations} \label{YM:eoms}
 \begin{align} \label{YM:eoms1}
 \dot{A}_a &= \{ A_a , H \}
 =\frac{\pi_a}{\sqrt{g}}
 - D_a \lambda
 \,,\\
 \label{YM:eoms2}
 \dot{\pi}^a &= \{ \pi^a , H \}
 =\partial_b (\sqrt{g}\, F^{ba}) +\sqrt{g}\, \extprod{A_b}{F^{ba}}+\extprod{\lambda}{\pi^a} \,.
\end{align}
\end{subequations}
The presence of the Gauss constraints~(\ref{YM:gauss-constraint}) in the Hamiltonian~(\ref{YM:Hamiltonian}) causes the equations of motion above to include a transformation, whose parameter is the arbitrary function $\lambda (x)$, which is precisely an infinitesimal gauge transformation.
Thus, let us discuss gauge transformations in the next subsection.

\subsection{Gauge transformations} \label{subsec:gauge-transformations-intro}

The equations of motion~(\ref{YM:eoms}) of the canonical fields depend on the Lagrange multiplier $\lambda$, which, as we have mentioned, is an arbitrary function.
As a consequence, the equations of motion cannot have a well-defined Cauchy problem, since the solutions are most-certainly not unique.
This issue is what lead us to the definition of gauge symmetries.

Specifically, we define \emph{gauge transformations} as those transformations generated by first-class constraints.
In other words, in the case of Yang-Mills, they are those transformations, whose canonical generator is
\begin{equation} \label{YM:gauge-generator}
 G[\lambda] \eqdef \int d^3 x\, \sprod{\lambda (x)}{\mathscr{G} (x)} \,,
\end{equation}
which is the Gauss constraints~(\ref{YM:gauss-constraint}) smeared with an arbitrary function $\lambda (x)$.
The above expression is precisely the last term appearing in the Hamiltonian~(\ref{YM:Hamiltonian}) and causing the equations of motion~(\ref{YM:eoms}) to depend on $\lambda$.

To see this fact, let us compute explicitly the vector field $X_\lambda$ associated to $G[\lambda]$ by means of the equation $\extderphase G [\lambda] = - \insertion_{X_{\lambda}} \Omega$.
In doing so, let us also take into consideration the potentially-problematic boundary terms, since this is an easy situation, in which to show the problems they may cause.
The exterior derivative of the gauge generator~(\ref{YM:gauge-generator}) is
\begin{equation} \label{YM:gauge-generator-variation}
 \extderphase G[\lambda] = \int d^3 x\, \Big[
 \sprod{-\extderphase \pi^a}{(\partial_a \lambda+\extprod{A_a}{\lambda})}
 -\sprod{\extderphase A_a}{\extprod{\lambda}{A_a}}
 \Big]+
 \oint_{S^2_{\infty}} d^2 \barr{x}_k \, \sprod{\lambda}{\pi^k} \,,
\end{equation}
where we have integrated by parts and obtained a boundary term, which has to be understood as an integral over a sphere whose radius is sent to infinity.
When the surface term in the expression above vanishes, the generator~(\ref{YM:gauge-generator}) is differentiable \`a la Regge-Teitelboim, which ensures the existence of the vector field $X_\lambda$ provided that the symplectic form is~(\ref{YM:symplectic-form}) without any boundary term.
In this case, we get the infinitesimal gauge transformations
\begin{subequations} \label{YM:gauge-transformations}
 \begin{align} \label{YM:gauge-transformations1}
 \delta_\lambda A_a &\eqdef \{ A_a, G[\lambda] \} = -D_a \lambda \,, \\
 \label{YM:gauge-transformations2}
 \delta_\lambda \pi^a &\eqdef \{ \pi^a, G[\lambda] \} = \extprod{\lambda}{\pi^a} \,,
\end{align}
\end{subequations}
which are, as expected, exactly the last terms appearing in the equations of motion~(\ref{YM:eoms}).
The infinitesimal transformations above can be integrated to get the gauge transformations with parameter $\mathcal{U}\eqdef \exp (-\lambda) \in \SU(N)$ or, in other words, the action $\Phi$ of $\SU(N)$ on the phase space.
Explicitly, we have 
\begin{equation} \label{YM:gauge-transformations-full}
 \Phi_\mathcal{U} (A_a)  = \mathcal{U}^{-1} A_a \,\mathcal{U}+\mathcal{U}^{-1} \partial_a  \,\mathcal{U}
 \qquad \text{and} \qquad
 \Phi_\mathcal{U} (\pi^a)  = \mathcal{U}^{-1} \pi^a \,\mathcal{U} \,,
\end{equation}
where the products on the right-hand sides are products among matrices.

Two field configurations related by gauge transformations are considered as \emph{physically equivalent}, although being mathematically distinguished.
In this way, the equations of motion have a well-posed Cauchy problem and, in particular, unique solutions for the congruence classes (or \emph{gauge orbits}) of physically-equivalent field configurations.
Therefore, the degrees of freedom in the mathematical description of the theory are redundant, since the same physical state can be described by all the elements in its gauge orbit equivalently.

Since gauge transformations are built from constraints, they are always symmetries of the theory.
Indeed, from the construction of subsection~\ref{subsec:YM:secondary-constraints}, their canonical generator must satisfy
\begin{equation}
 \{ G[\lambda] , H \} \weq 0 \,,
\end{equation}
so that they Poisson-commute with the Hamiltonian at least on the constraint surface, which include all the physical solutions.\footnote{
The equation $ \{ G[\lambda] , H \} \weq 0$ is valid even after extending the Hamiltonian to include all the constraints since, in the case of gauge theories, these are first class.
We thus see the importance of dealing with first-class constraints.
}
However, the group of gauge transformations $\Gau$ defines only a mathematical symmetry of the theory, since it maps solutions to the equations of motion to mathematically-different, but physically-equivalent, solutions.

The above considerations are valid at the condition that the boundary term in~(\ref{YM:gauge-generator-variation}) vanishes.
Whether or not this is actually the case depends on the asymptotic behaviour of the canonical fields and of the gauge parameter $\lambda(x)$, which is going to be thoroughly discussed in chapter~\ref{cha:Yang-Mills}.
For now, let us only mention that it is sometimes possible to extend the generator of gauge transformations~(\ref{YM:gauge-generator}) to include a boundary term so that it is differentiable for a larger class of function $\lambda$ than the original generator.
The larger group of transformations obtained in this way includes physical symmetries that have a non-trivial action on the physical state of a system, due to the boundary term at infinity.
We will refer to this transformations as \emph{improper gauge transformations} following the terminology of~\cite{Teitelboim-YM2}.
We will discuss in greater detail proper and improper gauge transformations in section~\ref{subsec:gauge-transformations} of the next chapter.
We will see that improper gauge transformations, together with the Poincar\'e transformations, will build the asymptotic-symmetry group of the theory.

\subsection{Reduced phase space} \label{subsec:reduced-ps}

Let us conclude this discussion about gauge theories with some general considerations.
As we have seen, the physical solutions belongs to the hypersurface defined by the constraints, which is in general a submanifold of the phase space $Z = T^* Q$ which we have considered.
Thus, one might try to work using the constraint-hypersurface $Z'$ instead of the full $Z$.
Since $Z' \subset Z$, the symplectic form $\Omega$ defines a two-form $\Omega'$ by means of the pull-back.
The two-form $\Omega'$ on $Z'$ is closed, since the pull-back of a closed form is closed, but it is also degenerate.
In particular, the gauge transformations defines Hamiltonian vectors fields on $Z$ that are tangent to $Z'$, since the constraints are first class.
Once restricted to $Z'$, these vector fields $(Y_a)_{a \in A}$  are such that $\insertion_{Y_a} \Omega' = 0$, so that the two-form $\Omega'$ is degenerate.
As a consequence, many of the tools discussed in this chapter, including the Poisson brackets, are not well-defined.

When this happens and $\Omega'$ is only closed (but not non-degenerate), we say that it is a \emph{pre-symplectic form} and that $Z'$ a \emph{pre-symplectic manifold}.
From this, one can define the \emph{reduced phase space} $\bar Z$ as the quotient of the pre-symplectic manifold $Z'$ modulo the transformations generated by the vector fields $(Y_a)_{a \in A}$.
In this way, one obtains a manifold $\bar Z$ with a closed, weakly non-degenerate symplectic form $\bar \Omega$.
The reduced phase space is, thus, a weakly symplectic manifold allowing the use of all the tools discussed in this chapter.
In addition, contrary to $Z$, it is not vastly redundant.

It would seem that the best thing to do would be to work with the reduced phase space $\bar Z$ instead of the original $Z$.
However, although being mathematically well-defined, it is very difficult, in fact, to find explicitly the manifold $\bar Z$ with its symplectic form $\bar \Omega$ and it is even more difficult to perform actual computations using it.
As a consequence, we will use the phase space $Z = T^* Q$ with the canonical symplectic form (eventually complemented by boundary terms) in the discussion of the next chapters.
For proofs and more details about this topic, see the discussion in~\cite{Henneaux-Teitelboim}.

\section{Hamiltonian approach to the study of asymptotic symmetries} \label{sec:principles}

We are now in a position to state precisely which are the first principles and the methods used in the Hamiltonian approach to the study of asymptotic symmetries of field theories.
Some of this discussion is redundant and has been already mentioned in previous parts of this thesis.
However, we find it useful to keep all these principles to be listed together in one section.
First and foremost, we require the following four basic structure of the Hamiltonian formulation to exist and be well-defined.
\begin{enumerate}[wide=\parindent]
\item \textsc{Phase space.} The phase space consists of all the field configurations that are allowed.
Other than identifying the canonical field of the theory, e.g. by means of the $3+1$ decomposition starting from the Lagrangian picture, it is necessary to impose conditions on the regularity and on the asymptotic behaviour of the fields.
Only in this way, the following conditions will be met in general.
\item \textsc{Symplectic form.} The symplectic form must be well-defined, as a closed (weakly) non-degenerate two-form on phase space.
Since we will often start from the canonical symplectic form, which is given in terms of an integral over the space manifold, we need to require that this integral is actually convergent.
We leave open the possibility to complement the symplectic form with a boundary term if this is useful to achieve the following conditions.
\item \textsc{Hamiltonian.} The Hamiltonian must be well-defined, which usually amounts to two conditions.
First, since we will often find the Hamiltonian as a formal expression involving integrals over the space manifold, we need to make sure that these integrals are convergent or, equivalently, that the Hamiltonian is finite.
Second, there must be a vector field on phase space associated to the Hamiltonian.
If the symplectic form is the canonical one without boundary terms, this fact is equivalent to the differentiability \`a la Regge-Teitelboim.
As a consequence, even in the case of a more general symplectic form, we will often refer to this property (of the existence of $X_H$) by saying that the Hamiltonian is differentiable.
\item \textsc{Poincar\'e group.} Since we will deal with relativistic field theories on a flat Minkowski background, we demand a Hamiltonian action of the Poincar\'e group on phase space.
This condition has to be met asymptotically in the case of field theories on an asymptotically-flat background.
\end{enumerate}

Regarding the last point, we recall that, if a symplectic action of the Poincar\'e group exists, then this action is also Hamiltonian.
For general Lie groups, there may be obstructions to turn a symplectic action into a Hamiltonian action, i.e. against the existence of a momentum map, and even if the latter exists, it need not be unique.
These issues of existence and uniqueness are classified by the Lie algebra's second and first cohomology group, respectively.
In case of the Poincar\'e group, these cohomology groups are both trivial, and these issues do not arise; compare,  e.g., \cite[Chap.\,3.3]{Woodhouse:GQ}.
Thus, in the case of the Poincar\'e transformations, it is sufficient to demand a symplectic action.
 
It should be clear that the possibility to simultaneously meet the requirements listed above will delicately depend on the precise characterisation of phase space.
For field theories, this entails mostly to characterise the canonical fields in terms of fall-off conditions and, as it turns out, also parity conditions.
The former ones tell us how quickly the fields vanish as one approaches spatial infinity, whereas the latter ones tell us the parity of the leading term in the asymptotic expansion of the fields as functions on the two-sphere at spatial infinity. 
In the context of Hamiltonian General Relativity it has long been realised that parity conditions are necessary in order to ensure the existence of integrals that represent Hamiltonian generators of symmetries that one wishes to include on field 
configurations that are asymptotically Minkowskian and represent isolated systems, as we shall briefly review in the next section.

Quite generally, the task is to find a compromise between two competing aspects: the size of phase space and the implementation of symmetries.
On the one hand, phase space should be large enough to contain sufficiently many interesting states, in particular those being represented by fields whose asymptotic fall-off is slow enough to allow globally ``charged'' states, like electric charge for the Coulomb solution in Electrodynamics,  or mass for the Schwarzschild solution in General Relativity.
On the other hand, for the  symmetry generators to exist as (differentiable) Hamiltonian functions, phase space cannot be too extensive. 
Since we are dealing with relativistic theories, the compatible symmetries should contain the Poincar\'e group, but might likely turn out to be a non-trivial extension thereof if we are dealing  with gauge or diffeomorphism-invariant theories.

Let us illustrate this last point in a somewhat more mathematical language.
In any gauge- or diffeomorphism-invariant theory, there is a large, infinite-dimensional group acting on the fields which transforms solutions of the equations of motions to solutions (of the very same equations).
For example, in ordinary gauge theories, these are certain (infinite-dimensional) groups of bundle automorphisms, or, in General Relativity, the group of diffeomorphisms of some smooth manifold.
Let us call it the ``symmetry group'' $\Sym$.
Now, inside $\Sym$, there is a normal subgroup of ``gauge transformations'', denoted by $\Gau$.
They, too, are symmetries in the sense that they map solutions of the field equations to solutions, but they are distinguished by their interpretation as ``redundancies in description''.
This means that any two phase-space points connected by the action of $\Gau$ are physically indistinguishable; they are two mathematical representatives of the same physical state. 
Accordingly, physical observables cannot distinguish between these two representatives, which means that physical observables are constant on each $\Gau$-orbit in phase space.
As we have seen, in the Hamiltonian setting the subset $\Gau\subset\Sym$ is characterised as the group that is generated by first-class constraints.
Accordingly, the space of physical observables is then defined to be the subset of phase-space functions that cannot separate points connected by $\Gau$, i.e. that Poisson-commute with the constraints on the set of points in phase-space allowed by the constraints.
Following \cite{Teitelboim-YM2}, elements of $\Gau$ are also called \emph{proper gauge transformations}.

The crucial observation is that $\Sym$ is strictly larger than $\Gau$, so that the quotient group $\Asym:=\Sym/\Gau$ is again a group of symmetries, now to be interpreted as \emph{proper physical symmetries}, in the sense of mapping states and solutions to \emph{new}, physically-different states and solutions.
It is this quotient group that one should properly address as group of \emph{asymptotic symmetries} and which should somehow contain the Poincar\'e group and, possibly, more.
Note that $\Asym$ contains residuals of those ``gauge transformations'' whose fall-off is too weak in order to be generated by constraints.
These are often called \emph{improper gauge transformations}~\cite{Teitelboim-YM2} and we will deal explicitly with them in section~\ref{subsec:gauge-transformations}. 

It has long been realised the insufficient distinction between proper and improper gauge transformations may result in apparently paradoxical conclusions, like that of an apparent violation of conservation of global non-abelian charges which follows as consequence if long-ranging (and hence improper) gauge 
transformations are taken for proper ones; see, e.g.,~\cite{Schlieder:1981}.
Strictly speaking, the improper gauge transformations do not only contain those with insufficient fall off, but they also may contain  those of rapid fall-off which are not in the component of the identity.
This is because the group $\Gau$ that is generated by the constraints is, by definition, connected.
Elements outside the component of the identity are sometimes referred to as \emph{large gauge transformations}. 

Quite generally, improper gauge transformations will combine with other symmetries, like the Poincar\'e group, into the group $\Asym$.
That combination need not be a direct product.
Often it is a semi-direct product or, more generally, an extension of one group by the other.
In fact, non-trivial extensions already appear when large gauge transformations are properly taken into account, with potentially interesting consequences for the physical content of the theory. 
For example, it may happen that the electromagnetic $\mathrm{U}(1)$ is extended to its (non-compact) universal cover $\mathbb{R}$, or that the spatial $\mathrm{SO}(3)$ is extended to its universal cover $\mathrm{SU}(2)$; see \cite{Giulini:1995}.

In the remained of this thesis, we will follow this principles in order to determine the group of asymptotic symmetries $\Asym$ of various physically-relevant situations.

\section{The situation in General Relativity} \label{sec:GR}

In this section, we briefly review the situation concerning asymptotically flat space in General Relativity.
We will only mention the main results and provide references for the full details.
The Hamiltonian formulation of General Relativity was achieved for the first time by and Arnowitt, Deser, and Misner~\cite{ADM} --- from whose initials takes the name of (ADM) formalism --- and independently by Dirac~\cite{Dirac:Hamiltonian1,Dirac:Hamiltonian2}.
A detailed review can be found in Chap. 21 of~\cite{Gravitation} and in~\cite{Giulini:Hamiltonian-GR}.

Contrary to the case of field theories on a fixed background, in General Relativity, the spacetime metric ${}^4 g$ itself is the dynamical field.
By means of the $3+1$ decomposition one finds that the Hamiltonian degrees of freedom of the theory are the Riemannian metric $g$ of the three-dimensional manifold $\Sigma$ and a canonical momentum $\pi$ associated to it.
Since ${}^4 g$ has ten independent components, while $g$ has only six, there will also be four independent constraints in the Hamiltonian formulation.
Specifically, the symplectic form is the canonical one
\begin{equation} \label{ADM:symplectic-form}
 \Omega = \int_\Sigma \extderphase \pi^{ij} \wedge \extderphase g_{ij}
\end{equation}
and the Hamiltonian takes the form
\begin{equation} \label{ADM:Hamiltonian}
 H[g,\pi; N, \vect{N}] = \int_\Sigma d^3 x \, \Big[
 N (x) \mathscr{H} (x) + N^i (x) \mathscr{H}_i (x) \Big] \,,
\end{equation}
where
\begin{equation} \label{ADM:super-Hamiltonian}
 \mathscr{H} = \frac{1}{\sqrt{g}} \left(
 \pi_{ij} \pi^{ij} - \frac{1}{2} \tilde{\pi}^2  \right)
 - \sqrt{g} \, R
\end{equation}
is called \emph{super-Hamiltonian} and
\begin{equation} \label{ADM:super-momentum}
 \mathscr{H}_i = -2 \nabla^j \pi_{ij}
\end{equation}
is called \emph{super-momentum}.
In the expressions above indices are raised and lowered using the three-dimensional metric $g$, $\tilde{\pi} \eqdef \pi^{ij} g_{ij}$, $R$ is the Ricci scalar of $g$ and $\nabla$ its Levi-Civita connection.
In addition, the constraints are
\begin{equation} \label{ADM:constraints}
 \mathscr{H} \weq 0
 \qquad \text{and} \qquad
 \mathscr{H}_i \weq 0 \,,
\end{equation}
which are precisely the super-Hamiltonian and the super-momentum.

For this reason, one often says that the Hamiltonian~(\ref{ADM:Hamiltonian}) of General Relativity is pure constraint, a fact that is not completely correct.
Indeed, as we have often mentioned in this section, the Hamiltonian needs to be complemented with boundary terms in general.
As first noted by Regge and Teitelboim~\cite{RT}, these boundary term are not only needed in the specific case of asymptotically-flat spacetimes, but they also carry an important physical significance.
Regge and Teitelboim assumed that there were some asymptotic coordinates $(x^i)_{i=1,2,3}$ such that the three-dimensional metric $g$ and the canonical momenta $\pi$ satisfied the fall-off condition
\begin{subequations} \label{RT:fall-off}
\begin{equation} \label{RT:fall-off1}
 g_{ij} = \delta_{ij} + \frac{1}{r} \barr{h}_{ij} + \bigo\big( 1/r^2 \big)
 \qquad \text{and} \qquad
 \pi^{ij} = \frac{1}{r^2} \barr{\pi}^{ij} + \bigo\big( 1/r^3 \big)
\end{equation}
where $\delta = \text{diag} (1,1,1)$, $r \eqdef x_i x^i$, and the quantities with a bar above, such as $\barr{h}$ and $\barr{\pi}$, do not depend on $r$ but only on the angles.
Thus, $\barr{h}$ and $\barr{\pi}$ are functions on the sphere at infinity.
The above equations need to be preserved by first-order spatial derivative.\footnote{
In the sense that, e.g., $\partial_r \bigo \big( 1/r^2 \big) = \bigo \big( 1/r^3 \big)$.
}
In addition, the lapse and the shift were also required to satisfy the fall-off conditions
\begin{equation} \label{RT:fall-off2}
 N = 1 + \bigo\big( 1/r \big)
 \qquad \text{and} \qquad
 N^i = \bigo\big( 1/r \big)
\end{equation}
\end{subequations}
up to first-order derivatives.

The fall-off conditions~(\ref{RT:fall-off}) do not suffice to make the canonical symplectic form~(\ref{ADM:symplectic-form}) well-defined, since, indeed, the integral can be verified to be logarithmically divergent.
For this reasons, they need to be complemented with some parity conditions.
The parity conditions chosen by Regge and Teitelboim were
\begin{equation} \label{RT:parity}
 \barr{h}_{ij} (-\barr{x}) = \barr{h}_{ij} (\barr{x})
 \qquad \text{and} \qquad
 \barr{\pi}^{ij} (-\barr{x}) = - \barr{\pi}^{ij} (\barr{x}) \,,
\end{equation}
where $\barr{x}$ are coordinates on the sphere at infinity and $\barr{x} \mapsto - \barr{x}$ is the antipodal map.\footnote{
\label{footnote:antipodal-map}
We remind that the antipodal map, denoted in this footnote by $\Phi \colon\barr{x} \mapsto - \barr{x}$, consists of the explicit transformation
$(\theta, \phi) \mapsto (\pi - \theta, \phi+\pi)$ in terms of the standard spherical coordinates.
A generic tensor field $T$ (or a density) is said to be \emph{even} under the antipodal map if $\Phi^* T = T$, being $\Phi^*$ the pull back of the antipodal map.
Analogously, $T$ is \emph{odd} if $\Phi^* T = -T$.
To see how this translate into the exact parity of the components of a tensor field (or density) expressed in some coordinates like the standard $(\theta,\phi)$ spherical coordinates, see footnote 2 of~\cite{Henneaux-ED}.
}
Hence, the former are even functions of the sphere and the latter are odd functions, so that the logarithmically-divergent part of~(\ref{ADM:symplectic-form}) becomes an odd function integrated over the sphere and, thus, vanishes.\footnote{
One needs to assume, as it is usually done and as we will do in this thesis, that the integral over $\Sigma$ is first performed in the angular coordinates and only then over the radial coordinate.
}
Note that the fall-off conditions~(\ref{RT:fall-off}) and the parity conditions~(\ref{RT:parity}) include the Schwarzschild and Kerr solutions.

In order to make it differentiable \`a la Regge-Teitelboim, the Hamiltonian~(\ref{ADM:Hamiltonian}) has to be complemented by the boundary term
\begin{equation}
 E [g ] = \oint d^2 \barr{x}_j (\partial_i g_{ij} - \partial_{j} g_{ii}) \,,
\end{equation}
as noted for the first time by Regge and Teitelboim.
In the above expression, which is valid only in the coordinates chosen above, the sum over $i$ and $j$ has to be understood.
Due to the constraints~(\ref{ADM:constraints}), the Hamiltonian takes the value of the boundary term above on the solutions of the equations of motion.
In particular, for the Schwarzschild spacetime, $E$ equates the Schwarzschild mass $M$ and, in general, it equates the ADM mass.
We thus see the importance of including the correct boundary term in the Hamiltonian (and, as a consequence, in the action) both from a mathematical and from a physical point of view.
Indeed, on the one hand, the boundary term makes the Hamiltonian differentiable, so that the equations of motion are well-defined.
On the other hand, $E[g]$ provides a definition for the global mass of an isolated system.
The fact that boundary terms are needed not only in gravity but also in other gauge theories was first discussed by Gervais, Sakita and Wadia~\cite{Gervais-Sakita-Wadia:surface-terms}.
See also the subsequent studies by Wadia and Yoneya~\cite{Wadia-Yoneya:surface-terms}, by Wadia~\cite{Wadia:surface-terms}, and by Gervais and Zwanziger~\cite{Gervais-Zwanziger:surface-terms}.

The fall-off~(\ref{RT:fall-off}) and parity conditions~(\ref{RT:parity}) allow a canonical action of the Poincar\'e transformations, as showed in~\cite{RT}.
The correct canonical generators of the Poincar\'e transformations, together with an ample discussion about the topic, can be found in the paper by Beig and \'o Murchadha~\cite{Beig}, that corrected a small mistake contained in~\cite{RT}.
However, no extension of the Poincar\'e group was possible with these conditions and, for this reason, it was long though that the BMS group was a feature of null infinity only, and not of spatial infinity.
This apparent clash between null and spatial infinity was solved only recently by Henneaux and Troessaert~\cite{Henneaux-GR}.
Notably, they showed that, using different parity conditions (which still include the Schwarzschild and Kerr solutions) one can achieve a Hamiltonian formulation with a well-defined symplectic form, a differentiable Hamiltonian, and a canonical action of the BMS group.
All the details can be found in~\cite{Henneaux-GR}. 

This concludes our small review about the Hamiltonian formulation of General Relativity and its asymptotic structure.
This also concludes this chapter about the Hamiltonian methods in field theories.
Starting from the next chapter, we will put this method into action to derive the results that constitute the original contribution of the author of this thesis.

\chapter{Non-abelian gauge theories} \label{cha:Yang-Mills}

We now put the machinery described in the previous chapters finally into action, in order to present the first original results of this thesis.
In particular, we will first complete the derivation of the Hamiltonian formulation of the $\SU(N)$-Yang-Mills theory, which was started in section~\ref{sec:gauge-theories}.
Then, we will analyse, in this framework, the asymptotic symmetries of the theory.
This chapter is taken from the paper~\cite{Tanzi-Giulini:YM} with minor changes, the only exception being section~\ref{sec:YM-higher-dim}, which contains new results.

The treatment of asymptotic symmetries of gauge theories using the Hamiltonian formulation was pioneered by Henneaux and Troessaert for the case of free electrodynamic~\cite{Henneaux-ED} and of General Relativity~\cite{Henneaux-GR}.
Their subsequent analyses covered also the cases of electrodynamics in higher dimensions~\cite{Henneaux-ED-higher} and of the couple Maxwell-Einstein theory~\cite{Henneaux-ED-higher}.
In this context, the goal of our investigations is to extend their analyses to include more and more physically-interesting theories, which we study following closely their strategy.
Therefore, in this chapter, we will discuss the $\SU(N)$-Yang-Mills theory on a flat Minkowski spacetime, which is one of the building blocks of the Standard Model of particle physics.
Indeed, it is both used to readily describe chromodynamics, in which case $N = 3$ represents the number of ``colours'' of the quarks, and it appears in the electroweak sector.
In greater detail, the electroweak sector is described by a $\SU(2) \times \U(1)$-Yang-Mills --- where the former factor represents the isospin and the latter the hypercharge --- coupled to a Higgs field.
In the next chapter, we will analyse the abelian Higgs model, a simple prototype of the electroweak Higgs, whose study is in preparation to a future complete study of the Standard Model of particle physics. 

The strategy pursued by Henneaux and Troessaert can be highlighted schematically as follows.
First, one starts from a theory in Lagrangian picture and, neglecting the boundary terms, derives candidates for the fundamental objects of the Hamiltonian picture.
To be more precise, one identifies a candidate for the phase space, for the symplectic form, for the Hamiltonian, and, in the case of supposedly Poincar\'e-invariant theories, for the action of the Poincar\'e group on the phase space.
Constraints and gauge transformations are also identified generically at this point.
The phase space, in this step, is merely identified as the space of functions needed to build the canonical fields and satisfying some weak regularity conditions, in order for the formal expressions of the equations of motion and of the Poincar\'e transformations to make sense.
This archetype of the phase space is, in general, too big.
Indeed, as we have seen in the general discussion of chapter~\ref{cha:Hamiltonian-ft}, many relevant physical objects, including the symplectic form and the Hamiltonian, are defined as integrals over the space manifold, which are not finite in general unless some conditions on the asymptotic behaviour of the canonical fields are imposed.

Thus, the second step is to restrict the phase space to those field configurations satisfying some fall-off and parity conditions.
The former conditions tell us how quickly the fields have to vanish while approaching spatial infinity, while the latter ones tell us how the parity of the leading terms in the asymptotic expansion as functions on the sphere at infinity, requiring that they are either even or odd functions of the sphere.
The combination of this conditions must be such that the defining objects of the Hamiltonian pictures are well-defined.
In particular, the symplectic form and the Hamiltonian must be finite, the latter must admit a Hamiltonian vector field (e.g. by being differentiable \`a la Regge-Teitelboim), and the Poincar\'e transformations must be canonical (which is ensured if they are symplectic).
Specifically, the fall-off conditions lead to a well-defined action of the Poincar\'e group (potentially non-canonical) and to a symplectic form which is, at most, logarithmically divergent.
The parity conditions, at this point, make sure that all the other mentioned conditions are satisfied, eventually with the inclusion of some boundary term in the Hamiltonian.

The reason why a combination of fall-off and parity conditions is employed instead of only relying on the fall-off conditions is that, in this way, the phase space turn out to be slightly larger and potentially more solutions to the equations of motion are included.
Indeed, if we tried to make the Hamiltonian formulation of free electrodynamics well-defined only by using fall-off conditions, these would be so strong that they would exclude the Coulomb solution from the phase space and, thus, from the theory.
Due to the great physical importance of this solution, we need to find a way of restricting the phase space that leads to a well-defined Hamiltonian formulation, but does not exclude this solution.
This is precisely achieved by the combination of fall-off and parity conditions and is needed, other than in electrodynamics, in other theories with long-range interactions, such as gravity and Yang-Mills.

The (strict) parity conditions --- found by requiring that the leading terms in the asymptotic expansions of the fields are either even or odd functions on the sphere at infinity --- are, in general, too strong.
Indeed, they exclude improper gauge transformations, since these would not have a well-defined action on the phase space, i.e. they would violates the parity conditions of the fields.\footnote{
Improper gauge transformations were briefly discussed in section~\ref{subsec:gauge-transformations-intro}.
Contrary to the (proper) gauge transformations, they are true, physical symmetries of the theory, whose canonical generator is obtained by complementing the one of proper gauge transformations with non-trivial boundary terms.
}
Therefore, the last step of the procedure is to relax the parity  conditions, in order to include the improper gauge transformations in the theory, but without loosing the good properties of a well-defined Hamiltonian formulation.
At this stage, it may be necessary to add a non-trivial boundary term to the symplectic form, as first noted by Henneaux and Troessaert for the case of electrodynamic~\cite{Henneaux-ED}.

The structure of this chapter is as follows.
Since part of the analysis was already carried in section~\ref{sec:gauge-theories}, we will directly start with the derivation of Poincar\'e transformation in section~\ref{sec:Lorentz-fall-off} and we will infer the fall-off conditions of the fields. 
In section~\ref{sec:parity}, we will find the strict parity conditions, which, in combination with the fall-off conditions, make the theory have a finite symplectic structure, a finite and functionally-differentiable Hamiltonian, and a canonical action of the Poincar\'e group.
However, as anticipated, these parity conditions seem too strong in that they exclude the possibility of non-trivial asymptotic symmetries and they prevent us from having a non-zero total colour charge.
For this reason, in section~\ref{sec:relax-parity-ED}, we review how this issue was resolved for electrodynamics in~\cite{Henneaux-ED}, which leads us to try a similar strategy in the Yang-Mills case in 
section~\ref{sec:relax-parity-YM}.
Interestingly, in the non-abelian case, this strategy now seems 
to manifestly fail for reasons that we outline in detail.
Finally, in section~\ref{sec:YM-higher-dim}, we will present the situation in higher spacetime dimensions, in order to show that this obstruction seems to be a peculiarity of the physically-relevant four-dimensional case.

\section{Poincar\'e transformations and fall-off conditions} \label{sec:Lorentz-fall-off}

The symplectic form and the Hamiltonian were derived in the general discussion about gauge theories of section~\ref{sec:gauge-theories}, where we made two simplifications.
First, we completely neglected if the relevant objects of the Hamiltonian formulation were actually well-defined, which usually require also a careful inspection of the potential boundary terms.
Secondly, we set the lapse and shift to the special values $N=1$ and $\vect{N} = 0$, which are enough to infer the time evolution of the fields, but not their behaviour under the action of the Poincar\'e group.

In this section, we will first provide the Poincar\'e transformation of the fields and, then, discuss their fall-off conditions.
These conditions should be strong enough, so that the Hamiltonian is finite and the symplectic form is, at most, logarithmically divergent, as already mentioned in the introduction of this chapter.
At the same time, they should be weak enough not to exclude any potentially interesting solution of the equations of motion.
Moreover, since one wishes to include the Poincar\'e transformations as symmetries of the theory, one also needs to impose that the fall-off conditions are preserved by Poincar\'e transformations.
For, otherwise, the transformations would map allowed filed configurations to non-allowed ones.

\subsection{Poincar\'e transformations of the fields} \label{subsec:poincare-transformations}

Let us begin by determining how the fields transform under Poincar\'e transformations.
The general method, described in section~\ref{sec:Poincare-general}, would require us to re-compute of the Hamiltonian, but this time for a foliation with arbitrary lapse and shift.
Then, replacing $N$ and $\vect{N}$ with $\xi^\perp$ and $\vect{\xi}$, respectively, we find a candidate for the generator of the Poincar\'e transformations.
A calculation of this kind will be presented in the next chapter for the case of a complex scalar field minimally-coupled to electrodynamics.

In this section, we follow a more intuitive and less cumbersome procedure, which will lead us, nevertheless, to the correct result.
To this end, let us write the Hamiltonian generator for arbitrary lapse and shift as
\begin{subequations} \label{YM:diffeo-generator-total}
\begin{equation} \label{YM:diffeo-generator-generic}
 H[N,\vect{N}]=
 \int d^3 x \, \big[
 N \, \mathscr{H} (A,\pi;g)
 + N^i \, \mathscr{H}_i (A,\pi;g)
 \big]
 +(\text{boundary}) \,.
\end{equation}
For the moment, let us neglect issues related to the fact that the generator above may be neither finite nor differentiable \`a la Regge-Teitelboim.
As we have already said, this is only a candidate and we will check \textit{a posteriori} in section~\ref{sec:parity} if these properties are fulfilled for the lapse and shift of the Poincar\'e transformations, after specifying the fall-off and parity conditions of the canonical fields.

At this point, let us note that the generator~(\ref{YM:diffeo-generator-generic}) needs to reduce to the already-computed Hamiltonian~(\ref{YM:Hamiltonian}) when we set $N=1$ and $\vect{N} = 0$.
Thus, we infer
\begin{equation} \label{YM:super-Hamiltonian}
 \mathscr{H} = \frac{\sprod{\pi^a}{\pi_a}}{2\sqrt{g}} + \frac{\sqrt{g}}{4} \sprod{F_{ab}}{F^{ab}} + \sprod{\lambda}{\mathscr{G}} \,. 
\end{equation}
Note that, due to the last term in~(\ref{YM:super-Hamiltonian}), the Hamiltonian generator~(\ref{YM:diffeo-generator-generic}) includes a gauge transformation with gauge parameter $\zeta \eqdef N \lambda$.
The tangential part of the generator $\mathscr{H}_i$ can be determined by geometrical reasons.
One simply requires that $A_a$ behaves like a covector field and $\pi^a$ like a density-one vector field under tangential deformations.
In other words, we ask that, if $N = 0$, the transformation of the fields is given by their Lie derivative along $\vect{N}$.
As a results, one finds
\begin{equation} \label{YM:super-momentum}
 \mathscr{H}_i = \sprod{\pi^a}{\partial_i A_a} - \partial_a (\sprod{\pi^a}{A_i}) \,.
\end{equation}
\end{subequations}
This determines completely the form of the generator~(\ref{YM:diffeo-generator-total}), up to boundary terms, that have been trivially neglected so far.

With this knowledge, we can compute the transformation of the fields under the Poincar\'e transformations by computing the equations of motion ensuing from~(\ref{YM:diffeo-generator-total}) and replacing the lapse and shift with $\xi^\perp$ and $\vect{\xi}$, as illustrated in section~\ref{sec:Poincare-general}.
Explicitly, we find
\begin{subequations} \label{YM:poincare-transformations}
\begin{align} \label{YM:poincare-transformations1}
 \delta_{\xi,\zeta} A_a &\eqdef \big\{ A_a, H[\xi^\perp,\vect{\xi}] \big\}
 =\xi^\perp \frac{\pi_a}{\sqrt{g}}+\xi^i \partial_i A_a+\partial_a \xi^i A_i-D_a \zeta \,,	\\
 \label{YM:poincare-transformations2}
 \delta_{\xi,\zeta} \pi^a &\eqdef \big\{ \pi^a, H[\xi^\perp,\vect{\xi}] \big\}
 = \sqrt{g} \, D_b(\xi^\perp F^{ba})
 +\partial_i (\xi^i \pi^a) - \partial_i \xi^a \pi^i+\extprod{\zeta}{\pi^a} \,.
\end{align}
\end{subequations}
Note that the Poincar\'e transformations contain an arbitrary gauge transformation, whose gauge parameter is $\zeta$.

\subsection{Fall-off conditions of the fields}
In this subsection, we determine the fall-off conditions of the fields.
To this end, we will work in radial-angular coordinate $(r,\barr{x})$, where the three metric can be written as~(\ref{3-metric-radial-angular}).

In order to derive the fall-off conditions of the fields, we demand the following requirements to be satisfied.
First of all, the canonical symplectic form~(\ref{YM:symplectic-form}) should be, at most, logarithmically divergent.
Second, the fall-off conditions of the fields should be preserved by the Poincar\'e transformations~(\ref{YM:poincare-transformations}).
Third, the asymptotic expansion of the fields should be of the form
\begin{subequations} \label{YM:fall-off-conditions-preliminary}
\begin{align}
  A_r (r,\barr{x}) &= \frac{\barr{A}_r (\barr{x})}{r^\alpha}
  +\bigo(1/r^{\alpha+1}) \,, \quad
  & \pi^r (r,\barr{x}) &= \frac{\barr{\pi}^r (\barr{x})}{r^{\alpha'}} 
  +\bigo(1/r^{\alpha'+1})\,, \\
  A_{\bar{a}} (r,\barr{x}) &= \frac{\barr{A}_{\bar{a}} (\barr{x})}{r^\beta}
  +\bigo(1/r^{\beta+1}) \,, \quad
  & \pi^{\bar{a}} (r,\barr{x}) &= \frac{\barr{\pi}^{\bar{a}} (\barr{x})}{r^{\beta'}}
  + \bigo(1/r^{\beta'+1}) \,.
\end{align}
\end{subequations}
As usual, the dependence of the fields on the time $t$, though present, is not denoted explicitly in the expressions above, nor in the following ones. 
Note that we require the leading term in the expansion to be an integer power of $r$ and the first subleading term in the expansion to be the power of $r$ with exponent reduced by one.
Functions whose fall-off behaviour is between the two next powers of $r$, such as those one could build using logarithms, are excluded at the first subleading order.
Fourth, the fall-off conditions should be the most general ones compatible with the previous three requirements, so that the space of allowed field configurations is as big as possible.
In addition, we expand also the gauge parameter appearing in~(\ref{YM:poincare-transformations}) according to
\begin{equation} \label{YM:fall-off-conditions-gauge-preliminary}
 \zeta= \frac{1}{r^\delta} \barr{\zeta} (\barr{x}) +\bigo(1/r^{\delta+1}) \,.
\end{equation}

To begin with, the requirement that the canonical symplectic form~(\ref{YM:symplectic-form}) is, at most, logarithmically divergent implies the relations
\begin{equation} \label{YM:fall-off-from-symplectic}
 \alpha+\alpha' \ge 1
 \qquad \text{and} \qquad
 \beta+\beta' \ge 1
\end{equation}
among the exponents defined in~(\ref{YM:fall-off-conditions-preliminary}).
If the two inequalities above are satisfied strictly, then the symplectic form is actually finite.

Then, one checks when the fall-off conditions~(\ref{YM:fall-off-conditions-preliminary}) and~(\ref{YM:fall-off-conditions-gauge-preliminary}) are preserved by the Poincar\'e transformations.
To do so, one considers the transformation of the fields~(\ref{YM:poincare-transformations}) and inserts, into these expressions, the asymptotic expansions~(\ref{YM:fall-off-conditions-preliminary}) and~(\ref{YM:fall-off-conditions-gauge-preliminary}).
As a result, one finds that the fall-off conditions are preserved by the Poincar\'e transformations if
\begin{align} \label{fall-off-from-poincare}
 1 \le \alpha < 2 \,, &&
 \alpha' = \alpha -1 \,, &&
 \beta=0 \,, &&
 \beta'=1 \,, &&
 \delta \ge 0 \,.
\end{align}
Note that these equations already imply~(\ref{YM:fall-off-from-symplectic}).
Finally, requiring that the fall-off conditions are the most general ones of all the possible ones, one obtains that the fields behave asymptotically as
\begin{subequations} \label{YM:fall-off-conditions}
\begin{align} \label{YM:fall-off-conditions-fields}
  A_r (r,\barr{x}) &= \frac{1}{r} \barr{A}_r (\barr{x})
  +\bigo\big(1/r^2\big) \,,
  & \pi^r (r,\barr{x}) &= \barr{\pi}^r (\barr{x}) +\bigo(1/r)\,, \\
  A_{\bar{a}} (r,\barr{x}) &=\barr{A}_{\bar{a}} (\barr{x}) +\bigo(1/r) \,, 
  & \pi^{\bar{a}} (r,\barr{x}) &= \frac{1}{r} \barr{\pi}^{\bar{a}} (\barr{x})
  + \bigo \big(1/r^2\big)
\end{align}
and the gauge parameter behaves as
\begin{equation} \label{YM:fall-off-conditions-gauge}
 \zeta (r,\barr{x})= \barr{\zeta} (\barr{x}) +\bigo(1/r) \,.
\end{equation}
\end{subequations}
Of course, the gauge parameter $\lambda$ appearing in~(\ref{YM:Hamiltonian}) and~(\ref{YM:gauge-generator}) needs to satisfy the same fall-off behaviour of $\zeta$, so that gauge transformations~(\ref{YM:gauge-transformations}) preserve the fall-off conditions~(\ref{YM:fall-off-conditions}) of the canonical fields.

To sum up,  we have determined the most general fall-off conditions of the fields and of the gauge parameter, under the requirements that they are preserved by the Poincar\'e transformations and that they make the symplectic form, at most, logarithmically divergent.
Specifically, the fall-off conditions~(\ref{YM:fall-off-conditions}) imply that the symplectic form is precisely logarithmically divergent and not yet finite.
We will solve this issue in section~\ref{sec:parity} by means of parity conditions.
But before we do that, we spend the remainder of this section to work out the explicit expressions for the Poincar\'e transformations of the asymptotic part of the fields.

\subsection{Asymptotic Poincar\'e transformations}

We will now write explicitly the action of the Poincar\'e transformations on the asymptotic part of the fields.
The results of this subsection will be used when discussing the parity conditions in the next section.

The procedure to obtain the Poincar\'e transformations of the asymptotic part of the fields is straightforward, although a little cumbersome.
One inserts the asymptotic expansions~(\ref{YM:fall-off-conditions}) into the transformations~(\ref{YM:poincare-transformations}) and write the explicit values~(\ref{poincare-xi}) of $\xi^\perp$ and $\vect{\xi}$.
After neglecting all the subleading contributions in the so-found expressions, one finds
\begin{subequations} \label{YM:poincare-asymptotic}
\begin{align}
 \label{YM:poincare-asymptotic1}
 \delta_{\xi,\zeta} \barr{A}_r ={}&
 \frac{b\, \barr{\pi}^r}{\sqrt{\barr{\gamma}}}
 + Y^{\bar m} \partial_{\bar m} \barr{A}_r
 + \extprod{\barr{\zeta}}{\barr{A}_r} \,,\\
 \label{YM:poincare-asymptotic2}
 \delta_{\xi,\zeta} \barr{A}_{\bar a} ={}&
 \frac{b \, \barr{\pi}_{\bar a}}{\sqrt{\barr{\gamma}}}
 +Y^{\bar m} \partial_{\bar m} \barr{A}_{\bar a} + \partial_{\bar a} Y^{\bar m} \barr{A}_{\bar m}
 -\barr{D}_{\bar a} \barr{\zeta}
 \,, \\
 \label{YM:poincare-asymptotic3}
 \delta_{\xi,\zeta} \barr{\pi}^r ={}&
 \barr{D}^{\bar m} \big( b\, \sqrt{\barr{\gamma}}\, \barr{D}_{\bar m} \barr{A}_r \big)
 +\partial_{\bar m} (Y^{\bar m} \barr{\pi}^r)
 +\extprod{\barr{\zeta}}{\barr{\pi}^r} \,, \\
 \label{YM:poincare-asymptotic4}
 \delta_{\xi,\zeta} \barr{\pi}^{\bar a} ={}& 
 \barr{D}_{\bar m} \big( b \, \sqrt{\barr{\gamma}}\, \barr{F}^{\bar m \bar a})
 +b \sqrt{\barr{\gamma}} \, \extprod{\barr{D}^{\bar a} \barr{A}_r}{\barr{A}_r}
 +\partial_{\bar m} (Y^{\bar m}\, \barr{\pi}^{\bar a})
  + \\
 &-\partial_{\bar m} Y^{\bar a} \, \barr{\pi}^{\bar m}
 +\extprod{\barr{\zeta}}{\barr{\pi}^{\bar a}} \,, \nonumber
\end{align}
\end{subequations}
where angular indices are lowered and raised with the use of the metric of the unit two-sphere $\barr{\gamma}_{\bar a \bar b}$ and its inverse $\barr{\gamma}^{\bar a \bar b}$, respectively.\footnote{
See section~\ref{sec:Poincare-general} and equation~(\ref{3-metric-radial-angular}) for comparison.
}
Furthermore,
\begin{equation}
 \barr{F}_{\bar m \bar n} \eqdef 
\partial_{\bar m} \barr{A}_{\bar n}-\partial_{\bar n} \barr{A}_{\bar m}+\extprod{\barr{A}_{\bar m}}{\barr{A}_{\bar n}}
\end{equation}
and  $\barr{D}_{\bar a} \eqdef \barr{\nabla}_{\bar a} + \extprod{\barr{A}_{\bar a}}{}$ is the asymptotic gauge-covariant derivative, being $\barr{\nabla}_{\bar a}$ the covariant derivative on the unit round two-sphere.

One sees immediately that the asymptotic transformations above are affected only by the boost $b$ and the rotations $Y^{\bar m}$, but not by the translations $T$ and $W$.
Moreover, these transformations exhibit two main differences with respect to the analogous transformations in electrodynamics~\cite{Henneaux-ED}.\footnote{
The asymptotic Poincar\'e transformations of free electrodynamics can also be read from the transformations~(\ref{SEDAH:poincare-asymptotic})  setting $\barr{\varphi}$ and $\barr{\Pi}$ to zero.
}
First, the radial and angular components of the fields do \emph{not} transform independently, due to the mixing terms in the transformation of the momenta.
Secondly, none of the asymptotic fields are gauge invariant.
Both these properties are a consequence of the non-abelian nature of the gauge group and will play an important role in the discussion of parity conditions in the next section.

\section{A well-defined Hamiltonian formulation and parity conditions} \label{sec:parity}

The fall-off conditions~(\ref{YM:fall-off-conditions}) are not sufficient to ensure the finiteness of the symplectic form~(\ref{YM:symplectic-form}), which is, indeed, still logarithmically divergent.
This problem can be fixed in the following way.
First, one assigns, independently to one another, a definite parity to the asymptotic part of the fields, $\barr{A}_r(\barr{x})$ and $\barr{A}_{\bar{a}}(\barr{x})$, so that they are either odd or even functions on the two-sphere.
Secondly, one imposes the opposite parity on the asymptotic part of the corresponding conjugated momenta, $\barr{\pi}^r(\barr{x})$ and $\barr{\pi}^{\bar a}(\barr{x})$.
This way, the logarithmically divergent term in the symplectic form is, in fact, zero once integrated on the two-sphere.

Specifically, let us assume that $\barr{A}_r$ has parity $s\in \mathbb{Z}_2$ and that $\barr{A}_{\bar{a}}$ has parity $\sigma \in \mathbb{Z}_2$, i.e., they behave under the antipodal map $\barr{x} \mapsto -\barr{x}$,\footnote{
See footnote~\ref{footnote:antipodal-map} on page~\pageref{footnote:antipodal-map}.
}
as 
\begin{equation} \label{YM:parity-conditions-undetermined}
 \barr{A}_r (-\barr{x}) = (-1)^s \, \barr{A}_r (\barr{x})
 \qquad \text{and} \qquad
 \barr{A}_{\bar{a}} (-\barr{x}) = (-1)^\sigma \, \barr{A}_{\bar{a}} (\barr{x}) \,.
\end{equation}
Then, the symplectic form is made finite by assuming that $\barr{\pi}^r$ has parity $s+1$ and that $\barr{\pi}^{\bar a}$ has parity $\sigma+1$.
The key observation is that the values of $s$ and $\sigma$ are unequivocally determined by the requirement that the Poincar\'e transformations are canonical and that they preserve the parity transformations.
In electrodynamics, it is possible to relax these strict parity conditions leaving the symplectic form still finite~\cite{Henneaux-ED}.
We will review how this procedure works in electrodynamics in section~\ref{sec:relax-parity-ED} and attempt to apply it to the Yang-Mills case in section~\ref{sec:relax-parity-YM}.

\subsection{Proper and improper gauge transformations} \label{subsec:gauge-transformations}

Before we determine the parity conditions, let us extend the discussion about gauge transformations, which was started in section~\ref{subsec:gauge-transformations-intro} of the previous chapter.
As we have already mentioned there, gauge transformations are generated by
\begin{equation} \label{YM:gauge-generator-again}
 G[\lambda] \eqdef \int d^3 x\, \sprod{\lambda (x)}{\mathscr{G} (x)} \,,
\end{equation}
which consists of the Gauss constraint~(\ref{YM:gauss-constraint}) smeared with a (Lie-algebra valued) function $\lambda$.
The generator above is differentiable \`a la Regge-Teitelboim --- whose conditions ensure the existence of an associated Hamiltonian vector field if the canonical symplectic form does not contain boundary terms --- if, and only if, the surface term
\begin{equation} \label{YM:gauge-generator-surface}
 \oint_{S^2_{\infty}} d^2 \barr{x}_k \, \sprod{\lambda}{\pi^k}
 = \oint_{S^2} d^2 \barr{x} \; \sprod{\barr{\lambda}}{\barr{\pi}^r} 
\end{equation}
vanishes.
In the right-hand side of the above expression, we have inserted the fall-off behaviour of the fields and of the gauge parameter~(\ref{YM:fall-off-conditions}).
Note that the integral on the right-hand side is an integral over a unit sphere, since the dependence on the radial coordinate $r$ disappears after taking the limit to an infinite-radius two-sphere in the left-hand side.
One sees immediately that the surface term vanishes for every allowed $\barr{\pi}^r$ if, and only if, the asymptotic gauge parameter $\barr{\lambda}$ has parity $s$, which is the opposite parity of $\barr{\pi}^r$.

There is an alternative way to make the generator~(\ref{YM:gauge-generator-again}) differentiable.
Precisely, one defines the extended generator
\begin{equation} \label{YM:gauge-generator-extended}
 G_{\text{ext.}}[\epsilon] \eqdef \int d^3 x \, \sprod{\epsilon(x)}{\mathscr{G}(x)}
 -\oint d^2 \barr{x}\; \sprod{\barr{\epsilon}(\barr{x})}{\barr{\pi}^r(\barr{x})} \,,
\end{equation}
where the function $\epsilon(x)$ is required to satisfy the same fall-off behaviour~(\ref{YM:fall-off-conditions-gauge}) of $\lambda (x)$ and $\zeta(x)$, but its asymptotic part $\barr{\epsilon}$ is \emph{not} restricted to have a definite parity.
One can easily verify that $G_{\text{ext.}}[\epsilon]$ is  differentiable \`a la Regge-Teitelboim for all the $\epsilon$ in this larger set of functions and that it generates the infinitesimal transformations
\begin{subequations} \label{YM:gauge-transformations-improper}
 \begin{align} \label{YM:gauge-transformations-improper1}
 \delta_\epsilon A_a &\eqdef \{ A_a, G_{\text{ext.}}[\epsilon] \} = -\partial_a \epsilon +\extprod{\epsilon}{A_a} \,, \\
 \label{YM:gauge-transformations-improper2}
 \delta_\epsilon \pi^a &\eqdef \{ \pi^a, G_{\text{ext.}}[\epsilon] \} = \extprod{\epsilon}{\pi^a} \,.
\end{align}
\end{subequations}
Moreover, one can also verify that $\big\{ G_{\text{ext.}}[\epsilon], H \big\}=0$, so that $G_{\text{ext.}}[\epsilon]$ is the generator of a symmetry.
The infinitesimal transformations above can be integrated to get the transformations with parameter $\mathcal{U}\eqdef \exp (-\epsilon) \in \SU(N)$
\begin{subequations} \label{YM:gauge-transformations-improper-full}
\begin{align} \label{YM:gauge-transformations-improper-full1}
 \Gamma_\mathcal{U} (A_a) & = \mathcal{U}^{-1} A_a \,\mathcal{U}+\mathcal{U}^{-1} \partial_a  \,\mathcal{U} \,, \\
 \label{YM:gauge-transformations-improper-full2}
 \Gamma_\mathcal{U} (\pi^a)  &= \mathcal{U}^{-1} \pi^a \,\mathcal{U} \,,
\end{align}
\end{subequations}
where the products on the right-hand sides are products among matrices.
In the expressions above, we have denoted the action of $\SU(N)$ on the fields with $\Gamma$, instead of the usual $\Phi$, in order not to make confusion with the field $\Phi$ introduced later in this chapter.

Note that, when $\barr{\epsilon}$ has parity $s$, the surface term in~(\ref{YM:gauge-generator-extended}) vanishes and $G_{\text{ext.}}[\epsilon]$ coincides with $G[\epsilon]$.
In this case, the symmetries generated by $G_{\text{ext.}}[\epsilon]$ are precisely a gauge transformations connecting physically-equivalent field configurations, as discussed in section~\ref{subsec:gauge-transformations-intro}.
We will refer to them in a rather pedantic way as \emph{proper gauge transformations}, in order to avoid any possible misunderstanding in the following discussion.

When $\barr{\epsilon}$ has parity $s+1$, the surface term in~(\ref{YM:gauge-generator-extended}) does \emph{not} vanish any more.
The transformation generated by $G_{\text{ext.}}[\epsilon]$, in this case, connects physically-inequivalent field configurations.
We refer to this transformations as \emph{improper gauge transformations}, following~\cite{Teitelboim-YM2}.
These, on the contrary of proper gauge transformations, are true symmetry of the theory connecting physically-inequivalent field configurations.
A general transformation generated by $G_{\text{ext.}}[\epsilon]$ will be the combination of a proper gauge transformation and of an improper one.

The generator~(\ref{YM:gauge-generator-extended}) is made of two pieces.
The former consists of the Gauss constraints $\mathscr{G}$ smeared with the function $\epsilon (x)$.
As a consequence, this term vanishes when the constraints are satisfied.
The latter is a surface term.
One can compute the value of the generator when the constraints are satisfied, which is, in particular, the case for any solution of the equations of motion.
One obtains
\begin{equation} \label{YM:charge}
 G_{\text{ext.}}[\epsilon]
 \weq -\oint d^2 \barr{x}\; \sprod{\barr{\epsilon}(\barr{x})}{\barr{\pi}^r(\barr{x})}
 \defeq Q[\barr{\epsilon}] \,,
\end{equation}
where we have defined the charge $Q[\barr{\epsilon}]$.
When the Lie-algebra-valued function $\epsilon(\barr{x})$ is constant over the sphere, we can write
$Q[\barr{\epsilon}] = \sprod{\barr{\epsilon}}{Q_0}$
in terms of the total colour charge measured at spatial infinity
\begin{equation} \label{YM:colour-charge}
 Q_0 \eqdef - \oint d^2 \barr{x} \; \pi^r (\barr{x}) \,.
\end{equation}

Finally, let us determine the transformation of the asymptotic fields under proper and improper gauge transformations.
Expanding the equations~(\ref{YM:gauge-transformations-improper})) using the fall-off conditions~(\ref{YM:fall-off-conditions}), one finds
\begin{equation} \label{YM:gauge-transformations-improper-asymptotic}
 \delta_\epsilon \barr{A}_r = \extprod{\barr{\epsilon}}{\barr{A}_r} \,,
 \quad
 \delta_\epsilon \barr{A}_{\bar a} = -\partial_{\bar a} \barr{\epsilon} +\extprod{\barr{\epsilon}}{\barr{A}_{\bar a}} \,,
 \quad
 \delta_\epsilon \barr{\pi}^r = \extprod{\barr{\epsilon}}{\barr{\pi}^r} \,,
 \quad \text{and} \quad
 \delta_\epsilon \barr{\pi}^{\bar a} = \extprod{\barr{\epsilon}}{\barr{\pi}^{\bar a}} \,,
\end{equation}
whereas, expanding the equations~(\ref{YM:gauge-transformations-improper-full}), one finds
\begin{subequations} \label{YM:gauge-transformations-improper-full-asymptotic}
\begin{align} \label{YM:gauge-transformations-improper-full-asymptotic1}
 \Gamma_{\mathcal{U}} (\barr{A}_r) &= \barr{\mathcal{U}}^{-1} \barr{A}_r \, \barr{\mathcal{U}} \,,
 &
 \Gamma_{\mathcal{U}} ( \barr{A}_{\bar a}) &=\barr{\mathcal{U}}^{-1} \barr{A}_{\bar a} \, \barr{\mathcal{U}} +\barr{\mathcal{U}}^{-1} \partial_{\bar a} \, \barr{\mathcal{U}} \,,
 \\ \label{YM:gauge-transformations-improper-full-asymptotic2}
 \Gamma_{\mathcal{U}} ( \barr{\pi}^r) &= \barr{\mathcal{U}}^{-1} \barr{\pi}^r \, \barr{\mathcal{U}}\,,
 &
\Gamma_{\mathcal{U}} ( \barr{\pi}^{\bar a} ) &= \barr{\mathcal{U}}^{-1} \barr{\pi}^{\bar a} \, \barr{\mathcal{U}} \,,
\end{align}
\end{subequations}
where $\barr{\mathcal{U}}\eqdef \exp(-\barr{\epsilon})$.
Note that the total colour charge transforms non-trivially under proper and improper gauge transformations as
\begin{equation} \label{YM:colour-charge-transformation}
 \Gamma_{\mathcal{U}} (Q_0) =
 - \oint d^2 \barr{x} \; \barr{\mathcal{U}}^{-1} (\barr{x}) \,  \barr{\pi}^r (\barr{x}) \, \barr{\mathcal{U}} (\barr{x}) \,.
\end{equation}
We will complete this discussion once that we have determined the parity conditions in the next subsection.

\subsection{Poincar\'e transformations and parity conditions} \label{subsec:poincare-parity}
In this subsection, we elaborate on some aspects of the Poincar\'e transformations, that were left aside in the previous discussions in section~\ref{sec:Lorentz-fall-off} and we determine the parity conditions of the asymptotic fields, that is the values of $s$ and $\sigma$, which were introduced at the beginning of this section.
In order to do so, we require the Poincar\'e transformations to be canonical and to preserve the parity conditions.
The former condition, as discussed in section~\ref{sec:Poincare-general}, is equivalent to the fact that the transformations are symplectic, i.e., that their vector field $X$ satisfies $\liephase_X \Omega =  \extderphase (i_X \Omega)=0$.
The latter condition is necessary is order to make sure that the Poincar\'e transformation have an action on the phase space.
If, indeed, they were not preserving the parity conditions, they would transform some field configurations belonging to the phase space to field configurations outside the phase space.

Let us begin with the analysis of the condition that the Poincar\'e transformations preserve parity.
To this end, let us take into consideration the asymptotic Poincar\'e transformations~(\ref{YM:poincare-asymptotic}).
The parts of the transformations depending on $\barr{\zeta}$ are, in fact, a proper gauge transformation, which we will discuss below.
The rest of the transformations preserves parity conditions as long as $\sigma=1$, as one can easily check.

Let us now impose that the Poincar\'e transformations are symplectic and, as a consequence, canonical.
Using the canonical the symplectic form~(\ref{YM:symplectic-form}) and denoting the components of the vector field of the Poincar\'e transformations as $X = (\delta_{\xi,\zeta} A_a,\delta_{\xi,\zeta} \pi^a)$, one has the generic expression
\begin{equation}
 \extderphase (i_X \Omega)=
 \int d^3 x \Big[
 \sprod{\extderphase \big( \delta_{\xi,\zeta} \pi^a \big) \wedge}{\extderphase A_a}+
 \sprod{\extderphase  \pi^a  \wedge}{\extderphase \big( \delta_{\xi,\zeta} A_a \big) }
 \Big]
\end{equation}
This can be evaluated by inserting the explicit value of the transformations~(\ref{YM:poincare-transformations}).
After a few lines of calculations and after the use of the fall-off conditions~(\ref{YM:fall-off-conditions}), one finds
\begin{equation} \label{YM:poincare-canonical-final}
 \extderphase (i_X \Omega)= \oint d^2 \barr{x} \; b \, \sqrt{\barr{\gamma}} \,
 \sprod{\extderphase \barr{A}_{\bar m} \wedge}{\extderphase \bigl( \barr{D}^{\, \bar m} \; \barr{A}_r \bigr)} \,.
\end{equation}
One can note three things.
First, after the fall-off conditions have been imposed, the only part of the Poincar\'e transformations which could lead to some problem is the boost sector.
Secondly, the above expression is precisely the non-abelian analogous of the one derived in~\cite{Henneaux-ED} for electrodynamics.
Lastly, if $\sigma = 1$, the right-hand side vanishes so long as $s=0$, which fully determines the parity conditions.

In short, the asymptotic fields need to satisfy the parity conditions
\begin{subequations} \label{YM:parity-conditions}
\begin{align} 
 \barr{A}_r (-\barr{x}) &= \barr{A}_r (\barr{x}) \,,
 &
 \barr{A}_{\bar{a}} (-\barr{x}) &= -\barr{A}_{\bar{a}} (\barr{x}) \,, \\
 \barr{\pi}^r (-\barr{x}) &= - \barr{\pi}^r (\barr{x}) \,,
 &
 \barr{\pi}^{\bar{a}} (-\barr{x}) &=  \barr{\pi}^{\bar{a}} (\barr{x}) \,.
\end{align}
\end{subequations}
Moreover, the gauge parameter of proper gauge transformations satisfies
\begin{equation} \label{YM:parity-condition-gauge}
 \barr{\epsilon}_{\text{proper}} (-\barr{x})=
 \barr{\epsilon}_{\text{proper}} (\barr{x}) \,.
\end{equation}
It is easy to check that proper gauge transformations --- including the terms of the Poincar\'e transformations~(\ref{YM:poincare-asymptotic}) depending on $\barr{\zeta}$ --- preserve the parity conditions.

The parity conditions of the fields~(\ref{YM:parity-conditions}) and of the paramter of proper gauge transformations~(\ref{YM:parity-condition-gauge}) have a few consequences, other than making the symplectic form~(\ref{YM:symplectic-form}) finite.
First, the Hamiltonian~(\ref{YM:Hamiltonian}) is finite and  differentiable \`a la Regge-Teitelboim, as one can easily check.
With the exclusion of term containing the Gauss constraint, this was already true after that we had imposed the fall-off conditions~(\ref{YM:fall-off-conditions}).
The parity conditions make it true also for this last term.

Second, improper gauge transformations are, at this stage, \emph{not} allowed.
Indeed, they change the asymptotic fields as in~(\ref{YM:gauge-transformations-improper-asymptotic}) when the asymptotic part of the gauge parameter has parity
\begin{equation}
 \barr{\epsilon}_{\text{improper}} (-\barr{x})=
 -\barr{\epsilon}_{\text{improper}} (\barr{x}) \,.
\end{equation}
However, these transformations do not preserve the parity conditions~(\ref{YM:parity-conditions}).
Therefore, if they were allowed, they would transform one point of the space of allowed field configurations to a point that does not belong to this space any more.
In other words, they do not have a well-defined action on the phase space.
In section~\ref{sec:relax-parity-YM}, we will discuss whether or not it is possible to modify parity conditions in order to restore the improper gauge transformations into the theory.

Third, the Poincar\'e transformations are canonical.
Their canonical generator, which is presented in the next subsection, is finite and differentiable \`a la Regge-Teitelboim.
Note that, with the exception of the boost, the transformations were already canonical even before imposing the parity conditions.
The parity conditions presented in this section fix the behaviour of the boost.

Last but not least, since $\barr{\pi}^r (\barr{x})$ is an odd function of $\barr{x}$, all the charges $Q[\barr{\epsilon}]$ defined in~(\ref{YM:charge}) are vanishing when $\barr{\epsilon} (\barr{x})$ is an even function.
Notably, this includes the total colour charge $Q_{0}$, defined in~(\ref{YM:colour-charge}), which is therefore zero.
Note that, despite the colour charge is not a gauge-invariant quantity, the statement that it is actually equal to zero is a gauge-invariant statement.
Indeed, using equation~(\ref{YM:colour-charge-transformation}), one sees that the colour charge vanishes for every \emph{proper} gauge transformation
$\barr{\mathcal{U}} (\barr{x}) = \exp \big[-\epsilon_{\text{proper}} (\barr{x}) \big]$,
after imposing the parity conditions~(\ref{YM:parity-conditions}) and~(\ref{YM:parity-condition-gauge}).

The above considerations would suggest that there are some issues if one wants a well-defined Lorentz boost and a non-zero colour charge in the Yang-Mills theory. A similar suggestion, coming from a different approach, 
was already present in~\cite{Christodoulou.Murchadha:1981}, where Christodoulou 
and {\'o} Murchadha studied the boost problem in General Relativity and 
briefly commented, at the end of their section 5, that the boost problem
does not seem to have  solutions for charged configurations 
in the Yang-Mills case, quite in contrast to the behaviour of 
General Relativity.

\subsection{Poincar\'e generator and algebra}

Now that we know that the Poincar\'e transformations are canonical, we present their finite and functionally-differentiable canonical generator, included the needed boundary term.
This is obtained, up to boundary terms, from the Hamiltonian generator $H[N,\vect{N},\lambda]$ given in~(\ref{YM:diffeo-generator-total}), by replacing $N$ and $\vect{N}$ with $\xi^\perp$ and $\vect{\xi}$ given in~(\ref{poincare-xi}).
Let us denote it as $P[\xi^\perp,\vect{\xi},\zeta]$ to stress that it is the generator of the Poincar\'e transformations. 
We have
\begin{subequations} \label{YM:poincare-generator-total}
\begin{equation} \label{YM:poincare-generator-generic}
 P[\xi^\perp,\xi^i,\zeta]=
 \int d^3 x \, \big[
 \xi^\perp \, \mathscr{P}_0
 + \xi^i \, \mathscr{P}_i
 + \sprod{\zeta}{\mathscr{G}}
 \big]
 +\oint d^2 \barr{x} \; \mathscr{B} \,,
\end{equation}
where the generator of the normal component of the Poincar\'e transformations is
\begin{equation} \label{YM:P0}
 \mathscr{P}_0 = \mathscr{H}_0 = \frac{\sprod{\pi^a}{\pi_a}}{2\sqrt{g}} + \frac{\sqrt{g}}{4} \sprod{F_{ab}}{F^{ab}} \,,
\end{equation}
the generator of the tangential component is
\begin{equation} \label{YM:Pi}
 \mathscr{P}_i = \mathscr{H}_i = \sprod{\pi^a}{\partial_i A_a} - \partial_a (\sprod{\pi^a}{A_i}) \,,
\end{equation}
the generator of the proper gauge transformations $\mathscr{G}$ is the Gauss constraint~(\ref{YM:gauss-constraint}), and the explicit expression of the boundary term is
\begin{equation} \label{YM:poincare-boundary}
 \mathscr{B} = \sprod{\barr{\pi}^r}{Y^{\bar a} \barr{A}_{\bar a}} \,,
\end{equation}
\end{subequations}
which is needed to make the generator~(\ref{YM:poincare-generator-total}) differentiable \`a la Regge-Teitelboim.

Moreover, one can show that the Poincar\'e generator satisfy the algebra
\begin{subequations}
\begin{equation} \label{YM:Poincare-algebra-generic}
 \Big\{ P \big[\xi_1^\perp, \vect{\xi}_1, \zeta_1 \big], P \big[\xi^\perp_2, \vect{\xi}_2, \zeta_2 \big] \Big\}=
 P \big[\widehat \xi^\perp, \widehat{\vect{\xi}}, \widehat \zeta \big]\,,
\end{equation}
where
\begin{align}
 \widehat \xi^\perp &= \xi^i_1 \partial_i \xi^\perp_2 -\xi^i_2 \partial_i \xi^\perp_1 \,, \\
 \widehat \xi^i &= g^{ij} (\xi^\perp_1 \partial_j \xi^\perp_2-\xi^\perp_2 \partial_j \xi^\perp_1)
 + \xi^j_1 \partial_j \xi^i_2 - \xi^j_2 \partial_j \xi^i_1 \,, \\
 \label{YM:Poincare-algebra-gauge}
 \widehat \zeta &= A_i g^{ij} (\xi^\perp_1 \partial_j \xi^\perp_2-\xi^\perp_2 \partial_j \xi^\perp_1)
 +\xi^i_1 \partial_i \zeta_2 - \xi^i_2 \partial_i \zeta_1
 + \extprod{\zeta_1}{\zeta_2} \,.
\end{align}
\end{subequations}
This is precisely a Poisson-representation of the Poincar\'e Lie algebra up to gauge transformations.
Note that the gauge transformation on the right-hand sides of~(\ref{YM:Poincare-algebra-generic}) remains present even when the gauge parameters on the left-hand side vanish.
Indeed, from~(\ref{YM:Poincare-algebra-gauge}), one can check that setting $\zeta_1$ and $\zeta_2$ to zero is not enough to make $\widehat \zeta$ vanish.
However, this fact does not constitute a problem (see e.g. the discussion in~\cite[Sec. 2]{Beig}).

This concludes this section, in which we have shown that imposing the fall-off conditions~(\ref{YM:fall-off-conditions}) together with the parity conditions~(\ref{YM:parity-conditions}) lead to a well-defined symplectic form with a well-defined Hamiltonian and a well-defined canonical action of the Poincar\'e group on the fields.
Moreover, enforcing the parity conditions~(\ref{YM:parity-conditions}) has two consequences other than the ones listed above.
First, the improper gauge transformations are not allowed any more and, as a result, the asymptotic symmetry group is trivial.
Secondly, some of the charges~(\ref{YM:charge}) measured at spatial infinity, and in particular the $Q[\barr{\epsilon}]$ with even $\barr{\epsilon}$, are vanishing.
Notably, this includes the total colour charge $Q_0$.\footnote
{\label{foot:radial-parity}
In order to have a non-vanishing colour charge, we would need the radial components to satisfy the opposite parity conditions to the ones presented in this section.
However, these would make the Poincar\'e transformations non-canonical.
Whether or not there is a way to implement the different parity conditions leaving the Poincar\'e transformations canonical will be discussed in the next section.
In addition, these parity conditions would also exclude the possibility of making proper gauge transformations with a non-vanishing part at infinity, but would allow improper gauge transformations.
}
In the next section, we explore the possibility of modifying the parity conditions, in order to restore improper gauge transformations as symmetries of the theory.

\section{Relaxing parity conditions and asymptotic symmetries in electrodynamics} \label{sec:relax-parity-ED}
\sectionmark{Relaxing parity conditions in electrodynamics}

In the previous analysis, we have imposed fall-off and parity conditions on the canonical fields and we have obtained, as a result, a well-defined Hamiltonian picture. 
However, at least in the case of electrodynamics, it is possible to weaken the parity conditions so that the symplectic form is still finite and improper gauge transformations are allowed, as it was shown in~\cite{Henneaux-ED}.
Before we investigate this possibility in the case of Yang-Mills, let us briefly show, in this section, how the procedure works in the simpler case of electrodynamics.

\subsection{Relaxing parity conditions} \label{subsec:loosen-parity-ED}
To begin with, let us note that the equations of the electromagnetic case can be inferred from the equations of section~\ref{sec:gauge-theories} and of this chapter by replacing formally the one-form Yang-Mills potential $A_a$ with one-form electromagnetic potential $A_a^{\text{ED}}$ and the Yang-Mills conjugated momentum $\pi^a$ with the electromagnetic conjugated momentum $\pi^a_{ED}$.
In addition, one also needs to replace the Killing scalar product $\sprod{}{}$ with the product among real numbers and set to zero every term containing the non-abelian contributions given by the commutator $\extprod{}{}$.
In the remainder of this section, we will not write explicitly the subscript and the superscript ``ED'' on the fields, since we will consider only the electromagnetic case.

If we followed the same line of argument of section~\ref{sec:Lorentz-fall-off} in the case of electrodynamics, we would arrive at the same fall-off conditions~(\ref{YM:fall-off-conditions}) for the canonical fields and the gauge parameter.
These are precisely the fall-off conditions presented in~\cite{Henneaux-ED}.

Then, if we determined the parity conditions with the same reasoning of the section~\ref{sec:parity}, we would find out that any choice of definite parity for $\barr{A}_r$ and $\barr{A}_{\bar a}$ would be preserved by the Poincar\'e transformations.
However, these would be canonical only if the parity of $\barr{A}_r$ were opposite to that of $\barr{A}_{\bar a}$.
At this point, we choose the parity of $\barr{\pi}^r$ to be even, so that Coulomb is an allowed solution.
Therefore, we arrive at the parity conditions
\begin{subequations}
\begin{align} \label{YM:parity-conditions-ED}
 \barr{A}_r (-\barr{x})&= -\barr{A}_r (\barr{x}) \,,
 &
 \barr{A}_{\bar{a}} (-\barr{x})&= \barr{A}_{\bar{a}} (\barr{x}) \,,
 \\
 \barr{\pi}^r (-\barr{x})&= \barr{\pi}^r (\barr{x}) \,,
 &
 \barr{\pi}^{\bar{a}} (-\barr{x})&= -\barr{\pi}^{\bar{a}} (\barr{x}) \,.
\end{align}
\end{subequations}
One consequence of these parity conditions is that the improper gauge transformations are not allowed, since they would add an odd part to the even $\barr{A}_{\bar a}$.
However, this issue can be easily solved by requiring that the fields satisfy the parity conditions given above up to an improper gauge transformation.
That is, we ask the field to satisfy the slightly weaker parity conditions
\begin{equation} \label{YM:parity-conditions-loosen-ED}
 \barr{A}_r = \barr{A}_r^{\text{odd}} \,,
 \qquad
 \barr{\pi}^r = \barr{\pi}^r_{\text{even}} \,,
 \qquad
 \barr{A}_{\bar{a}} = \barr{A}_{\bar{a}}^{\text{even}}
 -\partial_{\bar a} \barr{\Phi}^{\text{even}} \,,
 \quad \text{and} \qquad
 \barr{\pi}^{\bar{a}} =  \barr{\pi}^{\bar{a}}_{\text{odd}} \,,
\end{equation}
where $\barr{\Phi}^{\text{even}}(\barr{x})$ is an even function on the sphere.
With these parity conditions, the symplectic form is not finite any more.
Indeed, it contains the logarithmically divergent contribution
\begin{equation}
 \begin{aligned}
  \int \frac{dr}{r}\oint_{S^2} d^2 \barr{x} \, \extderphase \barr{\pi}^a \wedge \extderphase \barr{A}_a &=
 \int \frac{dr}{r}\oint_{S^2} d^2 \barr{x}
 \Bigl[ - \extderphase \barr{\pi}^{\bar a} \wedge \extderphase \partial_{\bar a} \barr{\Phi}^{\text{even}} \Bigr]= \\
 &= \int \frac{dr}{r}\oint_{S^2} d^2 \barr{x} \, \extderphase \partial_{\bar a} \barr{\pi}^{\bar a} \wedge \extderphase  \barr{\Phi}^{\text{even}} \,,
 \end{aligned}
\end{equation}
where we have integrated by parts in the last passage.
As it was noted in~\cite{Henneaux-ED}, supplementing the parity conditions~(\ref{YM:parity-conditions-loosen-ED}) with the further condition
\begin{equation} \label{YM:asymptotic-Gauss-ED}
 \partial_{\bar a} \barr{\pi}^{\bar a}=0
\end{equation}
makes the symplectic form finite without excluding any potential solution of the equations of motion.
This further condition is nothing else than the asymptotic part of the Gauss constraint since
$
 \mathscr{G} = \partial_a \pi^a 
 = \partial_{\bar a} \barr{\pi}^{\bar a} /r +\bigo \big( 1/r^2\big)
$.
As a consequence, imposing this further condition does not exclude any potential solution to the equation of motion, since these have already to satisfy the full Gauss constraint.

Furthermore, one notes that also the alternative parity conditions
\begin{equation} \label{YM:parity-conditions-loosen-ED-alternative}
 \barr{A}_r = \barr{A}_r^{\text{odd}} \,,
 \qquad
 \barr{\pi}^r = \barr{\pi}^r_{\text{even}} \,,
 \qquad
 \barr{A}_{\bar{a}} = \barr{A}_{\bar{a}}^{\text{odd}}
 -\partial_{\bar a} \barr{\Phi}^{\text{odd}} \,,
 \quad \text{and} \qquad
 \barr{\pi}^{\bar{a}} =  \barr{\pi}^{\bar{a}}_{\text{even}} \,,
\end{equation}
supplemented with~(\ref{YM:asymptotic-Gauss-ED}) lead to a finite symplectic form while allowing improper gauge transformations.
Either the choice of~(\ref{YM:parity-conditions-loosen-ED}) for the parity conditions or that of~(\ref{YM:parity-conditions-loosen-ED-alternative}) supplemented with~(\ref{YM:asymptotic-Gauss-ED}) provides a theory of electrodynamics, in which the symplectic form is finite and improper gauge transformations are allowed.
The former choice of parity conditions is preferable since the latter excludes the possibility of magnetic sources and leads generically to divergences in the magnetic field as one approaches future and past null infinity, as pointed out in~\cite{Henneaux-ED}.

\subsection{Making Poincar\'e transformations canonical} \label{subsec:poincare-canonical-ED}

The extended parity conditions~(\ref{YM:parity-conditions-loosen-ED}) and~(\ref{YM:parity-conditions-loosen-ED-alternative}) come with the advantage of including improper gauge transformations as symmetries of the theory at the cost, however, of making the Poincar\'e transformations non-canonical.
Indeed, with these relaxed parity conditions, the left-hand side of~(\ref{YM:poincare-canonical-final}) does not vanish any more.
The solution to this issue, presented in full details by Henneaux and Troessaert in~\cite{Henneaux-ED}, works as follows.

One introduces a new scalar field $\Psi$ and its corresponding canonical momentum $\pi_\Psi$, which is a scalar density of weight one.
In radial-angular coordinates, the scalar field and its canonical momentum are required to satisfy the fall-off conditions
\begin{equation} \label{YM:fall-off-Psi}
 \Psi= \frac{1}{r} \barr{\Psi} (\barr{x}) +\bigo \big( 1/r^2 \big)
 \qquad \text{and} \qquad
 \pi_{\Psi}= \frac{1}{r} \pi^{(1)}_\Psi (\barr{x}) +\smallo \big( 1/r \big) \,.
\end{equation}
Note that one assumes that the subleading contributions of  scalar field $\Psi$ are $\bigo(1/r^2)$, i.e. vanishing as $r$ tends to infinity at least as fast as $1/r^2$.
At the same time, one assumes that the subleading contributions of the momentum $\pi_\Psi$ are only $\smallo(1/r)$, i.e. vanishing faster than $1/r$, but not necessarily as fast as $1/r^2$.
Moreover, one imposes the constraint
\begin{equation}
 \pi_\Psi \weq 0 \,,
\end{equation}
so that the scalar field $\Psi$ is pure gauge in the bulk.\footnote{
The gauge transformations generated by the constraint $\pi_\Psi \weq 0$ smeared with a gauge parameter $\mu$ amount to nothing else than $\delta_\mu \Psi = \mu$, neglecting boundary terms and, thus, focusing only on the situation in the bulk.
As a consequence the field $\Psi$ can always be trivialised in the bulk by means of a proper gauge transformation.
}
At this point, one modifies the symplectic form to
\begin{equation}
 \Omega= \int d^3 x \, \big[ \extderphase \pi^a \wedge \extderphase A_a
 +\extderphase \pi_\Psi \wedge \extderphase \Psi \big] +\omega \,,
\end{equation}
which contains the standard contributions in the bulk and, in addition, the non-trivial surface term
\begin{equation}
 \omega \eqdef \oint d^2 \barr{x} \, \sqrt{\barr{\gamma}} \, \extderphase \barr{\Psi} \wedge \extderphase \barr{A}_r \,.
\end{equation}
Finally, one extends the Poincar\'e transformations to
\begin{subequations}
\begin{align}
 \delta_{\xi,\zeta} A_a &
 =\xi^\perp \frac{\pi_a}{\sqrt{g}}+\xi^i \partial_i A_a+\partial_a \xi^i A_i+\partial_a ( \xi^\perp \Psi -\zeta) \,,	\\
 \delta_{\xi,\zeta} \pi^a &=
   \partial_b(\sqrt{g}\, \xi^\perp F^{ba}) +\xi^\perp \nabla^a \pi_\Psi
 +\partial_i (\xi^i \pi^a) - \partial_i \xi^a \pi^i \,, \\
 \delta_{\xi,\zeta} \Psi &= \nabla^a (\xi^\perp A_a) + \xi^i \partial_i \Psi \,, \\
 \delta_{\xi,\zeta} \pi_\Psi &= \xi^\perp \partial_a \pi^a + \partial_i (\xi^i \pi_\Psi) \,.
\end{align}
\end{subequations}
Note that, up to gauge transformations and to constraints, the first two equations are the usual Poincar\'e transformations of $A_a$ and $\pi^a$.
It is now straightforward to show that the symplectic form is finite, that the fall-off conditions are preserved under Poincar\'e transformations, and that these latter are canonical.

In this paper, we present also an alternative way to achieve the same result. 
First we introduce a one-form $\phi_a$ and the corresponding canonical momentum $\Pi^a$, which is a vector density of weight one.
In radial-angular coordinates, these new fields are required to satisfy the fall-off conditions
\begin{subequations} \label{YM:fall-off-phi}
\begin{align} 
 \phi_r &=  \barr{\phi}_r (\barr{x}) +\bigo(1/r) \,, &
 \phi_{\bar a} &= r \barr{\phi}_{\bar a} (\barr{x}) +\bigo(r^0) \,, \\
 \Pi^r &= \frac{1}{r^2} \Pi^r_{(1)} (\barr{x}) +\smallo(1/r^2) \,, &
 \Pi^{\bar a} &= \frac{1}{r^3} \Pi^{\bar a}_{(1)} (\barr{x}) +\smallo(1/r^3) \,.
\end{align}
\end{subequations}
Note, as before, the different requirements for the subleading contributions of the field ($\bigo$) and of the momentum ($\smallo$).
Furthermore, we also impose the constraints
\begin{equation}
 \Pi^a \approx 0 \,,
\end{equation}
so that the new field $\phi_a$ is pure gauge in the bulk, and we modify the symplectic form to
\begin{equation}
 \Omega'= \int d^3 x \, \big[ \extderphase \pi^a \wedge \extderphase A_a
 +\extderphase \Pi^a \wedge \extderphase \phi_a \big] +\omega' \,,
\end{equation} 
which contains the non-trivial surface term
\begin{equation}
 \omega' \eqdef \oint d^2 \barr{x} \, \sqrt{\barr{\gamma}}\, \extderphase  (2 \barr{\phi}_r + \barr{\nabla}^{\bar a} \barr{\phi}_{\bar a}) \wedge \extderphase \barr{A}_r \,.
\end{equation}
Finally, one extends the Poincar\'e transformations to
\begin{subequations}
\begin{align}
 \delta_{\xi,\zeta} A_a &
 =\xi^\perp \frac{\pi_a}{\sqrt{g}}+\xi^i \partial_i A_a+\partial_a \xi^i A_i+\partial_a ( \xi^\perp \nabla^i \phi_i -\zeta) \,,	\\
 \delta_{\xi,\zeta} \pi^a &=
   \partial_b(\sqrt{g}\, \xi^\perp F^{ba}) -\xi^\perp \Pi^a
 +\partial_i (\xi^i \pi^a) - \partial_i \xi^a \pi^i \,, \\
 \delta_{\xi,\zeta} \phi_a &= \xi^\perp A_a + \xi^i \partial_i \phi_a+\partial_a \xi^i \phi_i \,, \\
 \delta_{\xi,\zeta} \Pi^a &= -\nabla^a (\xi^\perp \partial_i \pi^i) + \partial_i (\xi^i \Pi^a) - \partial_i \xi^a \Pi^i \,.
\end{align}
\end{subequations}
Again, note that, up to gauge transformations and to constraints, the first two equations are the usual Poincar\'e transformations of $A_a$ and $\pi^a$.
Moreover, the symplectic form is finite, the fall-off conditions are preserved under Poincar\'e transformations, and these latter are canonical.

\subsection{Asymptotic algebra}
In this subsection, we compute the asymptotic algebras of the two cases presented in the previous section and we show that these are equivalent.

The first case, which introduces the scalar field $\Psi$ and its momentum $\pi_\Psi$, is the solution presented in~\cite{Henneaux-ED}.
The Poincar\'e transformations are generated by
\begin{subequations} \label{YM:poincare-generator-em1-total}
\begin{equation} \label{YM:poincare-generator-em1-generic}
 P^{(1)}[\xi^\perp,\xi^i]=
 \int d^3 x \, \big[
 \xi^\perp \, \mathscr{P}_0^{(1)}
 + \xi^i \, \mathscr{P}_i^{(1)}
 \big]
 +\oint d^2 \barr{x} \; \mathscr{B}^{(1)} \,,
\end{equation}
where the generator of the normal component is
\begin{equation} \label{YM:P0-em1}
 \mathscr{P}_0^{(1)} = \frac{\pi^a \pi_a}{2\sqrt{g}} + \frac{\sqrt{g}}{4} F_{ab}F^{ab}
 -\Psi \partial_a \pi^a -A_a \nabla^a \pi_\Psi \,,
\end{equation}
the generator of the tangential component is
\begin{equation} \label{YM:Pi-em1}
 \mathscr{P}_i^{(1)} = \pi^a\partial_i A_a - \partial_a (\pi^a A_i) +\pi_\Psi \partial_i \Psi \,,
\end{equation}
the generator of the proper gauge transformations  is the Gauss constraint $\mathscr{G} = \partial_a \pi^a$, and the explicit expression of the boundary term is
\begin{equation} \label{poincare-boundary-em1}
 \mathscr{B}^{(1)} = b \left( \barr{\Psi} \barr{\pi}^r + \sqrt{\barr{\gamma}} \, \barr{A}_{\bar a} \barr{\nabla}^{\bar a} \barr{A}_r \right)
 +Y^{\bar a} \left( \barr{\pi}^r \barr{A}_{\bar a} + \sqrt{\barr{\gamma}}\, \barr{\Psi} \partial_{\bar a} \barr{A}_r \right)
 \,,
\end{equation}
\end{subequations}
which is needed to make the generator~(\ref{YM:poincare-generator-em1-total}) differentiable \`a la Regge-Teitelboim.
In addition, the proper and improper gauge symmetries are generated by
\begin{equation}
 G_{\epsilon,\mu}^{(1)} = \int d^3 x\, \big( \epsilon \, \mathscr{G}
 + \mu \, \pi_\Psi \big)
 -\oint d^2 \barr{x} \,\big( \barr{\epsilon} \, \barr{\pi}^r +\sqrt{\barr{\gamma}} \, \barr{\mu} \, \barr{A}_r \big) \,,
\end{equation}
which, together with~(\ref{YM:poincare-generator-em1-total}), satisfies the algebra
\begin{subequations} \label{asymptotic-algebra-ED1}
\begin{equation} 
 \big\{ P_{\xi_1^\perp,\xi_1}^{(1)} , P_{\xi_2^\perp,\xi_2}^{(1)} \big\} = P^{(1)}_{\widehat \xi^\perp, \widehat \xi} \,,
 \qquad
 \big\{ G_{\epsilon,\mu}^{(1)}, P_{\xi^\perp,\xi}^{(1)} \big\}= G_{\widehat \epsilon, \widehat \mu}^{(1)} \,,
 \qquad
 \big \{ G_{\epsilon_1,\mu_1}^{(1)}, G_{\epsilon_2,\mu_2}^{(1)} \big\}=0 \,,
\end{equation}
where
\begin{align}
 \widehat \xi^\perp &=\xi^i_1 \partial_i \xi_2^\perp-\xi^i_2 \partial_i \xi_1^\perp \,,
 &
 \widehat \xi^i &= g^{ij} (\xi^\perp_1 \partial_j \xi^\perp_2-\xi^\perp_2 \partial_j \xi^\perp_1)
 + \xi^j_1 \partial_j \xi^i_2 - \xi^j_2 \partial_j \xi^i_1 \,, \\
 \widehat \mu &= \nabla^i (\xi^\perp \partial_i \epsilon) -\xi^i \partial_i \mu \,, &
 \widehat \epsilon &= \xi^\perp \mu - \xi^i \partial_i \epsilon \,.
\end{align}
\end{subequations}

In the second case presented in the previous subsection, which introduces the one-form $\phi_a$ and its momentum $\Pi^a$, the Poincar\'e transformations are generated by
\begin{subequations} \label{YM:poincare-generator-em2-total}
\begin{equation} \label{YM:poincare-generator-em2-generic}
 P^{(2)}[\xi^\perp,\xi^i]=
 \int d^3 x \, \big[
 \xi^\perp \, \mathscr{P}_0^{(2)}
 + \xi^i \, \mathscr{P}_i^{(2)}
 \big]
 +\oint d^2 \barr{x} \; \mathscr{B}^{(2)} \,,
\end{equation}
where the generator of the normal component is
\begin{equation} \label{YM:P0-em2}
 \mathscr{P}_0^{(2)} = \frac{\pi^a \pi_a}{2\sqrt{g}} + \frac{\sqrt{g}}{4} F_{ab}F^{ab}
 -\nabla^a \phi_a \partial_b \pi^b+\Pi^a A_a \,,
\end{equation}
the generator of the tangential component is
\begin{equation} \label{YM:Pi-em2}
 \mathscr{P}_i^{(2)} = \pi^a\partial_i A_a - \partial_a (\pi^a A_i) +
 \Pi^a\partial_i \phi_a - \partial_a (\Pi^a \phi_i) \,,
\end{equation}
the generator of the proper gauge transformations  is the Gauss constraint $\mathscr{G} = \partial_a \pi^a$, and the explicit expression of the boundary term is
\begin{equation} \label{YM:poincare-boundary-em2}
 \mathscr{B}^{(2)} = b \big[(2 \barr{\phi}_r + \barr{\nabla}^{\bar a} \barr{\phi}_{\bar a}) \barr{\pi}^r + \sqrt{\barr{\gamma}} \, \barr{A}_{\bar a} \barr{\nabla}^{\bar a} \barr{A}_r \big]
 +Y^{\bar a} \, \barr{\pi}^r\, \barr{A}_{\bar a}
 \,,
\end{equation}
\end{subequations}
which is needed to make the generator~(\ref{YM:poincare-generator-em2-total}) differentiable \`a la Regge-Teitelboim.
In addition, the proper and improper gauge symmetries are generated by
\begin{equation}
 G_{\epsilon,\chi}^{(2)} = \int d^3 x\, \big( \epsilon \, \mathscr{G}
 + \chi_a \, \Pi^a \big)
 -\oint d^2 \barr{x} \,\big[ \barr{\epsilon} \, \barr{\pi}^r +\sqrt{\barr{\gamma}} \, (2 \barr{\chi}_r +\barr{\nabla}^{\bar a} \barr{\chi}_{\bar a}) \, \barr{A}_r \big] \,,
\end{equation}
which can be combined with~(\ref{YM:poincare-generator-em2-total}) into the generator
\begin{equation}
 A^{(2)}[\xi^\perp,\xi,\epsilon,\chi_a] \eqdef
 P^{(2)}[\xi^\perp,\xi]+G^{(2)}[\epsilon,\chi] \,,
\end{equation}
satisfying the algebra
\begin{subequations} \label{asymptotic-algebra-ED2}
\begin{align} 
 \big\{
 A^{(2)}[\xi_1^\perp,\xi_1,\epsilon_1,\chi_1] ,
 A^{(2)}[\xi_2^\perp,\xi_2,\epsilon_2,\chi_2]
 \big\} =
 A^{(2)}[\hat{\xi}^\perp,\hat{\xi},\hat{\epsilon},\hat{\chi}] \,,
\end{align}
where
\begin{align}
 \widehat \xi^\perp &= \xi^i_1 \partial_i \xi_2^\perp-\xi^i_2 \partial_i \xi_1^\perp \,,
 \\
 \widehat \xi^i &= \tilde \xi^i + \xi^j_1 \partial_j \xi^i_2 - \xi^j_2 \partial_j \xi^i_1 \,,
 \\
 \tilde \xi^i &\eqdef g^{ij} (\xi^\perp_1 \partial_j \xi^\perp_2-\xi^\perp_2 \partial_j \xi^\perp_1) \,,
 \\
 \widehat \chi_a &= \xi_1^\perp \partial_a \epsilon_2-\xi_2^\perp \partial_a \epsilon_1
 + \xi_1^i \partial_i \chi^2_a - \xi_2^i \partial_i \chi^1_a
 +\tilde{\xi}_a \nabla^m \phi_m - \tilde{\xi}^m \partial_m \phi_a -\partial_a (\tilde{\xi}^m\phi_m)\,,
 \\
 \widehat \epsilon &= \xi_2^i \partial_i \epsilon_1 -\xi_1^i \partial_i \epsilon_2+
 \xi_2^\perp \nabla^a \chi^1_a -\xi_1^\perp \nabla^a \chi^2_a \,.
\end{align}
\end{subequations}

The asymptotic algebras~(\ref{asymptotic-algebra-ED1}) and~(\ref{asymptotic-algebra-ED2}) are equivalent.
To see this, one has to consider, in the two cases, the group of all the allowed transformations and take the quotient of it with respect to the proper gauge.
Only then, one can compare the brackets~(\ref{asymptotic-algebra-ED1}) and~(\ref{asymptotic-algebra-ED2}).
In the first case presented above, the proper gauge amount to those transformations for which $\barr{\epsilon}$ is odd and $\barr{\mu}$ is even.
In the second case presented above, the proper gauge amount to those transformations for which $\barr{\epsilon}$ is odd and
$\barr{\nabla \cdot \chi} \eqdef 2 \barr{\chi}_r + \barr{\nabla}^{\bar a} \barr{\chi}_{\bar a}$ is even.
The equivalence is then shown by identifying $\barr{\mu}$ with $\barr{\nabla \cdot \chi}$.

\section{Relaxing parity conditions and asymptotic symmetries in Yang-Mills} \label{sec:relax-parity-YM}
\sectionmark{Relaxing parity conditions in Yang-Mills}

In this section, we try to apply the methods of the previous section to the non-abelian Yang-Mills case.
The goal is to obtain a Hamiltonian formulation of Yang-Mills with canonical Poincar\'e transformations and with allowed improper gauge transformations.
As we shall see, this goal cannot be entirely fulfilled.

\subsection{Relaxing parity conditions in Yang-Mills} \label{subsec:YM:loosen-parity}

Let us now study the possibility of relaxing the parity conditions in Yang-Mills, in order to restore the improper gauge transformations also in this case.
Following the same line of argument of the electromagnetic case, we begin by requiring the asymptotic fields to satisfy the parity conditions~(\ref{YM:parity-conditions}) up to asymptotic improper gauge transformations~(\ref{YM:gauge-transformations-improper-full-asymptotic}), so that
\begin{subequations} \label{YM:parity-conditions-loosen}
\begin{align} \label{YM:parity-conditions-loosen1}
 \barr{A}_r &= \barr{\mathcal{U}}^{-1} \barr{A}_r^{\text{even}} \, \barr{\mathcal{U}}  \,,
 &
 \barr{\pi}^r &=  \barr{\mathcal{U}}^{-1} \barr{\pi}^r_{\text{odd}} \, \barr{\mathcal{U}} \,,
 \\ \label{YM:parity-conditions-loosen2}
 \barr{A}_{\bar{a}} &= \barr{\mathcal{U}}^{-1} \barr{A}_{\bar{a}}^{\text{odd}} \barr{\mathcal{U}}
 + \barr{\mathcal{U}}^{-1} \partial_{\bar a}\, \barr{\mathcal{U}}   \,,
 &
 \barr{\pi}^{\bar{a}} &= \barr{\mathcal{U}}^{-1} \barr{\pi}^{\bar{a}}_{\text{even}} \, \barr{\mathcal{U}} \,,
\end{align}
\end{subequations}
where $\barr{\mathcal{U}} (\barr{x})=\exp \big[-\barr{\Phi}^{\text{odd}} (\barr{x}) \big] \in \SU(N)$ and the Lie-algebra-valued function $\barr{\Phi}^{\text{odd}} (\barr{x})$ is odd under the antipodal map $\barr{x} \mapsto -\barr{x}$.
Therefore, the Lie-group-valued function $\barr{\mathcal{U}} (\barr{x})$ behaves as $\barr{\mathcal{U}} (-\barr{x})= \barr{\mathcal{U}} (\barr{x})^{-1}$ under the antipodal map.
These new parity conditions introduce the logarithmically divergent part
\begin{equation} 
\label{YM:log-div1}
\begin{aligned}
 &\int \frac{dr}{r} \oint_{S^2} d^2 \barr{x} \;
 \sprod{\extderphase \barr{\pi}^a \wedge}{ \extderphase \barr{A}_a}
 =\\
 ={}&\int \frac{dr}{r} \oint_{S^2} d^2 \barr{x} \;
 \Bigg\{ \!
 \sprod{ \Big(\extderphase \barr{\mathcal{U}} \, \barr{\mathcal{U}}^{-1}   \Big)\wedge}
 {
 \extderphase \Big(
 \partial_{\bar a} \barr{\pi}^{\bar a}_{\text{even}}
 +\extprod{\barr{A}_r^{\text{even}}}{\barr{\pi}^r_{\text{odd}}}
 +\extprod{\barr{A}_{\bar a}^{\text{odd}}}{\barr{\pi}^{\bar a}_{\text{even}}}
 \Big)
 }
 +\\
 &
 -\frac{1}{2}
 \sprod{
 \Big[ \extprod{\Big(\extderphase \barr{\mathcal{U}} \, \barr{\mathcal{U}}^{-1}   \Big) \wedge}{\Big(\extderphase \barr{\mathcal{U}} \, \barr{\mathcal{U}}^{-1}   \Big)} \Big]
 \! }{ \!
 \Big(
 \partial_{\bar a} \barr{\pi}^{\bar a}_{\text{even}}
 +\extprod{\barr{A}_r^{\text{even}}}{\barr{\pi}^r_{\text{odd}}}
 +\extprod{\barr{A}_{\bar a}^{\text{odd}}}{\barr{\pi}^{\bar a}_{\text{even}}}
 \Big)
 }
 \! \Bigg\}
\end{aligned}
\end{equation}
in the symplectic form, whose precise derivation is presented in appendix~\ref{app:symplectic-form}.

At this point, we note that the second factor in both summands of the right-hand side of~(\ref{YM:log-div1}) is nothing else than the asymptotic Gauss constraint $\barr{\mathscr{G}}_0$ evaluated when $\barr{\Phi}^{\text{odd}}=0$, which is related to the asymptotic Gauss constrain $\barr{\mathscr{G}}$ with non-vanishing $\barr{\Phi}^{\text{odd}}$ by the expression $\barr{\mathscr{G}} = \barr{\mathcal{U}}^{-1} \, \barr{\mathscr{G}}_0 \, \barr{\mathcal{U}}$, so that the one vanishes if, and only if, the other does.
Therefore, we can keep the symplectic form finite by restricting the phase space to those field configurations that satisfy, together with the fall-off conditions~(\ref{YM:fall-off-conditions}) and the parity conditions~(\ref{YM:parity-conditions-loosen}), also the asymptotic Gauss constraint
\begin{equation} \label{asymptotic-Gauss-YM}
 \partial_{\bar a} \barr{\pi}^{\bar a}
 +\extprod{\barr{A}_r}{\barr{\pi}^r}
 +\extprod{\barr{A}_{\bar a}}{\barr{\pi}^{\bar a}}
=0 \,.
\end{equation}
Note that imposing this further condition does not exclude any of the former solutions to the equations of motion, since every solution was already satisfying the Gauss constraint altogether.
This shows that it is possible to relax the parity conditions in order to allow improper gauge transformations, but nevertheless leaving the symplectic form finite.

In electrodynamics, one notes that it is possible to start with a different set of parity conditions and to relax them, so that the symplectic form is nevertheless finite.
These freedom, was used in section~\ref{subsec:loosen-parity-ED} in order to present two possibility for the parity of the angular components of the asymptotic part of the fields.\footnote
{
In principle, one could use the same freedom for the parity of the radial component of the asymptotic fields, but this was already fixed by the physical requirement that Coulomb is a solution.
}
One could wonder whether or not this freedom is present also in the Yang-Mills case.

First, one notes that picking the opposite parity for the angular part is problematic.
Specifically, the asymptotic part of the Poincar\'e transformations~(\ref{YM:poincare-asymptotic}) contains the term $\barr{F}_{\bar a \bar b}$ and the operator
$\barr{D}_{\bar a} \eqdef \barr{\nabla}_{\bar a} + \extprod{\barr{A}_{\bar a}}{}$.
If we took $\barr{A}_{\bar a}$ to be of even parity (up to asymptotic proper/improper gauge transformations) we would end up with terms of indefinite parity after applying the Poincar\'e transformations.

Secondly, one could try to pick the opposite parity conditions for the radial components of the asymptotic fields (up to asymptotic proper/improper gauge transformations).
This choice would have the advantage of allowing a non-vanishing value of the colour charge, as discussed in footnote~\ref{foot:radial-parity} on page~\pageref{foot:radial-parity}. 
However, for this choice, the method used above to make the symplectic form finite does not work any more even after imposing the asymptotic Gauss constraint.\footnotemark

To sum up, we have found a way of relaxing the strict parity conditions of section~\ref{sec:parity} in order to allow improper gauge transformations, but leaving the symplectic form finite.
We have also discussed why different choices for the parity conditions are less appealing and more problematic in Yang-Mills compared to electrodynamics.
As expected, the price to pay when relaxing the parity conditions is that the Poincar\'e transformations are not canonical any more.
We will discuss what can be done to fix this issue in the next subsection.

\subsection{Attempt to make the Poincar\'e transformations canonical} \label{subsec:attempt-Poincare-canonical}

In order to make the Poincar\'e transformations canonical the following expression, which is the Lie derivative of the symplectic form, has to vanish:
\begin{equation} 
\label{YM:poincare-not-canonical}
 \liephase_X \Omega =
 \extderphase (i_X \Omega)=
 \oint_{S^2} d^2 \barr{x} \; 
b \, \sqrt{\barr{\gamma}} \, 
\barr{\gamma}^{\bar m \bar n} \,
\sprod{\extderphase \barr{A}_{\bar m} \wedge}{\extderphase ( 
\barr{D}^{\bar m} \barr{A}_r )} \,,
\end{equation}
possibly adding a surface term to the symplectic form and introducing new fields, which are non-trivial only at the boundary.
One could try  to follow the line of reasoning of section~\ref{subsec:poincare-canonical-ED} also in Yang-Mills.
Since the Lie derivative of the symplectic form fails to vanish due to the Lorentz boost, we will focus on the Lorentz boost and neglect the rest of the Poincar\'e transformations in the following.
In other words, we will consider the case in which $\xi^\perp = rb$ and $\vect{\xi} = 0$.
Moreover, we discuss, separately, the possible implementation of each one of the two solutions presented in section~\ref{subsec:poincare-canonical-ED} and adapted to the Yang-Mills case.

\footnotetext{ 
The method used to make the symplectic form finite in this subsection works if $\barr{A}_r$ and $\barr{A}_{\bar a}$ are chosen so that they have opposite parity when $\barr{\Phi}^{\text{odd}} = 0$.  Therefore, the method presented in this section would still work if we chose, at the same time, the opposite parity conditions both for the radial and for the angular components, with respect to those presented in~(\ref{YM:parity-conditions-loosen}).
However, we have already discussed that changing the parity conditions of the angular components leads to other issues.
}

\subsubsection{Case 1} \label{subsubsec:ansatz-Psi}
First, let us consider the solution described in section~\ref{subsec:poincare-canonical-ED} which uses the scalar field $\Psi$ and its conjugated momentum $\pi_\Psi$, first found in~\cite{Henneaux-ED}.
Also in the case of Yang-Mills, we supplement the field with the fall-off conditions~(\ref{YM:fall-off-Psi}), the further constraint $\pi_\Psi \approx 0$, and the symplectic structure in the bulk
\begin{equation}
 \Omega= \int d^3 x \, \big[
 \sprod{\extderphase \pi^a \, \wedge }{ \extderphase A_a}
 +\sprod{ \extderphase \pi_\Psi \, \wedge }{ \extderphase \Psi }
 \big] \,.
\end{equation}
Moreover, we impose the action of the Lorentz boost on the fields to be
\begin{subequations} \label{YM:boost-Psi}
\begin{align}
 \label{YM:boost-Psi-in}
 \delta_{\xi^\perp} A_a &
 =\xi^\perp \frac{\pi_a}{\sqrt{g}}+ D_a ( \xi^\perp \Psi ) \,,	\\
 \delta_{\xi^\perp} \pi^a &=
   \partial_b(\sqrt{g}\, \xi^\perp F^{ba}) +\xi^\perp \nabla^a \pi_\Psi - \xi^\perp \extprod{\Psi}{\pi^a} \,, \\
 \delta_{\xi^\perp} \Psi &= \nabla^a (\xi^\perp A_a) \,, \\
 \label{YM:boost-Psi-end}
 \delta_{\xi^\perp} \pi_\Psi &= \xi^\perp \mathscr{G} \,,
\end{align}
\end{subequations}
which preserve both the fall-off conditions and the constraints.
Let us denote with $X'$ the vector field in phase space defined by the transformations above.

These transformations \emph{would be} generated by
\begin{equation}
\begin{aligned}
P [\xi^\perp] \eqdef{}&
 \int d^3 x \, \xi^\perp \left[
 \frac{\sprod{\pi^a}{\pi_a}}{2\sqrt{g}} + \frac{\sqrt{g}}{4} \sprod{F_{ab}}{F^{ab}}
 - \sprod{\Psi}{\mathscr{G}} - \sprod{A_a}{\nabla^a \pi_\Psi}
 \right] +\\
 &+ (\text{boundary}) \,,
 \end{aligned}
\end{equation}
\emph{if} a suitable boundary term could be found, so that the $\extderphase P = - \insertion_{X'} (\Omega + \omega)$ eventually complementing the symplectic form $\Omega$ with a boundary term $\omega$.
As we shall see, such boundary term does not exist.
To see this, let us define
\begin{equation}
 \omega_0 \eqdef
 \oint_{S^2} d^2 \barr{x} \, \sqrt{\barr{\gamma}} \, \sprod{\extderphase \barr{\Psi} \, \wedge}{ \extderphase \barr{A}_r} \,,
\end{equation}
such that one finds
\begin{equation} \label{YM:lie-Omega-omega0}
 \liephase_{X'} (\Omega + \omega_0) =
 \oint_{S^2} d^2 \barr{x} \, b \sqrt{\barr{\gamma}} \,
 \left[
 \sprod{\extderphase \barr{A}_{\bar m} \,\wedge}{ \extderphase( \extprod{ \barr{A}^{\bar m} }{\barr{A}_r} ) } -
 \sprod{\extderphase \barr{\Psi} \,\wedge}{ \extderphase( \extprod{ \barr{\Psi} }{\barr{A}_r} ) }
 \right]
\end{equation}
At this point, one needs to find a second boundary term $\omega_1$, whose phase-space Lie derivative $\liephase_{X'} \omega_1$ is the opposite of the expression above.
However, one immediately faces the issue that even the first term inside square brackets of the expression above cannot be compensated by some expression contained in $\liephase_{X'} \omega_1$, for any $\omega_1$ built from the canonical fields.
Indeed, the first term in~(\ref{YM:lie-Omega-omega0}) contains only the asymptotic part of the field $A$, without any derivative, but the asymptotic transformations of the fields under Lorentz boosts do not contain any such term.
In other words, one \emph{cannot} find an extra surface term to the symplectic structure $\omega \eqdef \omega_0 + \omega_1$, which is build from the given fields and satisfies $\liephase_{X'} (\Omega + \omega)=0$.
This implies that $\insertion_{X'} (\Omega + \omega)$ cannot be a closed form and, thus, there cannot be a $P$ satisfying $\extderphase P = - \insertion_{X'} (\Omega + \omega)$.

\subsubsection{Case 2} \label{subsubsec:ansatz}
Secondly, one could try to adapt to the Yang-Mills case the other solution described in section~\ref{subsec:poincare-canonical-ED}, namely the one introducing the one form $\phi_a$ and its conjugated momentum $\Pi^a$.
Also in this case, we supplement the fields with the fall-off conditions~(\ref{YM:fall-off-phi}), the further constraints $\Pi^a \approx 0$, and the symplectic form in the bulk
\begin{equation} \label{YM:Omega-bulk-phi}
 \Omega' = \int d^3 x \, \big[
 \sprod{ \extderphase \pi^a \, \wedge }{ \extderphase A_a}
 +\sprod{ \extderphase \Pi^a \, \wedge }{ \extderphase \phi_a }
 \big] \,.
\end{equation} 
Moreover, we impose the action of the Lorentz boost on the fields to be
\begin{subequations} \label{YM:boost-YM-phi}
\begin{align}
 \label{YM:boost-YM-phi-in}
 \delta_{\xi^\perp} A_a &
 =\xi^\perp \frac{\pi_a}{\sqrt{g}}+ D_a ( \xi^\perp \mathscr{D}^i \phi_i) \,,	\\
 \delta_{\xi^\perp} \pi^a &=
   \partial_b(\sqrt{g}\, \xi^\perp F^{ba}) -\xi^\perp \Pi^a
   + \xi^\perp \extprod{ \pi^a }{ \mathscr{D}^i \phi_i }
   + \xi^\perp \, c \, \extprod{ \phi^a }{ \mathscr{G} } \,, \\
 \delta_{\xi^\perp} \phi_a &= \xi^\perp A_a  \,, \\
 \label{YM:boost-YM-phi-end}
 \delta_{\xi^\perp} \Pi^a &= -\mathscr{D}^a (\xi^\perp \mathscr{G})  \,,
\end{align}
\end{subequations}
where $\mathscr{D}_a \eqdef \nabla_a +c_1 \, \extprod{A_a}{}+ c_2 \, \extprod{\pi_a}{}$ and $c_1,c_2 \in \real$ are free parameters that one can set later to suitable values in order to make the Lorentz boost canonical.
One can verify that the above transformations preserve both the fall-off conditions and the constraints.
Moreover, they \emph{would be} generated by
\begin{equation} \label{YM:boost-generator-phi}
\begin{aligned}
 P' [\xi^\perp] \eqdef{}&
 \int d^3 x \, \xi^\perp \left[
 \frac{\sprod{\pi^a}{\pi_a}}{2\sqrt{g}} + \frac{\sqrt{g}}{4} \sprod{F_{ab}}{F^{ab}}
 - \sprod{\mathscr{D}^a \phi_a }{\mathscr{G}} + \sprod{A_a}{\Pi^a}
 \right] + \\
 &+ (\text{boundary}) \,,
\end{aligned}
\end{equation}
\emph{if} a suitable boundary term existed, as discussed in the previous case (as we shall see in the following, such boundary term does not exist also in this case).
One can easily compute that
\begin{equation} \label{YM:lie-symplectic-bulk}
 \liephase_{X'} \Omega' = 
 \oint_{S^2} d^2 \barr{x} \, b  
 \Big[
 \sqrt{\barr{\gamma}} \, \sprod{\extderphase \barr{A}_{\bar m} \,\wedge}{ \extderphase \big( D^{\bar m} \barr{A}_r \big) }
 + \sprod{ \extderphase \barr{\pi}^r \, \wedge}{ \extderphase \barr{\mathscr{D} \phi }}
 \Big]  \,,
\end{equation}
where
\[
\barr{\mathscr{D} \phi}\eqdef 2 \barr{\phi}_r + \barr{\nabla}^{\bar m} \barr{\phi}_{\bar m}
+ c_1 \, \big( \extprod{\barr{A}_r}{\barr{\phi}_r} + \extprod{ \barr{A}^{\bar m} }{ \barr{\phi}_{\bar m} } \big)
+ c_2 \, \big( \extprod{\barr{\pi}^r}{\barr{\phi}_r} + \extprod{ \barr{\pi}^{\bar m} }{ \barr{\phi}_{\bar m} } \big) 
\]
is the leading contribution in the expansion of $\mathscr{D}^a \phi_a =  \barr{\mathscr{D}\phi} /r + \bigo \big( 1/r^2 \big)$
and $X'$ is the vector field on phase space that defines the Lorentz boost~(\ref{YM:boost-YM-phi}).

One hopes that, with respect to the previous case concerning $\Psi$ and $\pi_\Psi$, one can now tackle the problem more efficiently, since there are now fields transforming asymptotically as the asymptotic part of $A_a$ without derivatives.
Namely, the one form $\phi_a$ transforms asymptotically under Lorentz boosts like
$\delta_{\xi^\perp} \barr{\phi}_{a} = b \, \barr{A}_a$.

In order to compensate for the terms contained in~(\ref{YM:lie-symplectic-bulk}), we use the following ansatz for the boundary term of the symplectic form:
\begin{equation} \label{YM:ansatz-omega}
 \begin{aligned}
 \omega' = \oint_{S^2} d^2 \barr{x} \; \sqrt{\barr{\gamma}} \,
 \Big[
 & a_0 \, \sprod{\extderphase \big( \barr{\nabla}^{\bar m}  \barr{\phi}_{\bar m} \big) \wedge}{\extderphase \barr{A}_r } +
 \\
 +& a_1 \, \sprod{\extderphase \barr{\phi}_{r} \wedge}{\extderphase \barr{A}_r } 
 + a_2 \, \sprod{ \barr{A}_{r} }{ \extprod{ \extderphase \barr{\phi}^{\bar m} \, \wedge}{ \extderphase \barr{A}_{\bar m} } } + \\
 +& a_3 \, \sprod{ \barr{\phi}_{r} }{ \extprod{ \extderphase \barr{A}^{\bar m} \, \wedge}{ \extderphase \barr{A}_{\bar m} } } 
 + a_4 \, \sprod{ \barr{A}_{\bar m} }{ \extprod{ \extderphase \barr{A}^{\bar m} \, \wedge}{ \extderphase \barr{\phi}_{r} } } + \\
 +& a_5 \, \sprod{ \barr{A}_{\bar m} }{ \extprod{ \extderphase \barr{\phi}^{\bar m} \, \wedge}{ \extderphase \barr{A}_{r} } } 
 + a_6 \, \sprod{ \barr{\phi}_{\bar m} }{ \extprod{ \extderphase \barr{A}^{\bar m} \, \wedge}{ \extderphase \barr{A}_{r} } } 
 \Big] \,,
\end{aligned}
\end{equation}
where $a_0, \dots, a_6 \in \real$ are free parameters that can be set to a suitable value in order to achieve $\liephase_{X'} (\Omega' + \omega') = 0$.
Note that one has to restrict the possible values of the parameters $a_0, \dots, a_6$, in order to ensure that the two-form $\omega'$ is closed.
In any case, one can show that no value of the parameters $a_0, \dots, a_6$, $c_1$, and $c_2$ can be found in order to make the Lorentz boost canonical.
A more detailed discussion about the reasons why we used the ansatz above and the computations needed to show that no value of the free parameters make the Lorentz boost canonical can be found in appendix~\ref{app:details-computations}.

In conclusion, we were not able to find a solution to the problem of making the Poincar\'e transformations canonical after having relaxed the parity conditions in the Yang-Mills case.

\section{The situation in higher dimensions} \label{sec:YM-higher-dim}

In this section, we will briefly comment on the situation in higher dimensions.
Although this case is not so interesting on a purely-physical side (we do live in four dimensions after all), it is nevertheless of mathematical interest and it shows that the four-dimensional case is somewhat special.
So, in this section, let us assume that the spacetime is an $(n+1)$-dimensional manifold, being $n > 3$.
The case of free electrodynamics in higher dimensions was already studied in~\cite{Henneaux-ED-higher} and the case of scalar electrodynamics will be briefly commented in section~\ref{subsec:Lorentz-Lorenz-SED}.

Without redoing all the derivations, let us begin by stating that the fall-off conditions of the fields in radial-angular coordinates\footnote{We use coordinates such that $r$ is the radial distance and $\barr{x}$ are coordinates on the $(n-1)$-unit-sphere.} are
\begin{subequations} \label{YM:fall-off-higher-dim}
\begin{align}
 A_r (r,\barr{x}) &= \frac{1}{r^{n-2}} \barr{A}_r (\barr{x})
 + \bigo \big(1/r^{n-1} \big) \,, \\
 \label{YM:fall-off-A-angular-higher-dim}
 A_{\bar a} (r,\barr{x}) &= \barr{\mathcal{U}}^{-1} (\barr{x}) \partial_{\bar a} \barr{\mathcal{U}} (\barr{x}) + \frac{1}{r^{n-3}} \barr{A}_{\bar a} (\barr{x})
 + \bigo \big( 1/r^{n-2} \big) \,, \\
 \pi_r (r,\barr{x}) &= \barr{\pi}_r (\barr{x})
 + \bigo \big( 1/r^{2} \big) \,, \\
 \pi_{\bar a} (r,\barr{x}) &= \frac{1}{r} \barr{\pi}_{\bar a} (\barr{x})
 + \bigo \big( 1/r^{2} \big) \,.
\end{align}
\end{subequations}
Note that the fall-off condition of $A_{\bar a}$ includes now two contributions.
One is a generic term vanishing at infinity as $1/r^{n-3}$, while the other is a zeroth-order contribution depending on $\barr{\mathcal{U}} \in \SU(N)$.
The latter is needed, for otherwise the improper gauge transformations would not be allowed.
Anyway, even with this zeroth-order term, the two-form curvature $F_{\bar m \bar n}$ is of order $1/r^{n-3}$.
Furthermore, in the case $n = 3$, these two contributions to the fall-off condition~(\ref{YM:fall-off-A-angular-higher-dim}) are of the same order, so that we can reabsorb $\barr{\mathcal{U}}^{-1} \partial_{\bar a}\barr{\mathcal{U}}$ into the definition of $\barr{A}_{\bar a}$, but this is not possible in higher dimensions.
The above fall-off conditions are such that they are preserved by the Poincar\'e transformations and that they allow, in principle, a finite, non-trivial value of the charges.

The action of the gauge transformations on the fields is well-defined so long as one requires that the infinitesimal gauge parameter $\zeta \in \su(N)$ falls off as
\begin{equation}
 \zeta (r, \barr{x}) = \barr{\zeta} (\barr{x}) +\bigo(1/r^{n-3}) \,.
\end{equation}
In principle, we could allow the gauge parameter to have also terms of order $\bigo(1/r)$. 
To do so, we would need to modify the fall-off condition of $A_{\bar a}$ by replacing $\barr{\mathcal{U}} (\barr{x})$ with a more generic $\mathcal{U}(x)$, whose leading order is $\barr{\mathcal{U}} (\barr{x})$.
In practice, this would bring some complications --- e.g. there is some ambiguity in the definition of $\barr{A}_{\bar a}$ and $\mathcal{U}$ --- without any real benefit, since the transformations that we add are proper gauge transformations and we are more interested in the possible implementation of the improper ones.

The fall-off conditions above are (almost) enough to ensure that the symplectic form is finite.
The only potentially-divergent term is the one built from $\barr{\mathcal{U}}$, i.e.
\begin{equation}
 \int \frac{dr}{r} \oint_{S^{n-1}} d^{n-1} \barr{x} \,
 \extderphase \barr{\pi}^{\bar a} \wedge \cdot
 \extderphase \left( \barr{\mathcal{U}}^{-1} \partial_{\bar a} \barr{\mathcal{U}} \right) \,.
\end{equation}
Let us define
$\barr{\pi}_0^{\bar a} \eqdef
\barr{\mathcal{U}} \, \barr{\pi}^{\bar a} \, \barr{\mathcal{U}}^{-1}$.
Then, following the same procedure used in appendix~\ref{app:symplectic-form} to compute $\Omega_3$, we find that the potentially divergent contribution to the symplectic form is
\begin{equation}
 \oint_{S^2} d^2 \barr{x} \;
  \bigg\{
  \sprod{\Big(
  \extderphase \barr{\mathcal{U}} \;  \barr{\mathcal{U}}^{-1}
  \Big) \wedge}
  {\extderphase \Big( \partial_{\bar a} \barr{\pi}^{\bar a}_{0} \Big)}-
  \sprod{
  \frac{1}{2} \Big[\extprod{ \Big(
  \extderphase \barr{\mathcal{U}} \; \barr{\mathcal{U}}^{-1} \Big) \wedge}{\Big(\extderphase \barr{\mathcal{U}} \, \barr{\mathcal{U}}^{-1} \Big)} \Big] 
  }
  { \partial_{\bar a} \barr{\pi}^{\bar a}_{0} }
  \bigg\} \,,
\end{equation}
so that it would vanish if $\partial_{\bar a} \barr{\pi}^{\bar a}_{0}= 0$.
Written in term of the original $\barr{\pi}^{\bar a}$, this condition becomes
\begin{equation}
 \barr{\mathcal{U}} \left[
 \partial_{\bar a} \barr{\pi}^{\bar a} +
 \left( \barr{\mathcal{U}}^{-1} \partial_{\bar a} \barr{\mathcal{U}} \right) \times \barr{\pi}^{\bar a}
 \right] \barr{\mathcal{U}}^{-1} = 0 \,.
\end{equation}
The term in square brackets in the expression above is the leading order of the Gauss constraint, so that we infer that the symplectic form is finite as long as we require the leading term of the Gauss constraint to vanish.

Note that, contrary to the situation in $3+1$ dimensions, we did not need to impose any parity conditions to the asymptotic part of the fields, in order to make the symplectic form finite.
This has some consequence on the proper and improper gauge transformations and on the charges.
In particular, as in the case of $3+1$ dimensions, the generator of the proper gauge transformations can be complemented by a boundary term, so that the new generator is differentiable also when the gauge parameter is non-vanishing at infinity.
In this way, one gets the extended generator
\begin{equation}
 G_{\text{ext.}}[\epsilon] = \int d^n x \, \epsilon (x) \cdot \mathscr{G} (x) -
 \oint d^{n-1} \barr{x} \, \barr{\epsilon} (\barr{x}) \cdot \barr{\pi}^r (\barr{x}) \,,
\end{equation}
where $\mathscr{G} = \partial_a \pi^a + A_a \times \pi^a$ is the Gauss constraint.
The boundary term, which defines the charges, is in general different from zero, since the function $\barr{\pi}^r$ is not restricted to have any definite parity.

The only part that we are left to check is whether or not the Poincar\'e transformations are canonical.
One can show that
\begin{equation} \label{YM:canonical-Poincare-YM-higher-dim}
\begin{aligned}
 \extderphase (\insertion_X \Omega) =
 &\oint d^{n-1} \barr{x} \, \sqrt{\barr{\gamma}} \, b \, \gamma^{\bar m \bar k}
 \extderphase \left( \barr{\mathcal{U}}^{-1} \partial_{\bar m} \barr{\mathcal{U}} \right) \wedge \cdot \\
 &\wedge \cdot \,\extderphase \left[ (n-3) \barr{A}_{\bar k} + \partial_{\bar k} \barr{A}_r
 +  \left( \barr{\mathcal{U}}^{-1} \partial_{\bar k} \barr{\mathcal{U}} \right) \times \barr{A}_r \right] \,.
\end{aligned}
\end{equation}
Although there is some improvement with respect to the Yang-Mills case in $(3+1)$-dimensions, we still need to make some assumptions in order to deal with the expression above.
In particular, we need to proceed as in the case of electrodynamics in higher dimensions~\cite{Henneaux-ED-higher}.
In that case, one needs to assume that $\barr{A}_{\bar m}^{\text{ED}}$, the term of order $1/r^{n-3}$ in the expansion of $A_{\bar m}^{\text{ED}}$, is the gradient of a function exactly as the leading order, i.e.,
\begin{equation}
 A_{\bar m}^{\text{ED}} = \partial_{\bar m} \left(
 \barr{\Phi} + \frac{\barr{\Theta}}{r^{n-3}}
 \right) + \bigo(1/r^{n-2}) \,.
\end{equation}
The zeroth-order term in $\Phi$ is needed in order to ensure the possibility of improper gauge transformations and has the same role of the term $\barr{\mathcal{U}}^{-1} \partial_{\bar a} \barr{\mathcal{U}}$ in~(\ref{YM:fall-off-A-angular-higher-dim}).
The above choice of $A_{\bar m}^{\text{ED}}$ leads to a vanishing $F_{\bar m \bar n}$ up to, and including, order $r^{n-3}$.
In order to be preserved by the Poincar\'e transformations, the above condition must be complemented with $\barr{\pi}^{\bar m} = 0$.

In the case of Yang-Mills in higher dimensions, we can follow a similar track.
Specifically, let us assume that, instead of~(\ref{YM:fall-off-A-angular-higher-dim}), we have the fall-off condition
\begin{equation}
 A_{\bar a} (r,\barr{x}) = \mathcal{U}^{-1} (x) \partial_{\bar a} \mathcal{U} (x) + \bigo(1/r^{n-2}) \,,
\end{equation}
where
\begin{equation}
 \mathcal{U} (x) = \barr{\mathcal{U}} (\barr{x})
 \left( \id + \frac{\barr{\Theta} (\barr{x})}{r^{n-3}} \right)
 +\bigo(1/r^{n-2}) \,.
\end{equation}
Using the expression
\begin{equation}
 \mathcal{U}^{-1} = 
 \left( \id - \frac{\barr{\Theta} (\barr{x})}{r^{n-3}} \right)
 \barr{\mathcal{U}}^{-1}
 +\bigo(1/r^{n-2}) \,,
\end{equation}
we find more explicitly the fall-off condition
\begin{equation} \label{YM:fall-off-A-angular-higher-dim-new}
 A_{\bar a} (r,\barr{x}) = \barr{\mathcal{U}}^{-1} \partial_{\bar a} \barr{\mathcal{U}}
 + \frac{1}{r^{n-3}} \barr{D}_{\bar a} \barr{\Theta} + \bigo(1/r^{n-2}) \,,
\end{equation}
where we have defined
$\barr{D}_{\bar a} \eqdef \barr{\nabla}_{\bar a} + \left( \barr{\mathcal{U}}^{-1} \partial_{\bar a} \barr{\mathcal{U}} \right) \times$, being $\barr{\nabla}$ the Levi-Civita covariant derivative on the unit $(n-1)$-sphere.
It is then trivial to see that $F_{\bar m \bar n} = \bigo (1/r^{n-2})$.
As in the case of electrodynamics, we need to complement the above conditions with the further requirement $\barr{\pi}^{\bar a} = 0$.
One can show that the new fall-off conditions are still preserved by the Poincar\'e transformations.

Let us now go back to check whether or not the Poincar\'e transformations are canonical, i.e., whether or not $\extderphase (\insertion_X \Omega)$ vanishes.
Inserting the new fall-off conditions~(\ref{YM:fall-off-A-angular-higher-dim-new}) into~(\ref{YM:canonical-Poincare-YM-higher-dim}), we find the simpler expression
\begin{equation}
 \extderphase (\insertion_X \Omega) =
 \oint d^{n-1} \barr{x} \, \sqrt{\barr{\gamma}} \, b \, \gamma^{\bar m \bar k}
 \extderphase \left( \barr{\mathcal{U}}^{-1} \partial_{\bar m} \barr{\mathcal{U}} \right) \wedge \cdot 
 \extderphase \left[ (n-3) \barr{D}_{\bar k} \barr{\Theta} + \barr{D}_{\bar k} \barr{A}_r \right] \,.
\end{equation}
In order to further simplifying the expression above, let us note that the term into square brackets can be written as
$\barr{D}_{\bar k} \mathcal{F}$ if we define $\mathcal{F} \eqdef (n-3) \barr{\Theta} + \barr{A}_r$.
Then, if we also define
$\mathcal{F}_0 \eqdef \barr{\mathcal{U}} \, \mathcal{F} \, \barr{\mathcal{U}}^{-1}$, so that
$\barr{D}_{k} \mathcal{F} = \barr{\mathcal{U}}^{-1} (\partial_{\bar k} \mathcal{F}_0 ) \,\barr{\mathcal{U}}$,
we reach the expression
\begin{equation} \label{YM:canonical-Poincare-YM-higher-dim-new}
 \extderphase (\insertion_X \Omega) =
 \extderphase \left[
 \oint d^{n-1} \barr{x} \, \sqrt{\barr{\gamma}} \, b \, \mathcal{F}_0 \cdot 
 \barr{\nabla}^{\bar m} \left( b \, \partial_{\bar m} \omega \right)
 \right]
 \,,
\end{equation}
where $\omega \eqdef \extderphase \barr{\mathcal{U}} \, \barr{\mathcal{U}}^{-1}$ is a one form which is neither close nor exact, since $\extderphase \omega = (\omega \times \wedge \omega)/2$.
Unfortunately, the right-hand side of expression above is, in general, non-vanishing.
As a consequence, also in the case of higher dimensions, the Poincar\'e transformations are not canonical if all the improper gauge transformations are allowed.

However, in contrast with the four-dimensional case, the right-hand side of~(\ref{YM:canonical-Poincare-YM-higher-dim-new}) is actually zero if we restrict the attention to the case in which $\barr{\mathcal{U}}(\barr{x})$ is constant, i.e. $\partial_{\bar m} \barr{\mathcal{U}}(\barr{x}) = 0$.
This actually corresponds to allowing, among all the possible improper gauge transformations, only the global $\SU (N)$ at infinity.
Thus, we see that, in higher dimensions, it is at least possible to have the global $\SU (N)$ as a symmetry of the theory, together with canonical Poincar\'e transformations.
One important consequence of this fact is that, now, the global colour charge $Q_{0}$ can take a non-trivial value.

This concludes this chapter in which we have provided a well-defined Hamiltonian formulation of Yang-Mills and discussed which asymptotic symmetries of the theory are present.
In particular, we have found that the situation is quite different from that of electrodynamic~\cite{Henneaux-ED} and of General Relativity~\cite{Henneaux-GR}.
Indeed, in the case of Yang-Mills in four dimensions, it is not possible to implement the improper gauge transformations in a way that does not prevent the Poincar\'e from being canonical.
Thus, the asymptotic symmetries of the theory are trivial in this case and the charges, including the global colour charge, have to vanish identically.
This situation is slightly improved in higher dimensions.
In this case, indeed, it is at least possible to implement the global $\SU(N)$ transformation at infinity without obstructing a canonical realisation of the Poincar\'e transformations.
In this way, the global colour charge is not-any-more vanishing and globally-charged states are allowed.

\chapter{Scalar electrodynamics and the abelian Higgs model} \label{cha:scalar-electrodynamics}

We now present the second original contribution of this thesis.
More precisely, we look at the case of electromagnetism coupled to a scalar field, following a similar strategy to that of the previous chapter.
The discussion contained in this chapter is taken from the paper~\cite{Tanzi-Giulini:abelian-Higgs} with minor changes.
 
In detail, we deal with two main cases, with two subcases in the first.
In the first main case, we consider what is commonly referred to as \emph{scalar electrodynamics}.
That is,  a scalar field endowed with a potential which, depending on its precise form, represents either a massless (first subcase) or a massive (second subcase) scalar field, minimally-coupled to the electromagnetic fields. 
Interestingly, the outcome of our analysis crucially depends 
on whether or not the scalar field  has a mass.
We show that a massive field has to decay at infinity faster than any power-like function in the affine coordinates, so that the behaviour of the electromagnetic fields, as well as the symmetry group, is the same as the one found by 
Henneaux and Troessaert in the case of free electrodynamics \cite{Henneaux-ED}.
On the other hand, a massless scalar field renders the boosts of the Poincar\'e transformations non-canonical in a way which is difficult to circumvent, leading either to trivial 
asymptotic symmetries or to a non-canonical action of the  Poincar\'e group.
We highlight a connection of this problem with the 
impossibility of a Lorenz gauge-fixing if the flux of  charge-current at null infinity is present, as  pointed out by Satishchandran and Wald \cite{Wald-Satishchandran}.
All this is derived in Section~\ref{sec:scalar-electrodynamics}.

As our second main case, we consider the \emph{abelian Higgs model}, i.e., a potential of the scalar field which leads to spontaneous symmetry breaking, thereby reducing the $\U(1)$ gauge-symmetry group to  the trivial group.
We show that the asymptotic symmetry group  reduces in a straightforward way to the Poincar\'e  transformations without any complication.
All this is derived in Section~\ref{sec:abelian-Higgs}. 

Section~\ref{sec:Lagrangian-Hamiltonian} sets up the Hamiltonian formalism for the present context and shows how to canonically implement the Poincar\'e action.
Section~\ref{sec:free-scalar} introduces the scalar-field models with a brief digression of the free scalar field for illustrative purposes.
Appendix~\ref{appendix:massive-fall-off} contains the proof of the statement that in the massive case the scalar field as well as its momentum fall off faster than any power in the 
affine coordinates.

\section{Hamiltonian and Poincar\'e transformations} 
\label{sec:Lagrangian-Hamiltonian}
In this section, we follow the methods illustrated in chapter~\ref{cha:Hamiltonian-ft} and establish the Hamiltonian formulation of an abelian gauge field $A$ minimally-coupled to a complex scalar field $\varphi$ on a flat Minkowski background.
The discussion about boundary terms is postponed to the next sections.
For now, we will assume that the analysed quantities are well-defined, in order to allow the following formal manipulations.

We start from the spacetime action in Lagrangian principle
\begin{equation} \label{SEDAH:action-Lagrangian}
\begin{aligned}
 S[A_\alpha,\dot A_\alpha, \varphi, \dot \varphi;g] =
  \int d^4 x \sqrt{-{}^4g} &\left[
  -\frac{1}{4} {}^4g^{\alpha \gamma} \, {}^4g^{\beta \delta} \, F_{\alpha \beta} F_{\gamma \delta}
  -{}^4g^{\alpha \beta} \big( D_{\alpha} \varphi \big)^* D_{\beta} \varphi +\right. \\
 &-V(\varphi^* \varphi)
 \Big]
 +(\text{boundary terms})\,,
\end{aligned}
\end{equation}
where $A$ is the one-form abelian potential,  $F \eqdef d A$ is the curvature (or field strength) two-form, $\varphi$ is the complex scalar field, and ${}^4 g$ is the four-dimensional \emph{flat} spacetime metric.
Moreover,
\begin{equation}
 D_\alpha \varphi \eqdef \partial_\alpha \varphi + i A_\alpha \varphi
\end{equation}
is the \emph{gauge-covariant derivative} in the fundamental representation and the potential $V(\varphi^* \varphi)$ is explicitly given by the expression
\begin{equation} \label{SEDAH:potential}
 V(\varphi^* \varphi) \eqdef -\mu^2 \varphi^* \varphi +\lambda (\varphi^* \varphi)^2 \,,
\end{equation}
where $\lambda$ and $\mu^2$ are two real parameters.
In this paper, we wish to analyse two specific situations, which arise depending on the value of these two parameters.

The former situation is \emph{scalar electrodynamics}.
Namely, it corresponds to the case in which the two parameters appearing in the potential~(\ref{SEDAH:potential}) are such that $m^2 \eqdef -\mu^2 \ge 0$ and $\lambda \ge 0$.
The so-defined parameter $m \ge 0$ is the \emph{mass} of the complex scalar field, while $\lambda \ge 0$ is the intensity of the self-interaction and we are leaving open the possibility for $\lambda$ to be non-zero, as this does not affect our analysis of the asymptotic structure.
As we shall see in section~\ref{sec:scalar-electrodynamics}, there are going to be some important differences in the asymptotic structure of the theory depending on whether we are dealing with a massless scalar field ($m^2 = 0$) or with a massive one ($m^2 > 0$).

The latter situation that we wish to analyse is the \emph{abelian Higgs model}.
This corresponds specifically to the case in which the two parameters appearing in the potential~(\ref{SEDAH:potential}) are such that $\mu^2 > 0$ and $\lambda >0$.
This choice leads to the well-known Mexican-hat shape of the potential and, ultimately, to the spontaneous symmetry breaking of the $\U(1)$ gauge symmetry.\footnote{
Note that the case $\lambda < 0$ needs to be excluded on physical grounds, as it would lead to a Hamiltonian which is not bounded from below.
For the same reason, we have to exclude the case $\lambda = 0$ when $\mu^2 > 0$, which is precisely the setup used in the abelian Higgs model.
}

Finally, let us point out again that we have included an undetermined boundary term in the action~(\ref{SEDAH:action-Lagrangian}), which ought to be chosen such that the variation principle is well-defined.
For now, we merely assume that a boundary leading to a well-defined action principle exists and postpone to the next sections a thorough discussion about whether or not this assumption is in fact correct.

\subsection{(3+1) decomposition}

The $3+1$ decomposition of the theory can be achieved following the general procedure of section~\ref{sec:3+1}.
We remind that, although we are on a flat Minkowski spacetime, it is better to consider a foliation with arbitrary lapse and shift, obtaining a Hamiltonian $H[N, \vect{N}]$, in order to readily infer the Poincar\'e transformations by replacing formally $N$ and $\vect{N}$ with $\xi^\perp$ and $\vect{\xi}$, as explained in section~\ref{sec:Poincare-general}.
The case of free electrodynamics was already discussed in~\cite[Sec.~11C]{Kuchar:hyperspace3} and in~\cite[Sec.~3]{Kuchar-Stone}, where the generator $H[N, \vect{N}]$ was derived on a spacetime manifold $M = \real \times \Sigma$, being $\Sigma$ a three-dimensional closed manifold.
Those results can be readily applied to our situations up to boundary terms, which are trivially absent in~\cite{Kuchar:hyperspace3,Kuchar-Stone}.
It is worth noting already at this point that obstructions to a well-defined Hamiltonian action of the Poincar\'e group are usually caused by the boost in the orthogonal deformation~$\xi^\perp$, so that one should usually pay more attention to the contribution due to $N$ rather than the one due to $\vect{N}$.
In addition, the transformation parametrised by $\vect{N}$ can be determined from geometrical considerations.
Specifically, one needs merely to require that the tangential transformations are given by Lie derivatives, as we did in chapter~\ref{cha:Yang-Mills}.
Nevertheless, we are going to do the explicit calculation with a generic $\vect{N}$ and obtain the mentioned fact as a result.

The complex scalar field $\varphi$ can be decomposed into
\begin{equation}
 \varphi = \frac{1}{\sqrt{2}} (\varphi_1 + i \varphi_2)
\end{equation}
where $\varphi_1$ and $\varphi_2$ are two real scalar fields.
Although this replacement makes some expressions less compact, it also makes clearer which are the actual degrees of freedom, with respect to which we have to vary the action.
In the following discussion, we will express the results either in terms of the complex scalar field $\varphi$ or in terms of the two real scalar fields $\varphi_1$ and $\varphi_2$, depending on which of the two approaches is more convenient in each situation.

Using the equations for the $3+1$ decomposition of a one-form~(\ref{3+1-decomp-A}) and of the metric~(\ref{4-metric-decomposition}) , the action~(\ref{SEDAH:action-Lagrangian}) becomes $S=\int dt \, L[A,\dot A,\varphi,\dot\varphi;g,N,\vect{N}]$, where the Lagrangian is
\begin{equation} \label{SEDAH:Lagrangian}
\begin{aligned}
  L
  ={}& \int d^3 x N \sqrt{g} \left\{
  \frac{1}{2 N^2} g^{ab} F_{0a}F_{0b}
  + \frac{g^{ab} N^c}{ N^2 } F_{0a} F_{bc}
  -\frac{1}{4} F_{ab} F^{ab}
  + \right. \\
  & + \frac{ g^{ac} N^b N^d}{2 N^2} F_{ab} F_{cd}
  +\frac{1}{2 N^2} \Big[ (\dot{\varphi}_1-A_0 \varphi_2)^2 + (\dot{\varphi}_2+A_0 \varphi_1)^2  \Big]  +\\
  & - \frac{N^a}{N^2} \left[
  (\dot \varphi_1  - A_0 \varphi_2)(\partial_a \varphi_1 -A_a \varphi_2)
  + (\dot \varphi_2 +A_0 \varphi_1)( \partial_a \varphi_2 + A_a \varphi_1)
  \right] +  \\
  & - \frac{1}{2} \left( g^{ab} -\frac{N^a N^b}{N^2} \right)
  \Big[ (\partial_a \varphi_1 - A_a \varphi_2)(\partial_b \varphi_1 - A_b \varphi_2) 
   +\\
  & + (\partial_a \varphi_2 + A_a \varphi_1)(\partial_b \varphi_2 + A_b \varphi_1) \Big]
  - V(\varphi^* \varphi) \bigg\}+(\text{boundary terms}) \,,
\end{aligned}
\end{equation}
where we have left $A_0 = N^m A_m - N A_\perp$, in order not to make the above expression even more involved.
The variation of the Lagrangian~(\ref{SEDAH:Lagrangian}) with respect to $\dot A_a$ yields the conjugate three-momenta
\begin{equation} \label{SEDAH:momenta}
 \pi^a \eqdef \frac{\delta L}{\delta \dot A_a} =\frac{\sqrt{g}}{N} \, g^{ab} \big( F_{0b} + N^m F_{bm} \big) \,,
\end{equation}
which are vector densities of weight $+1$, whereas the variation with respect to $\dot A_\perp$ returns the primary constraints
\begin{equation} \label{SEDAH:primary-constraints}
  \pi^\perp \eqdef \frac{\delta L}{\delta \dot A_\perp} = -N \, \frac{\delta L}{\delta \dot A_0} \weq 0 \,.
\end{equation}
Furthermore, the variation of the Lagrangian with respect to the time derivative of the real scalar fields gives the further three-momenta
\begin{subequations} \label{SEDAH:momenta-scalar}
\begin{align} 
 \label{SEDAH:momenta-scalar1}
 \Pi_1 \eqdef{}& \frac{\delta L}{\delta \dot{\varphi}_1} =
 \frac{\sqrt{g}}{N} \Big[\dot{\varphi}_1 - A_0 \varphi_2
 -N^m (\partial_m \varphi_1 - A_m \varphi_2)\Big] 
 \qquad \text{and}
 \\
 \label{SEDAH:momenta-scalar2}
 \Pi_2 \eqdef{}& \frac{\delta L}{\delta \dot{\varphi}_2} =
 \frac{\sqrt{g}}{N} \Big[\dot{\varphi}_2 + A_0 \varphi_1
 -N^m (\partial_m \varphi_2 + A_m \varphi_1)\Big] \,,
\end{align}
\end{subequations}
which are scalar densities of weight $+1$ and can be rewritten in the more compact complex form
\begin{equation}  \label{SEDAH:complex-field-momentum}
    \Pi \eqdef \frac{1}{\sqrt{2}} \big( \Pi_1 + i \Pi_2 \big) =
    \frac{\sqrt{g}}{N} \Big[D_0 \varphi -N^m D_m \varphi \Big] \,.
\end{equation}

Finally, the symplectic form is the canonical one
\begin{equation} \label{SEDAH:symplectic-form0}
\begin{aligned}
 \Omega[A,\pi,\varphi,\Pi] ={}&
 \int d^3 x \Big[ \extderphase \pi^\perp \wedge \extderphase A_\perp
 + \extderphase \pi^a \wedge \extderphase A_a
 + \\
 &+ \extderphase \Pi_1 \wedge \extderphase \varphi_1
 + \extderphase \Pi_2 \wedge \extderphase \varphi_2
 \Big] + (\text{boundary terms}),
\end{aligned}
\end{equation}
where the bold $\extderphase$ and $\wedge$ are, respectively, the exterior derivative and the wedge product in phase space.
Note that we are allowing the standard symplectic form to be complemented by a boundary term, which could emerge as a consequence of the boundary term included in the action.
A detailed discussion about boundary terms will be done in the next sections.
Before we complete the derivation of the generator $H[N, \vect{N}]$ and provide its explicit expression, let us briefly discuss the constraints, following section~\ref{sec:gauge-theories}.

\subsection{Constraints and constraints' algebra}

Following the general procedure highlighted in section~\ref{subsec:YM:secondary-constraints}, we need to check whether the primary constraint~(\ref{SEDAH:primary-constraints}) are preserved by time evolution.\footnote{
We remind that ``time evolution'' refers to the evolution with respect to the parameter $t$ of the foliation $(e_t)_{t \in I}$ and \emph{not} with respect to a physical clock.
}
Explicitly, we would need to derive a first version of the Hamiltonian, which includes the primary but not yet the secondary constraints.
From this Hamiltonian and from the symplectic form~(\ref{SEDAH:symplectic-form}), we would be able to compute the time derivative of $\pi^\perp$ as $\dot \pi^\perp = \{ \pi^\perp , H [N,\vect{N}] \}$.
Let us omit the details of this calculation\footnote{
The generator $H [N,\vect{N}]$ would be the same one presented in the next subsection with only the primary constraint included.
Thus, the reader could refer to the expressions of subsection~\ref{subsec:SEDAH-eoms}.
} and simply write the result
\begin{equation}
 \dot \pi^\perp = - N \big( \partial_a \pi^a + \varphi_1 \Pi_2 - \varphi_2 \Pi_1 \big)
\end{equation}
which shows that the primary constraint is not, in general, preserved under time evolution.
Thus, we impose the further constraint
\begin{equation} \label{SEDAH:gauss-constraint}
 \mathscr{G} \eqdef \partial_a \pi^a + \varphi_1 \Pi_2 - \varphi_2 \Pi_1 \weq 0 \,,
\end{equation}
which is the well-known \emph{Gauss constraint} in the presence of a charge density provided by the complex scalar field, so that the primary constraint is preserved for an arbitrary $N$.

As in the case illustrated in~\ref{subsec:YM:secondary-constraints}, the secondary constraints~(\ref{SEDAH:gauss-constraint}) is preserved under time evolution since
$\dot{\mathscr{G}}  = \{ \mathscr{G}, H [N, \vect{N}] \} = 0$.
This shows that we have found all the constraints of the theory, namely $\pi^\perp$ and the Gauss constraint $\mathscr{G}$.
Furthermore, since these expression were derived for arbitrary lapse and shift, the constraint are preserved also by the Poincar\'e transformations.
Finally, one can trivially verify that the constraints are first class and, more precisely, satisfy the abelian algebra
\begin{equation}
\label{SEDAH:constraint-algebra}
 \{ \pi^\perp (x), \pi^\perp (x') \}=0 \,, \qquad
 \{ \pi^\perp (x), \mathscr{G} (x') \}=0 \,,\qquad 
 \{ \mathscr{G} (x), \mathscr{G} (x') \}=0 \,.
\end{equation}
Having determined all the constraints of the theory, we can now complete the derivation of  $H[N, \vect{N}]$.

\subsection{Hamiltonian, equations of motion, and Poincar\'e transformations} \label{subsec:SEDAH-eoms}

The generator $H[N,\vect{N}]$ can be obtained in two steps, as discussed in section~\ref{sec:gauge-theories}.
First, one uses the definition 
$H \eqdef \int d^3 x \, (\pi^\alpha \dot{A}_\alpha + \Pi_1 \dot{\varphi}_1 +\Pi_2 \dot{\varphi}_2)-L$, in which one replaces $\dot A_a$ with $\pi^a$ by means of~(\ref{SEDAH:momenta}) and $\dot{\varphi}_{1,2}$ with $\Pi_{1,2}$ by means of~(\ref{SEDAH:momenta-scalar}), respectively.
Second, one includes the constraints multiplied by Lagrange multipliers.
In addition, as in the case described in section~\ref{sec:gauge-theories}, one can eliminate the degrees of freedom $\pi^\perp$ and $A_\perp$, since they do not carry any physical information.
Thus, we obtain the symplectic form
\begin{equation} \label{SEDAH:symplectic-form}
\begin{aligned}
 \Omega[A,\pi,\varphi,\Pi] ={}&
 \int d^3 x \Big[ \extderphase \pi^a \wedge \extderphase A_a
 + \extderphase \Pi_1 \wedge \extderphase \varphi_1
 + \extderphase \Pi_2 \wedge \extderphase \varphi_2
 \Big] \\ &+ (\text{boundary terms}),
\end{aligned}
\end{equation}
and the Hamiltonian
\begin{subequations} \label{SEDAH:Hamiltonian-total}
\begin{equation} 
\label{SEDAH:Hamiltonian-generic}
H[A,\pi,\varphi,\Pi;g,N,\vect{N};A_\perp]=
 \int  d^3 x \Big[ 
 N \mathscr{H}
 + N^i \mathscr{H}_i
 \Big] + (\text{boundary terms})\,,
\end{equation}
where
\begin{equation} 
\begin{aligned} \label{SEDAH:super-Hamiltonian}
 \mathscr{H} \eqdef{}& 
 \frac{\pi^a \pi_a + \Pi_1^2 + \Pi_2^2}{2\sqrt{g}}
 + \frac{\sqrt{g}}{4} F_{ab} F^{ab}
 + \frac{\sqrt{g}}{2} g^{ab} \big( \partial_a \varphi_1 \partial_b \varphi_1 + \partial_a \varphi_2 \partial_b \varphi_2 \big) + \\
  & \!+ \sqrt{g} A^a \big( \varphi_1 \partial_a \varphi_2 - \varphi_2 \partial_a \varphi_1 \big)
 +\frac{1}{2} A_a A^a \big( \varphi_1^2 + \varphi_2^2 \big)
 + \sqrt{g} \, V(\varphi^* \varphi)
 + A_\perp \, \mathscr{G}
\end{aligned}
\end{equation}
is responsible for the orthogonal transformations and
\begin{equation} \label{SEDAH:super-momentum}
 \mathscr{H}_i \eqdef
 \pi^a\partial_i A_a - \partial_a (\pi^a A_i)
 + \Pi_1 \partial_i \varphi_1 + \Pi_2 \partial_i \varphi_2
\end{equation}
\end{subequations}
is responsible for the tangential transformations.
Note that we named the Lagrange multiplier of the Gauss constraint~(\ref{SEDAH:gauss-constraint}) appearing in~(\ref{SEDAH:Hamiltonian-total}) as $N A_\perp$.
The reason for this is that, before the secondary constraint was included in the Hamiltonian, the Gauss constraint was already present in the Hamiltonian multiplied by $N A_\perp$.
Once the secondary constraint has been included, this term has  been reabsorbed in a redefinition of the Lagrange multiplier.
But, after the $\pi^\perp$ and $A_\perp$ are eliminated from the theory, we can rename the Lagrange multiplier as $N A_\perp$, in order to re-obtain the original term in form.
Note that, however, after all this steps, $A_\perp$ is not any more a degree of freedom of the theory, but only a Lagrange multiplier.

Finally, the knowledge of the symplectic form~(\ref{SEDAH:symplectic-form}) and of the Hamiltonian~(\ref{SEDAH:Hamiltonian-total}) in terms of arbitrary lapse and shift allows us to determine how the fields vary under time evolution and under the Poincar\'e transformations.
Specifically, let us compactly denote with $X = (\delta A_a \,, \delta \pi^a \,, \dots)$ the vector field associated to the infinitesimal change of the fields under a transformation parametrised by $N$ and $\vect{N}$, now not-any-more linked to a foliation.
In other words, $X$ need to satisfy $\extderphase H [N,\vect{N}]= - i_X \Omega$ in terms of the symplectic form~(\ref{SEDAH:symplectic-form}) and of the generator~(\ref{SEDAH:Hamiltonian-total}), which, now, depends on arbitrary $N$ and $\vect{N}$, even if they are not associated to a foliation.
Neglecting potential issues with boundary terms, as they will be thoroughly discussed in the next sections, we find
\begin{subequations} \label{SEDAH:eoms}
\begin{align} \label{SEDAH:eoms-begin}
 \delta A_a ={}& N \frac{\pi_a}{\sqrt{g}} 
 - \partial_a (N A_\perp ) + \lie_{\vect{N}} A_a \,,\\
 \delta \pi^a ={}& \partial_b (N \sqrt{g}\, F^{ba})
    - 2 \sqrt{g} \, N \,\imaginarypart \left( \varphi^* D^a \varphi \right) 
    + \lie_{\vect{N}} \pi^a\,, \\
 \delta \varphi ={}& N \frac{\Pi}{\sqrt{g}}
 + i (N A_\perp) \varphi +\lie_{\vect{N}} \varphi \,, \\
 \label{SEDAH:eoms-end}
 \delta \Pi ={}& D^a ( \sqrt{g} \, N D_a \varphi )
 + \sqrt{g} \, N \Big( \mu^2 - 2 \lambda |\varphi|^2  \Big) \varphi
 + i ( N A_\perp ) \Pi +\lie_{\vect{N}} \Pi \,,
\end{align}
\end{subequations}
where $\lie_{\vect{N}}$ is the three-dimensional Lie derivative on the space manifold $\Sigma$ with respect to $\vect{N}$ and we have chosen to use the more compact complex notation.
The above equations reduce to the usual equations of motion when $N = 1$ and $\vect{N}=0$ --- in which case the left-hand sides become the time derivative of the fields --- and to the Poincar\'e transformations when $N = \xi^\perp$ and $\vect{N} = \vect{\xi}$, as discussed in section~\ref{sec:Poincare-general}.

\subsection{Gauge transformations} \label{subsec:SEDAH-gauge-transformations}

As illustrated in section~\ref{subsec:gauge-transformations-intro}, the presence of the Gauss constraints~(\ref{SEDAH:gauss-constraint}) in the Hamiltonian~(\ref{SEDAH:Hamiltonian-total}) causes the transformations~(\ref{SEDAH:eoms}) to include a gauge transformation, whose gauge parameter is the arbitrary function $\zeta \eqdef  N A_\perp$.
In order to ensure the uniqueness of solutions despite the arbitrariness of $\zeta$, one needs to treat this transformations as mere relabelling of a physical state, i.e., a redundancy in the mathematical description of the theory.

The infinitesimal form of the gauge transformations, which we can read directly from the transformations~(\ref{SEDAH:eoms}), is
\begin{equation} \label{SEDAH:gauge-infinitesimal}
 \delta_\zeta A_a = - \partial_a \zeta \,,\qquad
 \delta_\zeta \pi^a = 0 \,, \qquad
 \delta_\zeta \varphi = i \zeta \varphi \,, \qquad \text{and} \qquad
 \delta_\zeta \Pi = i \zeta \Pi \,.
\end{equation}
Specifically, these are generated by
\begin{equation} \label{SEDAH:gauge-generator}
    G[\zeta] = \int d^3 x \, \zeta(x) \mathscr{G} (x)
\end{equation}
through the equation $\extderphase G[\zeta] = - i_{X_\zeta} \Omega$.
The left-hand side of this equation can be readily computed to be
\begin{equation}
\begin{aligned}
 \extderphase G[\zeta] ={}& \int d^3 x \, \Big[
   -\partial_a \zeta \extderphase \pi^a
   +\zeta \Pi_2 \extderphase \varphi_1
   -\zeta \Pi_1 \extderphase \varphi_2
   -\zeta \varphi_2 \extderphase \Pi_1
   +\zeta \varphi_1 \extderphase \Pi_2
   \Big] + \\ 
   & + \lim_{R \rightarrow \infty} \oint_{S^2_R} d^2 \barr{x}_k  \, \zeta \extderphase \pi^k \,.
\end{aligned}
\end{equation}
Assuming that the symplectic form~(\ref{SEDAH:symplectic-form}) does not contain any boundary term, the vector field $X_\zeta$ is ensured to exist so long as $G[\zeta]$ is differentiable \`a la Regge-Teitelboim, i.e., if the boundary term in the above expression vanishes.
Whether or not this is the case, and for which class of functions $\zeta (x)$ this happens, vastly depends on the asymptotic behaviour of the fields.
We will discuss this in the next sections and we will see that the asymptotic behaviour of the fields changes depending on the choice of parameters in the potentials~(\ref{SEDAH:potential}), i e., on whether we are dealing with scalar electrodynamics or with the abelian Higgs model.
For now, let us note that the generator $G[\zeta]$ can be in general extended to
\begin{equation} \label{SEDAH:gauge-generator-improper}
 G_{\text{ext.}}[\zeta] = G[\zeta] - \lim_{R \rightarrow \infty} \oint_{S^2_R} d^2 \barr{x}_k  \, \zeta \pi^k 
\end{equation}
whose variation, now, does not contain any boundary term.
In the case in which the boundary term in the above expression is non-trivial, the transformations corresponding to $G_{\text{ext.}}[\zeta]$ are not, in fact, proper gauge transformations.
Rather, they are true symmetries of the theory relating physically-different states and are commonly referred to as \emph{improper gauge transformations}, following~\cite{Teitelboim-YM2}. 
In addition, one can define the \emph{charge}
\begin{equation} \label{SEDAH:charge}
    Q[\zeta] \eqdef
    - \lim_{R \rightarrow \infty} \oint_{S^2_R} d^2 \barr{x}_k  \, \zeta \pi^k \,,
\end{equation}
which implies $G_{\text{ext.}}[\zeta] = G[\zeta] + Q[\zeta] \approx Q[\zeta]$, so that one can tell whether a transformation is a proper gauge or an improper one by checking whether the charge is zero or not, respectively.\footnote{
Note that, due to the limit in the definition~(\ref{SEDAH:charge}), the charge depends only on the asymptotic values of the fields and of the gauge parameter $\zeta$, which are going to be thoroughly discussed in the next sections.
Let us anticipate that the asymptotic part of $\zeta$ is going to be denoted by $\barr{\zeta}(\barr{x})$, which is a function on the two-sphere at infinity.
Thus, the charge can be written simply as $Q \left[ \,\barr{\zeta}\, \right]$, which is usually decomposed into spherical-harmonics components.
Specifically, one defines $Q_{\ell m} \eqdef Q[Y_{\ell m}]$ in terms of the spherical harmonics $Y_{\ell m}$.
The component $Q_{00}$ corresponds to the global (electric) charge up to a normalisation constant.
}
Whether or not there is a non-trivial class of functions $\zeta(x)$ such that improper gauge transformations exist and have a well-defined action on phase space depends, again, on the asymptotic behaviour of the fields, which will be discussed in the next sections.

Finally, let us point out that the expressions for the infinitesimal gauge transformations~(\ref{SEDAH:gauge-infinitesimal}) can be integrated to get the finite form of gauge transformations
\begin{equation} \label{SEDAH:gauge-finite}
 \Phi_\zeta ( A_a ) = A_a - \partial_a \zeta \,,\quad
 \Phi_\zeta ( \pi^a ) = \pi^a \,, \quad
 \Phi_\zeta ( \varphi ) = e^{i \zeta} \, \varphi \,, \quad 
 \text{and} \quad
 \Phi_\zeta ( \Pi ) = e^{i \zeta} \, \Pi \,,
\end{equation}
where $\Phi_\zeta$ denotes the action of $e^{i \zeta (x)} \in \U(1) $ on the fields.
The expressions above show clearly the $\U(1)$ nature of the gauge symmetry.

In the next sections, we are going to discuss the specific cases of scalar electrodynamics and of the abelian Higgs model.
Before that, in the next section, we are going to briefly discuss the case of a free scalar field with the potential~(\ref{SEDAH:potential}), as this simple situation let us highlight some of the features of the asymptotic structure.

\section{Free scalar field} 
\label{sec:free-scalar}
Let us study first the behaviour of the complex scalar field when it is not coupled to the gauge potential.
This can be achieved by considering the equations~(\ref{SEDAH:eoms}) and setting to zero the values of the gauge potential $A_a$, of the conjugated momenta $\pi^a$, and of the Lagrange multiplier $A_\perp$, obtaining
\begin{subequations} \label{SEDAH:eoms-complex}
\begin{align} \label{SEDAH:eoms-complex-begin}
 \delta \varphi ={}& N \frac{\Pi}{\sqrt{g}} +\lie_{\vect{N}} \varphi \,, \\
 \label{SEDAH:eoms-complex-end}
 \delta \Pi ={}&  \nabla^a ( \sqrt{g} N \partial_a \varphi ) + N \sqrt{g} \Big( \mu^2 - 2\lambda \, |\varphi|^2 \Big) \varphi
 + \lie_{\vect{N}} \Pi \,.
\end{align}
\end{subequations}
Finally, the Hamiltonian generator~(\ref{SEDAH:Hamiltonian-total}) reduces to
\begin{equation} \label{SEDAH:Hamiltonian-scalar}
 \begin{aligned}
  H_{\text{scalar}} 
  ={}&
  \int d^3 x \, \left\{
  N \left[
  \frac{\Pi^2}{\sqrt{g}}
  +\sqrt{g} g^{ab} \partial_a \varphi^* \partial_b \varphi
  + \sqrt{g} \Big( -\mu^2 |\varphi|^2 + \lambda |\varphi|^4 \Big)
  \right] + \right. \\
  &+  2 N^i \,\realpart \big( \Pi^* \partial_i \varphi \big)  \Big\} 
  +(\text{boundary terms}) \,,
 \end{aligned}
\end{equation}

Let us analyse separately the two different scenarios in the next two subsections.
First, we will consider the case in which $m^2 \eqdef - \mu^2 \ge 0$ and $\lambda \ge 0$.
This describes a massive ($m^2 >0$) or a massless ($m^2 = 0$) complex scalar field, with ($\lambda > 0$) or without ($\lambda = 0$) a self interaction.
This will be useful when studying scalar electrodynamics in section~\ref{sec:scalar-electrodynamics}.
Secondly, we will consider the case in which $\mu^2 > 0$ and $\lambda > 0$, so that the potential takes the well-known Mexican-hat shape.
This will be relevant in the analysis of the abelian Higgs model in section~\ref{sec:abelian-Higgs}.

\subsection{Massless and massive scalar field} \label{subsec:massless-massive-scalar}
Let us first consider the case of a complex scalar field with squared mass $m^2 \eqdef - \mu^2 \ge 0$.
The self interaction is either present or not, i.e., $\lambda \ge 0$.
Note that the equations of motion~(\ref{SEDAH:eoms-complex}) contain always the trivial solution
\begin{equation}
 \varphi^{(0)} (x) = 0
 \qquad \text{and} \qquad
 \Pi^{(0)} (x) =0 \,,
\end{equation}
which is also the solution that minimises the potential~(\ref{SEDAH:potential}) and the energy.
Indeed, neglecting the boundary, the value of the Hamiltonian for this solution is
\begin{equation}
 E^{(0)} \eqdef H_{\text{scalar}}[\varphi^{(0)},\Pi^{(0)};g,N=1, \vect{N} = 0] = 0 \,,
\end{equation}
whereas the value of the Hamiltonian for any other field configuration is easily seen to be positive.

At this point, we use a power-like ansatz for the fall-off behaviour of the field and the potential.
In detail, we assume that they behave as
\begin{equation}
 \varphi (x) = \frac{1}{r^\alpha} \barr{\varphi} (\barr{x}) +\bigo \big( 1/r^{\alpha+1} \big)
 \qquad \text{and} \qquad 
 \Pi (x) = \frac{1}{r^\beta} \barr{\Pi} (\barr{x}) +\bigo\big( 1/r^{\beta+1} \big)
\end{equation}
in radial-angular coordinates.
Whether or not $\alpha$ and $\beta$ can be found, such that the fall-off conditions are preserved by the Poincar\'e transformations, depends crucially on the value of the mass.
More precisely, in the \emph{massless} case, i.e. $m=0$, one finds the fall-off conditions
\begin{equation} \label{SEDAH:fall-off-scalar-massless}
 \varphi_\text{massless} (x) = \frac{1}{r} \barr{\varphi} (\barr{x}) +\bigo\big( 1/r^2 \big)
 \qquad \text{and} \qquad 
 \Pi_\text{massless} (x) = \barr{\Pi} (\barr{x}) +\bigo\big( 1/r \big) \,,
\end{equation}
which also make the symplectic form logarithmically divergent.

Before we discuss the massive case, let us point out that, in order to make the symplectic form actually finite, one needs to impose parity conditions on the asymptotic part of the fields in addition to the aforementioned fall-off conditions.
Specifically, it suffices, for instance, to require that $\barr{\varphi} (\barr{x})$ is either an even or an odd function of the sphere under the antipodal map\footnote{
See footnote~\ref{footnote:antipodal-map} on page~\pageref{footnote:antipodal-map}.
}
and, at the same time, that $\barr{\Pi}(\barr{x})$ has the opposite parity.
In this way, the potentially logarithmically divergent term in the symplectic form is, in fact, zero.
It is easy to check that these parity conditions are preserved by the Poincar\'e transformations, which take the asymptotic form
\begin{subequations}
\begin{align}
 \delta_{\xi} \barr{\varphi} &= \frac{b \, \barr{\Pi}}{\sqrt{\barr{\gamma}}} + Y^{\bar m} \partial_{\bar m} \barr{\varphi} \,,
 \\
 \delta_{\xi} \barr{\Pi} &= - b \sqrt{\barr{\gamma}} \, \barr{\varphi}
 + \barr{\nabla}^{\bar m} \left( \sqrt{\barr{\gamma}} \, b \partial_{\bar m} \barr{\varphi} \right)
 -2 b \sqrt{\barr{\gamma}} \, \lambda |\barr{\varphi}|^2 \, \barr{\varphi}
 + \partial_{\bar m} \left( Y^{\bar m} \barr{\Pi} \right) \,,
\end{align}
\end{subequations}
where we remind that $\barr{\nabla}$ denotes the covariant derivative of the round unit sphere, $b$ parametrises the Lorentz boost, and the Killing vector field $Y$ of the round-unit-two-sphere metric $\barr{\gamma}$ parametrises the rotations.
We will come back to the discussion about parity conditions in section~\ref{sec:scalar-electrodynamics} where we will consider the couple of the scalar field to electrodynamics.
Note that the Poincar\'e transformations of the asymptotic fields depends on the boosts and on the rotations, but \emph{not} on the translations, as we have already seen in the case of Yang-Mills.
This is actually a common feature of the asymptotic Poincar\'e transformations of field theories.

In the massive case, the appearance of a new term proportional to $m^2 > 0$ in the Poincar\'e transformations of the momentum, substantially modifies the fall-off behaviour of the fields, so that one does not find any power-like  solution.
One can show, as it is done in appendix~\ref{appendix:massive-fall-off}, that both $\varphi$ and $\Pi$ need to be function approaching zero at infinity faster than any power-like function.\footnote{
We will refer to this fall-off behaviour of the scalar field and its momentum and to similar behaviours encountered in the remainder of this paper by saying that the fields are ``quickly vanishing (at infinity)''.
}
Hence, we will restrict the phase space by requiring that both $\varphi$ and $\Pi$ are quickly-falling functions.
In details, we will require that the scalar field $\varphi$, as well as its spatial derivatives up to second order, and the momentum $\Pi$ vanish, in the limit to spatial infinity, faster than any power-like function (in Cartesian coordinates).
Note that, due to these fall-off conditions, the Hamiltonian and the generator of the Poincar\'e transformations of the massive scalar field are finite and functionally differentiable with respect to the canonical fields without any need of a boundary term.
In addition, also the symplectic form is finite without the need of parity conditions, contrary to the massless case.

Finally, let us note that the theory, both in the massless and in the massive case, possesses the global $\U (1)$ symmetry
\begin{equation} \label{SEDAH:global-symmetry-transformations}
 \big[ \Phi_\zeta ( \varphi ) \big] (x) = e^{i \zeta} \, \varphi (x)
 \qquad \text{and} \qquad
 \big[ \Phi_\zeta ( \Pi ) \big] (x) = e^{i \zeta} \, \Pi (x) \,,
\end{equation}
where $\Phi_\zeta$ denotes the action of $e^{i \zeta} \in \U (1)$ on the fields.
Note that, differently from~(\ref{SEDAH:gauge-finite}), the action is that of the global $\U (1)$, i.e., the parameter $e^{i \zeta} \in \U (1)$ is the same at each spacetime point.
The infinitesimal version of the above transformations is generated by
\begin{equation} \label{SEDAH:global-symmetry-generator}
    G[\zeta] = \int d^3 x \, \zeta \Big[
    \varphi_1 (x) \Pi_2 (x) - \varphi_2 (x) \Pi_1 (x)
    \Big] \,,
\end{equation}
where, again, $\zeta$ is independent of $x$.
Note that the above generator is always finite and differentiable, i.e. $\extderphase G[\zeta] = -i_{X_\zeta} \Omega$.\footnote{
In the massless case, the generator is finite thanks to the combination of the fall-off and parity conditions.
The former alone would make the generator logarithmically divergent.
}
In additions, it is \emph{not} proportional to a constraint, as the theory of a complex scalar field (with a potential) does not possess any constraint.
It can be easily verified that the generator above Poisson-commutes with Hamiltonian, i.e.
\begin{equation}
\Big\{ G[\zeta] , H[N=1,\vect{N} = 0] \Big\} \eqdef
i_{X_G} \big( i_{X_H} \Omega \big) =  0 \,,    
\end{equation}
showing that it generates, indeed, a physical symmetry.

To sum up, in this subsection, we have studied the fall-off conditions of a complex scalar field and its conjugated momentum with or without a quartic self-interaction.
The fall-off behaviour of the field and the momentum crucially depends on whether or not the mass is zero.
On the one hand, in the massless case, the fall-off conditions are power-like and, precisely, the ones in~(\ref{SEDAH:fall-off-scalar-massless}).
On the other hand, in the massive case, the scalar field --- as well as its spatial derivatives up to second order --- and its momentum need to vanish at spatial infinity faster than any power-like function.
In addition, we have seen that the theory possesses a global $\U (1)$ symmetry.

\subsection{Mexican-hat potential} \label{subsec:Mexican-hat-potential}
Let us now consider the case of a scalar field with a Mexican-hat potential $\mu^2 \ge 0$ and $\lambda > 0$.
Note that the parameter of the self interaction $\lambda$ has to be strictly positive for, otherwise, the potential and, as a consequence, the Hamiltonian are not bounded from below.
As in the cases of a massive and massless scalar field, the equations of motion~(\ref{SEDAH:eoms-complex}) contain the trivial solution
\begin{equation} \label{SEDAH:solution-zero}
 \varphi^{(0)} (x) = 0
 \qquad \text{and} \qquad
 \Pi^{(0)} (x) =0 \,.
\end{equation}
However, this is not-any-more the solution that minimises the potential $V(\varphi^* \varphi)$ and the Hamiltonian.
Indeed, this solution is found at a local maximum of the potential and gives the value of the Hamiltonian
\begin{equation}
 E^{(0)} \eqdef H_{\text{scalar}}[\varphi^{(0)},\Pi^{(0)};g,1,0] = 0 \,.
\end{equation}

In this case, the potential and the Hamiltonian are minimised by the constant and uniform solutions to the equations of motion
\begin{equation} \label{SEDAH:solutions-ssb}
 \varphi^{(\vartheta)} (x) = \frac{v}{\sqrt{2}} \, e^{i \vartheta}
 \qquad \text{and} \qquad 
 \Pi^{(\vartheta)} (x) = 0 \,,
\end{equation}
where the parameter $\vartheta \in \real/2\pi\integers$ and $v \eqdef \sqrt{\mu^2/ \lambda}$.
On all these solutions, neglecting eventual boundary terms, the Hamiltonian takes the same value
\begin{equation}
 E^{(\vartheta)} \eqdef H_{\text{scalar}}[\varphi^{(\vartheta)},\Pi^{(\vartheta)};g,1,0] =
 - \int d^3 x \, \sqrt{g} \, \lambda \, \frac{v^4}{4} \,,
\end{equation}
which diverges to $-\infty$, since it is the integral of a negative constant over a spatial slice $\Sigma \sim \real^3$.
This means that we would not be able to include the solutions~(\ref{SEDAH:solutions-ssb}) if we wished to have a well-defined, i.e. finite and functionally-differentiable, Hamiltonian.

The solution to this issue is quite simple.
We merely need to redefine the Hamiltonian generator~(\ref{SEDAH:Hamiltonian-scalar}) to
\begin{equation} \label{SEDAH:Hamiltonian-scalar-redefined}
 \begin{aligned}
  H'_{\text{scalar}} 
  ={}&
 \int d^3 x \, N \left\{
 \frac{\Pi^2}{\sqrt{g}}
 +\sqrt{g} g^{ab} \partial_a \varphi^* \partial_b \varphi
 + \sqrt{g} \Big( \lambda  \frac{v^4}{4}
 -\mu^2 |\varphi|^2 + \lambda |\varphi|^4 \Big)
  +  \right. \\
  &+  2 N^i \,\realpart \big( \Pi^* \partial_i \varphi \big)  \Big\} 
  +(\text{boundary terms}) \,.
 \end{aligned}
\end{equation}
This amounts to nothing else than the addition of the constant $\lambda v^4 / 4$ to the potential, without any impact on the equations of motion and on the Poincar\'e transformations.
Thus, neglecting the boundary, the value of the Hamiltonian evaluated on the solutions~(\ref{SEDAH:solutions-ssb}) is now
\begin{equation}
 E'{}^{(\vartheta)} \eqdef H'_{\text{scalar}}[\varphi^{(\vartheta)},\Pi^{(\vartheta)}; g,1,\vect{0}] =
 0 \,,
\end{equation}
whereas the value is positive for any other field configuration.
Note that, however, the value of the Hamiltonian evaluated on the trivial solution~(\ref{SEDAH:solution-zero}) is now divergent.
As a consequence, we need to remove this solution from the allowed field configuration, but this does not have a huge impact on the physical side, as~(\ref{SEDAH:solution-zero}) is on a local maximum of the potential and, thus, unstable under perturbations.

Let us now discuss the fall-off conditions of the field and its conjugated momentum.
Although most of the discussion does not differ much from the case of the massive scalar field discussed in the previous subsection, there are nevertheless a few subtleties that one should take into consideration.
We will work in radial-angular coordinates $(r,\barr{x})$.

First, let us focus on the terms in the Hamiltonian~(\ref{SEDAH:Hamiltonian-scalar-redefined}) containing the potential $V(\varphi^* \varphi)$ with the newly-added constant $\lambda v^4 / 4$.
If we wish this part to be finite upon integration, we need to require the absolute value of the field $|\varphi(x)|$ to approach the value $v / \sqrt{2}$ as $r \rightarrow \infty$.
In other words, this means that, if we write
\begin{equation}
 \varphi (x) = \frac{1}{\sqrt{2}} \rho (x) \, e^{i\vartheta (x)} \,,
\end{equation}
then $\rho (x) = v + h (x)$, where $h (x)$ vanishes in the limit $r \rightarrow \infty$.
Note that, in principle, we allow the phase $\vartheta(x)$ to be non-constant.
Nevertheless, we require that it has a well-defined limit $\barr{\vartheta} (\barr{x}) \eqdef \lim_{r \rightarrow \infty} \vartheta (x)$ as a possibly non-constant function on the sphere at infinity.

Secondly, let us note that, since $|\varphi (x)|$ converges to $v \ne 0$ at spatial infinity, $\varphi (x)$ is non vanishing at least in a neighbourhood of spatial infinity, so that we can always write, and it is convenient to do so,
\begin{equation}
    \Pi (x) = \Big[ u(x) + i w(x) \Big] \varphi(x) \,,
\end{equation}
where $u(x)$ and $v(x)$ are both real functions.
From the transformation of $\varphi$, one can easily find the transformations of its absolute value and phase as
\begin{equation} \label{SEDAH:transformations-abs-phase}
    \delta \rho = \rho \, \realpart \left( \frac{\delta \varphi}{\varphi} \right)
    \qquad \text{and} \qquad
    \delta \vartheta = \imaginarypart \left( \frac{\delta \varphi}{\varphi} \right) \,.
\end{equation}
Analogously, the transformations of $u$ and $w$ can be obtained from those of $\Pi$ and $\varphi$, as
\begin{equation} \label{SEDAH:transformations-u-w}
    \delta u = \realpart \left( \frac{\varphi \, \delta \Pi - \Pi \, \delta \varphi}{\varphi^2} \right)
    \qquad \text{and} \qquad
    \delta w = \imaginarypart \left( \frac{\varphi \, \delta \Pi - \Pi \, \delta \varphi}{\varphi^2} \right) \,.
\end{equation}

Thirdly, we can show that $h (x)$, $u (x)$, and $w (x)$ need to fall off at infinity faster than any power-like functions.
A precise proof of this statement would require us to proceed as in appendix~\ref{appendix:massive-fall-off} and is omitted here.
Instead, let us here provide a less-rigorous argumentation.
Specifically, let us assume the power-like behaviours
\begin{subequations} \label{SEDAH:power-like-ansatz}
\begin{align} 
    h(x) &= \barr{h} (\barr{x})/r^{\alpha} 
    + o \big( 1/r^\alpha \big) \,, \\
    u (x) &= \barr{u} (\barr{x})/r^{\beta} 
    + o \big( 1/r^\beta \big) \,, \\
    w (x) &= \barr{w} (\barr{x})/r^{\gamma} 
    + o \big( 1/r^\gamma \big) \,.
\end{align}
\end{subequations}
Note that $\alpha$ need to be greater than zero, since we requested $h$ to vanish as $r$ tends to infinity.
Now, let us consider only a part of the Poincar\'e transformations~(\ref{SEDAH:eoms-complex})
Namely,
\begin{equation} 
 \delta' \varphi = \xi^\perp \frac{\Pi}{\sqrt{g}}
 \qquad \text{and} \qquad
 \delta' \Pi = \sqrt{g} \, \xi^\perp \Big( \mu^2 - 2 \lambda |\varphi|^2  \Big) \varphi \,,
\end{equation}
from which we can derive the corresponding transformations of $h$, $u$, and $v$ using~(\ref{SEDAH:transformations-abs-phase}) and~(\ref{SEDAH:transformations-u-w}), obtaining
\begin{subequations}
\begin{align}
    \delta' h &= \frac{\xi^\perp}{\sqrt{g}} u \,, \\
    \delta' u &=  -2 v \xi^\perp \sqrt{g} \lambda \left( h - \frac{h^2}{2v} \right) - \frac{\xi^\perp}{\sqrt{g}} \big(u^2 -w^2 \big) \,,
    \quad \text{and} \\
    \delta' w &= - \frac{\xi^\perp}{ \sqrt{g}} 2 u w \,.
\end{align}
\end{subequations}
At this point, we insert~(\ref{SEDAH:power-like-ansatz}) in the above expressions and expand everything in powers of $r$, including $\sqrt{g} = r^2 \sqrt{\barr{\gamma}}$ and $\xi^\perp = r b +T$.
Requiring that the fall-off conditions~(\ref{SEDAH:power-like-ansatz}) are preserved, i.e., that the terms on the right-hand side of the above expressions do not fall off slower than the respective field, we find that the exponents in the power-like ansatz need to satisfy the non-trivial inequalities
\begin{equation} \label{SEDAH:inequalities-alpha-beta-gamma}
    \beta + 1 \ge \alpha \,, \qquad
    \alpha - 3 \ge \beta \,, \qquad
    2 \gamma + 1 \ge \beta \,, \qquad
    \beta + 2 \ge 0 \,,
\end{equation}
where the first inequality comes from the transformation of $h$, the last from that of $w$, and the remaining two from that of $u$.
One sees immediately that the first two inequalities lead to the contradiction
\begin{equation}
    \alpha \le \beta + 1 \le \alpha -2 \,, 
\end{equation}
which would lead to the conclusion that $h$ and $u$ are quickly vanishing at infinity, if one proceeded like in appendix~\ref{appendix:massive-fall-off}.
Furthermore, the third inequality in~(\ref{SEDAH:inequalities-alpha-beta-gamma}) would lead us to the conclusion that also $w$ is quickly vanishing.

Lastly, let us note that the conditions that we have determined so far show us that $\Pi$ and $h$ need to fall-off at infinity faster than any power-like function.
However, we have still to determine the fall-off behaviour of the phase $\vartheta (x)$.
To do so, it suffices to consider the transformation of $\Pi$ under time evolution, i.e., equation~(\ref{SEDAH:eoms-end}) at $N=1$ and $\vect{N}=0$.
Up to terms that are quickly vanishing at infinity, we find
\begin{equation}
    \delta \Pi = \varphi \left[
    - \sqrt{g} \, \partial_a \vartheta \, g^{ab} \, \partial_b \vartheta
    +i \partial_a \left( \sqrt{g} \, g^{ab} \partial_b \vartheta \right)
    \right]
   + (\text{quickly-vanishing terms}) \,.
\end{equation}
Thus, we have to impose that $\partial_a \vartheta$ is quickly vanishing in order to preserve the fall-off condition of $\Pi$.
This leads us to two fact.
First, the asymptotic part $\barr{\vartheta} (\barr{x})$ needs to be constant on the sphere at infinity.
We will simply denote it with $\barr{\vartheta}$.
Second, if we write $\vartheta (x) = \barr{\vartheta} + \chi (x) /v$, we will find out that $\chi (x) $ is quickly vanishing at infinity, as well as its derivatives up to second order.
We will see in section~\ref{sec:abelian-Higgs} that this situation changes when a gauge potential is present, as in the abelian Higgs model.
Finally, note that, from the Poincar\'e transformation of $\vartheta$
\begin{equation}
    \delta \vartheta = \imaginarypart \left( \frac{\delta \varphi}{ \varphi} \right) =
    \imaginarypart  \left( \frac{\xi^\perp \Pi}{\sqrt{g} \varphi}
    +\frac{\lie_{\vect{N}} \varphi }{\varphi}\right)  \,,
\end{equation}
we infer that $\barr{\vartheta}$ is invariant under the Poincar\'e transformations and, in particular, is time independent.
Indeed, the first summand on the left-hand side of the above expression is clearly quickly vanishing in the limit $r \rightarrow \infty$, while the second summand reduces to $\lie_{\vect{N}} \chi /v$ which, too, is quickly vanishing.

This concludes the derivation of the fall-off conditions of a complex scalar field with a Mexican-hat potential.
In short, we have shown that, when one considers the Mexican-hat potential as in the case of the Higgs mechanism, the Hamiltonian generator~(\ref{SEDAH:Hamiltonian-scalar}) needs to be modified to~(\ref{SEDAH:Hamiltonian-scalar-redefined}) by adding a constant to the potential, so that the minimal-energy solutions~(\ref{SEDAH:solutions-ssb}) to the equations of motion have finite energy.
The phase space is then defined by all those fields and momenta, whose difference from one of the minimum-energy solutions vanishes at infinity faster than any power-like function.
As in the case of the scalar massive field, one has to require the quick fall-off of the field up to the second-order spatial derivatives.
Note that the asymptotic part of the phase of the scalar field $\barr{\vartheta}$ needs to be constant on the sphere at infinity and is time-independent.
Moreover, the phase $\vartheta$ can differ from its constant value at infinity by a function $\chi/v$ that is quickly vanishing.
This will not be the case when we reintroduce the gauge potential $A_a$, as we shall see in section~\ref{sec:abelian-Higgs}.

Before we move to the study of scalar electrodynamics in section~\ref{sec:scalar-electrodynamics} and to that of the abelian Higgs model in section~\ref{sec:abelian-Higgs}, let us make the connection with the usual interpretation of $h$ and $\chi$ in high-energy physics.
To this end, let us consider the action in the Lagrangian picture, which can be obtained from~(\ref{SEDAH:action-Lagrangian}) by setting $A_\alpha = 0$ and adding the constant $\lambda v^4 /4$ to the potential.
Rewriting this action in terms of $h$ and $\chi$, we obtain
\begin{equation}
    S
    = \int d^4 x \left[
    -\frac{1}{2} \left( {}^4 g^{\alpha \beta} \partial_{\alpha} h \, \partial_\beta h + 2 \mu^2 h^2 \right)
    - \frac{1}{2} {}^4 g^{\alpha \beta} \partial_{\alpha} \chi \, \partial_\beta \chi
    +(\text{interactions})
    \right] \,,
\end{equation}
where the interactions include all the terms that are not quadratic in the fields.
From the above expression, we read that $h$ is a scalar field of squared mass $m_h^2 \eqdef 2 \mu^2 $, whereas $\chi$ is a massless scalar field.
The latter is precisely the Goldstone boson of the spontaneously broken global $\U (1)$ symmetry.
Indeed, as in the case analysed in the previous subsection, the theory possesses the symmetry~(\ref{SEDAH:global-symmetry-transformations}) generated by~(\ref{SEDAH:global-symmetry-generator}).
However, in this case, the minimum-energy solutions are not invariant under the action of the symmetry.
Rather, the vacuum solution
$\big( \varphi^{(\vartheta)} \,, \Pi^{(\vartheta)}  \big)$ is mapped to the different, physically-non-equivalent vacuum solution
$\big( \varphi^{(\vartheta+\zeta)} \,, \Pi^{(\vartheta+\zeta)}  \big)$ under the action of $\zeta \in \U (1)$.
In the abelian Higgs model analysed in section~\ref{sec:abelian-Higgs}, the Goldstone boson $\chi$ will turn out to be pure gauge, i.e. physically irrelevant, whereas $h$ will be the Higgs boson.

\section{Scalar electrodynamics} 
\label{sec:scalar-electrodynamics}

In this section, we will discuss the asymptotic symmetries of scalar electrodynamics, that is, the case of a complex scalar field minimally coupled to electrodynamics.
Specifically, this amount to consider the Hamiltonian~(\ref{SEDAH:Hamiltonian-total}) in the case in which the parameters in the potential~(\ref{SEDAH:potential}) are such that $m^2 \eqdef - \mu^2 \ge 0$ and $\lambda \ge 0$.
The former parameter represent the (squared) mass of the scalar field and distinguishes between the massive case ($m^2 > 0$) from the massless one ($m^2 = 0$).
The latter parameter regulates the magnitude of the self-interaction of the scalar field and we allow, in principle, $\lambda$ to be different from zero.

The ensuing discussion vastly differs depending on whether the scalar field is massive or massless.
Therefore, we will keep separated the analyses of these two different situations.
We will begin our discussion with the massive case, as this is significantly simpler and we will dedicate to it the first subsection, showing that a well-defined Hamiltonian formulation with non-trivial asymptotic symmetries can be found.

The rest of the section is devoted to the massless case, which presents subtle complications.
We will start the discussion of this second case by deriving the fall-off and (strict) parity conditions of the fields and their momenta, which are going to provide a theory with a finite symplectic form, a finite and functionally-differentiable Hamiltonian, and a symplectic action of the Poincar\'e group.
However, these conditions are a bit too strong, in the sense that they do not allow for non-trivial asymptotic symmetries.
We will attempt  to relax the strict parity conditions and discuss which issues arise during the process, that make either the asymptotic symmetry group trivial or the Lorentz boost non-canonical.
Finally, we will make the connection between these issues at spatial infinity and some problems concerning the Lorenz gauge fixing encountered in analyses at null infinity.
In making this connection, we will analyse also the situation in higher dimensions.

\subsection{Massive case}

Let us begin with the derivation of the fall-off conditions of the fields.
As in the case of a free complex scalar field of section~\ref{sec:free-scalar} and in the Yang-Mills case of section~\ref{sec:Lorentz-fall-off}, we are going to derive the fall-off conditions by demanding that they are the most general ones preserved by the action of the Poincar\'e group, whose infinitesimal form is given by~(\ref{SEDAH:eoms}) setting $N = \xi^\perp$ and $\vect{N} = \vect{\xi}$, as illustrated in section~\ref{sec:Poincare-general}.
The so-found fall-off conditions will be a natural generalisation of those discussed in~\cite{Henneaux-ED} for the case of free electrodynamics.

Focusing on the transformations of $\varphi$ and $\Pi$ and proceeding as in section~\ref{subsec:massless-massive-scalar}, one can show that the massive scalar field needs to vanish at infinity faster than any power-like function, as it happens in the free case.
It is easy to verify, at this point, that the fall-off conditions of $A$ and $\pi$ are exactly those of free electrodynamics discussed in~\cite{Henneaux-ED}.
Explicitly, they are
\begin{subequations} \label{SEDAH:fall-off-free-ED}
\begin{align} 
  A_r (r,\barr{x}) &= \frac{1}{r} \barr{A}_r (\barr{x}) +\bigo(1/r^2) \,,
  & \pi^r (r,\barr{x}) &= \barr{\pi}^r (\barr{x}) +\bigo(1/r)\,, \\
  A_{\bar{a}} (r,\barr{x}) &=\barr{A}_{\bar{a}} (\barr{x}) +\bigo(1/r) \,, 
  & \pi^{\bar{a}} (r,\barr{x}) &= \frac{1}{r} \barr{\pi}^{\bar{a}} (\barr{x}) + \bigo(1/r^2) \,,
\end{align}
\end{subequations}
where the results are expressed in radial-angular coordinates $(r,\barr{x})$.
In addition, the gauge parameter is required to fall off as
\begin{equation} \label{SEDAH:fall-off-gauge}
 \zeta (x) = \barr{\zeta} (\barr{x}) + \bigo (1/r) \,,
\end{equation}
so that the gauge transformations~(\ref{SEDAH:gauge-infinitesimal}) preserve the fall-off conditions of the canonical fields.
Note that this last expression, together with the fact that $N A_\perp$ is the gauge parameter in the generator~(\ref{SEDAH:Hamiltonian-total}) and that $N = rb +T$  for the Poincar\'e transformations, implies the fall-off condition
\begin{equation}
    A_\perp (r, \barr{x}) = \frac{1}{r} \barr{A}_\perp (\barr{x}) + \bigo(1/r^2) \,,
\end{equation}
so that the gauge transformations parametrised by $A_\perp$ do not violate the fall-off conditions~(\ref{SEDAH:fall-off-free-ED}).

Since the scalar field and its momentum vanish quickly at infinity, the asymptotic structure of the theory is effectively the same as in the free electrodynamics case.
This means that proceeding as in~\cite{Henneaux-ED}, one would find a well-defined Hamiltonian formulation of massive-scalar electrodynamics with a canonical action of the Poincar\'e group, and with non-trivial asymptotic symmetries, corresponding to an extension of the Poincar\'e group by the angle-dependent-$\U (1)$ transformations at infinity.
We redirect the reader to~\cite{Henneaux-ED} for all the details and calculations.

\subsection{Massless case: fall-off and parity conditions} \label{subsec:massless-scalar-ed-fall-off-parity}

As in the massive case, we  begin with the derivation of the fall-off conditions of the fields.
In this case, it is possible to find a power-law ansatz which is preserved by the Poincar\'e transformations~(\ref{SEDAH:eoms}).
Specifically, this corresponds to merging the fall-off conditions of the free massless scalar field~(\ref{SEDAH:fall-off-scalar-massless}) and of free electrodynamics~(\ref{SEDAH:fall-off-free-ED}).
Also in this case, the gauge parameter is required to fall-off as in~(\ref{SEDAH:fall-off-gauge}), so that the gauge transformations~(\ref{SEDAH:gauge-infinitesimal}) preserve the fall-off conditions of the fields.
The asymptotic Poincar\'e transformations of the fields are then found to be
\begin{subequations} \label{SEDAH:poincare-asymptotic}
\begin{align}
 \label{SEDAH:poincare-asymptotic-begin}
 \delta_{\xi,\zeta} \barr{A}_r ={}&
 \frac{b\, \barr{\pi}^r}{\sqrt{\barr{\gamma}}}
 + Y^{\bar m} \partial_{\bar m} \barr{A}_r
 \,,\\
 \delta_{\xi,\zeta} \barr{A}_{\bar a} ={}&
 \frac{b \, \barr{\pi}_{\bar a}}{\sqrt{\barr{\gamma}}}
 +Y^{\bar m} \partial_{\bar m} \barr{A}_{\bar a} + \partial_{\bar a} Y^{\bar m} \barr{A}_{\bar m}
 -\partial_{\bar a} \barr{\zeta}
 \,, \\
 \delta_{\xi,\zeta} \barr{\pi}^r ={}&
 \barr{\nabla}^{\bar m} \big( b\, \sqrt{\barr{\gamma}}\,
 \partial_{\bar m} \barr{A}_r \big)
 -2 b \sqrt{\barr{\gamma}} \, |\barr{\varphi}|^2 \barr{A}_r 
 +\partial_{\bar m} (Y^{\bar m} \barr{\pi}^r) \,, \\
 \label{SEDAH:poincare-asymptotic-pi-a}
 \delta_{\xi,\zeta} \barr{\pi}^{\bar a} ={}& 
 \partial_{\bar m} \big( b \, \sqrt{\barr{\gamma}}\, \barr{F}^{\bar m \bar a})
 -2 b \sqrt{\barr{\gamma}} \, \imaginarypart \left( \barr{\varphi}^* \barr{D}^{\bar a} \barr{\varphi} \right)
 +\partial_{\bar m} (Y^{\bar m}\, \barr{\pi}^{\bar a})
 -\partial_{\bar m} Y^{\bar a} \, \barr{\pi}^{\bar m} \,, \\
 \delta_{\xi,\zeta} \barr{\varphi} ={}& \frac{b \, \barr{\Pi}}{\sqrt{\barr{\gamma}}}
 + Y^{\bar m} \partial_{\bar m} \barr{\varphi}
 + i \barr{\zeta} \, \barr{\varphi} \,, \\
 \label{SEDAH:poincare-asymptotic-end}
 \delta_{\xi,\zeta} \barr{\Pi} ={}&
 - b\sqrt{\barr{\gamma}} \left( 1 + \barr{A}_r^2 \right) \barr{\varphi}
 + \barr{D}^{\bar m} \left( b\sqrt{\barr{\gamma}} \, \barr{D}_{\bar m} \barr{\varphi} \right)
 -2 b \sqrt{\barr{\gamma}} \, \lambda |\barr{\varphi}|^2 \, \barr{\varphi} + \\
 \nonumber
 &+ \partial_{\bar m} \left( Y^{\bar m} \barr{\Pi} \right)
 + i \barr{\zeta} \, \barr{\Pi}\,,
\end{align}
\end{subequations}
where we remind that $\barr{\nabla}$ is the covariant derivative of the round unit two sphere and we have defined $\barr{D}_{\bar m} \eqdef \barr{\nabla}_{\bar m} + i \barr{A}_{\bar m}$.

The fall-off conditions are not enough to provide a finite symplectic form and a symplectic action of the Poincar\'e group.
In particular, the symplectic form~(\ref{SEDAH:symplectic-form}) still contains two logarithmically-divergent contributions: The first is due to the fall-off conditions of $A$ and $\pi$, while the second is due to the fall-off conditions of $\varphi$ and $\Pi$.
One possible solution to this issue is quite simple.
One merely requires that the asymptotic part of the fields have one definite parity (either even or odd) as functions on the two-sphere at infinity and, then, imposes the opposite parity on their conjugated momenta.
This way, the potentially logarithmically-divergent contributions to the symplectic form are actually zero.
We will see that the presence of the massless scalar field will cause the parity conditions to be slightly more involved.

To fully determine the exact form of the parity conditions, let us remind that they should be such that, not only do they make the symplectic form finite, but also the Poincar\'e transformations symplectic and, thus, canonical.\footnote{
See the general discussion in section~\ref{sec:principles}.
}
Specifically, this happens when $\liephase_X \Omega = 0$, being $\liephase_X$ the Lie derivative in phase space with respect to the vector field $X$ defining the Poincar\'e transformations~(\ref{SEDAH:eoms}).
Using Cartan magic formula and the fact that the symplectic form is closed, one gets
\begin{equation} \label{SEDAH:poincare-canonical-final}
 \liephase_X \Omega =
 \extderphase (i_X \Omega)= \extderphase \oint d^2 \barr{x} \;  \sqrt{\barr{\gamma}} \,
 \barr{A}_{r} \left[
 \extderphase \barr{\nabla}^{\bar m} \left( b \, \barr{A}_{\bar m} \right)
 + 2 b\, \text{Im} \left( \barr{\varphi}^* \extderphase \barr{\varphi} \right)
 \right] \,,
\end{equation}
after having simplified the expression.
Note that the first summand in the right-hand side of the above expression is precisely the term already appearing in free electrodynamics~\cite{Henneaux-ED}, while the second summand appears due to the presence of the massless scalar field.
We wish to impose parity conditions that make the above expression to vanish identically.
To this end, let us decompose the complex scalar field as
\begin{equation}
 \varphi (x) = \frac{1}{\sqrt{2}} \, \rho (x) e^{i \vartheta (x)} \,.
\end{equation}
The newly-introduced absolute value and the phase of the scalar field need to satisfy the fall-off conditions
\begin{equation} 
 \rho (x) = \frac{1}{r} \barr{\rho} (\barr{x})
 + \bigo \big( 1/r^2 \big)
 \qquad \text{and} \qquad
 \vartheta (x) = \barr{\vartheta}  (\barr{x})
 + \bigo \big( 1/r \big) \,,
\end{equation}
in order to be consistent with~(\ref{SEDAH:fall-off-scalar-massless}).
Rewriting~(\ref{SEDAH:poincare-canonical-final}) in terms of these new fields, we see that the Poincar\'e transformations are canonical if
\begin{equation} \label{SEDAH:poincare-canonical-phase}
 \liephase_X \Omega =
 \extderphase \oint d^2 \barr{x} \;  \sqrt{\barr{\gamma}} \,
 \barr{A}_{r} \left[
 \extderphase \barr{\nabla}^{\bar m} \left( b \, \barr{A}_{\bar m} \right)
 + b\, \rho^2 \extderphase \vartheta
 \right]
\end{equation}
vanishes.
This can be achieved in the following way.
First, we require the parity conditions
\begin{equation}
 \barr{A}_r = \barr{A}_r^{\,\text{odd}}
 \qquad \text{and} \qquad
 \barr{\pi}^r = \barr{\pi}^r_{\text{even}} \,,
\end{equation}
so that the related part in the symplectic form is finite and the Coulomb solution\footnote{
The Coulomb solution in radial coordinates is simply given by  $\pi^r = - C$ (where $C$ is constant and equates the electric charge up to a normalisation constant) and all the other fields are zero up to gauge transformations.
}
is included in the allowed fields configurations.
Note that this choice of parity for $\barr{\pi}^r$ implies that gauge transformations are proper if $\barr{\zeta}$ is an odd function on the sphere and are improper if it is an even function, following the discussion of section~\ref{subsec:gauge-transformations-intro}.
Second, one makes~(\ref{SEDAH:poincare-canonical-phase}) to be finite by requiring that
\begin{equation}
 \barr{A}_{\bar m} = \barr{A}_{\bar m}^{\,\text{even}} \,,
 \qquad
 \barr{\pi}^{\bar m} = \barr{\pi}^{\bar m}_{\text{odd}} \,,
 \qquad
 \barr{\vartheta} = \barr{\vartheta}^{\,\text{odd}} \,,
\end{equation}
and that $\barr{\rho}$ is of definite parity, either even or odd.
Note that the parity conditions of $\barr{\vartheta}$ excludes the  improper gauge transformations, such as the constant $\U(1)$ at infinity, as these would shift $\barr{\vartheta}$ by an even function.
Finally, in order to make the symplectic form finite, we decompose also the momentum $\Pi$ as
\begin{equation}
 \Pi (x) = \frac{1}{\sqrt{2}} \, R (x) e^{i \Theta (x)} \,,
\end{equation}
which needs to satisfy the fall-off conditions
\begin{equation}
 R (x) = \frac{1}{r} \barr{R} (\barr{x})
 + \bigo \big( 1/r^2 \big)
 \qquad \text{and} \qquad
 \Theta (x) = \barr{\Theta}  (\barr{x})
 + \bigo \big( 1/r \big) \,.
\end{equation}
In terms of the absolute values and the phases, the logarithmically-divergent contribution to the symplectic form is
\begin{equation}
\begin{aligned}
    \int \frac{dr}{r} \int d^2 \barr{x} \,
    &\Big[ 
    \cos (\barr{\vartheta} - \barr{\Theta}) \Big(\extderphase \barr{R} \wedge \extderphase \barr{\rho} + \barr{\rho} \barr{R} \, \extderphase \barr{\vartheta} \wedge \extderphase  \barr{\Theta} \Big) + \\
    &- \sin (\barr{\vartheta} - \barr{\Theta}) \Big(\barr{\rho} \, \extderphase \barr{R} \wedge \extderphase \barr{\vartheta} + \barr{R} \, \extderphase \barr{\rho} \wedge \extderphase \barr{\Theta} \Big)
    \Big] \,,
\end{aligned}
\end{equation}
which vanishes identically once we require that $\barr{R}$  has the opposite parity of $\barr{\rho}$ and that $\barr{\Theta}$ is odd.
Note that also the parity of $\barr{\Theta}$, other than that of $\barr{\vartheta}$, is such that improper gauge transformations are not allowed.
Indeed, in order to preserve these parity conditions, we need to restrict the gauge parameters such that $\barr{\zeta}$ is an odd function.
In turn, this implies that the generator~(\ref{SEDAH:gauge-generator}) is finite and differentiable without the need of a surface term.

Finally, note that the parity conditions that we have just found are preserved by the Poincar\'e transformations.
To see this, one only need to use the asymptotic form of the transformations~(\ref{SEDAH:poincare-asymptotic}) and the equations
\begin{subequations}
\begin{align}
    \delta \barr{\vartheta} &=  \imaginarypart \left( \frac{\delta \barr{\varphi}}{\barr{\varphi}} \right) \,, &
    \delta \barr{\rho} &= \barr{\rho} \, \realpart \left( \frac{\delta \barr{\varphi}}{\barr{\varphi}} \right) \,, \\
    \delta \barr{\Theta} &=  \imaginarypart \left( \frac{\delta \barr{\Pi}}{\barr{\Pi}} \right) \,,  &
    \delta \barr{R} &=  \barr{R} \, \realpart \left( \frac{\delta \barr{\Pi}}{\barr{\Pi}} \right) \,.
\end{align}
\end{subequations}

To sum up, we have seen that, in the massless case, the fields satisfy power-like fall-off conditions.
In order to have a finite symplectic form and a canonical action of the Poincar\'e group, the fall-off conditions need to be complemented with some parity conditions.
We have shown that it is possible to find (strict) parity conditions leading to a well-defined Hamiltonian formulation.
Specifically, the strict parity conditions of $A$ and $\pi$ are the same as those in free electrodynamics~\cite[Sec. 5]{Tanzi-Giulini:YM}.
The parity conditions of the complex scalar field and its momentum have been found after decomposing them into an absolute value and a phase.
The absolute values of $\varphi$ and $\Pi$ are required to have opposite parity, while the phases need to be both of odd parity.
Notably, the parity conditions imposed on the phases, as well as those on $\barr{A}_{\bar a},$ exclude the improper gauge transformations from the theory and reduces the asymptotic symmetry group to the Poincar\'e group.
In the next subsection, we will try to solve this problem by relaxing the parity conditions.

\subsection{Relaxing the parity conditions}

The solution to reintroduce the possibility of performing improper gauge transformations is quite simple in theory.
Specifically, since the improper gauge transformations are excluded due to the (strict) parity conditions, we simply need relax them so that they are satisfied up to an improper gauge transformations.
Therefore, we require the asymptotic part of the fields that transform non-trivially under gauge transformations to be such that
\begin{subequations} \label{SEDAH:parity-conditions-loosen-ED}
\begin{equation}
 \barr{A}_{\bar{a}} = \barr{A}_{\bar{a}}^{\text{even}}
 -\partial_{\bar a} \barr{\Phi}^{\text{even}} \,,  \qquad
 \barr{\vartheta} = \barr{\vartheta}^{\text{odd}} + \barr{\Phi}^{\text{even}} \,,
 \qquad \text{and} \qquad
 \barr{\Theta} = \barr{\Theta}^{\text{odd}} + \barr{\Phi}^{\text{even}} \,,
\end{equation}
where $\barr{\Phi}^{\text{even}}(\barr{x})$ is an even function on the sphere.
At the same time, the other fields are required to satisfy the same parity conditions as before, that is
\begin{equation}
 \barr{A}_r = \barr{A}_r^{\text{odd}} \,, \qquad
 \barr{\pi}^r = \barr{\pi}^r_{\text{even}} \,, \qquad
 \barr{\pi}^{\bar{a}} =  \barr{\pi}^{\bar{a}}_{\text{odd}} \,,
\end{equation}
\end{subequations}
while $\barr{R}$ and $\barr{\rho}$ are of definite, and opposite, parity.

These relaxed parity conditions allow for certain the possibility of performing improper gauge transformations, thus extending the asymptotic symmetry group.
However, they also reintroduce back in the theory two issues.
First, the symplectic form is not finite any more.
Indeed, it contains now the logarithmically divergent contribution
\begin{equation}
    \Omega=
    \int \frac{dr}{r}\oint_{S^2} d^2 \barr{x} \;
    \extderphase \left[ \partial_{\bar a} \barr{\pi}^{\bar a}
    -2 \, \imaginarypart \left( \barr{\Pi}^* \barr{\varphi} \right ) \right]
    \wedge \extderphase  \barr{\Phi}^{\text{even}}
    + (\text{finite terms})\,.
\end{equation}
To solve this issue, we need merely to note that the term in square brackets in the expression above is nothing else than the leading contribution in the asymptotic expansion of the Gauss constraint~(\ref{SEDAH:gauss-constraint}).
Indeed, it is easy to verify that
\begin{equation}
    \mathscr{G} =
    \frac{1}{r} \left[ \partial_{\bar a} \barr{\pi}^{\bar a}
    -2 \, \imaginarypart \left( \barr{\Pi}^* \barr{\varphi} \right ) \right]
    + \bigo(1/r^2) \defeq
    \frac{1}{r} \barr{\mathscr{G}} + \bigo(1/r^2) \,.
\end{equation}
As a consequence, the symplectic form can be made finite by restricting the phase space to those fields configurations satisfying the further condition $\barr{\mathscr{G}} = 0$.
This does not exclude any solution to the equations of motion, since they already need to satisfy the full Gauss constraint $\mathscr{G} \approx 0$.

The second issue reintroduced after relaxing the parity condition is that the Poincar\'e transformations are not canonical any more.
This is due to the fact that
\begin{equation} \label{SEDAH:poincare-canonical-scalar-ed}
 \liephase_X \Omega =
 \extderphase \oint d^2 \barr{x} \;  \sqrt{\barr{\gamma}} \,
 \barr{A}_{r} \left[
 \extderphase \barr{\nabla}^{\bar m} \left( b \, \barr{A}_{\bar m} \right)
 +2 b\, \text{Im} \left( \barr{\varphi}^* \extderphase \barr{\varphi} \right)
 \right]
\end{equation}
does not vanish identically any more.
In the expression above $\liephase_X \Omega$ is the Lie derivative (in phase space) of the symplectic form $\Omega$ with respect to the vector field $X$, which identifies the Poincar\'e transformations.
In the case of \emph{free} electrodynamics, it was shown that it is possible to make the Poincar\'e transformations canonical once again, by introducing a new boundary degree of freedom $\barr{\Psi}$ and complementing the symplectic form with a boundary term $\omega$~(see section~\ref{sec:relax-parity-ED} and~\cite{Henneaux-ED}).
Specifically, this works as follows.
First, one requires that $\barr{\Psi}$ transform under the Poincar\'e transformations as
$\delta_X \barr{\Psi} = \barr{\nabla}^{\bar m} \left( b \, \barr{A}_{\bar m} \right) + Y^{\bar m} \partial_{\bar m} \barr{\Psi}$ and chooses the boundary term to be
\begin{equation} \label{symplectic-boundary}
    \omega = \oint d^2 \barr{x} \, \sqrt{\barr{\gamma}} \, \extderphase \barr{\Psi} \wedge \extderphase \barr{A}_r \,,
\end{equation}
so that $\liephase_X (\Omega + \omega) = 0$.
Second, one extends the new field $\barr{\Psi}$ in the bulk and makes $\Psi_{\text{bulk}}$ pure gauge.
A detailed discussion can be found in~\cite{Henneaux-ED}, where this method was presented for the first time.
Here, we are interested in pointing out that a similar attempt in this case would not be as successful.
Indeed, on the one hand, one would still be able to compensate the first summand in square brackets of~(\ref{SEDAH:poincare-canonical-scalar-ed}).
On the other hand, one would not be able to compensate also the second summand in square brackets, as this is not an exact form.\footnote{
To see precisely that the form is not exact one could either rely on the decomposition into phase and absolute value as done in~(\ref{SEDAH:poincare-canonical-phase}) or on the decomposition into $\varphi_{1,2}$.
We will come back to this point in section~\ref{subsec:Lorentz-Lorenz-SED}.
}

While studying a similar issue in Yang-Mills, we have shown that it is in general not-easily possible to circumvent this type of problems (see section~\ref{sec:relax-parity-YM}).
In that case, we used a general ansatz with quite a few free parameters for the boundary degrees of freedom, for the boundary term of the symplectic form, and for the Poincar\'e transformations of the boundary degrees of freedom and showed that no choice of free parameters was yielding a solution.
In this chapter, we will not pursue a similar tedious path.
Rather, we will point out a possible connection between obstructions to a canonical Lorentz boost and some issues in the Lorenz gauge fixing when a flux of charge-current at null infinity is present~\cite{Wald-Satishchandran}, as in the case of a charged massless scalar field with the weakest possible fall-off conditions compatible with the Poincar\'e transformations. 
To this end, we will first analyse some aspects of the free electrodynamics case and, then, deal with the scalar electrodynamics one.

\subsection{The Lorentz boost and the Lorenz gauge: free electrodynamics} \label{subsec:Lorentz-Lorenz-free-ED}

In this subsection, we focus on free electrodynamics and highlight the relation between canonical Poincar\'e transformations and the Lorenz gauge fixing (at infinity).
We will once again start our analysis from the action in Lagrangian picture, but use the knowledge that we have gained so far in the discussion concerning fall-off and parity conditions.

Before we begin the analysis, let us remind that the action in Lagrangian picture can be written when we have a foliation of the spacetime, as described in section~\ref{sec:3+1}.
In order to study the time evolution of the fields, it suffices to consider a foliation satisfying $N = 1$ and $\vect{N} = 0$.
More generally, following Regge and Teitelboim, we could require the lapse and the shift to satisfy the fall-off conditions~(\ref{RT:fall-off2}).
These conditions, together with the fall-off~(\ref{SEDAH:fall-off-free-ED}) and parity~(\ref{SEDAH:parity-conditions-loosen-ED}) conditions of the fields, are enough to ensure, first, that the canonical symplectic~(\ref{SEDAH:symplectic-form}) form is finite and, second, that the Hamiltonian~(\ref{SEDAH:Hamiltonian-total}) is finite and differentiable \`a la Regge-Teitelboim without the addition of any boundary term.
In turn, when this result is translated back in the language of the Lagrangian picture, it implies that the variation of the spacetime action~(\ref{SEDAH:action-Lagrangian}) is well-behaved without the need of any boundary term.

However, we have no guarantees that the variation of the action keeps to be well-behaved if the lapse and shift are asked to satisfy weaker fall-off conditions than~(\ref{RT:fall-off2}). 
In particular, as illustrated in section~\ref{sec:Poincare-general}, the Poincar\'e transformations are obtained by formally replacing the generic lapse $N$ and shift $\vect{N}$ with $\xi^\perp$ and $\vect{\xi}$, respectively.
The so-found lapse and shift are both violating the fall-off~(\ref{RT:fall-off2}) and, specifically, they grow linearly in the radial coordinate $r$ when boosts and rotations are considered.
Of course, the lapse and shift of the Poincar\'e transformations are, in general, not coming from a foliation, but rather from a generic path in the space of embeddings, since $\xi^\perp$ may be zero.
For this reason, it would make little sense to consider the spacetime action in this case, as it is not well-defined in those regions with vanishing lapse.

Nevertheless, we may consider a generic foliation whose lapse and shift are allowed to grow at spatial infinity in a way similar to $\xi^\perp$ and $\vect{\xi}$. 
If the variation of the spacetime action were well-behaved also at the boundary, the ensuing symplectic form and Hamiltonian would already include the boundary terms needed to make them well-defined, even for the linearly-growing lapse and shift.
Thus, when we replace formally the lapse and shift with those of the Poincar\'e transformations, we will not have to worry about boundary terms since these will be already there.

Before we begin, let us point out, that the Poincar\'e transformations are obtained by the formal replacement of $N$ and $\vect{N}$ with $\xi^\perp$ and $\vect{\xi}$ at the level of the Hamiltonian formulation.
The space manifold $\Sigma$ (which is $\real^3$ in our case) is not affected by this formal replacement, nor is its asymptotic boundary $\partial \Sigma$ (which is $S^2_\infty$ in our case).
Thus, we will write the boundary terms as integrals on $\real \times \partial \Sigma$, adding a boundary term at spatial infinity, as it is more fitted for the ensuing discussion, although different types of boundary terms can be considered and they have been considered in the literature, as we will comment  later.

Let us begin with the action of free electrodynamics, which is given by~(\ref{SEDAH:action-Lagrangian}) when setting the scalar field $\varphi$ to zero.
Note that, in principle, the action in the bulk can be complemented by a boundary term.
Let us write it now explicitly as
\begin{equation} \label{SEDAH:boundary-generic}
    (\text{boundary term}) =
    \int dt \oint_{S^2_\infty} d^2 \barr{x} \, \mathcal{B} (\barr{x}) \,.
\end{equation}
The function $\mathcal{B} (\barr{x})$ depends on (the asymptotic part of) the fields, of the lapse $N$, and on the shift $\vect{N}$.
The variation of the bulk action, in general, will produce other boundary terms due to the necessity of performing some integration by parts while deriving the equations of motion.
In order to have a well-defined action principle, we need to require that, not only do the bulk part of the variation  vanishes producing the bulk equations of motion, but also that the boundary term (which contains also the contribution due to $\mathcal{B}$) of the variation is zero.
One way to deal with the boundary term in the variation of the action is to make it vanish identically by imposing some suitable (fall-off and parity) conditions on the asymptotic behaviour of the fields.
Another way is to make sure that, even if it is not identically zero, it produces boundary equations of motion that do not contain any new information with respect to the bulk ones.
If neither one of the two said situations happens, we end up with some non-trivial equations of motion at the boundary, which could affect the physics of the theory, for instance by trivialising some symmetry and making some charge identically zero.

Before we actually show explicitly the situation in electrodynamics, let us stress that whether or not boundary terms are produced during the variation of the action in the bulk depends on the asymptotic behaviour of the fields, of the lapse $N$, and of the shift $\vect{N}$.
However, since we wish to obtain in the end a well-defined action of the Poincar\'e group as well, we need to allow the lapse and the shift to behave asymptotically in a way similar to those of the Poincar\'e transformations.
Thus, we will assume that $N = r b (\barr{x}) + \bigo (1)$ and we will set $\vect{N} = 0$ for simplicity, since we are interested in issues arising due to the boost.
Note that $b (\barr{x})$ here represents an arbitrary function on the sphere at infinity and not the specific $b$ of a boost.
Moreover, in the following expressions, a dot above a quantity represents the change of that quantity under the parameter of the foliation, as in the case of section~\ref{sec:3+1}.

Explicitly, the variation of the action $\extderphase S$ evaluated on an arbitrary vector field $(\delta A)$ is given by the expression
\begin{equation}
\begin{aligned}
    \int_\real dt \biggl\{
    &\int_\Sigma d^3 x \left[
    \frac{\sqrt{g}}{N} g^{ab} F_{0b} \, \delta \dot{A}_a
    + \partial_b \left( N \sqrt{g} F^{ba} \right) \delta A_a
    - N \partial_a \left( \frac{\sqrt{g}}{N} g^{ab} F_{0b} \right) \delta A_\perp
    \right] 
     \\
    + &\oint_{S^2_\infty} d^2 \barr{x} \, \Big[
    \sqrt{g} \, g^{rm} F_{0m} \, \delta A_\perp
    + N \sqrt{g} F^{a r} \, \delta A_a
    + \delta \mathcal{B}
    \Big]
    \biggr\} \,,
\end{aligned}
\end{equation}
where the surface integral on $S^2_\infty$ has to be understood as a surface integral over a sphere of radius $R$ followed by the limit $R \rightarrow \infty$.
In the above expression, we have replaced $A_0$ with $A_\perp$ using~(\ref{3+1-decomp-A}) and we have already performed the needed integration by parts.
The bulk part of the variation lead to the usual equations of motion and symplectic form, which we have already discussed.
Thus, let us focus on the boundary part.
Inserting the usual fall-off conditions of the field and $N = rb +\bigo(1)$, the boundary part of the variation becomes
\begin{equation} \label{SEDAH:variation-boundary}
    \int_\real dt \oint_{S^2} d^2 \barr{x} \, \Big[
    \sqrt{\barr{\gamma}} \, \dot{\barr{A}}_r \, \delta \barr{A}_\perp 
    + b \sqrt{\barr{\gamma}} \, \partial_{\bar m}  \barr{A}_r \barr{\gamma}^{\bar m \bar n} \, \delta \barr{A}_{\bar n}
    + \delta \mathcal{B}
    \Big] \,,
\end{equation}
where the limit $R \rightarrow \infty$ has already been taken, so that the remaining surface integral is effectively on a unit two-sphere. 
Let us assume for the moment that $\mathcal{B} = 0$, i.e., the action is the usual action of Maxwell electrodynamics without any boundary term.

On the one hand, we could try to make the expression in~(\ref{SEDAH:variation-boundary}) identically zero by imposing parity conditions on the fields similarly as in section~\ref{subsec:massless-scalar-ed-fall-off-parity}, but this choice would exclude the improper gauge transformations from the theory trivialising the asymptotic symmetries.
On the other hand, in the absence of parity conditions, the above expression would produce the boundary equations of motion
\begin{equation} \label{SEDAH:boundary-eoms-bad}
    \dot{\barr{A}}_r = 0
    \qquad \text{and} \qquad
    \partial_{\bar m} \barr{A}_r = 0
\end{equation}
in order for the variation to be well-defined for any value of $b$.
The first one of the above equations implies that the asymptotic part of the electric field vanishes and, as a consequence, so do the charges.
One way to see this is to expand the expression in~(\ref{SEDAH:momenta}) to get $\barr{\pi}^r = \dot{\barr{A}}_r \sqrt{\barr{\gamma}} / b = 0$.
In addition, the second equation implies that $\barr{A}_r$ is constant on the sphere.
Since it is also required to be an odd function --- in order to have a finite symplectic form --- we must conclude that $\barr{A}_r = 0$.
Thus, also in this case, we end up with trivial asymptotic symmetries.

This shows that, if we wish to have a well-defined action principle for Maxwell electrodynamics with the Poincar\'e transformations and non-trivial asymptotic symmetries, we must include a boundary term in the original action.
A suitable choice is
\begin{equation} \label{SEDAH:boundary}
    \mathcal{B} =
    2 \, b \, \sqrt{\barr{\gamma}} \, \barr{A}_r^2
    - \sqrt{\barr{\gamma}} \, \dot{\barr{A}}_r  \barr{A}_\perp 
    - b \, \sqrt{\barr{\gamma}} \, \partial_{\bar m}  \barr{A}_r \barr{\gamma}^{\bar m \bar n} \barr{A}_{\bar n} \,.
\end{equation}
The first summand of the boundary term above is chosen for later convenience and the latter two because they move all the variations in~(\ref{SEDAH:variation-boundary}) to $\barr{A}_r$ and $\dot{\barr{A}}_r$, so that we obtain one, single boundary equations of motion rather than the two, very-restrictive ones of~(\ref{SEDAH:boundary-eoms-bad}).
A similar boundary term was already considered by Henneaux and Troessaert in~\cite[App. B]{Henneaux-ED}, in order to solve a similar issue.
Note that, if $b$ were exactly the one of a boost and, thus, an odd function on the sphere, the first summand in the boundary term above would vanish upon integration on the sphere due to its odd parity.

At this point, the boundary part of the variation of the action becomes
\begin{equation} \label{SEDAH:variation-boundary-new}
    \int_\real dt \oint_{S^2_\infty} d^2 \barr{x} \, \Big[
    - \sqrt{\barr{\gamma}} \, \barr{A}_\perp \, \delta \dot{\barr{A}}_r 
    + \sqrt{\barr{\gamma}} \, \barr{\nabla}^{\bar m} \left( b \barr{A}_{\bar m} \right)  \, \delta \barr{A}_r
    + 2 \, b \, \sqrt{\barr{\gamma}} \, \barr{A}_r \delta \barr{A}_r
    \Big] \,,
\end{equation}
Two things can be noted.
First, when going from the Lagrangian to the Hamiltonian picture as in section~\ref{sec:Lagrangian-Hamiltonian}, $\barr{A}_r$ has now a conjugated momentum on the boundary, namely $-\sqrt{\barr{\gamma}} \, \barr{A}_\perp$.
This means that the usual bulk symplectic form needs to be complemented with the boundary term
\begin{equation} \label{SEDAH:symplectic-boundary-new}
    \omega = - \oint_{S^2_\infty} d^2 \barr{x} \, \sqrt{\barr{\gamma}} \, \extderphase \barr{A}_\perp \wedge \extderphase \barr{A}_r \,,
\end{equation}
which coincides with the boundary term~(\ref{symplectic-boundary}) used by Henneaux and Troessaert in~\cite{Henneaux-ED}, after identifying $- \barr{A}_\perp$ with $\Psi$.
Note that the need of such a boundary term was first pointed out by Campiglia and Eyheralde in~\cite[Sec. 4]{Campiglia-U1}.
Second, the single boundary equation of motion ensuing from~(\ref{SEDAH:variation-boundary-new}) is
\begin{equation}
 \dot{\barr{A}}_\perp + 2\, b \, \sqrt{\gamma} \, \barr{A}_r +
 \barr{\nabla}^{\bar m} \left( b \barr{A}_{\bar m} \right) = 0 \,.
\end{equation}
This is nothing else than the leading term in the asymptotic expansion of the Lorenz gauge condition ${}^{4} \nabla^\mu A_\mu = 0$, where ${}^4 \nabla$ is the Levi-Civita connection of the four-metric ${}^4 g$ given in~(\ref{4-metric-decomposition}).
To see this let us first compute
\begin{equation} \label{Lorenz-gauge-derivation}
    \begin{aligned}
        {}^{4} \nabla^\mu A_\mu ={}&
        {}^{4} \nabla_\mu \left( {}^4 g^{\mu \nu} A_\nu \right) 
        = \frac{1}{\sqrt{|{}^4 g|}} \, \partial_\mu \left( \sqrt{|{}^4 g|} \; {}^4 g^{\mu \nu} A_\nu \right) = \\
        ={}& \frac{1}{N \sqrt{g}} \left\{
        \partial_0 \left[ N \sqrt{g} \left( - \frac{1}{N^2}  \right) A_0 \right]
        + \partial_m \left( N \sqrt{g} \, g^{mn} A_m \right)
        \right\} = \\
        ={}& \frac{1}{N} \left[ \dot A_\perp
        + \nabla^m (N A_m) \right] \,,
    \end{aligned}
\end{equation}
where $\nabla$ is the Levi-Civita connection of the three-dimensional metric $g$.\footnote{
In deriving the expression, we have used twice, once for the four metric ${}^4 g$ and once for the three metric $g$, the fact that, if $\gamma$ is a non-singular metric of any signature on a manifold of any dimension and if $V$ is a vector field, then
\[
 \nabla_m V^m =
 \frac{1}{\sqrt{|\gamma|}} \, \partial_m \left(  \sqrt{|\gamma|} \, V^m \right) \,,
\]
where $\nabla$ is the Levi-Civita connection of $\gamma$.
}
The expansion in powers of $r$ of the expression in square brackets on the last line is then
\begin{equation}
    \frac{1}{r} \left[
    \dot{\barr{A}}_\perp + 2 b \, \barr{A}_r +
    \barr{\nabla}^{\bar m} \left( b \, \barr{A}_{\bar m} \right) 
    \right] + \bigo(1/r^2) \,,
\end{equation}
which shows the validity of the claim above.

It is possible, although not strictly necessary, to extend the boundary equation of motion into the bulk.
To do so, one can proceed as in~\cite[Sec. 4-5]{Kuchar-Stone} and introduce a new contribution to the action
\begin{equation}
    \tilde S [A,\psi] = \int d^4 x \sqrt{- {}^4 g } \; \partial_\alpha \psi \, g^{\alpha \beta} \, A_\beta \,,
\end{equation}
where $\psi$ is a new scalar field.\footnote{
In the analysis done in~\cite{Kuchar-Stone}, the spatial slices of the spacetime are closed manifolds, i.e., compact and without boundary.
Nevertheless, the results of the paper can be applied to our situation as well and are correct up to boundary terms.
}
The variation of the action with respect to $\psi$ yields the desired Lorenz gauge condition in the bulk.
When passing to the Hamiltonian picture, one finds that the conjugated momentum of $\psi$ is $\pi_\psi = \sqrt{g} A_\perp$, which is the Lagrange multiplier $A_\perp$ up to the density weight.
In order to make sure that the equations of motion in the bulk are not physically affected in the procedure, one needs merely to impose the constraint $\psi \approx 0$.
Without redoing all the computations, let us simply note that, in this way, we obtain exactly the same solution proposed in~\cite{Henneaux-ED}, after identifying $\Psi = - A_\perp =  \pi_\psi /\sqrt{g}$ and $\pi_\Psi = - \sqrt{g} \, \psi \approx 0$.
Note that the constraint $\pi_\Psi \approx 0$ induces gauge transformations that shift the value of $\Psi = A_\perp$ by an arbitrary function in the bulk, so that the bulk part of the Lorenz condition ${}^{4} \nabla^\mu A_\mu = 0$ can be violated arbitrarily.
However, on the boundary this is not the case since shifting $\barr{\Psi}$ by an odd function is not a gauge transformation, but rather a true symmetry of the theory.
Finally, note that this procedure introduces two new canonical degrees of freedom: the orthogonal component of the vector potential $A_\perp$, which has been elevated from being a mere Lagrange multiplier to a true degree of freedom, and a momentum conjugated to it.\footnote{
Due to the constraint $\pi_\Psi$ and the gauge symmetry ensuing from it, however, the only physically-relevant degree of freedom that has be introduced is the odd component of $\barr{A}_\perp$.
}

The above consideration are true for an arbitrary foliation with a lapse $N = rb +\bigo(1)$.
After, the Hamiltonian formulation is achieved, we can remove the restriction of $N$ being associated to a foliation and consider a generic path in the space of embeddings.
As a consequence, the derived results apply for the case of a Lorentz boost, which was our main interest.

To sum up, we have shown that the action of Maxwell electrodynamics needs to be complemented by a boundary term, if one wishes to have a well-defined action principle, which works also with the lapse and shift given by the Poincar\'e transformations, featuring non-trivial asymptotic symmetries.
A suitable choice for the boundary term is~(\ref{SEDAH:boundary}), which, once added to the original action, leads to two consequences.
First, when deriving the symplectic form, one finds that it contains the boundary term~(\ref{SEDAH:symplectic-boundary-new}) as in~\cite{Henneaux-ED}.
Second, one gets a new, non-trivial boundary equation of motion, which is nothing else than the leading term in the asymptotic expansion of the Lorenz gauge condition.
In addition, changing this gauge fixing at infinity by shifting $\barr{\Psi}$ by an odd function is not a proper gauge transformation, but rather a true symmetry of the theory, as thoroughly explained in~\cite{Henneaux-ED}.

\subsection{More details about the situation at the boundary}

Before we actually deal with the case in which a charged massless scalar field is present, let us characterise in greater details the situation at the boundary described in the last subsection.
To begin with, let us point out that, instead of starting from the action in the Lagrangian picture, we could have performed a similar analysis beginning with the action in the Hamiltonian picture, in which case we would have not needed to restrict the initial part of the analysis to a foliation.
In the Hamiltonian picture, one piece of the boundary term in the action would have been interpreted as being part of the Hamiltonian and the other piece as coming from the symplectic form.

In any case, independently of the chosen path, we end up with an action containing a boundary term, whose effect is to impose a Lorenz condition at infinity.
In order to better describe the situation at the boundary, let us remind that, once the passage to the Hamiltonian formulation is done, we deal with an abstract space manifold $\Sigma$ on which the canonical fields live.
The space manifold $\Sigma$ is complemented with an asymptotic boundary $\partial \Sigma = S^2_\infty$, i.e., the sphere at infinity.
It is to this sphere that the boundary terms in the integrals of the previous subsection referred.

As discussed in section~\ref{sec:Poincare-general}, a Lorenz boost can be seen as a one-parameter family of embeddings $(e_\lambda)_{\lambda \in \real}$ acting on an initial hypersurface.
For instance, using normal coordinates $(x^\alpha)$ on $M$, a boost along the $x^1$-axis of the hypersurface
$\Sigma_0 \eqdef \{ x \in M : x^0 = 0 \}$ would be described by the embeddings~(\ref{boost-along-x1}).
For this subsection, let us reintroduce temporarily the notation that the points on $\Sigma$ are denoted by bold letters and let us use coordinates $(\vect{x}^a)$ on $\Sigma$.
The coordinates are chosen so that the four-metric ${}^4 g = \text{diag} (-1,1,1,1)$, the three-metric $g = \text{diag} (1,1,1)$, and the boost is given precisely by~(\ref{boost-along-x1}).
   
We now claim that a Lorentz boost maps the sphere at infinity (in $\Sigma$) to a portion of the hyperboloid at infinity (in $M$).
To see this, let us replace $S^2_\infty$ with a sphere of finite radius $R >0$, of which we will take the limit to infinity only at the end.
This sphere contains all the points of $\Sigma$ satisfying
\begin{equation}
 \big( \vect{x}^1 \big)^2 +
 \big( \vect{x}^2 \big)^2 +
 \big( \vect{x}^3 \big)^2 = R^2 \,.
\end{equation}
Using this condition and the explicit expressions~(\ref{boost-along-x1}) for the boost, it is easy to see that, for every $\lambda \in \real$, every point $x$ in the image $e_\lambda \big( S^2_R \big)$ satisfies the condition
\begin{equation}
- \big( x^0 \big)^2 +
 \big( x^1 \big)^2 +
 \big( x^2 \big)^2 +
 \big( x^3 \big)^2 = R^2 \,.
\end{equation}
Thus, we see that the image of the sphere $S^2_R \subset \Sigma$ under a boost is contained in the hyperboloid $\mathcal{H}_R \subset M$ and that, varying the embedding parameter $\lambda$ in the real numbers, the image of $S^2_R$ sweeps a subset of $\mathcal{H}_R$.
That this is actually only a proper subset can be deduced from the fact that, for instance, any point at $\vect{x}^1 = 0$ in $S^2_R$ is mapped to the same point in $\Sigma_0$ for every value of $\lambda$.
On the contrary, any point with $\vect{x}^1 \ne 0$ sweeps completely (one branch of) an hyperbola.
   
One can show that, varying the embedding parameter $\lambda$ in the real numbers, the image of the sphere at infinity sweeps the portion of the hyperboloid, whose coordinates satisfy the further condition $\big( x^2 \big)^2 + \big( x^3 \big)^2 \le R^2$.
For instance, suppressing the dimension $x^3$, the portion of hyperboloid would be the one between the two planes $x^2 = \pm R$.
Thus, adding the generic boundary~(\ref{SEDAH:boundary-generic}) --- given explicitly in terms of~(\ref{SEDAH:boundary}) and expressed using $S^2_\infty$ --- corresponds to adding a boundary term in $M$ on a portion of the hyperboloid at infinity $\mathcal{H}_\infty$.
This boundary term imposes the asymptotic Lorenz condition discussed in the previous section.
With regard to this point, let us stress that equation~(\ref{Lorenz-gauge-derivation}) is a local expression derived using foliation-induced coordinates, which are locally well-defined so long as $N \ne 0$, as it is in a neighbourhood of the said portion of $\mathcal{H}_\infty$.

Finally, let us note that the boundary term and the Lorenz condition can be written in terms of the four-dimensional quantities, so that their formal expression do not contain the boost parameter $b_1$, as in the case of the left-hand side of~(\ref{Lorenz-gauge-derivation}).
The only dependence on the chosen boost is in selecting the portion of $\mathcal{H}_\infty$ on which the boundary term is integrated and on which the Lorenz condition holds asymptotically.
However, if we wish to include well-defined boosts in any direction, we need to extend this portion to the whole $\mathcal{H}_\infty$.
Proceeding in this way, we see that our discussion at the boundary is equivalent to the one already contained in~\cite[App. B]{Henneaux-ED}.
For convenience, we will work in terms of the sphere at infinity in the following analysis.

\subsection{The Lorentz boost and the Lorenz gauge: scalar electrodynamics} \label{subsec:Lorentz-Lorenz-SED}

Let us now reintroduce the massless scalar field minimally-coupled to electrodynamics.
Proceeding as in the previous subsection, we consider the variation of the action~(\ref{SEDAH:action-Lagrangian}) and split it into a bulk part and a boundary part.
The part in the bulk provides the equations of motion  in the bulk, after some integration by parts.
The boundary part of the variation, at this point, reads
\begin{equation} \label{SEDAH:variation-boundary-scalar-ed}
\begin{aligned}
    \int dt \oint d^2 \barr{x} \, \Big[&
    -\sqrt{\barr{\gamma}} \, \dot{\barr{A}}_r \, \delta \barr{A}_\perp 
    + b \sqrt{\barr{\gamma}} \, \partial_{\bar m}  \barr{A}_r \barr{\gamma}^{\bar m \bar n} \, \delta \barr{A}_{\bar n} + \\
    &+2 b \sqrt{\barr{\gamma}} \, \realpart \left( \barr{\varphi}^* \delta \barr{\varphi} \right)
    -2 b \sqrt{\barr{\gamma}} \, \barr{A}_r \imaginarypart \left( \barr{\varphi}^* \delta \barr{\varphi} \right)
    + \delta \mathcal{B}
    \Big] \,,
\end{aligned}
\end{equation}
where, again, we are allowing the presence of a boundary term $\mathcal{B}$ in the action.
The first line of the expression above contains the contribution due to free electrodynamics, which was amply discussed in the previous subsection.
The second line appears due to the presence of the scalar field and is made up of two contributions.
The former does not bring any issue, as it can be readily absorbed into $\delta \mathcal{B}$.
Indeed,
\begin{equation}
    2 \realpart \left( \barr{\varphi}^* \delta \barr{\varphi} \right) = 
    \barr{\varphi}_1 \delta \barr{\varphi}_1 +
    \barr{\varphi}_2 \delta \barr{\varphi}_2 =
    \delta \left( \frac{\barr{\varphi}_1^2 + \barr{\varphi}_2^2}{2} \right)
    = \delta \left( \barr{\varphi}^* \, \barr{\varphi} \right)\,.
\end{equation}
The second term is the one causing all the troubles.
Indeed, not only cannot it be rewritten as a total variation, but also it cannot be written as $\barr{A}_r$ times the total variation of something, since
\begin{equation}
    2 \imaginarypart \left( \barr{\varphi}^* \delta \barr{\varphi} \right) = 
    \barr{\varphi}_1 \delta \barr{\varphi}_2 -
    \barr{\varphi}_2 \delta \barr{\varphi}_1 \,,
\end{equation}
which is not even a closed one-form if $\delta$ is formally replaced with $\extderphase$.\footnote{
Actually, one would need to consider two independent variations, but this operation behaves formally like applying two exterior derivatives.
}
If we had been able to rewrite the term in the variation as $\barr{A}_r \delta \mathcal{B}'$, we could have included a term $- \barr{A}_r \mathcal{B}'$ into $\mathcal{B}$ and we would have obtain a single, more relaxed equation of motion at the boundary, as in the previous subsection.
This equation would have been the Lorenz gauge condition at infinity modified by a contribution coming from $\mathcal{B}'$.
When passing to the Hamiltonian formalism, the above issue translate in the fact that the Lorentz boost fails to be canonical due to the presence of a boundary term in $\liephase_X \Omega$, unless strict parity conditions are impose, \textit{de facto} trivialising the asymptotic algebra.

We propose a connection between the impossibility of having a canonical Lorentz boost when asymptotic symmetries are allowed, i.e. when we impose the relaxed parity conditions, and some issues related to the Lorenz gauge fixing when a flux of charge-current at null infinity is present~\cite{Wald-Satishchandran}.
Although we will not provide a formal proof of this statements, we will provide two indications that this is the case.

Before we present the two arguments, let us summarise the relevant results described by Wald and Satishchandran in~\cite{Wald-Satishchandran}.
Specifically, they have analysed the case of electrodynamics in four and higher dimensions and shown that, due to the fall-off conditions of the fields, it is not possible to find a Lorenz gauge fixing --- i.e. it does not exist a gauge parameter satisfying the correct fall-off conditions and bringing the four potential in Lorenz gauge --- if the dimension of the spacetime is four and a flux of charge-current at null infinity is present.
This setup is expected in our situation, due to the presence of a charged massless scalar field satisfying the most general fall-off at spatial infinity, compatible with the Poincar\'e transformations.
Note that no obstruction to the Lorenz gauge fixing is present in higher dimension, even in the presence of a charge-flux at null infinity.

The first argument which we provide is that, in the case of free electrodynamics, the Lorenz gauge fixing at infinity appears as a boundary equations of motion that needs to be imposed if we wish, at the same time, a well-defined action principle (also for the lapse and shift of the Poincar\'e transformations) and non-trivial asymptotic symmetries.
The same equation cannot be derived if a massless scalar field is present, as we have seen in the first part of this subsection.

The second argument is that no issue arises in the variation of the action (or, equivalently, in the Lorentz boost being canonical) in higher dimensions.
So far, we have worked exclusively in $3+1$ dimensions, as this is the physically-relevant case.
However, it is possible to repeat the same analyses in higher dimensions, as in the case of free electrodynamics, which was already studied by Henneaux and Troessaert in~\cite{Henneaux-ED-higher}.
So, let us assume for the remainder of this subsection that the spacetime dimension is $n+1$, being $n \ge 3$ an odd number.\footnote{
The case of even $n \ge 4$ would lead to the same results but to slightly different expression in the following analysis.
Thus, we will limit the discussion to the odd case.
}

We can derive also in this case the fall-off conditions of the fields by requiring that they are power-like and that they are preserved by the Poincar\'e transformations, obtaining\footnote{
We work in radial angular components $(r, \barr{x})$, where $\barr{x}$ are coordinates on the unit $(n-1)$-sphere.
}
\begin{subequations} \label{SEDAH:fall-off-higher-dim}
\begin{equation} \label{SEDAH:fall-off-higher-dim-radial}
 A_r (r,\barr{x}) = \frac{\barr{A}_r (\barr{x})}{r^{n-2}}  +\bigo \left( 1/r^{n-1} \right)
 \quad \text{and} \quad
 \pi^r (r,\barr{x}) = \barr{\pi}^r (\barr{x}) +\bigo(1/r)
\end{equation}
for the radial components of the canonical fields of electrodynamics,
\begin{equation} \label{SEDAH:fall-off-higher-dim-angular}
  A_{\bar{a}} (r,\barr{x}) = \partial_{\bar a} \barr{\Phi} (\barr{x}) +
  \frac{\barr{A}_{\bar{a}} (\barr{x})}{r^{n-3}} +\bigo \left( 1/r^{n-2} \right)
  \quad \text{and} \quad
  \pi^{\bar{a}} (r,\barr{x}) = \frac{\barr{\pi}^{\bar{a}} (\barr{x})}{r}  + \bigo \left( 1/r^2 \right)
\end{equation}
for their radial components, and
\begin{equation} \label{SEDAH:fall-off-higher-dim-scalar}
  \varphi (r, \barr{x}) = \frac{\barr{\varphi} (\barr{x})}{r^{(n-1)/2}}  + \bigo \left( 1/r^{(n+1)/2} \right)
  \quad \text{and} \quad
  \Pi (r, \barr{x}) = \frac{\barr{\Pi} (\barr{x})}{r^{(3-n)/2}}  + \bigo\left( 1/r^{(5-n)/2} \right)
\end{equation}
for the scalar field and its angular momentum.
\end{subequations}
The fall-off conditions of the radial~(\ref{SEDAH:fall-off-higher-dim-radial}) and of the angular~(\ref{SEDAH:fall-off-higher-dim-angular}) components of the canonical fields of electrodynamics are precisely those already discussed in~\cite{Henneaux-ED-higher}.
Two things can be noted about them.
First, if $n > 3$, the fall-off conditions are enough to ensure that the symplectic form is finite (see~\cite{Henneaux-ED-higher} for a detailed discussion), so that no parity condition is needed.
Second, $A_{\bar a}$ contains two relevant asymptotic parts: a zeroth-order contribution, which is a gradient of a function on the $(n-1)$-sphere, and a contribution of order $1/r^{n-3}$.
If $n = 3$, as it is in the previous part of this section, the gradient can be reabsorbed in $\barr{A}_{\bar a}$, but this is not possible if $n > 3$.
Finally, the fall-off conditions of free electrodynamics need to be complemented with the fall-off conditions of the scalar field and its momentum~(\ref{SEDAH:fall-off-higher-dim-scalar}).
These lead to a logarithmic divergence in the symplectic form which can be dealt with by means of parity conditions.

Ignoring the details about these subtleties, let us show directly that the scalar field does not bring any obstruction to a canonical Lorentz boost.
To this end, let us compute the Lie derivative of the symplectic form with respect to the vector field of the Poincar\'e transformations.
After a few passages, we find
\begin{equation} \label{SEDAH:canonical-poincare-higher-dim}
    \liephase_X \Omega =
    \oint_{S^{n-1}_\infty} d^{n-1} \barr{x} \left[
    \xi^\perp \sqrt{g} \, \extderphase F^{ra} \wedge \extderphase A_a
    + 2 \sqrt{g} \, \xi^\perp \, \realpart \left(
    \extderphase D_r \varphi \wedge \extderphase \varphi^*
    \right)
    \right] \,,
\end{equation}
where the integration over $S^{n-1}_\infty$ has to be understood as an integration over an $(n-1)$-sphere of radius $R$ followed by the limit $R \rightarrow \infty$.
Also in the case of higher dimensions, we see that the Poincar\'e transformations fail to be canonical due to a boundary contribution.

The first term in square brackets in equation~(\ref{SEDAH:canonical-poincare-higher-dim}) is the contribution due to free electrodynamics.
Using the fall-off conditions~(\ref{SEDAH:fall-off-higher-dim}), it reduces to
\begin{equation}
     \oint d^{n-1} \barr{x} \left\{
    - b \sqrt{\barr{\gamma}} \, \barr{\gamma}^{\bar m \bar n} \, \extderphase
    \left[ (n-3) \barr{A}_{\bar m} + \partial_{\bar m} \barr{A}_r \right]
    \wedge \extderphase \partial_{\bar n} \barr{\Phi}
    \right\} \,,
\end{equation}
where the integration is now performed on a unit $(n-1)$-sphere, as the limit $R \rightarrow \infty$ has been already taken.
This contribution has been already thoroughly analysed in~\cite{Henneaux-ED-higher} and, basically, can be dealt with by introducing a new boundary degree of freedom, similarly to the case of free electrodynamics in four dimension, which we have discussed in the previous subsection.
The second term, on the contrary, is the new contribution due to the massless scalar field.
Expanding it with the use of the fall-off conditions~(\ref{SEDAH:fall-off-higher-dim}), it reduces to
\begin{equation}
     \lim_{R \rightarrow \infty} \oint_{S^{n-1}_R} d^{n-1} \barr{x} \left\{
    - \frac{1}{R^{n-3}} \, 2 b \sqrt{\barr{\gamma}} \; \imaginarypart \left[
    \extderphase \left( \barr{\varphi} \, \barr{A}_r \right) \wedge \extderphase \barr{\varphi}^*
    \right]
    \right\} \,,
\end{equation}
which vanishes if $n > 3$ and produces the problematic term of~(\ref{SEDAH:poincare-canonical-scalar-ed}) if $n = 3$.
Thus, we have shown that no issue is present if $n > 3$ even if there is a massless charged field.

In summary, we have studied the situation of scalar electrodynamics in this section.
We have seen that is the scalar field is massive, the analysis and the asymptotic symmetries do not differ from the free electrodynamics case, which is already well known~\cite{Henneaux-ED}.
A massless scalar field, however, brings some complications.
Specifically, despite it is possible to provide a well-defined Hamiltonian formulation of the theory with canonical Poincar\'e transformations, this does not include any non-trivial asymptotic symmetry, due to the strict parity conditions required.
Relaxing the parity conditions in order to allow improper gauge transformation and keeping the symplectic form finite is possible, but at the cost of making the Lorentz boost non-canonical.
This is a second example of an incompatibility between improper gauge transformations and canonical Poincar\'e transformations, the first one being the non-abelian Yang-Mills case discussed in chapter~\ref{cha:Yang-Mills}.

Furthermore, we have identified a possible explanation for the failure of having, at the same time, a canonical action of the Poincar\'e group and non-trivial asymptotic symmetries in the impossibility of imposing a Lorenz gauge condition if there is a flux of charge-current at null infinity~\cite{Wald-Satishchandran}. Interestingly, 
this fact is a peculiarity of the physically-relevant
four-dimensional spacetime and does not happen in 
higher dimensions.
We have provided two evidences in support of this hypothesis.
First, the importance of the Lorenz gauge condition at infinity in free electrodynamics with canonical Poincar\'e transformations and non-trivial asymptotic symmetries.
Second, the fact that neither the obstruction to the Lorenz gauge fixing nor the issues in having a canonical action of the Poincar\'e group are present in higher dimensions.
This concludes our analysis of the asymptotic structure of scalar electrodynamics using the Hamiltonian formulation.
In the next section, we will focus on the abelian Higgs model.

\section{Abelian Higgs model} 
\label{sec:abelian-Higgs}

In this section, we wish to study the asymptotic symmetries of the theory described by the Hamiltonian~(\ref{SEDAH:Hamiltonian-total}), when $\mu^2 > 0$ and $\lambda > 0$.
This choice of the parameters leads to the Mexican-hat potential for  the scalar field and to the abelian Higgs mechanism.
Let us begin by determining the fall-off behaviour of the fields.

\subsection{Fall-off conditions of the fields} \label{subsec:Higgs-fall-off}

Let us begin the discussion about the abelian Higgs model by studying the asymptotic behaviour of the fields and, in particular, their fall-off conditions.
As usual, we wish to find the ``largest'' phase space which is stable under the action of the Poincar\'e transformations.
The derivation of the fall-off conditions is very similar to that presented in section~\ref{subsec:Mexican-hat-potential} and differs only in the last steps and in the fact that one needs to take into consideration a greater number of fields, as we have to include the abelian one-form potential $A_a$ and its conjugated momentum $\pi^a$ in the discussion.
This will have an effect also on the fall-off conditions of the phase of the scalar field, which will turn out to be a bit different from those of section~\ref{subsec:Mexican-hat-potential}.

First of all, let us note that we need the phase space to contain the minimum-energy solutions to the equations of motion, as these are physically-relevant solutions.
Specifically, this means that the phase space needs to include at least the solutions
\begin{equation}
 A_a (x) = 0 \,, \qquad
 \pi^a (x) = 0 \,, \qquad
 \varphi (x) = \varphi^{(\vartheta)} (x) \,, \qquad \text{and} \qquad
 \Pi (x) = 0 \,,
\end{equation}
where the constant solution $\varphi^{(\vartheta)} (x) \eqdef v/ \sqrt{2} \exp (i \vartheta)$ was already defined in equation~(\ref{SEDAH:solutions-ssb}).
We already know that one consequence of this fact is that the potential~(\ref{SEDAH:potential}) needs to be corrected by the addition of the constant $\lambda v^4 /4 $, being $v \eqdef \sqrt{ \mu^2 / \lambda }$, so that it becomes
\begin{equation} \label{SEDAH:potential-new}
    V(\varphi^* \varphi) = \lambda \left(\frac{v^2}{2} -\varphi^* \varphi \right)^2 \,.
\end{equation}
Another consequence is that we have to exclude the trivial solution to the equation of motion --- i.e. all fields and momenta equal to zero --- from phase space, for otherwise the Hamiltonian would not be finite.

Secondly, the fall-off conditions are expressed more effectively when the $\varphi(x)$ is expressed in terms of its absolute value and phase.
So, let us write
\begin{equation} \label{SEDAH:scalar-abs-phase}
 \varphi (x) = \frac{1}{\sqrt{2}} \rho (x) \, e^{i \vartheta (x)}
\end{equation}
At this point, we only need to proceed in the same way as in section~(\ref{subsec:Mexican-hat-potential}) excluding the last step, in which the behaviour of $\vartheta (x)$ was determined.
With the same arguments, we conclude also in this case that $\rho (x) = v + h (x)$, where $h (x)$ is quickly vanishing up to the second derivative order, and that $\Pi (x)$ is quickly vanishing.
\enlargethispage{-\baselineskip}

Thirdly, let us determine the fall-off behaviour of the phase $\vartheta (x)$.
As in section~\ref{subsec:Mexican-hat-potential}, let us consider the transformation of $\Pi$ under time evolution, i.e., equation~(\ref{SEDAH:eoms-end}) at $N=1$ and $\vect{N}=0$.
Up to terms that are quickly vanishing at infinity, we find
\begin{equation}
\begin{aligned}
    \delta \Pi ={}& \,\varphi \, \Big\{
    - \sqrt{g} \, g^{ab} ( \partial_a \vartheta + A_a ) (\partial_b \vartheta + A_b)
    +i D_a \big[ \sqrt{g} \, g^{ab} (\partial_b \vartheta + A_b) \big]
    \Big\}\\
    &+ (\text{quickly-vanishing terms}) \,.
\end{aligned}
\end{equation}
The above transformation preserves the fall-off condition of $\Pi$ so long as $\partial_a \vartheta + A_a$ is quickly vanishing together with its first-order derivatives.
So, let us write $A = - d \vartheta + \tilde A$, where $\vartheta$ is only required to have a well-defined limit $\barr{\vartheta} (\barr{x}) = \lim_{r \rightarrow \infty} \vartheta (x) $ as a function on the sphere at infinity, whereas $\tilde A$ is a quickly-vanishing function together with its first derivatives.
Note that the Lagrange multiplier $N A_\perp$ needs to satisfy the same fall-off conditions of $\vartheta$.

Lastly, we need to determine the fall-off behaviour 
of $\pi^a$.
To do so, one merely need to demand that the fall-off behaviour of $A= -d \vartheta +\tilde A$ is preserved by a generic Lorentz boost.
One sees that the only possibility is to require that $\pi^a$ is quickly vanishing.
In turn, this fall-off behaviour is preserved by the Poincar\'e transformations so long as the second derivatives of $\tilde A$, too, are quickly vanishing.
This concludes the discussion about the fall-off conditions of the fields in the abelian Higgs model.

To sum up, we have shown that, if one splits the scalar field into an absolute value and a phase as in~(\ref{SEDAH:scalar-abs-phase}), the former has to be \mbox{$\rho (x) = v + h (x)$}, where $h(x)$ is quickly vanishing up to its second-order derivatives.
The phase $\vartheta (x)$, on the contrary, is merely required to have a well-defined limit
$\barr{\vartheta} (\barr{x}) = \lim_{r \rightarrow \infty} \vartheta (x)$
as a function on the sphere at infinity and the same holds true for the Lagrange multiplier $N A_\perp$.
In addition, the one-form $A$ can be written as $A= -d \vartheta + \tilde A$, where $\tilde A$ is quickly vanishing up to its second-order derivatives.
Finally, the momenta $\pi^a$ need to be quickly vanishing.
In particular, note that the fall-off behaviour of the one-form $A_a$ and of its momentum $\pi^a$ is substantially different from that of electrodynamics, either in the free case~\cite{Henneaux-ED,Tanzi-Giulini:YM} or when coupled to a scalar field (see section~\ref{sec:scalar-electrodynamics}).
Indeed, an important consequence of the presence of a Higgs field is that it makes both $\tilde{A}$ and $\pi$ quickly vanishing at infinity, affecting in a non-trivial way the physics of the system, as we shall see in greater detail in the next subsections.
These fall-off conditions ensure that the Poincar\'e transformations have a well-defined action on the phase space.
\enlargethispage{-\baselineskip}

\subsection{Well-defined Hamiltonian formulation and symmetries} \label{subsec:Higgs-Hamiltonian-symmetries}

Having derived the fall-off conditions of the fields, we can now provide the well-defined Hamiltonian formulation of the abelian Higgs model.
In particular, we will provide the exact form of the Hamiltonian, of the generator of the Poincar\'e transformations, and of the generator of the gauge transformations.
Furthermore, we will also identify the asymptotic symmetries of the theory.

To begin with, let us note that the symplectic form~(\ref{SEDAH:symplectic-form}) is finite, thanks to the quick fall-off of the fields.
For the same reason,  both the Hamiltonian and the generator of the Poincar\'e transformations are finite and differentiable.
These can be inferred from the generator
\begin{subequations} \label{SEDAH:Hamiltonian-Higgs-total}
\begin{equation} \label{SEDAH:Hamiltonian-Higgs-generic}
H[A,\pi,\varphi,\Pi;g,N,\vect{N};A_\perp]=
 \int  d^3 x \Big[ 
 N \mathscr{H}
 + N^i \mathscr{H}_i
 \Big] \,.
\end{equation}
Indeed, the Hamiltonian is obtained by setting $N=1$ and $\vect{N} = 0$, while the generator of the Poincar\'e transformations is obtained by setting $N=\xi^\perp$ and $\vect{N} = \vect{\xi}$.
In the above generator
\begin{equation} 
\begin{aligned} \label{SEDAH:superHamiltonian-Higgs}
 \mathscr{H} \eqdef{}& 
 \frac{\pi^a \pi_a + \Pi_1^2 + \Pi_2^2}{2\sqrt{g}}
 + \frac{\sqrt{g}}{4} F_{ab} F^{ab}
 + \frac{\sqrt{g}}{2} g^{ab} \big( \partial_a \varphi_1 \partial_b \varphi_1 + \partial_a \varphi_2 \partial_b \varphi_2 \big) + \\
 & \! + \sqrt{g} A^a \big( \varphi_1 \partial_a \varphi_2 - \varphi_2 \partial_a \varphi_1 \big)
 +\frac{1}{2} A_a A^a \big( \varphi_1^2 + \varphi_2^2 \big)
 + \sqrt{g} \, V(\varphi^* \varphi)
 + A_\perp \, \mathscr{G}
\end{aligned}
\end{equation}
is responsible for the orthogonal transformations and
\begin{equation} 
 \mathscr{H}_i \eqdef
 \pi^a\partial_i A_a - \partial_a (\pi^a A_i)
 + \Pi_1 \partial_i \varphi_1 + \Pi_2 \partial_i \varphi_2
\end{equation}
is responsible for the tangential transformations.
Note that the potential is 
\begin{equation}
    V(\varphi^* \varphi) = \lambda \left( \frac{v^2}{2} - \varphi^* \varphi \right)^2 \,,
\end{equation}
\end{subequations}
which differs from the original potential~(\ref{SEDAH:potential}) due to the addition of the constant $\lambda v^4 /4$, so that the energy of the vacuum solutions to the equations of motion is finite (and, actually, zero).

Furthermore, let us note that the Gauss constraint $\mathscr{G} (x)$ --- which has the same expression as in~(\ref{SEDAH:gauss-constraint}) --- appears in the generator~(\ref{SEDAH:Hamiltonian-Higgs-total}) multiplied by the Lagrange multiplier $N A_\perp$.
More generally, gauge transformations are generated by
\begin{equation} \label{SEDAH:generator-gauge-Higgs}
 G[\zeta] \eqdef \int d^3 x \,  \zeta (x) \mathscr{G} (x) \weq 0 \,,
\end{equation}
which is finite and differentiable \`a la Regge-Teitelboim without the need of any surface term.
Note that, in the above generator, $\zeta$ is only required to have a well-defined limit $\barr{\zeta} (\barr{x}) = \lim_{r \rightarrow \infty} \zeta (x)$, so that the transformations~(\ref{SEDAH:gauge-infinitesimal}) preserve the fall-off conditions identified in the previous subsection.

Two things can be noted at this point.
First, the phase $\vartheta$ can always be trivialised by a proper gauge transformation, so that it carries no physical meaning.
Specifically, from~(\ref{SEDAH:gauge-finite}), we see that $\Phi_{-\vartheta} (\varphi) = \rho /\sqrt{2}$, without any phase.\footnote{
Note that this is a complete gauge fixing.
}
Second, since~(\ref{SEDAH:generator-gauge-Higgs}) is already finite and differentiable without the need of any boundary term, it cannot be extended to a generator of improper gauge transformations, contrary to the case of electrodynamics~\cite{Henneaux-ED}.
As a consequence, the asymptotic symmetries of the theories are trivially the Poincar\'e transformations.
Indeed, the only generator of asymptotic symmetries is $H [\xi, \vect{\xi}]$ which satisfies the algebra
\begin{subequations} \label{SEDAH:HH-bracket-total}
\begin{equation}
\label{SEDAH:HH-bracket-generic}
 \big\{ H[\xi^\perp_1,\vect{\xi}_1] , H[\xi^\perp_2,\vect{\xi}_2] \big\}
 = H[\hat \xi^\perp,\hat{\vect{\xi}}] + G [\hat \zeta] \,,
\end{equation}
where the parameters of the Poincar\'e transformations on the right-hand sides are given by
\begin{equation}
 \hat \xi^\perp = \lie_{\vect{\xi}_1} \xi^\perp_2 - \lie_{\vect{\xi}_2} \xi^\perp_1
 \qquad \text{and} \qquad
 \hat \xi^m = \tilde \xi^m + [\vect{\xi}_1,\vect{\xi}_2]^m \,,
\end{equation}
while the gauge parameter is given by
\begin{equation}
 \hat \zeta = A_m \tilde \xi^m
 + \xi_1 \lie_{\vect{\xi_2}} A_\perp - \xi_2 \lie_{\vect{\xi_1}} A_\perp \,.
\end{equation}
\end{subequations}
Here, we have defined
$\tilde \xi^i \eqdef g^{ij} (\xi^\perp_1 \partial_j \xi^\perp_2 - \xi^\perp_2 \partial_j \xi^\perp_1)$,
which simplifies the expressions above and the following discussion.
In addition, $\lie$ is the Lie derivative on space manifold $\Sigma$ and $[\vect{\xi}_1,\vect{\xi}_2]$ is the Lie-Jacobi commutator of the vector fields $\vect{\xi}_1$ and $\vect{\xi}_2$.
The above algebra is easily seen to be a Poisson-representation of the Poincar\'e algebra up to (proper) gauge transformations, due to the presence of the constraint on the right-hand side of~(\ref{SEDAH:HH-bracket-generic}).
The fact that the Poincar\'e algebra is recovered up to proper gauge transformations is not in general a problem (see e.g. the discussion in~\cite[Sec. 2]{Beig}).

Before we conclude this section, let us note that the $\hat \zeta$ in the expressions above depends on the canonical fields and, in particular, on $A_m$.\footnote{We remind that $A_\perp$ is not a canonical field, but only a Lagrange multiplier.}
As a consequence the transformation generated by $G[\hat \zeta]$ slightly differs from the usual gauge transformations.
Specifically, it induces the transformations
\begin{equation}
 \delta A_a = - \partial_a \hat \zeta
 \,,
 \qquad
 \delta \pi^a = - \tilde \xi^a \, \mathscr{G} \approx 0  \,,
 \qquad
 \delta \varphi =  i \hat \zeta \, \varphi  \,,
 \qquad \text{and} \qquad
 \delta \Pi = i \hat \zeta \, \Pi \,.
\end{equation}
It is useful to compare the above transformations with those caused by $\zeta$ in equations~(\ref{SEDAH:gauge-infinitesimal}).
Two things emerge.
First, $A$, $\varphi$, and $\Pi$ transform in the same way, with the only difference being that the parameter $\hat \zeta$ is field-dependent.
Secondly, the transformation of $\pi$ due to $\hat \zeta$ is not trivial any more.
Nevertheless, it is proportional to the Gauss constraint and, thus, vanishes on the constraint hypersurface.
Note that the transformations above deserve by all means the title of gauge transformations, as one part of them is generated by the constraints
\begin{equation}
 A_m \mathscr{G} \weq 0
\end{equation}
smeared with $\tilde \xi^m$, while the other part of them is generated by the usual Gauss constraint $\mathscr{G}$ smeared by $\xi_1 \lie_{\vect{\xi_2}} A_\perp - \xi_2 \lie_{\vect{\xi_1}} A_\perp$.
Let us neglect this second part, as it is of a well-known shape, and focus on the first one, whose generator
$\tilde G[\tilde{\vect{\xi}}] \eqdef G[\tilde \xi^m A_m]$ is easily seen to be well-defined and functionally differentiable \`a la Regge-Teitelboim, as the fields are rapidly vanishing while approaching spatial infinity.
Furthermore, it satisfies the algebra
\begin{equation}
 \big\{ \tilde G[\tilde{\vect{\xi}}_1] , \tilde G[\tilde{\vect{\xi}}_2] \big\} = \tilde G[\tilde{\vect{\xi}}] \,,
 \qquad \text{where} \qquad
 \tilde{\vect{\xi}} = [\tilde{\vect{\xi}}_1 , \tilde{\vect{\xi}}_2] \,.
\end{equation}

This concludes the discussion about the asymptotic symmetries of the abelian Higgs model.
To sum up, we have shown that the fall-off conditions derived in the previous subsection are enough to ensure a well-defined Hamiltonian formulation with a canonical action of the Poincar\'e group.
Moreover, we have seen that the phase $\vartheta$ can always be trivialised by a proper gauge transformation and that the asymptotic symmetries of the abelian Higgs model are trivial, in the sense that the asymptotic-symmetry group is the Poincar\'e group.
This was shown by computing the Poisson-algebra of $H[\xi,\vect{\xi}]$, which is a Poisson representation of the Poincar\'e algebra up to proper gauge transformations.
Before we draw our conclusions, let us briefly comment on the fate of the Goldstone boson, which emerged as a consequence of the spontaneous symmetry break of the global $\U (1)$ in section~\ref{subsec:Mexican-hat-potential}.

\subsection{The fate of the Goldstone boson}

At the end of section~\ref{subsec:Mexican-hat-potential}, we discussed that, in the case of the spontaneous symmetry break of the global $\U (1)$ symmetry, the action could be rewritten in terms of two real scalar fields: the massive $h$ and the massless $\chi$.
The former was identified to be the candidate Higgs field in the abelian Higgs model, while the latter was recognised as the Goldstone boson.

Let us repeat that analysis for the abelian Higgs model using the fall-off conditions of section~\ref{subsec:Higgs-fall-off}.
Proceeding as in section~\ref{subsec:Mexican-hat-potential}, let us consider the action in the Lagrangian picture, which can be obtained from~(\ref{SEDAH:action-Lagrangian}) by adding the constant $\lambda v^4 /4$ to the potential.
Rewriting this action in terms of $h$, $\vartheta$, and $\tilde A$, we obtain
\begin{equation} \label{SEDAH:action-expansion-Higgs}
\begin{aligned}
    S[h,\vartheta,\tilde A] ={}& \int d^4 x \left\{
    - \frac{1}{2} \left( {}^4 g^{\alpha \beta} \partial_{\alpha} h \, \partial_\beta h + 2 \mu^2 h^2 \right) + \right. \\
    &- \left. \left( \frac{1}{4} {}^4 g^{\alpha \gamma} \, {}^4 g^{\beta \delta}  \tilde{F}_{\alpha \beta} \tilde{F}_{\gamma \delta}
    + \frac{v^2}{2} \, {}^4 g^{\alpha \beta} \tilde{A}_{\alpha} \tilde{A}_{\beta} \right)
    +(\text{interactions})
    \right\} \,,
\end{aligned}
\end{equation}
where the interactions include all the terms that are not quadratic in the fields.
In the expression above, we have introduced $\tilde{A}_0 \eqdef A_0 +\dot \vartheta$, whose quickly-falling asymptotic behaviour can be inferred from that of the momentum $\Pi$, and $\tilde{F} \eqdef d \tilde{A}$.

Three things can be noted from the expression above.
First, there is a real scalar fields $h$ of squared mass $m_h^2 \eqdef 2 \mu^2$, which corresponds to the Higgs field.
Second, the spin-one field $\tilde{A}$ becomes massive with a squared mass $m_A^2 \eqdef v^2$.
The mass $m_A$ of the spin-one field depends on the vacuum expectation value $v$ of the complex scalar field $\varphi$ and, in general, on the coupling of the Higgs to the original gauge potential $A$ (in this chapter, it was set to the value of $1$).
Last, but not least, there is no trace of a massless scalar field, which could play the role of the Goldstone boson.

The disappearance of the Goldstone boson can be tracked down precisely to the choice $A = - d \vartheta + \tilde A$, which we did in section~\ref{subsec:Higgs-fall-off}.
On the one hand, this choice makes the fall-off condition of the momentum $\Pi$ to be preserved by the Poincar\'e transformations.
On the other hand, it makes the gauge-covariant derivative of $\varphi$ to be independent of $\vartheta$, so that the action~(\ref{SEDAH:action-expansion-Higgs}) is also independent of the phase $\vartheta$.\footnote{We remind that in section~\ref{subsec:Mexican-hat-potential}, the role of the Goldstone boson was played by the part $\chi$ of the phase $\vartheta$ which was quickly falling at infinity.
}
Therefore, if we wished to reintroduce the Goldstone boson, we would have to modify slightly the fall-off conditions of section~\ref{subsec:Higgs-fall-off}.

To this end, let us write the phase $\vartheta = \vartheta' + \chi/v$ as the sum of two parts.
The former of the two, $\vartheta'$, is the ``power-like'' part of $\vartheta$, while the latter, $\chi /v$, is the ``quickly-falling'' part.
The only requirement while performing this split is that $\chi /v$ is actually a quickly-falling function.
Two things can be noted.
First, the fall-off behaviour of $\Pi$ is preserved by the Poincar\'e transformations so long as $A = -d \vartheta' +\tilde A$, being $\tilde A$ quickly falling.
When this choice is introduced in the action~(\ref{SEDAH:action-Lagrangian}), we get the expression
\begin{equation} \label{SEDAH:action-expansion-Higgs-Goldstone}
\begin{aligned}
    S[h,\chi,\tilde A] ={}& \int d^4 x \left\{
    - \frac{1}{2} \left( {}^4 g^{\alpha \beta} \partial_{\alpha} h \, \partial_\beta h + 2 \mu^2 h^2 \right)
    - \frac{1}{2} {}^4 g^{\alpha \beta} \partial_{\alpha} \chi \, \partial_\beta \chi +  \right. \\
    & \left. - \left( \frac{1}{4} {}^4 g^{\alpha \gamma} \, {}^4 g^{\beta \delta}  \tilde{F}_{\alpha \beta} \tilde{F}_{\gamma \delta}
    + \frac{v^2}{2} \, {}^4 g^{\alpha \beta} \tilde{A}_{\alpha} \tilde{A}_{\beta} \right) 
    +(\text{interactions})
    \right\} \,,
\end{aligned}
\end{equation}
rather than~(\ref{SEDAH:action-expansion-Higgs}).
The expression above does indeed contain the massless Goldstone boson $\chi$, other than the already-present Higgs field $h$ and massive spin-one field $\tilde A$.
Second, the split of $\vartheta$ into a power-like part $\vartheta'$ and a quickly-falling part $\chi /v$ is obviously ambiguous.
This was not the case in section~\ref{subsec:Mexican-hat-potential}, since, in that case, the only allowed power-like part of $\vartheta$ was its asymptotic value $\barr{\vartheta}$ on the sphere at infinity, which could be unequivocally identified by $\barr{\vartheta} \eqdef \lim_{r \rightarrow \infty} \vartheta$.
The main consequence of this ambiguity in the splitting of $\vartheta$ into $\vartheta'$ and $\chi$ is that $\Omega$ becomes degenerate, so that it is a pre-symplectic form rather than a symplectic one.
Indeed, one can easily check that $i_Y \Omega = 0$, if $Y$ is chosen so that
\begin{equation}
    \delta_Y \vartheta' = \zeta \,, \qquad
    \delta_Y \chi = - v \zeta \,, \qquad
    \delta_Y \tilde A = d \zeta \,, \quad \text{and} \qquad
    \delta_Y (\text{other fields}) = 0 \,,
\end{equation}
for any quickly-falling $\zeta$.
At this point, one would need to deal with this issue as in~\cite{Henneaux:Rarita-Schwinger}.

In this chapter, we have preferred not to pursue this path, since it would introduce some mathematical complications without any advantage on the physical side.
Indeed, as we have seen in section~\ref{subsec:Higgs-Hamiltonian-symmetries}, the phase $\vartheta$ can always be set to zero by means of a proper gauge transformation (with gauge parameter $-\vartheta$).
As a consequence, neither $\vartheta'$ nor $\chi$ are physically-relevant fields.

This concludes the discussion concerning the abelian Higgs model.
To summarise this section, we have derived the fall-off conditions of the fields and shown that these lead to a well-defined Hamiltonian formulation of the theory with a canonical action of the Poincar\'e group.
As a consequence of the quick fall-off behaviour of the fields, the proper gauge transformations cannot be extended to improper ones and the asymptotic symmetry group trivially coincide with the Poincar\'e group.
Furthermore, we have seen that the various fields can be interpreted as a massive spin-one field, a Higgs field, and a Goldstone boson.
The latter, although absent due to the chosen fall-off conditions, can be reintroduced by a slight modification of these.
Nevertheless, it is physically irrelevant, since it can be trivialised by means of a proper gauge transformation.

\chapter{Discussion and conclusions} \label{cha:conclusions}

We are now in a position to look back and see how the Hamiltonian treatment of the asymptotic symmetries of gauge theories pursued in this thesis proceeded.
Although the Hamiltonian formulation of classical field theory is an old and well-know subject, its consistent and systematic application to the study of asymptotic symmetries is rather recent. 
The general techniques, which were first used successfully by Henneaux and Troessaert to study the asymptotic symmetries of General Relativity~\cite{Henneaux-GR} and of electrodynamics~\cite{Henneaux-ED}, relies on the careful analysis of four essential elements of the Hamiltonian formulation.
\begin{enumerate}[wide=\parindent] 
 \item \textsc{Phase space.} \emph{The phase space, i.e. the space of allowed field configurations, is defined in terms of (canonical) fields on the space manifold $\Sigma$, satisfying conditions that allow the features described in the following points.}
 One usually starts with a fairly large phase space, in which many-physically relevant quantities are only given as formal expressions, and then imposes conditions on the regularity of the fields and on their asymptotic behaviour, in order to make these formal expressions actually well-defined.
 In particular, the asymptotic behaviour of the fields is usually expressed in terms of fall-off and parity conditions.
 The former ones specify how quickly the fields approach a certain fixed value (often zero) at infinity, whereas the latter ones indicate the parity (under the antipodal map) of the leading term in the asymptotic expansion of the canonical fields.
 In any case, the phase space should be big enough to include physically-relevant solutions to the field equations --- such as the Coulomb solution in electrodynamics or the Schwarzschild one in General Relativity --- and, among all the possible choices featuring the following qualities, it should be the ``biggest'' one.
 It is possible to introduce new fields on the phase space (that were not originally present in the theory) under the condition that, only at the boundary, the new fields are allowed to be physically non-trivial and to affect the original fields non-trivially.
 \item \textsc{Symplectic form.} \emph{There must be a well-defined symplectic form, which is a closed weakly non-degenerate two-form on the phase space.}
 The symplectic form of local field theories is defined in terms of an integral over the space manifold and an integral over its (asymptotic) boundary, which is often, but not always, zero.
 Thus, in order for it to be well-defined, the greatest effort goes in making sure that the integral over the space manifold converges, obtained usually by carefully imposing fall-off and parity conditions on the canonical fields.
 Generalisations to the pre-symplectic case, i.e. to the degenerate case, are possible and work similarly to the situation described here, although they were not used to derive the results contained in this thesis.
 \item \textsc{Hamiltonian.} \emph{The Hamiltonian must be well-defined, which means that it must be finite and must admit a  Hamiltonian vector field associated to it.}
 The latter condition is usually stated by saying that the Hamiltonian must be functionally-differentiable with respect to the canonical fields, extending the original definition of differentiability \`a la Regge and Teitelboim~\cite{RT}, used when the symplectic form does not have boundary terms.
 \item \textsc{Poincar\'e group.} Since we have restricted our attention to relativistic field theories on a flat Minkowski background, we have added the further requirement that \emph{the Poincar\'e group must be a canonical symmetry of the theory}.
 In particular, there must be a canonical generator and a Hamiltonian vector field associated to the Poincar\'e group, which is true as long as there is a symplectic vector field associated to the Poincar\'e group, due to the particular structure of the group itself.
 In the cases in which the spacetime is not the flat Minkowski one, a similar requirement should be imposed, eventually replacing the Poincar\'e group with a different symmetry group.
 For instance, in the asymptotically-flat case, one uses the asymptotic Poincar\'e transformations, i.e., the ones with respect to the asymptotic metric; whereas, in an FLRW spacetime, one would use the group of spatial translations and rotations. 
\end{enumerate}

The first three requirements provide the minimal structure to set up the Hamiltonian formulation.
Only once these conditions are met, one is allowed to study the (asymptotic) symmetries of the theory.
The fourth requirement ensures that the asymptotic symmetry group is, in general, an extension of the Poincar\'e group. 
Whether or not this extension is non-trivial and how big it turns out to be depends amply on the choice of phase space, e.g. on the fall-off and parity conditions imposed on the fields.
The case of General Relativity is emblematic in this regard: If parity conditions are chosen as in~\cite{RT}, the symmetry group is merely the Poincar\'e one; however, if they are chosen as in~\cite{Henneaux-GR}, the entire BMS group is recovered.
It is important to mention that the correct characterisation of the phase space, despite being described at the first point of the list above, is a process that actually extends throughout the other points and requires a certain amount of trial and error, as we have seen in the explicit cases treated in this thesis.

Specifically, we have exploited the method described above, in order to study two main cases.
The first one consisted of non-abelian gauge theories, while the second one of a complex scalar fields minimally coupled to an abelian gauge potential.
The latter was specialised in two situations of interest: scalar electrodynamics and the abelian Higgs model.
We will present the discussions and conclusions concerning each one of these two main cases separately in the following two sections.
These are partially taken and adapted from~\cite{Tanzi-Giulini:YM} and~\cite{Tanzi-Giulini:abelian-Higgs}, respectively.

\section{Non-abelian gauge theories}

Non-abelian gauge theories and, more precisely, the $\SU(N)$-Yang-Mills theories have been discussed in section~\ref{sec:gauge-theories} and in chapter~\ref{cha:Yang-Mills}.
We have seen that the fall-off conditions can be unequivocally determined from a power-law ansatz if one requires that the usual action of the Poincar\'e transformations leaves them invariant.

The discussion on the parity conditions is more involved, as it was expected from the experience with the electromagnetic case, already studied by Henneaux and Troessaert~\cite{Henneaux-ED}. 
We started by showing that strict parity conditions can be employed which allow the theory to meet all the required Hamiltonian requirements, though they turned out to not allow for improper gauge transformations and non-zero global charges.
We certainly did expect some additional constraints on the range of such conditions, over and above those already known from the electrodynamics case. 
After all, there are additional terms from the non-vanishing commutators in the covariant derivatives, which one needs to take care of.
But we did not quite expect these constraints to be as restricting 
as they finally turned to be.  

In a second step we investigated the possibility to regain non-trivial asymptotic symmetries and colour charges by carefully 
relaxing the parity conditions.
We found that it is possible to relax the parity conditions
so that they are still preserved under Poincar\'e transformations, that the symplectic form is still finite, and that non-trivial improper gauge transformations exist.
But this possibility had two independent drawbacks.
First, the Poincar\'e transformations ceased to be canonical.
We originally expected to be able to fix this issue in a manner similar to that employed in the electromagnetic case 
in~\cite{Henneaux-ED}, but this turned out not to work.
Second, the relaxed parity conditions allowing non-zero colour charge fail to ensure the existence of a symplectic 
form. 
Furthermore, the impossibility of having canonical Poincar\'e transformations and a non-vanishing colour charge at the same time is peculiar to the four-dimensional case, since we have shown that these complications disappear in higher dimensions.

Let us clearly state that we do not pretend to have proven the impossibility of non-trivial asymptotic symmetries and non-vanishing global charges in an entirely rigorous sense,
taking full account of functional-analytic formulations of infinite-dimensional symplectic manifolds.
However, the constraints we encountered are not of the kind that one can expect to simply disappear through proper identifications 
of function spaces.
We believe that the obstructions we encountered point towards a deeper structural property of non-abelian Yang-Mills theory that has hitherto not been taken properly into consideration,
despite the fact that similar concerns were already raised  
several years ago in~\cite[Sec.~5]{Christodoulou.Murchadha:1981} based on a careful asymptotic analysis of the field equations.
Given that this view is correct, it is tempting to speculate that further clarification of that structure might tell us something relevant in connection with the problem of confinement.
After all, the general idea that confinement might be related to structures already seen at a purely classical level is not new;
see, e.g., \cite{Feynman:1981}.  

An important further step would be to reconcile the Hamiltonian treatment at spacelike infinity with the already existing study at null infinity~\cite{Strominger-YM,Strominger-YM2,Barnich-YM}.
Here, too, a confirmation of the obstructions we have seen would highlight a clear difference between non-abelian Yang-Mills theory on one hand, and electrodynamics and gravity on the other.
In particular, it would be of interest to learn whether such a reconciliation is possible only at the price of allowing certain symmetries to act non-canonically.

\section{Scalar electrodynamics and abelian Higgs model}

The second main case considered in this thesis was that of a complex scalar field minimally coupled to an abelian gauge potential, discussed in chapter~\ref{cha:scalar-electrodynamics}.
The complex scalar field was provided with a quartic potential which, depending on the value of its parameters, enabled us to study two relevant cases: scalar electrodynamics and the abelian Higgs model.

In the case of scalar electrodynamics, the discussion and the results extensively depend on whether the scalar field is massive or massless.
On the one hand, when the scalar field is massive, its asymptotic behaviour is such that it decays quickly enough not to affect asymptotically the electromagnetic fields.
Therefore, the equations reduce effectively to the ones of free electrodynamics~\cite{Henneaux-ED} and so does the discussion about the asymptotic symmetries.
On the other hand, when the scalar field is massless, this fact is no longer true and the asymptotic structure of the theory is non-trivially affected.
In this case, we have seen that it is not possible to have a canonical Poincar\'e boost if the fall-off condition of the scalar field is as general as allowed.
In addition, we have pointed out a connection between this issue and similar concerns which arose in studies at null infinity~\cite{Wald-Satishchandran}, in which it was shown that the Lorenz gauge cannot be imposed asymptotically if a non-zero flux of charged particle is present at null infinity.

The situation of the abelian Higgs model is actually quite simpler.
In this case, we have seen that the complex scalar field and the abelian potential need to approach certain values at infinity and that they differ from these values by quickly vanishing functions.
The physical degrees of freedom are contained in a real scalar field --- the Higgs field --- and in a spin-one field, both of which are effectively massive as a consequence of the spontaneous symmetry breaking.
Their quickly-vanishing behaviour at infinity is precisely that expected from any massive field.
In addition, the asymptotic-symmetry group trivialises to the Poincar\'e group as a consequence of the spontaneous symmetry breaking, as it was expected from this model, in which the $\U(1)$-group is broken to the trivial group.

We consider these results to be both interesting and encouraging.
The results concerning massive scalar fields were clearly expected and it is encouraging to see that this expectation was correct, thereby providing further confidence into the Hamiltonian method for the analysis of asymptotic structures 
and symmetries.
As already discussed at length, the obvious and characteristic 
advantage of this method is to embed the discussion on asymptotic symmetries into a formalism of clear-cut rules and interpretation.
The result for the massless case was not a surprise, though we had no firm intuition whether we should expect it.
In that sense, we consider it interesting.

Although the models considered here are not at the forefront of physical phenomenology, the abelian model does provides good insight into what to expect in other Higgs models, such as the physically-relevant case of the  electroweak sector.
We provided ample discussion of these expectations.  
In fact, one may speculate that similar results hold in the case of the abelian mechanism of the electroweak theory, that is 
$\SU(2)_L\times\U(1)_Y \rightarrow \U(1)_{\text{e.m.}}$, where $\SU(2)_L$ is the isospin acting on the left-handed fermions, $U(1)_Y$ is generated by hypercharge, and  $\U(1)_{\text{e.m.}}$ by electric charge.
Let $W^I$ (with $I=1,2,3$) be the standard components of the connection associated to $\SU(2)_L$ and $B$ the one associate to $\U(1)_Y$.
Then, one can rewrite
\begin{subequations}
\begin{align}
 A &= \sin \theta_W W^3 + \cos \theta_W B \,, \\
 Z &= \cos \theta_W W^3 - \sin \theta_W  B \,,\\
 W^{\pm} &= \frac{1}{\sqrt{2}} \left( W^1 \mp i W^2 \right) \,,
\end{align}
\end{subequations}
where $\theta_W$ is the Weinberg angle.
Due to the Higgs mechanism, the equations of motion (and the Poincar\'e transformations) of $Z$ and $W^\pm$ will contain an effective mass term, while no such term will be present in the equations of the electromagnetic $A$.

Given that, one may expect to find that $A$ has a power-like behaviour, while $Z$ and $W^\pm$ are quickly vanishing.
This expectation is due to the fact that the behaviour at infinity seems to depend on whether or not a mass term (or an effective mass term) is  present, and \emph{not} on the specific 
field under consideration.
In this thesis, this is what happens to the free scalar field and to the one form $A$ in the abelian Higgs mechanism.
A consequence of the above-mentioned fall-off conditions is that the equations of motion (and the Poincar\'e transformations) of $A$ and its conjugated momentum $\pi$ become those of free electrodynamics near spatial infinity.
Therefore, one may be led to the conjecture that the discussion on parity conditions simply reduces to that already presented by Henneaux and Troessaert.

\section{Outlook}

There are several possible future developments in the Hamiltonian analysis of asymptotic symmetries, building on the results contained in this thesis.
At least three possibilities are of quick realisation, in principle.
First, as already mentioned in the discussion about non-abelian gauge theories, it would be interesting to reconcile our analysis at spatial infinity with those at null infinity.
In particular, it would be interesting to verify whether obstructions to non-trivial asymptotic symmetries can be found at null infinity as well.
Secondly, a natural way to generalise the results of this thesis is to analyse the situation of the Standard Model of particle physics and we have already provided ample discussion of what we expect to find in this case.
Thirdly, another way to generalise the field content, that would also be of particular interest in view of our results in free Yang-Mills, is to consider the $\SU(2)$-Yang-Mills-Higgs case.
This is currently under investigation and will be presented soon.

All the aforementioned extensions to this work still rely on a flat non-dynamical background.
Therefore, an important and interesting medium-term possibility for future investigations consists in  turning on the gravitational interaction and see what happens, for instance, on an asymptotically-flat spacetime.
In addition, another compelling possibility is to consider the quantum effects and to investigate on a solid theoretical basis the role that asymptotic symmetries and conserved charges play in gravity and black-hole physics.

Finally, in the future, similar methods should also be applied to models beyond General Relativity, including 
$f(R)$, scalar-tensor theories (which are used to describe inflation, e.g.), and teleparallel alternative formulation of General Relativity.
In these cases, analyses of asymptotic symmetries, pursued in a similar fashion to those highlighted in this thesis, may state something about the viability of these theories, or maybe, even serve as arguments as to why reformulations (or extensions) of General Relativity are better or worse behaved as General Relativity itself.

\appendix

\noappendicestocpagenum\addappheadtotoc
\chapter{Some detailed computations of chapter~\ref{cha:Yang-Mills}}
\label{appendix:Yang-Mills}

In this appendix, we provide two detailed computations, which were omitted in the text of chapter~\ref{cha:Yang-Mills}.
The former concerns the logarithmically-divergent contribution to the symplectic form of the free $\SU(N)$-Yang-Mills theory.
As explained in section~\ref{subsec:YM:loosen-parity}, this contribution emerges when relaxing the strict parity conditions.
Moreover, the second computation provides the details about the ansatz presented in section~\ref{subsec:attempt-Poincare-canonical}.

\section{The logarithmically-divergent contribution to the symplectic form} \label{app:symplectic-form}

We show a step-by-step computation of the logarithmically-divergent contribution to the symplectic form, which arises once we relax the  parity conditions to match~(\ref{YM:parity-conditions-loosen}), as discussed in subsection~\ref{subsec:YM:loosen-parity}.
In short, we will evaluate
\begin{equation} \label{YM:log-div-symplectic}
 \begin{split}
  \oint_{S^2} d^2 \barr{x} \;
  \sprod{\extderphase \barr{\pi}^a \wedge}{ \extderphase \barr{A}_a} = 
  \oint_{S^2} d^2 \barr{x} \;
  \bigg[
  &\sprod{\extderphase \Big( \barr{\mathcal{U}}^{-1} \barr{\pi}^r_{\text{odd}} \, \barr{\mathcal{U}} \Big) \wedge}
  {\extderphase \Big( \barr{\mathcal{U}}^{-1} \barr{A}_r^{\text{even}} \, \barr{\mathcal{U}} \Big)}
  + \\
  +&\sprod{\extderphase \Big( \barr{\mathcal{U}}^{-1} \barr{\pi}^{\bar{a}}_{\text{even}} \, \barr{\mathcal{U}} \Big) \wedge}
  {\extderphase \Big( \barr{\mathcal{U}}^{-1} \barr{A}_{\bar{a}}^{\text{odd}} \barr{\mathcal{U}} \Big)}
  + \\
  +&\sprod{\extderphase \Big( \barr{\mathcal{U}}^{-1} \barr{\pi}^{\bar{a}}_{\text{even}} \, \barr{\mathcal{U}} \Big) \wedge}
  { \extderphase \Big( \barr{\mathcal{U}}^{-1} \partial_{\bar a}\, \barr{\mathcal{U}} \Big)}
  \bigg] \,.
 \end{split}
\end{equation}
Let us call $\barr{\Omega}_1$, $\barr{\Omega}_2$, and $\barr{\Omega}_3$ the contributions of the first, the second, and the third summand of the above expression, respectively.
In the following, we compute these three contributions separately.
\enlargethispage{-\baselineskip}

\subsection{Preliminaries}
In order to make the ensuing computation of the three contributions easier to follow, let us evaluate in advance a few useful quantities.
To begin with, we note that most of the contributions in~(\ref{YM:log-div-symplectic}) are of the form
\begin{equation} \label{YM:div-gen1}
 \begin{aligned}
  \extderphase \Big( \barr{\mathcal{U}}^{-1} \eff \, \barr{\mathcal{U}} \, \Big) ={}&
  \, \extderphase \big( \barr{\mathcal{U}}^{-1} \big) \eff \, \barr{\mathcal{U}}+
  \barr{\mathcal{U}}^{-1} \extderphase \eff \, \barr{\mathcal{U}}+
  \barr{\mathcal{U}}^{-1} \eff \, \extderphase \barr{\mathcal{U}}= \\
  ={}& \, \barr{\mathcal{U}}^{-1}
  \Big( \extderphase \eff +
  \eff \, \extderphase \barr{\mathcal{U}} \, \barr{\mathcal{U}}^{-1}
  -\extderphase \barr{\mathcal{U}} \, \barr{\mathcal{U}}^{-1} \eff
  \Big) \,\barr{\mathcal{U}} = \\
  ={}& \, \barr{\mathcal{U}}^{-1}
  \Big( \extderphase \eff +
  \extprod{\eff}{\big(\extderphase \barr{\mathcal{U}} \, \barr{\mathcal{U}}^{-1} \big) }
  \Big)  \,\barr{\mathcal{U}} \,,
 \end{aligned}
\end{equation}
where $\eff$ needs to be replaced by one of definite-parity parts appearing in the canonical fields.
In the above expression, we have made use of the identity
$\extderphase \big( \barr{\mathcal{U}}^{-1} \big) =
- \barr{\mathcal{U}}^{-1} \extderphase \barr{\mathcal{U}} \; \barr{\mathcal{U}}^{-1}$, in order to obtain the expression on the second line.
Moreover, let us also compute
\begin{equation} \label{YM:div-gen2}
 \begin{aligned}
  \extderphase \Big( \barr{\mathcal{U}}^{-1} \partial_{\bar a}\, \barr{\mathcal{U}} \Big) ={}&
  \, \extderphase \big( \barr{\mathcal{U}}^{-1} \big) \partial_{\bar a} \, \barr{\mathcal{U}}+
  \barr{\mathcal{U}}^{-1} \partial_{\bar a} \big( \extderphase \barr{\mathcal{U}} \big)= \\
  ={}& \, \barr{\mathcal{U}}^{-1} \Big[
  -\extderphase \barr{\mathcal{U}} \; \barr{\mathcal{U}}^{-1} \partial_{\bar a}\, \barr{\mathcal{U}} \; \barr{\mathcal{U}}^{-1} +
  \partial_{\bar a} \big( \extderphase \barr{\mathcal{U}} \big) \, \barr{\mathcal{U}}^{-1}
  \Big] \,\barr{\mathcal{U}} = \\
  ={}& \, \barr{\mathcal{U}}^{-1} \Big[
  \extderphase \barr{\mathcal{U}} \; \partial_{\bar a} \big( \barr{\mathcal{U}}^{-1} \big) +
  \partial_{\bar a} \big( \extderphase \barr{\mathcal{U}} \big) \, \barr{\mathcal{U}}^{-1}
  \Big] \,\barr{\mathcal{U}} = \\
  ={}& \, \barr{\mathcal{U}}^{-1}
  \partial_{\bar a} \Big(
  \extderphase \barr{\mathcal{U}} \;  \barr{\mathcal{U}}^{-1}
  \Big) \,\barr{\mathcal{U}} \,,
 \end{aligned}
\end{equation}
where we have made use of the further identity
$\partial_{\bar a} \big( \barr{\mathcal{U}}^{-1} \big) =
- \barr{\mathcal{U}}^{-1} \partial_{\bar a} \, \barr{\mathcal{U}} \; \barr{\mathcal{U}}^{-1}$,
in order to obtain the expression on the third line.

Finally, let us evaluate
$\sprod{ \big( \extprod{U}{\omega} \big) \wedge }
{\big( \extprod{V}{\omega} \big)} $,
where $U$ and $V$ are $\su(N)$-valued functions 
and $\omega$ is a $\su(N)$-valued one-form on phase 
space to which the exterior product refers.
From our definition \eqref{eq:CompModifiedKillingProduct} 
of the inner product, we get
\begin{equation}
 \sprod{ \big( \extprod{U}{\omega} \big) \wedge }{\big( \extprod{V}{\omega} \big)} =
 - \tr \Big( 
 \extprod{U}{\omega} \wedge \extprod{V}{\omega}
 \Big) =
 -\tr \Big( 
 \big[ U , \omega \big] \wedge 
 \big[ V , \omega \big]
 \Big) \,,
\end{equation}
where next to the exterior product of $\su(N)$-valued 
one-forms matrix multiplication in $\su(N)$ is also understood. In the following we shall also temporarily 
drop the wedge-product symbol. We only need to remember 
to insert an extra minus sign every time we invert 
the order of the two $\omega$.
Expanding the commutators and the composition, we get
\begin{equation}
 \sprod{ \big( \extprod{U}{\omega} \big) \wedge }{\big( \extprod{V}{\omega} \big)} =
 -\tr \Big( 
 U \omega V \omega +
 \omega U \omega V -
 U \omega \omega V -
 \omega UV\omega
 \Big) \,.
\end{equation}
Using the cyclicity of the trace and taking
into account the minus sign whenever the order of the two one-forms 
$\omega$ changes, we immediately see that the first two terms cancel. 
Therefore, applying the same rules, we get
\begin{equation}
 \sprod{ \big( \extprod{U}{\omega} \big) \wedge }{\big( \extprod{V}{\omega} \big)} =
 - \tr \Big( 
 - U \omega \omega V
 - \omega U V \omega
 \Big) =
 - \tr \Big( 
 - \omega \omega V U
 + \omega \omega U V
 \Big)
 \,.
\end{equation}
This can be factorised in the form 
\begin{equation}
 \sprod{ \big( \extprod{U}{\omega} \big) \wedge }{\big( \extprod{V}{\omega} \big)} =
 - \tr \Big( 
 \omega \omega \big[ U , V \big]
 \Big) =
 - \tr \left( 
 \frac{1}{2} \big[ \omega , \omega \big]
 \, \big[ U , V \big]
 \right)
 \,,
\end{equation}
where we have replaced the product $\omega\omega$ with the commutator divided by two using the antisymmetry of the exterior product.\footnote{Note that the commutator of the two $\omega$ does not identically vanish because it is combined with the (antisymmetric) exterior product.}
Finally, recalling our definition \eqref{eq:CompModifiedKillingProduct} of the inner 
product and again displaying the exterior product, 
we arrive at the desired identity
\begin{equation} \label{YM:double-product-identity}
 \sprod{ \big( \extprod{U}{\omega} \big) \wedge }{\big( \extprod{V}{\omega} \big)} =
 \frac{1}{2}
 \sprod{
 \big( \extprod{\omega \wedge}{\omega} \big)
 }{
 \big( \extprod{U}{V} \big)
 } \,.
\end{equation}
We are now ready to present the actual computation of the three terms $\barr{\Omega}_1$, $\barr{\Omega}_2$, and $\barr{\Omega}_3$, whose sum gives the divergent contribution~(\ref{YM:log-div-symplectic}) to the symplectic form.

\subsection{Computation of the divergent contribution}
Let us define $\barr{\varepsilon} \eqdef \extderphase \barr{\mathcal{U}} \, \barr{\mathcal{U}}^{-1}$, which is a one-form in phase space, in order to write the following expressions in a more-compact way.
First, let us compute $\barr{\Omega}_1$, the first line of the right-hand side of~(\ref{YM:log-div-symplectic}).
Using~(\ref{YM:div-gen1}), we get
\begin{equation}
 \begin{aligned}
  \barr{\Omega}_1 \eqdef
  \oint_{S^2} \!\! d^2 \barr{x} \; &
  \sprod{\extderphase \Big( \barr{\mathcal{U}}^{-1} \barr{\pi}^r_{\text{odd}} \, \barr{\mathcal{U}} \Big) \wedge}
  {\extderphase \Big( \barr{\mathcal{U}}^{-1} \barr{A}_r^{\text{even}} \barr{\mathcal{U}} \Big)} = \\
  = \oint_{S^2} \!\! d^2 \barr{x} \; &
  \sprod
  {\barr{\mathcal{U}}^{-1}
  \Big[ \extderphase \barr{\pi}^r_{\text{odd}} +
  \extprod{\barr{\pi}^r_{\text{odd}}}{\barr{\varepsilon}}
  \Big]  \,\barr{\mathcal{U}} \, \wedge}{
  \barr{\mathcal{U}}^{-1}
  \Big[ \extderphase \barr{A}_r^{\text{even}} +
  \extprod{\barr{A}_r^{\text{even}}}{\barr{\varepsilon}}
  \Big]  \,\barr{\mathcal{U}}
  } = \\
  = \oint_{S^2} \!\! d^2 \barr{x} \; &
  \sprod
  {\Big[ \extderphase \barr{\pi}^r_{\text{odd}} +
  \extprod{\barr{\pi}^r_{\text{odd}}}{\barr{\varepsilon} }
  \Big]  \wedge}
  {\Big[ \extderphase \barr{A}_r^{\text{even}} +
  \extprod{\barr{A}_r^{\text{even}}}{\barr{\varepsilon} }
  \Big]  } \,,
 \end{aligned}
\end{equation}
where, on the last step, we have simplified $\barr{\mathcal{U}}$ with $\barr{\mathcal{U}}^{-1}$ using the cyclicity of the trace, which appears in the definition of the Killing inner product.
At this point, we can expand the product of the two terms in square brackets.
The term
$\sprod{\extderphase \barr{\pi}^r_{\text{odd}} \wedge}{\extderphase \barr{A}_r^{\text{even}}}$
vanishes upon integration because it is an odd function on the sphere.
Using the symmetries of the triple product and being careful in putting an extra minus sign every time we change the order of the forms in the exterior product, we can rearrange the terms as
\begin{equation} \label{YM:omega1-partial}
 \begin{aligned}
  \barr{\Omega}_1
  = \oint_{S^2} d^2 \barr{x} \;
  &\bigg[
  \sprod{\barr{\varepsilon} \, \wedge}
  { \Big(
  \extprod{ \barr{A}_r^{\text{even}} }{\extderphase \barr{\pi}^r_{\text{odd}}}+
  \extprod{\extderphase \barr{A}_r^{\text{even}} }{ \barr{\pi}^r_{\text{odd}}}
  \Big) }
  + \\
  &+\sprod
  {\Big(
  \extprod{\barr{\pi}^r_{\text{odd}}}{\barr{\varepsilon} }
  \Big)  \wedge}
  {\Big(
  \extprod{\barr{A}_r^{\text{even}}}{\barr{\varepsilon} }
  \Big)}
  \bigg] \,.
 \end{aligned}
\end{equation}
The second factor in the first summand can be rewritten as
$\extderphase \big(
\extprod{ \barr{A}_r^{\text{even}} }{ \barr{\pi}^r_{\text{odd}}}
\big)$, simply using the Leibniz rule.
Moreover, the second summand can be rewritten using the identity~(\ref{YM:double-product-identity}).
Hence, we arrive at the expression
\begin{equation} \label{YM:div-line1}
  \barr{\Omega}_1
  =\oint_{S^2} d^2 \barr{x} \;
  \bigg[
  \sprod{\barr{\varepsilon} \, \wedge }
  {\extderphase \Big(
  \extprod{\barr{A}_r^{\text{even}}}{\barr{\pi}^r_{\text{odd}}}
  \Big)}-\sprod{
  \frac{1}{2} \Big(\extprod{ \barr{\varepsilon} \, \wedge }{ \barr{\varepsilon} } \Big) 
   }
  { \Big( \extprod{\barr{A}_r^{\text{even}}}{\barr{\pi}^r_{\text{odd}}} \Big) }
  \bigg] \,.
\end{equation}

Second, let us note that the second line of~(\ref{YM:log-div-symplectic}) is analogous to the first line, so that we can get the value of $\barr{\Omega}_2$ with a computation almost identical to the one for $\barr{\Omega}_1$, obtaining
\begin{equation} \label{YM:div-line2}
  \barr{\Omega}_2
  = \oint_{S^2} d^2 \barr{x} \;
  \bigg[
  \sprod{\barr{\varepsilon} \, \wedge }
  {\extderphase \Big(
  \extprod{\barr{A}_{\bar a}^{\text{odd}}}{\barr{\pi}^{\bar a}_{\text{even}}}
  \Big)} -\sprod{
  \frac{1}{2} \Big(\extprod{ \barr{\varepsilon} \, \wedge }{\barr{\varepsilon}} \Big) 
   }
  { \Big( \extprod{\barr{A}_{\bar a}^{\text{odd}}}{\barr{\pi}^{\bar a}_{\text{even}}} \Big) }
  \bigg] \,.
\end{equation}

Third, let us compute the last contribution $\barr{\Omega}_3$.
Using~(\ref{YM:div-gen1}) and~(\ref{YM:div-gen2}), we get
\begin{equation}
 \begin{aligned}
  \barr{\Omega}_3 \eqdef
  & \oint_{S^2} d^2 \barr{x} \;
  \sprod{\extderphase \Big( \barr{\mathcal{U}}^{-1} \barr{\pi}^{\bar{a}}_{\text{even}} \, \barr{\mathcal{U}} \Big) \wedge}
  { \extderphase \Big( \barr{\mathcal{U}}^{-1} \partial_{\bar a}\, \barr{\mathcal{U}} \Big)}
  = \\
  ={}& \oint_{S^2} d^2 \barr{x} \; 
  \sprod
  {\barr{\mathcal{U}}^{-1}
  \Big( \extderphase \barr{\pi}^{\bar a}_{\text{even}} +
  \extprod{\barr{\pi}^{\bar a}_{\text{even}}}{\barr{\varepsilon} }
  \Big)  \,\barr{\mathcal{U}} \;\wedge}
  {\barr{\mathcal{U}}^{-1}
  \partial_{\bar a} \barr{\varepsilon} \; \barr{\mathcal{U}} } \,.
 \end{aligned}
\end{equation}
Once again, we can simplify $\barr{\mathcal{U}}$ and $\barr{\mathcal{U}}^{-1}$ using the cyclicity of the trace employed in the definition of the Killing  inner product.
Expanding afterwards the expression, we get
\begin{equation}
 \begin{aligned}
  \barr{\Omega}_3
  ={}& \oint_{S^2} d^2 \barr{x} \;
  \bigg[
  \sprod{\extderphase \barr{\pi}^{\bar a}_{\text{even}} \, \wedge}
  {\partial_{\bar a} \barr{\varepsilon} } +
  \sprod{
  \Big(\extprod{\barr{\pi}^{\bar a}_{\text{even}}}{\barr{\varepsilon}} \Big) \wedge
  }
  {\partial_{\bar a} \barr{\varepsilon}} \,
  \bigg] = \\
  ={}& \oint_{S^2} d^2 \barr{x} \;
  \bigg[
  - \sprod{\extderphase \Big( \partial_{\bar a} \barr{\pi}^{\bar a}_{\text{even}} \Big) \, \wedge}
  {\barr{\varepsilon}} +
  \sprod{
  \Big(\extprod{\partial_{\bar a} \barr{\varepsilon} \, \wedge}{\barr{\varepsilon}} \Big) 
  }
  {\barr{\pi}^{\bar a}_{\text{even}}}
  \bigg] \,,
 \end{aligned}
\end{equation}
where we have integrated by part the first summand.
Moreover, in the second summand, we have used the symmetries of the triple product and inserted an extra minus sign due to the ordering of the forms in the exterior product.
The above expression can be easily rewritten as
\begin{equation} \label{YM:div-line3}
 \begin{aligned}
  \barr{\Omega}_3 =
  & \oint_{S^2} d^2 \barr{x} \;
  \bigg[
  \sprod{\barr{\varepsilon} \, \wedge}
  {\extderphase \Big( \partial_{\bar a} \barr{\pi}^{\bar a}_{\text{even}} \Big)}+
  \sprod{
  \frac{1}{2} \partial_{\bar a} \Big(\extprod{ \barr{\varepsilon} \, \wedge}{\barr{\varepsilon}} \Big)
  }
  {\barr{\pi}^{\bar a}_{\text{even}}}
  \bigg] = \\
  ={}& \oint_{S^2} d^2 \barr{x} \;
  \bigg[
  \sprod{\barr{\varepsilon} \, \wedge}
  {\extderphase \Big( \partial_{\bar a} \barr{\pi}^{\bar a}_{\text{even}} \Big)}-
  \sprod{
  \frac{1}{2} \Big( \extprod{ \barr{\varepsilon} \, \wedge}{\barr{\varepsilon}} \Big) 
  }
  { \partial_{\bar a} \barr{\pi}^{\bar a}_{\text{even}} }
  \bigg] \,,
 \end{aligned}
\end{equation}
where we have integrated by part the second summand.

Finally, we find the logarithmically-divergent contribution to the symplectic form by summing the three contributions $\barr{\Omega}_1$, $\barr{\Omega}_2$, and $\barr{\Omega}_3$, given by the expressions~(\ref{YM:div-line1}), (\ref{YM:div-line2}), and~(\ref{YM:div-line3}), respectively.
Reminding that $\barr{\varepsilon} = \extderphase \barr{\mathcal{U}} \, \barr{\mathcal{U}}^{-1}$, we see that the result coincides exactly with the expression~(\ref{YM:log-div1}) presented in subsection~\ref{subsec:YM:loosen-parity}.

\section{Details about the computations of section~\ref{subsec:attempt-Poincare-canonical}} \label{app:details-computations}
In this appendix, we provide a more detailed discussion about the attempts to make the Poincar\'e transformations canonical after having relaxed the parity conditions in the Yang-Mills case.
In particular, we extend the information of subsection~\ref{subsubsec:ansatz}.
There, some assumptions were made in the behaviour of the fields under Poincar\'e transformations and in the ansatz~(\ref{YM:ansatz-omega}) for the boundary term of the symplectic form.
In the following, we comment on the fact that these assumptions are actually not so restrictive.

\subsection{Poincar\'e transformations of the fields}
We remind that, in section~\ref{subsubsec:ansatz}, we introduced a one form $\phi_a$ and the conjugated momenta $\Pi^a$.
These new canonical fields were required to satisfy the fall-off conditions~\ref{YM:fall-off-phi} and the further constraint $\Pi^a \approx 0$.
At this point, we have to specify how the fields transform under the Poincar\'e transformations and, in particular, the Lorentz boost.
In order to do so, let us make a few assumptions.

First, we wish that, ultimately, the Poincar\'e transformations would be generated by the Poisson brackets with a function $P$ on phase space, as it is in the case of General Relativity and electrodynamics.
So, let us write the candidate for the generator of the boost as
\begin{equation} \label{YM:boost-generator-new0}
 P' [\xi^\perp] \eqdef
 \int d^3 x \, \xi^\perp \left[
 \frac{\sprod{\pi^a}{\pi_a}}{2\sqrt{g}} + \frac{\sqrt{g}}{4} \sprod{F_{ab}}{F^{ab}}
 +\mathscr{P}'_{(1)}
 \right]
 + (\text{boundary}) \,,
\end{equation}
where the first two summands in the square brackets are responsible for the usual transformations~(\ref{YM:poincare-transformations}), while $\mathscr{P}'_{(1)}$ takes into account the transformation of $\phi_a$ and $\Pi^a$, as well as some possible new contributions to the transformation of $A_a$ and $\pi^a$.
For now, we ignore any issue concerning the existence of a boundary term which makes the generator above well-defined.
We pretend that it exists, in order to allow the following formal manipulations, and check at the end whether or not this is consistent.
It is the goal of this appendix to show that such boundary term does \emph{not} actually exist.
Note that, due to the presence of an unspecified boundary term in the expression above, $\mathscr{P}'_{(1)}$ is defined up to a total derivative.
We will implicitly make use of this fact in some of the following equalities.

Secondly, the attempt done in section~\ref{subsubsec:ansatz-Psi} failed because it was not possible to compensate for the term containing
$\sprod{\extderphase \barr{A}_{\bar m} \,\wedge}{ \extderphase( \extprod{ \barr{A}^{\bar m} }{\barr{A}_r} ) }$
in $\liephase_{X} \Omega$.
Indeed, there was no field transforming (asymptotically) as (the asymptotic part of) $A_a$ without any derivative.
Therefore, as a further assumption, we ask that $\phi_a$ transforms exactly as $\delta_{\xi^\perp} \phi_a = \xi^\perp A_a$, thus finding
\begin{equation} \label{YM:boost-generator-new1}
  \mathscr{P}'_{(1)} = \sprod{A_a}{\Pi^a} + \mathscr{P}'_{(2)} \,,
\end{equation}
where $\mathscr{P}'_{(2)}$ does \emph{not} depend on $\Pi^a$.

Thirdly, we ask that the transformations of $A_a$ and $\pi^a$ differ from the ones in~(\ref{YM:poincare-transformations}) by, at most, gauge transformations and constraints.
Since $\mathscr{P}'_{(2)}$ cannot depend on the constraints $\Pi^a \approx 0$, this implies that $\mathscr{P}'_{(2)} = \sprod{F}{\mathscr{G}}$, for some function $F$ of the canonical fields (except for $\Pi^a$) and their derivatives.
Note that, since $\mathscr{G}$ is a weight-one scalar density, $F$ needs to be a scalar in order for the integral in~(\ref{YM:boost-generator-new0}) to make sense.

Finally, we require that the transformations of $A_a$ and $\pi^a$ are exactly the ones in~(\ref{YM:poincare-transformations}), when the new fields $\phi_a$ and $\Pi^a$ are set to zero.
Therefore, we can write, up to boundary terms, $F = \mathscr{D}^a \phi_a$, where the operator $\mathscr{D}^a$ is built using the fields $A_a$, $\pi^a$, and $\phi_a$, as well as an arbitrary number of derivatives and $\su(N)$ commutators.
At the lowest order in the derivatives and in the fields, we find
\begin{equation}
 \mathscr{D}^a \phi_a = c_0 \nabla^a \phi_a + c_1 \extprod{A^a}{\phi_a} + c_2 \extprod{\pi^a}{\phi_a} \,,
\end{equation}
where $c_0, c_1, c_2 \in \real$ are three free parameters.
After noting that $c_0$ can be set to $1$ by redefining $\phi_a$, we find exactly the transformations~(\ref{YM:boost-YM-phi}), that were assumed in section~\ref{subsubsec:ansatz}.
\enlargethispage{2\baselineskip}

\subsection{The boundary term of the symplectic form}

Before we can verify whether or not the Poincar\'e transformations are canonical, we need to specify how the symplectic form is affected by the introduction of the new fields $\phi_a$ and $\Pi^a$.
We assume that the contribution in the bulk is of the usual form $\sprod{\extderphase \Pi^a \, \wedge}{\extderphase \phi_a}$.
Therefore, the symplectic form in the bulk $\Omega'$ is given by~(\ref{YM:Omega-bulk-phi}).
To this, we add a boundary term $\omega'$ built using the asymptotic part of the fields.
There is potentially an infinite number of possibilities when one writes contributions to $\omega'$.
However, a few things need to be taken into consideration.

First, we want to achieve $\liephase_{X'} (\Omega' + \omega') = 0 $.
Now, $\liephase_{X'} \Omega'$ contains non-zero boundary contributions as shown in~(\ref{YM:lie-symplectic-bulk}).
In order for $\liephase_{X'} (\Omega' + \omega')$ to be actually zero, we need that terms in~(\ref{YM:lie-symplectic-bulk}) are compensated by some terms in $\liephase_{X'} \omega'$.
The ansatz~(\ref{YM:ansatz-omega}) is designed exactly in this spirit.
In particular, the terms with coefficients $a_0$ and $a_1$ should compensate those parts of~(\ref{YM:lie-symplectic-bulk}) containing derivatives of $\barr{A}$ and those containing $\barr{\pi}^r$, whereas the terms with coefficients $a_2,\dots,a_6$ should tackle the part in~(\ref{YM:lie-symplectic-bulk}) containing
$\sprod{\extderphase \barr{A}_{\bar m} \,\wedge}{ \extderphase( \extprod{ \barr{A}^{\bar m} }{\barr{A}_r} ) }$.

Secondly, introducing contributions to $\omega'$ built using the momenta $\barr{\pi}^a$ does not help, since these would introduce terms with at least two derivatives of $\barr{A}_a$ in $\liephase_{X'} \omega'$, due to their asymptotic transformations under boost --- see~(\ref{YM:poincare-asymptotic3}) and~(\ref{YM:poincare-asymptotic4}) --- while $\liephase_{X'} \Omega'$ only contains terms with at most one derivative.

Thirdly, having terms in $\omega'$ containing a great number of fields and of their commutators would introduce a big complication in the problem.
Furthermore, it would be difficult to justify such terms when comparing the theory at spatial and at null infinity.

In conclusion, we consider the ansatz~(\ref{YM:ansatz-omega}) for the boundary term of the symplectic form for the aforementioned reasons.
Although it is not the most general ansatz, it is general enough to show that the Yang-Mills case is substantially different from electrodynamics and General Relativity.
\enlargethispage{2\baselineskip}

\subsection{The Poincar\'e transformations are not canonical}
We finally show that no value of the free parameters $a_0,\dots,a_6$, $c_1$, and $c_2$ makes the Poincar\'e transformations canonical.
To begin with, the symplectic form $\Omega' + \omega'$ must be a closed two-form on phase space.
Since $\extderphase \Omega' = 0$, one need to impose that also $\extderphase \omega' = 0$.
One can easily check that this amount to consider the general ansatz~(\ref{YM:ansatz-omega}) with the free parameters $a_0,\dots,a_6$ restricted by the two conditions
\begin{equation} \label{YM:conditions-param-1}
 a_3 + a_4 = 0
 \qquad \text{and} \qquad
 a_2 + a_5 + a_6 = 0 \,.
\end{equation}
The two conditions above imply that one can rewrite the boundary term $\omega'$ of the symplectic form as
\begin{equation} 
\label{YM:ansatz-omega-close}
 \begin{aligned}
 \omega' = \oint_{S^2} d^2 \barr{x} \; \sqrt{\barr{\gamma}} \,
 \Big[ &
  a_0 \, \sprod{\extderphase \big( \barr{\nabla}^{\bar m}  \barr{\phi}_{\bar m} \big) \wedge}{\extderphase \barr{A}_r } 
 +a_1 \, \sprod{\extderphase \barr{\phi}_{r} \wedge}{\extderphase \barr{A}_r } + \\
 + & {\tilde a}_2 \, \sprod{ \extderphase \barr{A}_{\bar m} \, \wedge }{ \extderphase \big( \extprod{ \barr{A}^{\bar m} }{ \barr{\phi}_r } \big) }
 + {\tilde a}_3 \, \sprod{ \extderphase \barr{A}_{\bar m} \, \wedge }{ \extderphase \big( \extprod{ \barr{A}_r }{ \barr{\phi}^{\bar m} } \big) }
 + \\
 +&{\tilde a}_4 \, \sprod{ \extderphase \barr{A}_{r} \, \wedge }{ \extderphase \big( \extprod{ \barr{A}_{\bar m} }{ \barr{\phi}^{\bar m} } \big) }
 \Big] \,,
\end{aligned}
\end{equation}
where the three parameters ${\tilde a}_2$, ${\tilde a}_3$, and ${\tilde a}_4$ are related to $a_2, \dots, a_5$ by
\begin{equation} \label{YM:conditions-param-2}
 \tilde a_2 = a_3 \,, \qquad
 \tilde a_3 = -a_2 \,,
 \qquad \text{ and} \qquad
 \tilde a_4 = -a_5 \,.
\end{equation}
Note that~(\ref{YM:ansatz-omega-close}) is not only close but also exact.

It is now not difficult to show that the Poincar\'e transformations are not canonical for any value of the free parameters $a_0$, $a_1$, ${\tilde a}_2$, ${\tilde a}_3$, ${\tilde a}_4$, $c_1$, and $c_2$.
Indeed, the Poincar\'e transformations would be canonical if, and only if,
\begin{equation} \label{YM:lie-Omega-omega}
 \liephase_{X'} \Omega' = - \liephase_{X'} \omega' \,.
\end{equation}
The left-hand side of the above expression was already computed in~(\ref{YM:lie-symplectic-bulk}).
It contains a first summand with the term
$\sprod{\extderphase \barr{A}_{\bar m} \,\wedge}{ \extderphase \big( D^{\bar m} \barr{A}_r \big) }$,
which would appear also on the right-hand side of the above expression if we imposed
\begin{equation} \label{YM:conditions-param-3}
 a_0 = 1
 \qquad \text{and} \qquad
 {\tilde a}_3 = {\tilde a}_2 +1 \,.
\end{equation}
Moreover, the left-hand side of~(\ref{YM:lie-Omega-omega}) contains a second summand with the term
$\sprod{ \extderphase \barr{\pi}^r \, \wedge}{ \extderphase \barr{\mathscr{D} \phi }}$.
This contribution would be compensated by a similar contribution on the right-hand side of~(\ref{YM:lie-Omega-omega}) if we imposed the further conditions
\begin{equation} \label{YM:conditions-param-4}
 a_1 = 1 \,, \qquad
 c_1 = 0 \,, 
 \qquad \text{ and} \qquad
 c_2 = 0 \,.
\end{equation}
After restricting the free parameters to those satisfying~(\ref{YM:conditions-param-3}) and~(\ref{YM:conditions-param-4}), every term in the left-hand side of~(\ref{YM:lie-Omega-omega}) appears also on the right-hand side.
However, the latter contains also other terms, which one has to set to zero with an appropriate choice of the remaining parameters, if this is actually possible.
In particular, the right-hand side still contains, among others, some contribution proportionate to $\extderphase \barr{\pi}^r$.
These would vanish, if we set
\begin{equation} \label{YM:conditions-param-5}
 {\tilde a}_2 = -1
 \qquad \text{and} \qquad
 {\tilde a}_4 = 0 \,.
\end{equation}
These conditions, together with the previous ones~(\ref{YM:conditions-param-1}), (\ref{YM:conditions-param-2}), (\ref{YM:conditions-param-3}), and (\ref{YM:conditions-param-4}), completely fix the values of the free parameters, so that
\begin{equation} \label{YM:ansatz-omega-final}
 \omega' = \oint_{S^2} d^2 \barr{x} \; \sqrt{\barr{\gamma}} \,
 \Big[
 \sprod{\extderphase \big( 2 \barr{\phi}_{r} + \barr{\nabla}^{\bar m}  \barr{\phi}_{\bar m} \big) \wedge}{\extderphase \barr{A}_r } 
 - \sprod{ \extderphase \barr{A}_{\bar m} \, \wedge }{ \extderphase \big( \extprod{ \barr{A}^{\bar m} }{ \barr{\phi}_r } \big) }
 \Big]
\end{equation}
does not depend any more on any free parameter, nor do the Poincar\'e transformations (\ref{YM:boost-YM-phi}).
One can now easily verify by direct computation that $\liephase_{X'} (\Omega' + \omega') \ne 0$, i.e., the Poincar\'e transformations are not canonical, as we wanted to show.
In particular, this also shows that the boundary term in~(\ref{YM:boost-generator-new0}) cannot exist.

\chapter{Fall-off behaviour of massive fields} \label{appendix:massive-fall-off}

We would like to show that, in the massive case, the fall-off behaviour of the field and the momentum needs to be decreasing more rapidly than any power-like function.
Specifically, let us denote with $Z$ the phase space consisting of all the allowed field configurations $(\varphi,\Pi)$ and with $\Poi$ the Poincar\'e group.
In order to have a well-defined relativistic field theory, we need to require that the action of any Poincar\'e transformation maps points belonging to the phase space into points belonging to the phase space.\footnote{
The phase space $Z$ should be though of as 
a sub-manifold of some infinite-dimensional manifold 
of functions which are sufficiently regular so that 
the explicit expressions of the Poincar\'e transformations~(\ref{SEDAH:eoms}) make sense.
}
In other words, we have to impose the condition that, for all $g \in \Poi$, one has  $g \, Z \subseteq Z$.\footnote{Note that we 
require the group $\Poi$ to operate on 
$Z$ by a \emph{group action}, which means 
that the map 
$\Poi\times Z \rightarrow Z$, 
$(g,p)\mapsto gp$, satisfies $g(hp)=(gh)p$ and 
$ep=p$ (where $e$ is the group identity) for 
all $g,h\in\Poi$ and all $p \in Z$. 
This immediately implies that, for 
any $g\in\Poi$, the map $Z \rightarrow Z$, $p\mapsto gp$ is a bijection. Hence $g Z \subseteq Z$ is, in fact, equivalent to 
$g Z = Z$.}
In this appendix, we wish to show that this requirement, together with the finiteness of the Hamiltonian and the at-most-logarithmic divergence of the symplectic form, implies that,
\begin{equation} \label{SEDAH:thesis-fall-off}
 \forall (\varphi,\Pi) \in Z \,, \;
 \forall \alpha,\beta \in \integers \,, \quad
 r^\alpha \, \varphi (x) \rightarrow 0
 \quad \text{and} \quad
 r^\beta \, \Pi (x) \rightarrow 0
\end{equation}
in the limit $r \eqdef |x| \rightarrow \infty$.\footnote{
In order to avoid issues in the ensuing proof, we need to assume that the phase space $Z$ is not empty.
This is easily achieved by assuming that $(\varphi^{(0)},\Pi^{(0)}) \in Z$, being $\varphi^{(0)} (x) = 0$ and $\Pi^{(0)} (x) = 0$.
This field configuration, other than being the minimum-energy solution to the equations of motion, is also invariant under the action of the Poincar\'e group and satisfies the statement in~(\ref{SEDAH:thesis-fall-off}).
}

To this end, let us focus only on a part of the full Poincar\'e transformations~(\ref{SEDAH:eoms}) and, specifically on
\begin{equation} 
 \delta' \varphi = \xi^\perp \frac{\Pi}{\sqrt{g}}
 \qquad \text{and} \qquad
 \delta' \Pi = - \xi^\perp \sqrt{g} \, m^2 \varphi \,.
\end{equation}
When considering only a Lorentz boost, i.e. $\xi^\perp = r \, b(\barr{x})$, and writing explicitly the dependence on the radial and angular coordinates, the above expressions become
\begin{equation} \label{SEDAH:boost-special-case}
 \delta' \varphi (r,\barr{x}) =
 \frac{b(\barr{x})}{\sqrt{\barr{\gamma}(\barr{x})}} \, \frac{\Pi(r,\barr{x})}{r} 
 \quad \text{and} \quad
 \delta' \Pi (r,\barr{x}) =
 - b (\barr{x}) \sqrt{\barr{\gamma}(\barr{x})} \;m^2 \, r^3 \, \varphi (r,\barr{x}) \,.
\end{equation}

Let us define for all $(\varphi,\Pi) \in Z$ the quantities
\begin{equation}
 \alpha_\varphi \eqdef
 \sup \big\{ \alpha \in \integers \colon r^\alpha \varphi (x) \rightarrow 0 \big\}
 \quad \text{and} \quad
 \beta_\Pi \eqdef
 \sup \big\{ \beta \in \integers \colon r^\beta \Pi (x) \rightarrow 0 \big\} \,.
\end{equation}
First of all, let us note that these quantities are well-defined.
Indeed, the finiteness of the mass term in the Hamiltonian (proportional to $\varphi^* \varphi$) implies that $\varphi (x) \rightarrow 0$, whereas the finiteness of the kinetic term (proportional to $\Pi^2$) implies that $r^{-1} \Pi (x) \rightarrow 0$.
Therefore, the sets on the right-hand sides of the definitions above are not empty and the suprema exist.
Note that, with the same argument, we can also conclude that
$\alpha_\varphi \ge 0$ and $\beta_\Pi \ge -1$ for every $(\varphi,\Pi)$ belonging to the phase space.

There are two possibilities for $\alpha_\varphi$ and, analogously, for $\beta_\Pi$.
First, the value of $\alpha_\varphi$ may be $+\infty$, in which case $\varphi$ is according to the statement~(\ref{SEDAH:thesis-fall-off}) that we wish to prove.
Secondly, it may happen that $\alpha_\varphi$ is a finite integer number, in which case the supremum is actually a maximum and $r^{\alpha_\varphi + 1} \varphi$ converges to some function on the sphere.
In principle, this function on the sphere can be divergent, but need not be identically zero.

In order to prove the original statement~(\ref{SEDAH:thesis-fall-off}), we need to show that, for all $(\varphi, \Pi) \in Z$, both $\alpha_\varphi$ and $\beta_\Pi$ are infinite.
To this purpose, let us define
\begin{equation}
 \barr{\alpha} \eqdef \min \{ \alpha_\varphi \colon \varphi \in Z_{|\varphi} \}
 \qquad \text{and} \qquad
 \barr{\beta} \eqdef \min \{ \beta_\Pi \colon \Pi \in Z_{|\Pi}\} \,,
\end{equation}
which are well-defined quantities, since $\alpha_\varphi \ge 0$ and $\beta_\Pi \ge -1$ for every $(\varphi,\Pi) \in Z$, so that the sets on the right-hand sides of the definitions above are non-empty subsets of $\integers \cup \{ +\infty \}$ bounded from below and, as a consequence, the minima exist.
Note that the value of $\barr{\alpha}$ and $\barr{\beta}$ can actually be infinite.
This happens, respectively, when $\alpha_\varphi = +\infty$ for all $\varphi$ and when $\beta_\Pi = +\infty$ for all $\Pi$.
It is easy to see that the statement~(\ref{SEDAH:thesis-fall-off}) is equivalent to the case in which both $\barr{\alpha}$ and $\barr{\beta}$ are infinite.

Let us assume, \emph{ad absurdum}, that at least one among $\barr{\alpha}$ and $\barr{\beta}$ is finite.
To begin with, we note that also the other quantity need to be finite.
This can be seen as follows.
Let us assume that $\barr{\alpha} \in \integers$ and let $(\varphi, \Pi) \in Z$ be such that $\alpha_\varphi = \barr{\alpha}$.\footnote{
The existence of $(\varphi, \Pi) \in Z$ satisfying
$\alpha_\varphi = \barr{\alpha}$  is guaranteed by the fact that $\barr{\alpha}$ is a minimum.
}
After applying a Poincar\'e transformation, we reach the field configuration $(\varphi',\Pi')$ which still belongs to the phase space $Z$ due to the hypothesis.
From the second equation in~(\ref{SEDAH:boost-special-case}), it follows that $\Pi'$ contains the term
\begin{equation}
 \delta' \Pi (r,\barr{x}) =
 - b (\barr{x}) \sqrt{\barr{\gamma}(\barr{x})} \; m^2 \, r^3 \, \varphi (r,\barr{x}) \,,
\end{equation}
which is easily seen to satisfy
$r^{\barr{\alpha} -3} \, \delta' \Pi \rightarrow 0$, while
$r^{\barr{\alpha} -2} \, \delta' \Pi$ does not converge to zero.
Since this is only one of the terms composing $\Pi'$, we cannot make an exact statement about the value of $\beta_{\Pi'}$, but we can nevertheless conclude that $\beta_{\Pi'} \le \barr{\alpha} - 3$, which implies
\begin{equation} \label{SEDAH:inequality-beta}
 \barr{\beta} \le \barr{\alpha} -3 \,,
\end{equation}
showing that $\barr{\beta}$ is finite if $\barr{\alpha}$ is finite.
Analogously, one can show that, if $\barr{\beta}$ is finite, also $\barr{\alpha}$ is finite and satisfies the inequality
\begin{equation} \label{SEDAH:inequality-alpha}
 \barr{\alpha} \le \barr{\beta} + 1 \,.
\end{equation}
The combination of the two inequalities~(\ref{SEDAH:inequality-beta}) and~(\ref{SEDAH:inequality-alpha}) readily yields us the contradiction
\begin{equation}
 \barr{\alpha} \le
 \barr{\beta} + 1 \le
 (\barr{\alpha} - 3) +1 =
 \barr{\alpha}-2 \,.
\end{equation}
Hence, we must conclude that both $\barr{\alpha}$ and $\barr{\beta}$ are infinite, which proves the statement~(\ref{SEDAH:thesis-fall-off}), as we wished.
\hfill$\blacksquare$

\bibliography{biblio}{}
\bibliographystyle{jhep-fullnames-erratum-arxiv}
\end{document}